
\input lanlmac
\writedefs
\input epsf


\noblackbox

\def\frac#1#2{{\textstyle{#1\over#2}}}

\def\tr{{\rm tr}\,}
\def\gc{g\dup_c}
\def\ud{\half}
\def\ee#1{{\rm e}^{^{\textstyle#1}}}
\def\d{{\rm d}}
\def\e{{\rm e}}
\def\CP{{\cal P}}
\def\Res{{\rm\ Res}\,}
\def\QT{Q}

\def\Re{{\rm Re\,}}
\def\Im{{\rm Im\,}}
\def\Pit{\,\Pi}
\def\gs{\gamma_{\rm str}}
\font\cmss=cmss10 \font\cmsss=cmss10 at 7pt

\def\inbar{\,\vrule height1.5ex width.4pt depth0pt}
\def\IC{\relax\hbox{$\inbar\kern-.3em{\rm C}$}}
\def\IR{\relax{\rm I\kern-.18em R}}
\def\IZ{\relax\ifmmode\mathchoice
{\hbox{\cmss Z\kern-.4em Z}}{\hbox{\cmss Z\kern-.4em Z}}
{\lower.9pt\hbox{\cmsss Z\kern-.4em Z}}
{\lower1.2pt\hbox{\cmsss Z\kern-.4em Z}}\else{\cmss Z\kern-.4em Z}\fi}
\def\refsubsec#1{subsec.~{\it #1\/}}
\def\refapp#1{app.~{\it #1\/}}
\def\refsec#1{sec.~#1}
\def\figin{\epsfcheck\figin}\def\figins{\epsfcheck\figins}
\def\epsfcheck{\ifx\epsfbox\UnDeFiNeD
\message{(NO epsf.tex, FIGURES WILL BE IGNORED)}
\gdef\figin##1{\vskip2in}\gdef\figins##1{\hskip.5in}
\else\message{(FIGURES WILL BE INCLUDED)}%
\gdef\figin##1{##1}\gdef\figins##1{##1}\fi}
\def\figinsert{\goodbreak\midinsert}
\def\ifig#1#2#3{\DefWarn#1\xdef#1{fig.~\the\figno}
\writedef{#1\leftbracket fig.\noexpand~\the\figno}%
\figinsert\figin{\centerline{#3}}\medskip\centerline{\vbox{\baselineskip12pt
\advance\hsize by -1truein\noindent\footnotefont{\bf Fig.~\the\figno:} #2}}
\bigskip\endinsert\global\advance\figno by1}

\lref\rpoly{A. M. Polyakov, Phys. Lett. 103B (1981) 207, 211.}
\lref\rdOne{E. Br\'ezin, V. A. Kazakov, and Al. B. Zamolodchikov,
Nucl. Phys. B338 (1990) 673\semi
G. Parisi, Phys. Lett. B238 (1990) 209, 213;
Europhys. Lett. 11 (1990) 595\semi
D. J. Gross and N. Miljkovic Phys. Lett. B238 (1990) 217.}%
\lref\rGZ{P. Ginsparg and J. Zinn-Justin, Phys. Lett. B240 (1990) 333.}%
\lref\rdOneGK{D.J. Gross and I. Klebanov, Nucl. Phys. B344 (1990) 475.}
\lref\rDS{M. Douglas and S. Shenker, Nucl. Phys. B335 (1990) 635.}%
\lref\rBK{E. Br\'ezin and V. Kazakov, Phys. Lett. B236 (1990) 144.}%
\lref\rGM{D. Gross and A. Migdal, Phys. Rev. Lett. 64 (1990) 127;
Nucl. Phys. B340 (1990) 333.}%
\lref\rbdss{T. Banks, M. Douglas, N. Seiberg, and S. Shenker,
Phys. Lett. B238 (1990) 279.}
\lref\rGZlob{P. Ginsparg and J. Zinn-Justin,
Phys. Lett. B255 (1991) 189.}
\lref\rGZaplob{P. Ginsparg and J. Zinn-Justin,
``Action principle and large order behavior of non-perturbative gravity'',
LA-UR-90-3687/SPhT/90-140 (1990), published in
proceedings of 1990 Carg\`ese workshop.}
\lref\rthooft{G. 't Hooft, Nucl. Phys. B72 (1974) 461.}
\lref\rFDi{F. David, Mod. Phys. Lett. A5 (1990) 1019.}
\lref\rDavidetal{D. Weingarten, Nucl. Phys. B210 [FS6] (1982) 229\semi
F. David, Nucl. Phys. B257[FS14] (1985) 45, 543\semi
J. Ambj{\o}rn, B. Durhuus and J. Fr\"ohlich, Nucl. Phys. B257[FS14]
(1985) 433; J. Fr\"ohlich, in: Lecture Notes in Physics, Vol. 216,
ed. L. Garrido (Springer, Berlin, 1985)\semi
V. A. Kazakov, I. K. Kostov and A. A. Migdal, Phys. Lett. 157B (1985) 295;
D. Boulatov, V. A. Kazakov, I. K. Kostov and A. A. Migdal,
Phys. Lett. B174 (1986) 87; Nucl. Phys. B275[FS17] (1986) 641.}
\lref\rBIPZ{E. Br\'ezin, C. Itzykson, G. Parisi and J.-B. Zuber,
Comm. Math. Phys. 59 (1978) 35.}
\lref\rBIZ{D. Bessis, C. Itzykson, and J.-B. Zuber,
Adv. Appl. Math. 1 (1980) 109.}
\lref\rKPZ{V. G. Knizhnik, A. M. Polyakov, and A. B. Zamolodchikov,
Mod. Phys. Lett. A3 (1988) 819\semi
F. David, Mod. Phys. Lett. A3 (1988) 1651\semi
J. Distler and H. Kawai, Nucl. Phys. B321 (1989) 509.}
\lref\rKM{V. A. Kazakov and A. A. Migdal, Nucl. Phys. B311 (1988) 171.}
\lref\rBPZ{A. A. Belavin, A. M. Polyakov and A. B. Zamolodchikov,
Nucl. Phys. B241 (1984) 333.}
\lref\rD{M. R. Douglas, Phys. Lett. B238 (1990) 176.}
\lref\rNotes{L. Alvarez-Gaum\'e, ``Random surfaces, statistical mechanics,
and string theory'', Lausanne lectures, winter 1990\semi
P. Ginsparg, ``Matrix Models of 2d Gravity,'' lectures given at Trieste Summer
school,
July, 1991, LA-UR-91-4101 (hep-th/\-9112013)\semi
A. Bilal, ``2d gravity from matrix models,'' Johns Hopkins
Lectures, CERN TH5867/90\semi
V. Kazakov, ``Bosonic strings and string field theories in one-dimensional
target space,'' LPTENS 90/30, in {\it Random surfaces and
quantum gravity\/}, proceedings of 1990 Carg\`ese workshop, edited by
O. Alvarez, E. Marinari, and P. Windey, NATO ASI Series B262\semi
E. Br\'ezin, ``Large $N$ limit and discretized two-dimensional quantum
gravity'', in {\it Two dimensional quantum gravity and random surfaces\/},
proceedings of Jerusalem winter school (90/91),
edited by D. Gross, T. Piran, and S. Weinberg\semi
D. Gross, ``The c=1 matrix models'', in proceedings of Jerusalem winter school
(90/91)\semi
I. Klebanov, ``String theory in two dimensions'',
Trieste lectures, spring 1991, Princeton preprint PUPT--1271
(hep-th/9108019)\semi
D. Kutasov, ``Some properties of (non) critical Strings'',
Trieste lectures, spring 1991, Princeton preprint PUPT--1277
(hep-th/9110041)\semi
J. Ma\~nes and Y. Lozano, ``Introduction to Nonperturbative 2d quantum
gravity'', Barcelona preprint UB-ECM-PF3/91\semi
E. Martinec, ``An Introduction to 2d Gravity and Solvable String
Models'' (hep-th/9112019), lectures at 1991 Trieste spring school, Rutgers
preprint RU-91-51\semi
F. David, ``Simplicial quantum gravity and random lattices''
(hep-th/9303127), Lectures given at Les Houches Summer School, July 1992,
Saclay T93/028\semi
A. Marshakov,  ``Integrable structures in matrix models
and physics of 2-d gravity'' (hep-th/9303101), NORDITA-93-21,
Based on lectures given at the Niels Bohr Inst.\semi
A. Morozov, ``Integrability and Matrix Models'' (hep-th/9303139), ITEP-M2/93.}
\lref\rGiMo{P. Ginsparg and G. Moore, ``Lectures on 2D gravity and 2D string
theory'' (hep-th/9304011), TASI lectures (summer, 1992), to appear in
proceedings\semi
P. Ginsparg and G. Moore, ``{\it Lectures on 2d gravity and 2d string
theory\/}, {\bf the book},'' Cambridge University Press, to appear
later in 1993.}
\lref\rDijkW{R. Dijkgraaf and E. Witten, 
``Mean field theory, topological field theory, and multimatrix models'',
Nucl. Phys. B342 (1990) 486}
\lref\rAW{O. Alvarez and P. Windey, Nucl. Phys. B348 (1991) 490.}
\lref\rGD{I. M. Gel'fand and L. A. Dikii, Russian Math. Surveys
30:5 (1975) 77\semi
I. M. Gel'fand and L. A. Dikii, Funct. Anal. Appl. 10 (1976) 259.}
\lref\rGGPZ{P. Ginsparg, M. Goulian, M. R. Plesser, and J. Zinn-Justin,
Nucl. Phys. B342 (1990) 539.}
\lref\rpglh{P. Ginsparg, ``Applied conformal field theory'' Les Houches
Session XLIV, 1988, {\it Fields, Strings, and Critical Phenomena\/}, ed.\ by
E. Br\'ezin and J. Zinn-Justin (1989).}
\lref\rkazcon{V. Kazakov, Mod. Phys. Lett. A4 (1989) 2125.}
\lref\rfrdc{F. David, ``Nonperturbative effects in 2D gravity and matrix
models,'' Saclay-SPHT-90-178, in proceedings of Carg\`ese workshop
(1990), edited by
O. Alvarez, E. Marinari, and P. Windey, NATO ASI Series B262.}
\lref\FF{B. Feigin and D. Fuchs, Funct. Anal. and Appl. {\bf 16} (1982) 114.}
\lref\NATI{N. Seiberg, ``Notes on Quantum Liouville Theory and
Quantum Gravity,'' in {\it Common Trends in Mathematics and Quantum Field
Theory\/}, Proc. of the 1990 Yukawa International Seminar, Prog. Theor. Phys.
Suppl 102, and in proceedings of
1990 Carg\`ese workshop.}
\lref\SEL{A. Selberg, Norsk Matematisk Tidsskrift {\bf 26} (1944) 71.}
\lref\DKU{ P. Di Francesco and D. Kutasov, Phys. Lett. {\bf 261B} (1991) 385;
``World sheet and space time physics in two
dimensional (super) string theory'' (hep-th/9109005),
Nucl. Phys. B375 (1992) 119.}
\lref\WSTR{E. Witten, ``On the structure of the topological phase of
two dimensional gravity,'' Nucl. Phys. B340 (1990) 281.}
\lref\KON{M. Kontsevich,
``Intersection theory on the moduli space of curves,''
Funk. Anal.\& Prilozh., {\bf 25} (1991) 50-57\semi
``Intersection theory on the moduli space of curves and the matrix Airy
function,'' lecture at the Arbeitstagung, Bonn, June 1991 and
Comm. Math. Phys. {\bf 147} (1992) 1.}
\lref\Wun{E. Witten,
``Two dimensional gravity and intersection theory on moduli space,''
Surv. in Diff. Geom. {\bf 1} (1991) 243-310.}
\lref\WIT{E. Witten,
``On the Kontsevich model and other models of two dimensional gravity,''
preprint IASSNS-HEP-91/24}
\lref\Wdep{E. Witten, ``The $N$ matrix model and gauged WZW models,''
Nucl.Phys.B371 (1992) 191.}
\lref\Wtr{E. Witten,
``Algebraic geometry associated with matrix models of two dimensional
gravity,''
 preprint IASSNS-HEP-91/74.}
\lref\CJB{ C. Itzykson and J.-B. Zuber,
``Combinatorics of the Modular Group II: The \KK\ integrals,''
Int. J. Mod. Phys. A7 (1992) 5661.}
\lref\INT{Harish-Chandra,
{``Differential operators on a semisimple Lie algebra},
Amer. J. Math. {\bf 79} (1957) 87\semi
C. Itzykson and J.-B. Zuber,
``The planar approximation II,''  J. Math. Phys. {\bf 21} (1980) 411-421.}
\lref\DIJ{R. Dijkgraaf, ``Intersection theory, integrable hierarchies
and topological field theory'' (hep-th/9201003), 1991 Carg\`ese lectures,
published in  NATO ASI: Cargese 1991:95-158 (QC174.45:N2:1991).}
\lref\KMM{S. Kharchev, A. Marshakov, A. Mironov, A. Morozov and A. Zabrodin,
``Towards unified theory of 2D gravity'' (hep-th/9201013),
Nucl. Phys. B380 (1992) 181\semi
A. Marshakov, A. Mironov, and A. Morozov,
``From Virasoro Constraints in Kontsevich's Model
to $\cal W$-constraints in 2-matrix Models'' (hep-th/9201010),
Mod. Phys. Lett. A7 (1992) 1345\semi
S. Kharchev, A. Marshakov, A. Mironov, and A. Morozov,
``Generalized Kontsevich model versus Toda hierarchy and discrete matrix
models'' (hep-th/9203043), FIAN-TD-03-92.}
\lref\DWIT{R. Dijkgraaf and E. Witten, ``Mean field theory, topological
field theory and multi--matrix models,'' Nucl. Phys. {\bf B342}(1990) 486.}
\lref\KEKE{K. Li, ``Topological gravity with minimal matter,''
Nucl. Phys. B354 (1991) 711.}
\lref\EY{T. Eguchi and S.-K. Yang, ``$N=2$ superconformal models as topological
field theories,'' Mod. Phys. Lett. A5 (1990) 1693.}
\lref\DVV{R. Dijkgraaf, E. Verlinde and H. Verlinde, ``Topological strings
in $d<1$,'' Nucl. Phys. {\bf B352} (1991) 59.}
\lref\PCJB{P. Di Francesco, C. Itzykson and J.--B. Zuber, ``Polynomial
averages in the Kontsevich model,'' Comm. Math. Phys. {\bf 151} (1993) 193.}
\lref\PPK{P. Di Francesco, ``Observables in the Kontsevich model,'' to
appear in  {\it Low dimensional topology and gravity\/},
Cambridge University Press (1993).}
\lref\rWtp{E. Witten, Nucl. Phys. B340 (1990) 281.}
\lref\DS{V. Drinfeld and V. Sokolov, J. Sov. Math. {\bf 30}(1985) 1975.}
\lref\DIFK{P. Di Francesco and D. Kutasov,
``Integrable models of two-dimensional quantum gravity'',
Princeton preprint PUPT-1206 (1990), published in proceedings of Carg\`ese
workshop  on Random Surfaces, Quantum Gravity and Strings(1990)\semi
P. Di Francesco and D. Kutasov, Nucl. Phys. {\bf B342} (1990) 589.}
\lref\rKBK{V. Kazakov, Phys. Lett. 119A (1986) 140\semi
D. Boulatov and V. Kazakov, Phys. Lett. 186B (1987) 379.}
\lref\rising{E. Br\'ezin, M. Douglas, V. Kazakov, and S. Shenker, Phys. Lett.
B237 (1990) 43\semi
D. Gross and A. Migdal, Phys. Rev. Lett. 64 (1990) 717.}
\lref\rIYL{C. Crnkovi\'c, P. Ginsparg, and G. Moore,
Phys. Lett. B237 (1990) 196.}
\lref\DOTFAT{V. Dotsenko and V. Fateev, Phys. Lett. {\bf 154B}
(1985) 291.}
\lref\rEyZJii{B.Eynard and J. Zinn-Justin,
{\it Phys. Lett.} B302 (1993) 396.}
\lref\ZAMO{A. Zamolodchikov, Sov. J. Nucl. Phys.{\bf 44} (1986) 529.}
\lref\rBout{P. Boutroux, {\it
Ann. Ecole Normale} 30 (1913) 256; E. Hille, {\it Ordinary Differential
Equations in the complex domain\/}, Pure and Applied Mathematics, J. Wiley and
Sons, (1976).}
\lref\rDoDKK{M. Douglas, ``The two-matrix model'', published in proceedings of
1990 Carg\`ese workshop\semi
E. Martinec, ``On the origin of integrability in matrix models,''
Commun. Math. Phys. 138 (1991) 437\semi
T. Tada, Phys. Lett. B259 (1991) 442\semi
J.M.~Daul, V.A.~Kazakov and I. K. Kostov,
``Rational theories of 2D gravity from the two-matrix model''
(hep-th/9303093), CERN-TH.6834/93.}
\lref\rnonor{ G. Harris and E. Martinec, Phys. Lett. B245 (1990) 384\semi
E. Brezin and H. Neuberger,
Phys. Rev. Lett. 65 (1990) 2098; Nucl.Phys. B350 (1991) 513.}
\lref\rNeu{H. Neuberger, ``Regularized string and flow equations,''
preprint RU-90-50.}
\lref\rIK{I. Kostov, {\it Mod. Phys. Lett.} A4 (1989) 217, {\it Phys. Lett.}
B266 (1991) 312\semi
M. Gaudin and I. Kostov, {\it Phys. Lett.} B220 (1989) 200\semi
I. K. Kostov and M. Staudacher, {\it Nucl. Phys.} B384 (1992) 459.}
\lref\rZJ{J. Zinn-Justin, {\it Quantum Field Theory and Critical Phenomena},
Oxford Univ. Press (1989), pp. 915--928 in 2nd edition (1993).}
\lref\rScar{S. Shenker, ``The strength of nonperturbative effects in string
theory,'' in proceedings of Carg\`ese workshop (1990).}
\lref\rEyZJ{B. Eynard and J. Zinn-Justin, {\it Nucl. Phys.} B386 (1992) 558.}
\lref\rEyZJiii{B. Eynard and J. Zinn-Justin, {\it The $O(n)$ model on a random
surface: critical points and large order behavior (II)}, Saclay preprint
SPhT/93-029}
\lref\rbachpet{C. Bachas and P.M.S. Petropoulos,
Phys. Lett. B247 (1990) 363\semi
C. Bachas, ``On triangles and squares,''
in proceedings of Carg\`ese workshop (1990).}
\lref\rFDii{F. David, {\it Nucl. Phys.} B348 (1991) 507.}
\lref\rBMP{E. Br\'ezin, E. Marinari, and G. Parisi, Phys. Lett. B242 (1990)
35.}
\lref\rDSS{M. Douglas, N. Seiberg, and S. Shenker,
Phys. Lett. B244 (1990) 381.}
\lref\rMoore{G. Moore,
``Geometry of the string equations,'' Commun. Math. Phys. 133 (1990) 261.}
\lref\rMPsup{E. Marinari and G. Parisi, Phys. Lett. B240 (1990) 375\semi
M. Karliner and S. Migdal,
``Nonperturbative 2D quantum gravity via supersymmetric string,''
Mod. Phys. Lett. A5 (1990) 2565\semi
M. Karliner, A. Migdal and B. Rusakov,
``Ground State of 2D Quantum Gravity and Spectral Density
of Random Matrices'' (hep-th/9212114), Nucl. Phys. B399 (1993) 514.}
\lref\rdvvvir{R. Dijkgraaf, H. Verlinde, E. Verlinde,
``Loop equations and virasoro constraints in nonperturbative 2-d quantum
gravity,'' Nucl. Phys. B348 (1991) 435\semi
M. Fukuma, H. Kawai, and Ryuichi Nakayama,
``Continuum Schwinger-Dyson equations and universal structures
in two-dimensional quantum gravity,''
Int. J. Mod. Phys. A6 (1991) 1385\semi
Yu. Makeenko, A. Marshakov, A. Mironov, and A. Morozov,
``Continuum versus discrete Virasoro in one matrix models,''
Nucl. Phys. B356 (1991) 574.}
\lref\rpoperev{C. Pope, ``Review of W Strings'' (hep-th/9204093),
Contribution to Proceedings of HARC Meeting, CTP TAMU--30/92.}
\lref\rtric{M. Kreuzer and R. Schimmrigk, ``Tricritical
Ising model, generalized KdVs, and nonperturbative 2d-quantum gravity'',
Phys. Lett. B248 (1990) 51\semi
H. Kunitomo and S. Odake, ``Nonperturbative analysis of three matrix
model'', Phys. Lett. B247 (1990) 57\semi
K. Fukazawa, K. Hamada, and H. Sato, ``Phase structures of 3-Matrix chain
model'', Mod. Phys. Lett. A5 (1990) 2431.}
\lref\rMSeSt{G. Moore, N. Seiberg, and M. Staudacher,
``From loops to states in 2-d quantum gravity'',
Nucl. Phys. {\bf 362} (1991) 665.}
\lref\rstaud{Matthias Staudacher,
``The Yang-Lee edge singularity on a dynamical planar random surface,''
Nucl. Phys. B336 (1990) 349.}
\lref\rpqdual{S. Kharchev, A. Marshakov,
``On $p-q$ duality and explicit solutions in $c \le 1$ 2d gravity models''
hep-th/9303100;
``Topological versus Non-Topological Theories and $p-q$ Duality in $c \le 1$
2d Gravity Models'' (hep-th/9210072), FIAN-TD-15-92.}
\lref\rMardis{A. Gerasimov, A. Marshakov, A. Mironov, A. Morozov, and A. Orlov,
``Matrix models of 2-d gravity and Toda theory,''
Nucl. Phys. B357 (1991) 565\semi
A. Marshakov, A. Mironov, and A. Morozov,
``Generalized matrix models as conformal field theories: discrete case,''
Phys. Lett. B265 (1991) 99.}
\lref\rdhok{Eric D'Hoker, ``Lecture notes on 2-d quantum gravity and
Liouville theory,'' UCLA-91-TEP-35,
in Particle Physics VI-th Jorge Andre Swieca Summer School, eds. O.J.P. Eboli,
M. Gomes and A. Santoro, World Scientific Publishers (1992).}
\lref\rGLi{M. Goulian and M. Li, Phys. Rev. Lett. 66 (1991) 2051.}
\lref\rAoDhok{Ken-ichiro Aoki and Eric D'Hoker,
``On the liouville approach to correlation functions for 2-d quantum gravity''
(hep-th/910902417), Mod. Phys. Lett. A7 (1992) 235.}
\lref\rtorPF{Michael Bershadsky, Igor R. Klebanov,
``Genus one path integral in two-dimensional quantum gravity'',
Phys. Rev. Lett. 65 (1990) 3088\semi                  
N. Sakai and Y. Tanii, ``Compact boson coupled to two-dimensional gravity'',
Int. J. Mod. Phys. A6 (1991) 2743.}
\lref\grndrng{E. Witten,
Ground ring of two-dimensional string theory'' (hep-th/9108004),
Nucl. Phys. B373 (1992) 187.}
\lref\rmgrng{M. Bershadsky and D. Kutasov, (hep-th/9110034),
Phys. Lett. B274 (1992) 331;
also (hep-th/9204049) Nucl. Phys. B382 (1992) 213\semi
D. Kutasov, E. Martinec, and N. Seiberg,
``Ground rings and their modules in 2-d gravity with c <= 1 matter''
(hep-th/9111048), Phys. Lett. B276 (1992) 437.}

\def\stomm{\hyperref {}{section}{2}{2}}
\def\sstlnl{\hyperref {}{subsection}{3.1}{3.1}}
\def\sqmomm{\hyperref {}{section}{4}{4}}
\def\ssesgkf{\hyperref {}{subsection}{4.3}{4.3}}
\def\sTg{\hyperref {}{section}{5}{5}}
\def\stcalg{\hyperref {}{section}{6}{6}}
\def\ssDw{\hyperref {}{subsection}{6.2}{6.2}}
\def\ssPG{\hyperref {}{subsection}{7.2}{7.2}}
\def\ssIalobom{\hyperref {}{subsection}{7.7}{7.7}}
\def\sMccrda{\hyperref {}{section}{8}{8}}
\def\sstomxc{\hyperref {}{subsection}{8.1}{8.1}}
\def\ssmmdis{\hyperref {}{subsection}{8.3}{8.3}}
\def\stOnmm{\hyperref {}{section}{9}{9}}

\def\ssTrLcq{\hyperref {}{subsection}{\hbox {A.}1}{\hbox {A.}1}}
\def\sstlze{\hyperref {}{subsection}{\hbox {A.}3}{\hbox {A.}3}}
\def\eresolv{(\hyperref {}{equation}{\hbox {A.}7}{\hbox {A.}7})}
\def\sGkdv{\hyperref {}{appendix}{B}{B}}
\def\ssEc{\hyperref {}{subsection}{\hbox {B.}1}{\hbox {B.}1}}
\def\sspqqp{\hyperref {}{subsection}{\hbox {B.}2}{\hbox {B.}2}}
\def\smmaj{\hyperref {}{appendix}{C}{C}}
\def\ssDccr{\hyperref {}{subsection}{\hbox {D.}1}{\hbox {D.}1}}
\def\sAfur{\hyperref {}{subsection}{\hbox {D.}2}{\hbox {D.}2}}

\Title{\vbox{\vskip-10pt\baselineskip12pt\hbox{LA-UR-93-1722}\hbox{SPhT/93-061}
\hbox{hep-th/9306153}}}{2D Gravity and Random Matrices}

\vskip-.3in
\centerline{P. Di Francesco$^{1,2}$, P. Ginsparg$^4$
and J. Zinn-Justin$^{1,3}$}

\medskip{\baselineskip14pt
\centerline{$^1$Service de Physique Th\'eorique de Saclay}
\centerline{F-91191 Gif-sur-Yvette Cedex, FRANCE}
\centerline{$^2$philippe@amoco.saclay.cea.fr}
\centerline{$^3$zinn@amoco.saclay.cea.fr}}
\smallskip
{\baselineskip14pt\centerline{${}^4$MS-B285}
\centerline{Los Alamos National Laboratory}
\centerline{Los Alamos, NM \ 87545}
\centerline{ginsparg@xxx.lanl.gov}}


\smallskip
We review recent progress in 2D gravity coupled to $d<1$ conformal matter,
based on a representation of discrete gravity in terms of random matrices. We
discuss the saddle point approximation for these models, including a class of
related $O(n)$ matrix models. For $d<1$ matter, the matrix problem can be
completely solved in many cases by the introduction of suitable orthogonal
polynomials. Alternatively, in the continuum limit the orthogonal polynomial
method can be shown to be equivalent to the construction of representations of
the canonical commutation relations in terms of differential operators.
In the case of pure gravity or discrete Ising--like matter, the sum
over topologies is reduced to the solution of non-linear differential
equations (the Painlev\'e equation in the pure gravity case)
which can be shown to follow from an action principle.
In the case of pure gravity and more generally all unitary models,
the perturbation theory is not Borel summable and therefore alone does not
define a unique solution.
In the non-Borel summable case, the matrix model does not define the
sum over topologies beyond perturbation theory.
We also review the computation of correlation functions directly in the
continuum formulation of matter coupled to 2D gravity, and compare with the
matrix model results. Finally, we review the relation between matrix models
and topological gravity, and as well the relation to intersection theory of
the moduli space of punctured Riemann surfaces.

\smallskip
\Date{\qquad 6/93, submitted to Physics Reports}
\listtoc\writetoc

\vfill\eject

\secno-1
\newsec{Introduction} 
It was proposed some time ago \rDavidetal\ that the integral over the internal
geometry of a 2D surface can be discretized as a sum over randomly
triangulated surfaces. The use of such a lattice regularization allows the
partition function of 2D quantum gravity coupled to certain matter systems to
be expressed as the free energy of an associated hermitian matrix model. This
matrix realization can frequently be solved by means of large $N$ techniques
\rBIPZ\ (see also \rBIZ\ and references therein), and the solutions restricted
to fixed topology of the two dimensional spacetime are found to be in
agreement with the continuum Liouville results of \rKPZ\ (as we shall review
in detail later on here).

More recently \refs{\rDS\rBK{--}\rGM},
a continuum limit that includes the sum over
topologies of two dimensional surfaces was defined for certain matter
systems coupled to 2D quantum gravity.
The continuum limit specific heat for these models was moreover found
to satisfy an ordinary differential equation, in principle allowing a full
non-perturbative solution.

The above progress \refs{\rDS{--}\rGM} suggests hope for extracting
nonperturbative information from string theory, at least in some simple
contexts. A prime obstacle to our understanding of string theory has been an
inability to penetrate beyond its perturbative expansion. Our understanding of
gauge theory is enormously enhanced by having a fundamental formulation based
on the principle of local gauge invariance from which the perturbative
expansion can be derived. Symmetry breaking and nonperturbative effects such
as instantons admit a clean and intuitive presentation. In string theory, our
lack of a fundamental formulation is compounded by our ignorance of the true
ground state of the theory.

String theory is an attempt to overcome the difficulties encountered in the
quantization of 4D gravity by replacing particles by string-like one
dimensional objects, which describe some two dimensional ``worldsheet''
$\Sigma$ as they evolve in time (interactions are encoded in the genus
of the surface $\Sigma$). Polyakov showed that such theories
could be interpreted as theories of {\it two dimensional\/}
quantum gravity, in which the world--sheet is exchanged with
the space--time, and the string coordinate $X^{\mu}(\sigma)$,
$\sigma \in \Sigma$, is considered as a $D$--dimensional ``matter'' field
defined on $\Sigma$.

In string theory we thus wish to perform an integral over two dimensional
geometries and a sum over two dimensional topologies,
$$Z\sim\sum_{\rm topologies} \int\CD g\,\CD X\ \ee{-S}\ ,$$
where the spacetime physics (in the case of the bosonic string) resides in the
conformally invariant action
\eqn\eString{S\propto\int \d^2\xi\,\sqrt g\, g^{ab}\,\del_a X^\mu\,\del_b
X^\nu \,G_{\mu \nu}(X)\,.}
Here $\mu,\nu$ run from $1,\ldots,D$ where $D$ is the number of spacetime
dimensions, $G_{\mu \nu}(X)$ is the spacetime metric,
and the integral $\CD g$ is over worldsheet metrics. Typically we
``gauge-fix'' the worldsheet metric to $g\dup_{ab}=\e^{\ph}\delta_{ab}$, where
$\ph$ is known as the Liouville field.
Following the formulation of string theory in this form (and in particular
following the appearance of \rpoly), there was much work to develop the
quantum Liouville theory (some of which is reviewed in sec.~\stcalg\ here).

The method of \refs{\rDS{--}\rGM}, using a discretization of the string
worldsheet to incorporate in the continuum limit
simultaneously the contribution of 2d surfaces with any number of handles,
makes it possible not only to integrate over all possible
deformations of a given genus surface (the analog of the integral over Feynman
parameters for a given loop diagram), but also to sum over all genus
(the analog of the sum over all loop diagrams).
This progress, however, is limited in the sense that these methods only apply
currently for non-critical strings embedded in dimensions $D\le1$ (or
equivalently critical
strings embedded in $D\le2$), and the nonperturbative information even in this
restricted context has proven incomplete. Due to familiar problems with lattice
realizations of supersymmetry and chiral fermions, these methods have also
resisted extension to the supersymmetric case.

In addition an investigation of the large order behavior of the perturbative
(topological) expansion shows that terms at large
orders $k$ have a typical $(2k)!$ behavior. The perturbation series are
divergent, and for the most interesting models (pure gravity, unitary
models) the series are non-Borel summable because all terms are positive.
Perturbation theory does not define unique functions, and
in the pure gravity case it can be shown that the real solution of the
Painlev\'e equation has unphysical properties. It is thus conjectured that in
the non-Borel summable case, the matrix model does not define the sum over
topologies beyond perturbation theory.

The developments we shall describe here nonetheless provide at least a
half-step in the correct direction, if only to organize the perturbative
expansion in a most concise way. They have also prompted much useful evolution
of related continuum methods. Our point of view here is that string theories
embedded in $D\le1$ dimensions provide a simple context for testing ideas and
methods of calculation \ref\BrJZ{see for example: E. Br\'ezin and J.
Zinn-Justin, {\it Phys. Lett.} B288 (1992) 54.}. Just as we would encounter
much difficulty
calculating infinite dimensional functional integrals without some prior
experience with their finite dimensional analogues, progress in string theory
should be aided by experimentation with systems possessing a restricted number
of degrees of freedom.

Other review references on the same general subject can be found in
\rNotes. In this review we concentrate mainly on the properties of
$D<1$ systems coupled to 2D gravity. Many
interesting recent developments in the field, including study of
issues of principle such as topology change in 2D quantum gravity,
and as the relation to recent work on $d=2$ black holes in string theory,
are based on $D=1$ matter coupled to gravity\foot{which, together with
the Liouville field, results in a $d=2$
dimensional target space in the critical string interpretation}
(see e.g.\ \rGiMo).

\newsec{Discretized surfaces, matrix models, and the continuum limit}
\seclab\sDsmmcl

\subsec{Discretized surfaces}

We begin here by considering a ``$D=0$ dimensional string theory'',
i.e.\ a pure
theory of surfaces with no coupling to additional ``matter'' degrees of
freedom  on the string worldsheet. This is equivalent to the propagation of
strings in a non-existent embedding space. For partition function we take
\eqn\eZdo{Z=\sum_h\int\CD g\,\ee{-\beta A + \gamma \chi}\ ,}
where the sum over topologies is represented by the summation over $h$, the
number of handles of the surface, and the action consists of couplings to the
area $A=\int\sqrt g$, and to the Euler character
$\chi={1\over4\pi}\int\sqrt g\,R=2-2h$.

(Recall that the Einstein action for pure gravity
with cosmological term reads $S(g)=\int \d^d x \sqrt{g}(K R+ \Lambda)$,
in which $g\dup_{ij}$ is the metric tensor, $R$ the scalar curvature and
$K,\Lambda$ are two coupling constants. The cosmological constant $\Lambda$
multiplies the volume element. In two dimensions classical gravity is trivial
because the scalar curvature term $\int \sqrt{g}R$
is topological (Gauss--Bonnet theorem) and
thus does not contribute to the equations of motion.  In the quantum case,
however, even two dimensional gravity is non-trivial because large quantum
fluctuations may change the genus of the surface and the partition function
hence involves a sum over surfaces of all genus. In addition, on higher
genus surfaces there are non-trivial topological sectors.)

\ifig\frandtri{A piece of a random triangulation of a surface.
Each of the triangular faces is dual to a three point vertex of a quantum
mechanical matrix model.}{\epsfxsize3.25in\epsfbox{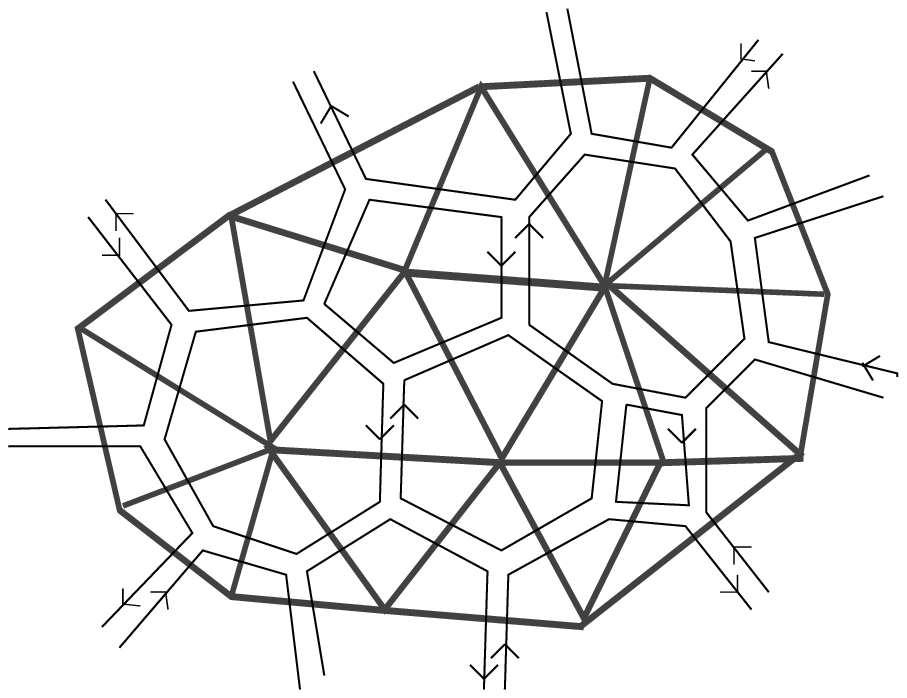}}

The integral $\int\CD g$ over the metric on the surface in \eZdo\ is
difficult to calculate in general. The most progress in the continuum has
been made via the Liouville approach which we briefly review in sec.~\stcalg.
If we discretize the surface, on the other hand, it turns out that \eZdo\
is much easier to calculate, even before removing the finite cutoff. We
consider in particular a ``random triangulation'' of the surface
\rDavidetal, in which the surface is constructed from triangles, as in
\frandtri. The triangles are designated to be equilateral,\foot{We point out
that this constitutes a basic difference from the Regge calculus, in which
the link lengths are geometric degrees of freedom. Here the geometry is
encoded entirely into the coordination numbers of the vertices.
This restriction of degrees of freedom roughly corresponds to fixing a
coordinate gauge, hence we integrate only over the gauge-invariant moduli
of the surfaces.} so that
there is negative (positive) curvature at vertices $i$ where the number
$N_i$ of incident triangles is more (less) than six, and zero curvature
when $N_i=6$. Indeed if we call $V$, $E$, and $F$ the total number of
vertices, edges, and faces respectively, of the triangulation, due to the
topological relations $2E=\sum_i N_i$ and $3F=2E$ (a relation obeyed by
triangulations of surfaces, since each face has three edges each of which is
shared by two faces), all quantities can be expressed in terms of the $N_i$'s.
The discrete counterpart to the
Ricci scalar $R$ at vertex $i$ is $R_i=2\pi(6-N_i)/ N_i$, so that
$$\int\sqrt g\,R\to \sum_i 4 \pi(1-N_i/6)
=4 \pi(V-\half F)=4 \pi(V-E+F)=4 \pi \chi\ ,$$
coincides with the simplicial definition which gives the Euler
character $\chi$.
The discrete counterpart to the
infinitesimal volume element $\sqrt g$ is $\sigma_i=N_i/3$,
so that the total area $|S|=\sum_i \sigma_i$ just counts the total number of
triangles, each designated to have unit area. (The factor
of $1/3$ in the definition of $\sigma_i$ is because
each triangle has three vertices and is counted three times.)
The summation over all such random triangulations is thus
the discrete analog to the integral $\int \CD g$ over all possible
geometries,
\eqn\ediscr{\sum_{{\rm genus}\ h}\ \int \CD g \quad
\to \ \sum_{\scriptstyle\rm random \atop \scriptstyle\rm triangulations}\ .}

In the above, triangles do not play an essential role and may be replaced
by any set of polygons. General random polygonulations of surfaces
with appropriate fine tuning of couplings may, as we shall see, have more
general critical behavior, but can in particular always reproduce the
pure gravity behavior of triangulations in the continuum limit.

\subsec{Matrix models}
\subseclab\ssMm

We now demonstrate how the integral over geometry in \eZdo\ may be
performed in its discretized form as a sum over random triangulations. The
trick is to use a certain matrix integral as a generating functional for
random triangulations. The essential idea goes back to work \rthooft\ on
the large $N$ limit of QCD, followed by work on the saddle point approximation
\rBIPZ.

We first recall the (Feynman) diagrammatic expansion of the (0-dimensional)
field theory integral
%
\eqn\esft{\int_{-\infty}^\infty
{\d\ph\over\sqrt{2\pi}}\, \ee{-\ph^2/2+ \lambda\ph^4/4!}\ ,}
where $\ph$ is an ordinary real number.\foot{The integral is understood to
be defined by analytic continuation to negative $\lambda$.} In a formal
perturbation series in $\lambda$, we would need to evaluate integrals such as
\eqn\epel{{\lambda^n\over n!}
\int_\ph\ee{-\ph^2/2}\left({\ph^4\over4!}\right)^n\ .}
Up to overall normalization we can write
\eqn\essft{\int_\ph \ee{-\ph^2/2}\ph^{2k}=
\left.{\del^{2k}\over\del J^{2k}}\int_\ph\ee{-\ph^2/2+J\ph}\right|_{J=0}
=\left.{\del^{2k}\over\del J^{2k}}\,\ee{J^2/2}\right|_{J=0}\ .}
Since ${\del\over\del J}\e^{J^2/2}=J \e^{J^2/2}$, applications of
$\del/\del J$ in the above need to be paired so that any factors of $J$
are removed before finally setting $J=0$. Therefore if we represent each
``vertex'' $\lambda\ph^4$ diagrammatically as a point with four emerging
lines (see fig.~\the\figno b), then \epel\ simply counts the number of ways
to group
such objects in pairs. Diagrammatically we represent the possible pairings
by connecting lines between paired vertices. The connecting line is known
as the propagator $\langle\ph\,\ph\rangle$ (see fig.~\the\figno a) and the
diagrammatic rule we have described for connecting vertices in pairs is
known in field theory as the Wick expansion.

$$\vbox{\hrule height .7pt width 30pt\vskip30pt
\hbox{\quad(a)}}\qquad\qquad\qquad
\vbox{\hbox{\hskip20pt\vrule width .7pt height 40pt}\vskip-20pt\hrule width
40pt height .7pt\vskip30pt\hbox{\quad\ (b)}}$$
\vglue5pt\nobreak
\centerline{\footnotefont{\bf Fig.~\the\figno:}
(a) the scalar propagator.   (b) the scalar four-point vertex.}
\bigbreak\xdef\fscpv{fig.~\the\figno}\global\advance\figno by1

When the number of vertices $n$ becomes large, the allowed diagrams begin
to form a mesh reminiscent of a 2-dimensional surface. Such diagrams
do not yet have enough structure to specify a Riemann surface. The
additional structure is given by widening the propagators to ribbons (to
give so-called ribbon graphs or ``fatgraphs''). From the standpoint of \esft,
the required extra structure is given by replacing the scalar
$\ph$ by an $N\times N$ hermitian matrix $M^i{}_j$. The analog of \essft\
is given by adding indices and traces:
\eqn\esmft{\eqalign{\int_M\ee{-\tr M^2/2} M^{i_1}{}_{j_1}\cdots
M^{i_n}{}_{j_n} &=\left.{\del\over\del J^{j_1}{}_{i_1}}
\cdots{\del\over\del J^{j_n}{}_{i_n}} \ee{-\tr M^2/2+\tr J M}\right|_{J=0}\cr
&=\left.{\del\over\del J^{j_1}{}_{i_1}}
\cdots{\del\over\del J^{j_n}{}_{i_n}}\, \ee{\tr J^2/2}\right|_{J=0}\ ,\cr}}
where the source $J^i{}_j$ is as well now a matrix. The measure in \esmft\ is
the invariant
$\d M=\prod_i\d M^i{}_i\,\prod_{i<j}\d{\rm Re} M^i{}_j\,\d{\rm Im} M^i{}_j$,
and the normalization is such that $\int_M \e^{-\tr M^2/2}=1$.
To calculate a quantity such as
\eqn\empel{{\lambda^n\over n!}\int_M \e^{-\tr M^2/2}(\tr M^4)^n\ ,}
we again lay down $n$ vertices (now of the type depicted in fig.~\the\figno b),
and connect the legs with propagators $\langle
M^i{}_j\,M^k{}_l\rangle=\delta^i_l\,\delta^k_j$ (fig.~\the\figno a).
The presence of upper and lower matrix indices is represented in
fig.~\the\figno\ by the
double lines\foot{This is the same notation employed in the large $N$ expansion
of QCD \rthooft.} and it is understood that the sense of the arrows is to be
preserved when linking together vertices. The resulting diagrams are similar to
those of the scalar theory, except that each external line has an associated
index $i$, and each internal closed line corresponds to a summation over an
index $j=1,\ldots,N$. The ``thickened'' structure is now sufficient to
associate a Riemann surface to each diagram, because the closed internal loops
uniquely specify locations and orientations of faces.

\font\bigarrfont=cmsy10 scaled\magstep 3
\def\extarr{\mathord-\mkern-6mu}
\def\vuline{\raise7.5pt\hbox{\textfont2\bigarrfont$\uparrow$}\hskip-4.75pt
\hbox{\vrule width.7pt depth -18 pt height 27pt}}
\def\vdline{\raise15pt\hbox{\textfont2\bigarrfont$\downarrow$}\hskip-4.75pt
\vrule width.7pt depth -4 pt height 27pt}
$$\hbox{\vbox{\hbox{\textfont2\bigarrfont$\extarr
\mathord\rightarrow\mkern-6mu\extarr\extarr$}
\vskip-10pt\hbox{$\textfont2\bigarrfont\extarr\extarr
\mathord\leftarrow\mkern-6mu\extarr$}\vskip21pt\hbox{\quad\ (a)}}
\qquad\qquad\qquad
\vbox{\hbox{$\textfont2\bigarrfont\mathord\rightarrow\mkern-6mu\extarr
\vuline\hskip2.5pt\vdline\mkern-1mu\mathord\rightarrow\mkern-6mu\extarr$}
\vskip-10pt
\hbox{$\textfont2\bigarrfont\extarr\mathord\leftarrow\mkern-6mu
\lower23pt\hbox{\vuline}\hskip2.5pt
\lower23pt\hbox{\vdline}\mkern-2mu\extarr\mathord\leftarrow$}
\vskip7pt\hbox{$\,$\qquad(b)}}}$$
\nobreak
\centerline{\footnotefont{\bf Fig.~\the\figno:}
(a) the hermitian matrix propagator.
(b) the hermitian matrix four-point vertex.}
\bigbreak\xdef\fmapv{fig.~\the\figno}\global\advance\figno by1

To make contact with the random triangulations discussed earlier,
we consider the diagrammatic expansion of the matrix integral
\eqn\ecmm{\ee{Z}=\int\d M\ \ee{-\ha\tr M^2+{g\over\sqrt N}\tr M^3}}
(with $M$ an $N\times N$ hermitian matrix, and the integral again
understood to be defined by analytic continuation in the coupling $g$.)
The term of order $g^n$ in a power series expansion counts the number of
diagrams constructed with $n$ 3-point vertices. The dual to such a diagram (in
which each face, edge, and vertex is associated respectively to a dual
vertex, edge, and face) is identically a random triangulation inscribed on
some orientable Riemann surface (\frandtri). We see that the matrix integral
\ecmm\ automatically generates all such random triangulations.\foot{Had we
used real symmetric matrices rather than the hermitian matrices $M$, the two
indices
would be indistinguishable and there would be no arrows in the propagators and
vertices of \fmapv. Such orientationless vertices and propagators generate an
ensemble of both orientable and non-orientable surfaces \rnonor.}
Since each triangle has unit area, the area of the
surface is just $n$. We can thus make formal identification with \eZdo\
by setting $g=\e^{-\beta}$. {\it Actually the matrix integral generates both
connected and disconnected surfaces, so we have written $\e^ Z$ on the
left hand side of \ecmm}. As familiar from field theory, the exponential of
the connected diagrams generates all diagrams, so $Z$ as defined above
represents contributions only from connected surfaces. We see that the
{\it free energy\/} from the matrix model point of view is actually the
{\it partition function\/} $Z$ from the 2d gravity point of view.

There is additional information contained in $N$, the size of the matrix.
If we change variables $M\to M\sqrt N$ in \ecmm, the matrix action becomes
$N\,\tr(-\half\tr M^2+g\,\tr M^3)$, with an overall factor of
$N$.\foot{Although we could as well rescale $M\to M/g$ to pull out an overall
factor of $N/g^2$, note that
$N$ remains distinguished from the coupling $g$ in the model since it enters
as well into the traces via the $N\times N$ size of the matrix.} This
normalization makes it easy to count the power of $N$ associated to any
diagram. Each vertex contributes a factor of $N$, each propagator (edge)
contributes a factor of $N\inv$ (because the propagator is the inverse of
the quadratic term), and each closed loop (face) contributes a factor of
$N$ due to the associated index summation.
Thus each diagram has an overall factor
\eqn\efN{N^{V-E+F}=N^\chi=N^{2-2h}\ ,}
where $\chi$ is the Euler character of the surface associated to the diagram.
We observe that the value $N=\e^{\gamma}$ makes contact with the coupling
$\gamma$ in \eZdo.  In conclusion, if we take
$g=\e^{-\beta}$ and $N=\e^{\gamma}$, we can formally identify the continuum
limit of the partition function $Z$ in \ecmm\ with the $Z$ defined in \eZdo.
The metric for the discretized formulation is not smooth, but one can imagine
how an effective metric on larger scales could arise after averaging over
local irregularities. In the next subsection, we shall see explicitly how this
works.

(Actually \ecmm\ automatically calculates \eZdo\ with the measure factor
in \ediscr\ corrected to $\sum_S {1\over|G(S)|}$, where $|G(S)|$ is the
order of the (discrete) group of symmetries of the triangulation $S$. This
is familiar from field theory where diagrams with symmetry result in an
incomplete cancellation of $1/n!$'s such as in \epel\ and \empel. The
symmetry group $G(S)$ is the discrete analog of the isometry group of a
continuum manifold.)

The graphical expansion of \ecmm\ enumerates graphs as shown in \frandtri,
where the triangular faces that constitute the random triangulation are dual to
the 3-point vertices. Had we instead used 4-point vertices as in \fmapv b, then
the dual surface would have square faces (a ``random squarulation'' of the
surface), and higher point vertices $(g_k/N^{k/2-1})\tr M^k$ in the matrix
model would result in more general ``random polygonulations'' of surfaces. (The
powers of $N$ associated with the couplings are chosen so that the rescaling
$M\to M\sqrt N$ results in an overall factor of $N$ multiplying the action. The
argument leading to \efN\ thus remains valid, and the power of $N$ continues to
measure the Euler character of a surface constructed from arbitrary polygons.)
The different possibilities for generating vertices constitute additional
degrees of freedom that can be realized as the coupling of 2d gravity to
different varieties of matter in the continuum limit.

\subsec{The continuum limit}
\subseclab\sstcl

{}From \efN, it follows that we may expand $Z$ in powers of $N$,
\eqn\elne{Z(g)
=N^2 Z_0(g)+Z_1(g)+N^{-2}Z_2(g) + \cdots=\sum N^{2-2h} Z_h(g)\ ,}
where $Z_h$ gives the contribution from surfaces of genus $h$.
In the conventional large $N$ limit, we take $N\to\infty$ and only $Z_0$,
the planar surface (genus zero) contribution, survives. $Z_0$ itself may be
expanded in a perturbation series in the coupling $g$, and for large order $n$
behaves
as (see \rBIZ\ for a review)
\eqn\eloln{Z_0(g)\sim \sum_n n^{\gamma-3} (g/\gc){}^n\sim (\gc-g)^{2-\gamma}\
.}
These series thus have the property that they diverge as $g$ approaches
some critical coupling $\gc$. We can extract the continuum limit of these
surfaces by tuning $g\to\gc$. This is because the expectation value of the
area of a surface is given by
$$\langle A\rangle=\langle n\rangle
={\del\over\del g}\ln Z_0(g)\sim {1\over g-\gc}$$
(recall that the area is proportional to the number of vertices $n$, which
appears as the power of the coupling in the factor $g^n$ associated to each
graph).
As $g\to\gc$, we see that $A\to\infty$ so that we may rescale the area of
the individual triangles to zero, thus giving a continuum surface with finite
area. Intuitively, by tuning the coupling to the point where the
perturbation series diverges the integral becomes dominated by diagrams
with infinite numbers of vertices, and this is precisely what we need to
define continuum surfaces.
In general, surfaces of large area are connected
with the large order behavior of the Taylor series expansion
in powers of $g$ and therefore to the singularity of $Z_h(g)$ closest to
the origin.

There is no direct proof as yet that this procedure for defining continuum
surfaces is ``correct'', i.e.\ that it coincides with the continuum
definition \eZdo. We are able, however, to compare properties of the
partition function and correlation functions calculated by matrix model
methods with those properties that can be calculated directly in the
continuum (as in the early work of \rKM, and which we shall review in later
sections here).  This gives implicit confirmation that
the matrix model approach is sensible and gives reason to believe other
results derivable by matrix model techniques (e.g.\ for higher genus) that
are not obtainable at all by continuum methods.

One of the properties of these models derivable via the continuum Liouville
approach is a ``critical exponent'' $\gamma_{\rm str}$,
defined in terms of the area dependence of the partition function for surfaces
of fixed large area $A$ as
\eqn\elpoa{Z(A)\sim A^{(\gamma_{\rm str}-2)\chi/2-1}\ .}
(Note that if we consider \elpoa\ restricted to genus zero,
i.e.\ with $\chi=2$, then we
see that $\gamma_{\rm str}$ coincides with $\gamma$ of \eloln.)
To anticipate some relevant results, we recall that the
unitary discrete series of conformal field theories is labelled by an integer
$m\ge2$ and has central charge $D=1-6/m(m+1)$
(for a review, see e.g.\ \rpglh), where the central charge is normalized such
that $D=1$ corresponds to a single free boson. If we couple conformal field
theories with these fractional values of $D$ to 2d gravity, the continuum
Liouville theory prediction for the exponent $\gamma_{\rm str}$ is (see
\refsubsec\ssDw)
\eqn\epbl{\gamma_{\rm str}
={1\over12}\bigl(D-1-\sqrt{(D-1)(D-25)}\,\bigr)=-{1\over m}\ .}
The case $m=2$, for example, corresponds to $D=0$ and hence
$\gamma_{\rm str}=-\ha$ for
pure gravity. The next case $m=3$ corresponds to $D=1/2$, i.e.\ to a
1/2--boson or fermion. This is the conformal field theory of the critical
Ising model, and we learn from \epbl\ that the Ising model coupled to 2d
gravity has $\gamma_{\rm str}=-{1\over3}$. Notice that \epbl\ ceases to be
sensible for
$D>1$. This is the first indication of a ``barrier'' at $D=1$.

In sections \stomm--\sqmomm,
we shall present the solution to the matrix model formulation
of the problem, and the value of the exponent $\gamma_{\rm str}$ provides a
coarse means
of determining which specific continuum model results from taking the
continuum limit of a particular matrix model. Indeed the coincidence of
$\gs$ and other scaling exponents (see sec.~\sqmomm)
calculated from the two points of view were originally the only evidence that
the continuum limit of matrix models was a suitable definition for the
continuum problem of interest (note however a subtlety in the comparison
for non-unitary models). The simplicity of matrix model results for
correlation functions has spurred a rapid evolution of continuum Liouville
technology so that as well many correlation functions can be computed in both
approaches and are found to coincide, as we shall review in sec.~\stcalg.

\subsec{The double scaling limit}
\subseclab\sdsl

Thus far we have discussed the naive $N\to\infty$ limit which retains only
planar surfaces. It turns out that the successive coefficient functions
$Z_h(g)$ in \elne\ as well diverge at the same critical value of the
coupling $g=\gc$ (this should not be surprising since the divergence of the
perturbation series is a local phenomenon and should not depend on global
properties such as the effective genus of a diagram).
As we shall see in the next section, for the higher genus contributions
\eloln\ is generalized to
\eqn\elolnhg{Z_h(g)\sim \sum_n n^{(\gamma_{\rm str}-2)\chi/2-1}
(g/\gc)^n\sim
(\gc-g)^{(2-\gamma_{\rm str})\chi/2}\ .}
We see that the contributions from higher genus ($\chi<0$)
are enhanced as $g\to\gc$.
This suggests that if we take the limits $N\to\infty$ and $g\to\gc$ not
independently, but together in a correlated manner, we may compensate the
large $N$ high genus suppression with a $g\to\gc$ enhancement. This would
result in a coherent contribution from all genus surfaces \refs{\rDS{--}\rGM}.

To see how this works explicitly,
we write the leading singular piece of the $Z_h(g)$ as
$$Z_h(g)\sim f_h(g-\gc)^{(2-\gamma_{\rm str})\chi/2}\ .$$
Then in terms of
\eqn\ekappa{\kappa\inv\equiv N(g-\gc)^{(2-\gamma_{\rm str})/2}\ ,}
the expansion \elne\ can be rewritten\foot{Strictly speaking the first two
terms here have additional non-universal pieces that need to be subtracted
off.}
\eqn\elner{Z=\kappa^{-2}f_0+f_1+\kappa^2 f_2+\cdots
=\sum_h \kappa^{2h-2}\,f_h\ .}
The desired result is thus obtained by taking the limits $N\to\infty$,
$g\to\gc$ while holding fixed the ``renormalized'' string coupling $\kappa$
of \ekappa. This is known as the ``double scaling limit''.

\newsec{The one--matrix model}
\seclab\stomm

In order to justify the claims made in the previous section,
we introduce some formalism to solve the matrix models. Since the integrand
in \ecmm\ depends only on the eigenvalues of the matrix $M$, we can factorize
the integration measure into the product of the Haar measure for unitary
matrices and an integration measure for eigenvalues. The integration over
unitary matrices is
then trivial and we can rewrite the partition function \ecmm\ in the form
\eqn\egpf{\ee{Z}=\int \d M\ \ee{-\tr V(M)}=\int\prod_{i=1}^N
\d \lambda_i\, \Delta^2(\lambda)\ \ee{-\sum_i V(\lambda_i)}\ ,}
where we now allow a general polynomial potential
$V(M)=M^2+\sum_{k\geq 3} \alpha_k M^k$. In \egpf, the
$\lambda_i$'s are the $N$ eigenvalues of the hermitian matrix $M$, and
\eqn\eVand{\Delta(\lambda)=\prod_{i<j}(\lambda_j-\lambda_i)}
is the Vandermonde determinant.\foot{\egpf\ may be derived via the usual
Fadeev-Popov method: Let $U_0$ be the unitary matrix such that
$M=U_0^\dagger \Lambda' U_0$, where $\Lambda'$ is a diagonal matrix with
eigenvalues $\lambda'_i$. The right hand side of \egpf\ follows by
substituting the definition
$1=\int \prod_i\d \lambda_i\,\d U\,\delta(U M U^\dagger-\Lambda)
\,\Delta^2(\lambda)$ (where $\int\d U\equiv1$).
We first perform the integration over $M$, and then $U$ decouples due to
the cyclic invariance of the trace so the integration over $U$ is trivial,
leaving only the integral over the eigenvalues $\lambda_i$ of $\Lambda$.
To determine $\Delta(\lambda)$, we note that only the infinitesimal
neighborhood $U=(1+T)U_0$ contributes to the $U$ integration, so that
$$1=\int \prod_{i=1}^N\d \lambda_i\,\d U\,
\delta^{N^2}\!\bigl(U M U^\dagger-\Lambda\bigr)\,\Delta^2(\lambda)
=\int\d T\ \delta^{N(N-1)}\bigl([T,\Lambda']\bigr)\Delta^2(\lambda')\ .$$
Now $[T,\Lambda']_{ij}=T_{ij}(\lambda'_j-\lambda'_i)$, so \eVand\ follows
(up to a sign) since the integration $\d T$ above is over
real and imaginary parts of the off-diagonal $T_{ij}$'s.}
(In appendix~\smmaj, we give a more formal derivation for the appearance of
the Vandermonde determinant based on the group metric.)
Due to antisymmetry in interchange of any two eigenvalues, \eVand\ can be
written $\Delta(\lambda)=\det\,\lambda^{j-1}_i$
(where the normalization is determined by comparing leading terms).
In the case $N=3$ for example we have
$$(\lambda_3-\lambda_2)(\lambda_2-\lambda_1)(\lambda_3-\lambda_1)=
\det\pmatrix{1&\lambda_1&\lambda_1^2\cr
1&\lambda_2&\lambda_2^2\cr
1&\lambda_3&\lambda_3^2\cr}\ .$$

\subsec{The large $N$ limit: steepest descent}
\subseclab\sSD
\def\zp{s}

The large $N$ limit of the matrix models considered here was originally solved
by saddle point methods in \rBIPZ. For this procedure,
it is convenient to change the normalization in the
integrand of \egpf\ and consider instead
\eqn\eparta{\ee{Z\left(g,\alpha_k,N\right)}
=\int \d M\, \ee{-(N/g)\tr V(M)}=
\int\prod_i\d \lambda_i\, \Delta^2(\Lambda)
\,\ee{-(N/g) \sum_i V (\lambda_i)}\ ,}
where the coupling constant $g$ plays the role of the cosmological constant.
To describe pure gravity, we recall that only one $\alpha_k$ is needed, for
instance one can use only triangles. More general models correspond to
additional degrees of freedom on the surface.

In the conventional large $N$ limit, in which according to \elne\
only surfaces with the topology of the sphere contribute,
the integral \eparta\ can be evaluated by steepest descent.
The Vandermonde determinant leads to a
repulsive force between eigenvalues which otherwise would accumulate at the
minimum of the potential $V$. The saddle point equations that come from
varying a single eigenvalue in \eparta\ are
\eqn\ecol{{2\over N}\sum_{j\ne i}{1\over\lambda_i-\lambda_j}=
{1\over g}V'\left(\lambda_i\right)\ .}

This equation can be solved by the following method: We introduce the trace
of the resolvent of the matrix $M$
\eqn\eresolM{\omega(z)={1\over N}\,\tr{1
\over M-z}={1\over N}\sum_i {1 \over \lambda_i -z}\ .}
Multiplying eq.~\ecol\ by $1/(\lambda_i -z)$ and summing over $i$,
we find
\eqn\eRicat{\omega^2(z)-{1\over N}\omega'(z)+{1 \over g}V'(z)\omega(z)=
-{1 \over N g}\sum_i {V'(z)-V'(\lambda_i) \over z-\lambda_i}\ .}
This equation is analogous to the Riccati form of the Schr\"odinger
equation, the wave function $\psi$ being related to $\omega$
by $N\omega(z)+NV'(z)/(2g)=\psi'/\psi$. The eigenvalues $\lambda_i$
are the zeros of the wave function. In the large $N$ limit, we can neglect
$N^{-1}\omega'(z)$ (this is the WKB approximation, but note that the equation
can more generally be used to study the convergence of the distribution of
zeros toward its limit).

In this limit the distribution of eigenvalues $\rho(\lambda)={1\over N}\sum_i
\delta(\lambda-\lambda_i)$ becomes continuous, and
\eqn\eanal{\omega(z)=\int {\rho(\lambda)\d\lambda\over\lambda-z}\ .}
Note that the normalization condition
$\int\rho(\lambda')\d\lambda'=1$ is the analogue of the Bohr--Sommerfeld
quantization condition. Eq.~\ecol\ can now be rewritten
$$2\int\kern-10.5pt-\kern7pt
{\rho(\lambda')\d\lambda'\over\lambda-\lambda'}=
{1\over g}V'\left(\lambda\right)\ ,$$
or equivalently
\eqn\ecolc{\omega(z+i0)+ \omega(z-i0)=-{1\over g}V'(z)\ .}
Finally, eq.~\eRicat\ becomes
\eqn\ecolom{\omega^2(z)+{1 \over g}V'(z)\,\omega(z)+{1 \over 4g^2}R(z)=0\ ,}
where
\eqn\edefR{R(z)=4g \int\d\lambda\,\rho(\lambda) {V'(z)-V'(\lambda) \over
z-\lambda}\ }
is a polynomial of degree $l-2$ when $V$ is of degree $l$.
Note that the coefficient of highest degree of $R$ is fixed by the
normalization of $\rho(\lambda)$ while the remaining coefficients
depend explicitly on the eigenvalue distribution.

The eigenvalue density $\rho(\lambda)$ is extracted from $\omega(z)$ via
the relation
\eqn\eanalinv{\rho(\lambda)={1\over2i\pi}
\bigl(\omega(z+i0)-\omega(z-i0)\bigr)\ ,}
and from \eparta, we can write the gravity partition function as
\eqn\esaddFa{Z=N^2\left(\int\d\lambda\,\d\mu\,
\rho(\lambda)\rho(\mu)\ln|\lambda-\mu| -{1\over g}\int
\d\lambda\,\rho(\lambda)V(\lambda)\right)\ .}

\medskip
{\it The solution.} The solution to eq.~\ecolom\ is\foot{Recall that this
solution could also be determined indirectly by first solving the homogeneous
equation, i.e.\ with the r.h.s.\ of \ecolc\ set to zero, by looking for a
function that has a cut on the support of $\rho$ and takes opposite values
above and below the cut. Such a function has square root branch points and is
therefore the square root of a polynomial. The inhomogeneous equation has
particular solution $-V'(z)/2 g$, and additional constraints arise from the
condition that $\omega(z)$ behaves like $-1/z$ for $|z|$ large. This
equivalently determines the solution given with the degree $l-2$ polynomial
$R(z)$ having fixed highest degree coefficient.}
\eqn\esigsol{\omega(z)={1 \over 2 g}\bigl(-V'(z)+\sigma(z)\bigr)\ ,}
where $\sigma(z)$, up to the normalization, is the singular part
$\omega_{\rm sing}(z)$ of $\omega(z)$,
\eqn\esigma{\sigma(z)=2g\,\omega_{\rm sing}(z)
=\sqrt{\bigl(V'(z)\bigr)^2-R(z)}\ .}
Generically $\omega(z)$ has $2(l-1)$
branch points corresponding to the roots of the polynomial $V'{}^2-R$.
Therefore the support of $\rho(\lambda)$ is formed of $l-1$ disconnected
pieces. In the simplest case, when
the potential has only one minimum, we expect a single
connected support and thus
only two branch points. It follows that the polynomial $V'{}^2-R$
must have $l-2$ double roots and this yields $l-2$ conditions that fully
determine $R$.

For later purposes, it is convenient to give a more explicit
representation of the one-cut solution. The simplest one-cut
solution of the homogeneous equation \ecolc\ is $\sqrt{(z-a_1)(z-a_2)}$.
Dividing $\omega(z)$ by this function, we can transform \ecolc\ into a
discontinuity equation,
$${\omega(z+i0)\over i\sqrt{(a_2-z)(a_1-z)}}-
{\omega(z-i0)\over -i\sqrt{(a_2-z)(z-a_1)}}
=-{1\over ig\sqrt{(a_2-z)(z-a_1)}}V'(z)\ .$$
It follows that
\eqn\esiggen{\omega(z)={\sqrt{(z-a_1)(z-a_2)}\over 2\pi g}\int^{a_2}_{a_1}
{\d\lambda \over \lambda-z}\,
{V'(\lambda) \over \sqrt{(a_2-\lambda)(\lambda-a_1)}}\ .}

The discontinuity equation defines a solution up to regular terms,
and the large $|z|$ behavior implies their absence. The large $|z|$
behavior yields also two other conditions which determine $a_1$ and
$a_2$: if we expand the r.h.s.\ of \esiggen\ for $|z|$ large we
find first a constant piece which must vanish, so that
\eqna\ecndct
$$\int^{a_2}_{a_1} \d\lambda\,
{V'(\lambda) \over \sqrt{(a_2-\lambda)(\lambda-a_1)}}  =0\ ,\eqno\ecndct a$$
and then we find a term proportional to $1/z$ whose residue is known, so that
$$\int^{a_2}_{a_1}\d\lambda\,
{\lambda V'(\lambda)\over\sqrt{(a_2-\lambda)(\lambda-a_1)}}
=2\pi g\ .\eqno\ecndct b$$

These equations can be written in
another convenient form by replacing the integrals over the cut with
contour integrals,
\eqn\econdcuti{\eqalign{\oint \d\lambda\, {V'(\lambda) \over
\sqrt{(\lambda-a_1)(\lambda-a_2)}}  &=0\, \cr
\oint\d\lambda\,\lambda {V'(\lambda)\over\sqrt{(\lambda-a_1)(\lambda-a_2)}}
 &= 4i\pi g\ .\cr}}
After the change of variable
$$\lambda=z+{1\over 2}\left(a_1+a_2\right)+{\left(a_1-a_2\right)^2 \over
16z}\ ,$$
eqs.~\ecndct{a,b}\ take the form
\eqna\econdcut
$$\eqalignno{&\oint{\d z\over2i\pi}
\,V'\bigl(\lambda(z)\bigr)=g\ ,&\econdcut a\cr
&\oint{\d z \over2i\pi z}\,V'\bigl(\lambda(z)\bigr)= 0\ .&\econdcut b\cr}$$

\medskip
{\it The partition function in the large $N$ limit.}
The partition function can be calculated from eq.~\esaddFa\ but it is more
convenient to work with the derivative
$${\del Z\over \del g}={N\over g^2}\bigl<\tr V(M)\bigr> \sim
{N^2\over g^2}\int\d\lambda\,\rho(\lambda)\,V(\lambda)\ ,$$
differentiated again after multiplying by $g^3$:
\eqn\embgda{{\del \over \del g}\left(g^3{\del Z\over \del g}\right)=N^2
\int\d\lambda\,V(\lambda)\,{\del \bigl(g\rho(\lambda)\bigr)\over \del g}\ .}
This leads us to consider the function
\eqn\eOmi{\Omega(z)={\del\bigl(g\omega(z)\bigr)\over \del g}
={1\over 2}{\del \sigma(z)\over \del g}
=\int{\d \lambda \over \lambda-z}  {\del
\bigl(g\rho(\lambda)\bigr)\over \del g},}
whose real part vanishes according to eq.~\ecolc.
Moreover since for $z$ large, $\omega$ behaves as $-1/z$,
$\Omega(z)$ also behaves  as $-1/z$.  Finally since $\omega(z)$ behaves near
$a_1,a_2$ as $\sqrt{z-a_i}$, its derivative behaves at most as $1/
\sqrt{z-a_i}$.  The unique solution is
\eqn\eOmii{\Omega(z)=-\bigl((z-a_1)(z-a_2)\bigr)^{-1/2}.}

Transforming the integral in \embgda\ into a contour integral,
\eqn\efreeu{{\del \over \del g}\left(g^3{\del Z\over \del g}\right)\equiv
gu(g)=N^2{1\over 2i\pi}
\oint\d\lambda\,{V(\lambda)\over\sqrt{(\lambda-a_1)(\lambda-a_2)}}\ ,}
and differentiating a last time with respect to $g$ gives
$$\eqalign{{\del \bigl(gu(g)\bigr)\over \del g}&={N^2\over 4i\pi}\left({\del
a_1\over \del g}
\oint\d\lambda\,{V(\lambda)\over(\lambda-a_1)^{3/2}(\lambda-a_2)^{1/2}}
\right. \cr
&\qquad\qquad \left. +{\del a_2\over \del g}
\oint\d\lambda\,{V(\lambda)\over(\lambda-a_1)^{1/2}(\lambda-a_2)^{3/2}}
\right)\ .\cr}$$
Integrating by parts in eqs.~\econdcuti, we can
generate the two needed integrals so that finally we have
\eqn\eusing{{\del \bigl(gu(g)\bigr)\over \del g}=2 N^2 g {\del
\ln|a_1-a_2|\over \del g}\ ,}
which we shall use to determine the singular part of the partition function.

\medskip
{\it Even potentials.} For an even potential the cut end-points take equal
and opposite values $\pm a$. Eq.~\econdcut{b}\
is automatically satisfied and \econdcut{a}\ becomes
\eqn\edetasq{g=\oint{\d z \over 2i\pi}V'\left(z+a^2/4z\right)\ .}
For the quartic potential
\eqn\equapot{V(\lambda)={1\over2}\lambda^2+{1\over4}\lambda^4\ ,}
for example, we find\foot{With respect to the
conventions in eqs.~$(17a,b)$ of \rBIPZ, we have $g\to g/4$,
$\lambda^2\to\lambda^2/g$, $a^2\to a^2/4g$, and $\omega=-F$.}
\eqn\erhodist{\omega(z)={1\over 2 g}\Bigl(-z-z^3+\bigl(z^2+1+\half a^2\bigr)
\sqrt{z^2-a^2}\Bigr)\ ,}
with
$$a^2={2\over3}\left(-1+\sqrt{1+12g}\right)\ .$$
{}From \eanalinv, we find that
\eqn\erhoiv{\rho(\lambda)={1\over2\pi g}\bigl(\lambda^2+1+\half a^2\bigr)
\sqrt{a^2-\lambda^2}\ .}

It is important to notice here that the partition function \eparta\
to leading order in
large $N$ (the spherical limit) has an analytic continuation\foot{Due to our
choice of normalization the average matrix is anti-hermitian when $g$ is
negative.}  from $g>0$ to $g<0$ and the first singularity arises at
$g=g\dup_c=-1/12$, at which point
$Z$ has a square root branch point.  The existence of the
$g<0$ region can be
understood as follows: the number of planar diagrams increases only
geometrically, while the barrier penetration effects for $g<0$
responsible for the divergence of perturbation theory behave as $\e^{-K(g)N}$.
The latter are therefore exponentially suppressed for $N$ large
(for details see \refsubsec\ssIalobom).

Note that, as should have been expected, the singularity in $g$ occurs at a
point where a zero of the singular part $\sigma(z)$ of $\omega(z)$ coalesces
with an end-point of the cut. At $g\dup_c$, $\sigma(z)$ becomes
$$\sigma(z)=\left(z^2-a_c^2\right)^{3/2},$$
with $a_c\equiv a(g\dup_c)$.
\medskip
{\it The continuum limit.} We want now to study the singular behavior of
functions near $g\dup_c$ (the continuum limit). Let us blow up the
neighborhood of the cut end-point. Due to the symmetry of the potential both
end-points play a role, for generic potentials only one end-point would be
relevant. We set $x=1-g/g\dup_c$ and $z=a_c(1-\zp)$. Then for $\zp$ and $x$
small with $\zp=O(\sqrt{x})$, $\sigma(z)$ has a scaling form:
$$\sigma(z)\sim (4/3)^{3/2}\left(\zp+{\textstyle{1\over4}}\sqrt{x}\right)
\sqrt{\half\sqrt{x}-\zp}\quad {\rm and}\quad  -a_c\,\omega_{\rm sing}(s)
=2\sqrt{6}\,\sigma(z)\ .$$
The singular part of the partition function can be calculated directly using
equations \efreeu\ and \eusing. Near $g\dup_c$ at leading order, we can
replace the explicit powers of $g$ by $g\dup_c$, and integrate to find
$$g_c^2 Z''\sim 2N^2 \ln|a_1-a_2| +\ \hbox{less singular terms}\ .$$
For the potential \equapot, this gives
%
$$g_c^2 Z''_{\rm sing.}\sim -N^2 x^{1/2}\ .$$
The partition function thus takes the form
\eqn\efreec{Z=-{\textstyle{4\over15}}N^2 x^{5/2}
+\ \hbox{less singular terms}\ ,}
which implies $\gamma_{\rm str}=-1/2$. (Note that the Legendre transform from
the fixed area partition function $Z(A)$ of \elpoa\ gives
$\int \d A\,A^{(\gamma_{\rm str}-2)\chi/2-1}\e^{-Ax}
\sim x^{(2-\gamma_{\rm str})\chi/2}$, i.e.\
for genus zero $Z(x)\sim x^{2-\gamma_{\rm str}}$.)

\subsec{Multicritical points}
\subseclab\ssMp

We have seen that a critical point is the result of a confluence of a
regular zero of the singular part $\sigma$ of $\omega(z)$ with a cut
end-point. By taking potentials of higher degree, which thus depend on more
parameters, we can adjust these parameters in such a way that $m-1$ zeros
of $\sigma$ reach a cut end-point for the same value $g=g\dup_c$. At the
critical point, $\sigma(z)$ will then have the form\foot{in a minimal
realization because it could have additional irrelevant zeros}
\eqn\esigc{\sigma(z)=z^{m-1/2}(z-b)^{1/2}\ .}
We have assumed here that at the critical point the end-point of the cut is at
$z=0$, and that the potential is generic (not even) such that the other
end-point of the cut is $z=b>0$.

Note that the form \esigc\ determines both the
critical potential and $g\dup_c$ since from \esigsol\ we have
\eqn\esVz{\sigma(z)=V'(z)-2g z^{-1}+O\left(z^{-2}\right)\ .}
It follows that
\eqn\epotcrit{V'(z)=\left(z^m(1-b/z)^{1/2}\right)_+\ ,}
where the subscript $+$ means the sum of the terms with non-negative powers
of $z$ in the large $z$ expansion. Equivalently, we can write
$$V'(z)={1\over B(m,1/2)}\left(b^m/m
+ (z-b)^{1/2} \int^z_b \d s\, (z-s)^{-1/2}s^{m-1}\right)\ ,$$
where $B$ is the usual ratio of $\Gamma$ functions
$$B(\alpha,\beta)={\Gamma(\alpha)\,\Gamma(\beta)\over\Gamma(\alpha+\beta)}\ .$$
Note that the minimal potential goes as $z^{m+1}$ for $z$ large,
and therefore the corresponding matrix integral can be defined
only by analytic continuation for $m$ even.
{}From \eqns{\esigc{--}\epotcrit}, we also have
\eqn\egcritm{g\dup_c={b^{m+1}\over 2\pi}B(m+1/2,3/2)\ .}

For $x=1-g/g\dup_c$ small (but not critical), the zeros at $z=0$
and the end-point of the cut, now at $z=a$, will split.
The other end-point (at $z=b$) also moves but it is easy to
verify that this effect is negligible at leading order in $x$. For $a$ and
$z$ small, with $z=O(a)$, $\sigma$ again assumes a scaling form.
For $z$ large, this scaling
form  must match the small $z$ behavior of the critical
form \esigc. Thus we find
$$\sigma_{\rm sc}= b^{1/2}(a-z)^{1/2}a^{m-1} P_{m-1}(z/a)\ ,$$
where $P_m(z)$ is a polynomial such that $P_m \sim z^m$ for $z$ large.
Moreover from \eOmi\ and \eOmii, we infer
\eqn\esigsci{{\del \sigma_{\rm sc}\over \del g}=-2b^{-1/2}(a-z)^{-1/2}\ .}

This implies that all terms with positive powers in the expansion of
$\sigma_{\rm sc}$ for $z$ large must vanish, except for the first term
which is independent of $a$. It follows that
$$P_m(z)=\left(z^m(1-1/z)^{-1/2}\right)_+$$
(where again the subscript means the sum of terms with non-negative powers
in the large $z$ expansion), and the term of order $z^{-1/2}$ in
$\sigma_{\rm sc}$ is proportional to $a^m$.
Comparing with \esigsci, we find that
$\del a^m/\del g$ is a constant. Eq.~\esigsci\ then determines
$\sigma(z)$ up to a multiplicative constant which is fixed by the large $z$
behavior, and we obtain the integral representation
\eqn\esigscii{\sigma_{\rm sc}(z)={b^{1/2}\over B(m,1/2)}\int_z^a\d s\,
(s-z)^{-1/2}s^{m-1}\ .}
Expanding this expression in powers of $z$ and comparing with the leading
term of expression \esigsci, we find
$${\del a^m \over \del g}=-2m b^{-1}B(m,1/2)\ ,$$
and thus the relation between $a$ and $x$ is
\eqn\ecutscale{2(m+1)\left(a \over b\right)^m\equiv x(a)=x\ .}

Combining this relation with eqs.~\eqns{\esigsol{,\ }\egcritm{,\ }\esigscii},
we obtain
a useful representation for the singular part $\omega_{\rm sing}$ of $\omega$
in the scaling limit:
\eqn\eomegsc{\omega_{\rm sing}
=b^{-1/2}\int^a_z \d s\,{\del x\over \del s}\,(s-z)^{-1/2}\ . }
{}From \efreeu\ and \eusing, we then immediately obtain the singular part of
the partition function:
\eqn\eZab{Z''_{\rm sing}(x)\sim -2N^2(a/b)\ ,}
and thus
\eqn\eZmsg{Z_{\rm sing}(x)\sim -N^2{2
m^2 \over (m+1)(2m+1)}\bigl(2(m+1)\bigr)^{-1/m}x^{2+1/m}\ .}
We conclude in particular that the susceptibility exponent
$\gamma_{\rm str}$ takes the value
$$\gamma_{\rm str}=-1/m$$
(see comment following eq.~\efreec\ for the identification of
$\gamma_{\rm str}$).
As anticipated at the end of \refsubsec\ssMm, we see that more general
polynomial matrix interactions provide the necessary degrees of freedom to
result in matter coupled to 2d gravity in the continuum limit.

To compare the result \eZmsg\ for $m=2$ with \efreec, it is necessary to
relate the normalizations of $x$ in both calculations. Note
that $\int\d z\,\omega(z)$ is normalization independent, because
the related quantity $\int\d\lambda\,\rho(\lambda)=1$, so that comparison
between the scaling forms of the singular part of $\omega(z)$ in both
calculations determines the relative normalizations of $x$ and $z$. We
find here for $Z_{\rm sing}$ one half of the result \efreec. This surprising
result has a simple explanation \rbachpet:
For even potentials both cut end-points
contribute to the partition function and this yields an additional factor of 2.
In what follows, we shall largely restrict for reasons of simplicity to even
potentials, so we need to keep this peculiarity in mind.
The relative normalization of $Z_{\rm sing}$ and $\sigma$ will also
be useful when discussing large order behavior of the topological expansion.
Finally we indicate in sec.~\stOnmm\
how the $O(n)$ gas loop on a random surface
\rIK\ can be investigated by a generalization of the method presented here
\rEyZJ.

\medskip
{\it The loop average.} Eventually we shall also consider correlation
functions of quantities of the form $L(s)={1\over N}\tr M^s$, which create in
the surface a loop of length $s$. Here we examine the
behavior of the loop average $\bigl<L(s)\bigr>$, for large loop length $s$,
in the case of a general critical point.
Such an average represents a sum over surfaces with a
loop of length $s$ as boundary.

In the large $N$ limit, we find
$$\bigl<L(s)\bigr>=\int\d\lambda\,\rho(\lambda)\lambda^s\ .$$
Using the analogue of eqs.~\eqns{\embgda{--}\efreeu}, we obtain
$${\del \over \del g}(g\bigl<L(s)\bigr>)={1\over 2i\pi}\oint\d\lambda{\lambda^s
\over\sqrt{(\lambda-a_1)(\lambda-a_2)}}\ .$$
For $s$ large, the integral is dominated by the neighborhood of
$\max\{|a_1|,|a_2|\}$. Let us assume $ |a_2|>|a_1|$. Thus we have
$${\del \over \del g}(g\bigl<L(s)\bigr>)
\mathop{\sim}_{s\to\infty}{1\over\sqrt{1-a_2/a_1}}{1\over \sqrt{\pi s}}
a_2^s\ .$$

We shall now assume that $a_2$ corresponds to the cut end-point where all zeros
of the resolvent coalesce in the critical limit (a condition satisfied by the
minimum potentials \epotcrit\ when $\lambda$ is shifted such that
$V'(\lambda)\sim \lambda$ when $\lambda\to 0$),
and we consider the case of an $m^{\rm th}$ order critical point.
It follows from eq.~\ecutscale\ that
$$a_2-[a_2]_c\propto x^{1/m}\ ,$$
and therefore
$$\ln \bigl<L(s)\bigr>\sim s\ln a_{2c}+\ {\rm const}\cdot s\,x^{1/m}\ .$$
The first term on the r.h.s.\ is a short distance effect and can be
cancelled by a matrix renormalization so that we have
\eqn\elooplaw{\bigl[\ln\bigl<L(s)\bigr>\bigr]_{\rm ren}\propto s\,x^{1/m}\ .}
This behavior shows that the distance should be rescaled
by a factor $x^{1/m}$ in the continuum limit. For $m=2$
(pure gravity), this agrees with the area scaling as $1/x$ as we argued
in \refsubsec\sstcl. For $m\ge3$, however, the result is different, leading
to a difficulty in identifying $x$ with the cosmological constant (which by
definition is coupled to the area). Let us define the length scale by
\elooplaw\ and call $\mu\sim x^{2/m}$ the cosmological constant. If we then
define the
exponent $\gamma$ by the behavior \elpoa\  of the fixed area partition
function $Z(A)$, we find (see comment following eq.~\efreec)
$$Z(\mu)\propto \mu^{2-\gamma}\sim x^{(2-\gamma)2/m}\ ,$$
and thus $\gamma=3/2-m$. We see therefore that there are several ways to
define a string exponent, depending on the reference parameter.
A similar problem will arise in general
for non-unitary models (of which the $m\ge3$ multicritical one matrix models
comprise a particular subclass), as will be discussed in sec.~\sqmomm.

Note finally that the normalized average
${1\over N}\langle\tr\,\e^{sM}\rangle$,
which yields a weighted superposition of loops of different length, has the
same behavior  in the continuum as $L(s)$. It is often used instead of $L(s)$
because it has simpler algebraic properties. It is in particular related
to the trace of the resolvent $\omega(z)$ by
\eqn\eWloop{\bigl<W(s)\bigr>
\equiv{1\over N}\bigl<\tr\,\e^{sM}\bigr>
=-{1\over 2i\pi}\oint\d z\,\e^{sz}\omega(z)\ .}

\subsec{The method of orthogonal polynomials}
\subseclab\sorth

The steepest descent method allows a general discussion of the large $N$
limit.
It is difficult however to calculate the subleading orders in the $1/N$
expansion and therefore to discuss perturbation theory to all orders. We now
present another method that allows us to recover previous results and to
extend them to all orders in $1/N$.

This alternative method for solving \egpf\ makes use of an infinite set of
polynomials $P_n(\lambda)$, orthogonal with respect to the measure
\eqn\eopoly{\int_{-\infty}^\infty
\d \lambda\ \e^{-V(\lambda)}\,P_n(\lambda)\,P_m(\lambda)
=s_n\,\delta_{nm}\ .}
The $P_n$'s are known as orthogonal polynomials and are functions of a single
real variable $\lambda$. Their normalization is given by having leading term
$P_n(\lambda)=\lambda^n+\ldots$, hence the constant $s_n$ on the r.h.s.\ of
\eopoly. Due to the relation
\eqn\evand{\Delta(\lambda)=\det\,\lambda^{j-1}_i=
\det\,P_{j-1}(\lambda_i)}
satisfied by the Vandermonde determinant \eVand\
(recall that arbitrary polynomials may be built up by adding
linear combinations of preceding columns, a procedure that leaves the
determinant unchanged),
the polynomials $P_n$ can be employed to solve \egpf. We substitute the
determinant $\det\,P_{j-1}(\lambda_i)=\sum(-1)^{\pi}
\prod_k P_{i_k-1}(\lambda_k)$
for each of the $\Delta(\lambda)$'s in \egpf\
(where the sum is over permutations $i_k$ and $(-1)^{\pi}$ is the parity of
the permutation).  The integrals over individual
$\lambda_i$'s factorize, and due to orthogonality the only contributions are
from terms with all $P_i(\lambda_j)$'s paired. There are $N!$ such terms so
\egpf\ reduces to
\eqn\egrt{\eqalign{\ee{Z}&=
\int \prod_\ell\d \lambda_\ell\,\e^{-V(\lambda_\ell)}
\,\sum_{\pi,\pi'}(-1)^{\pi}(-1)^{\pi'}\prod_k P_{i_k-1}(\lambda_k)
\prod_j P_{i_j-1}(\lambda_j)\cr
&=N!\prod_{i=0}^{N-1}s_i=N!\, s_0^N\,\prod_{k=1}^{N-1}f_k^{N-k}\ ,\cr}}
where we have defined $f_k\equiv s_k/s_{k-1}$. The solution of the original
matrix integral is thus reduced to the problem of determining the
normalizations $s_k$, or equivalently the ratios $f_k$.

In the naive large $N$ limit (the planar limit), the rescaled index $k/N$
becomes a continuous variable $\xi$ that runs from 0 to 1, and $f_k/N$ becomes
a continuous function $f(\xi)$. In this limit, the partition function
(up to an irrelevant additive constant) reduces to a simple one-dimensional
integral:
\eqn\enlnl{{1\over N^2} Z={1\over N}\sum_k(1-k/N)\ln f_k
\sim\int_0^1\d \xi(1-\xi)\ln f(\xi)\ .}

To derive the functional form for $f(\xi)$, we assume for simplicity that the
potential $V(\lambda)$ in \eopoly\ is even.
Since the $P_i$'s from a complete set of basis vectors in the space of
polynomials, it is clear that $\lambda P_n(\lambda)$ must be expressible as a
linear combination of lower $P_i$'s, $\lambda P_n(\lambda)=\sum_{i=0}^{n+1}
a_i\,P_i(\lambda)$ (with $a_i=s_i\inv\int\e^{-V}\lambda P_n\,P_i$). In fact,
the orthogonal polynomials satisfy the simple recursion relation,
\eqn\elpn{\lambda P_n=P_{n+1}+r_n\,P_{n-1}\ ,}
with $r_n$ a scalar coefficient independent of $\lambda$. This is because any
term proportional to $P_n$ in the above vanishes due to the assumption that
the potential is even, $\int\e^{-V}\lambda\,P_n\,P_n=0$. Terms proportional to
$P_i$ for $i<n-1$ also vanish since $\int \e^{-V}P_n\,\lambda\,P_i=0$ (recall
$\lambda P_i$ is a polynomial of order at most ${i+1}$ so is orthogonal to
$P_n$ for $i+1<n$).

By considering the quantity $P_n\lambda P_{n-1}$ with $\lambda$
paired alternately with the preceding or succeeding polynomial, we derive
$$\int \e^{- V} \,P_n\, \lambda\, P_{n-1}=r_n\,s_{n-1}=s_n\ .$$
This shows that the ratio $f_n=s_n/s_{n-1}$ for this simple case\foot{In
other models, e.g.\ multimatrix models, $f_n=s_n/s_{n-1}$ has a more
complicated dependence on recursion coefficients.} is identically the
coefficient defined by \elpn, $f_n=r_n$.
Similarly, since $P'_n=nP_{n-1}+O(\lambda^{n-2})$,
\eqn\enhn{n s_{n-1}=\int \e^{-V}\,P_n'\, P_{n-1}
= -\int P_n{\d \over \d\lambda}\left(\e^{-V}\,P_{n-1}\right)=\int\e^{-V}\,
V'\,P_n\,P_{n-1}\ .}
This is the key relation that will allow us to determine $r_n$.

\subsec{The genus zero partition function revisited}
\subseclab\sstgzpfr

Our intent now is to find an expression for $f_n=r_n$ and substitute into
\enlnl\ to calculate a partition function.
For definiteness, we take as example the potential
\eqn\epex{\eqalign{&V(\lambda)={N\over 2g}\Bigl(\lambda^2+\lambda^4
+b\lambda^6\Bigr)\ ,\cr
\llap{\rm with\ derivative\quad\qquad}
g &V'(\lambda)
=N\left(\lambda+2\lambda^3+3b\lambda^5\right)\ .\cr}}
The right hand side of \enhn\ involves terms of the form
$\int\e^{-V}\, \lambda^{2p-1}\,P_n\,P_{n-1}$. According to \elpn, these may be
visualized as ``walks'' of $2p-1$ steps ($p-1$ steps up and $p$ steps down)
starting at $n$ and ending at $n-1$, where each step down from $m$ to $m-1$
receives a factor of $r_m$ and each step up receives a factor of unity. The
total number of such walks is given by ${2p-1\choose p}$, and each results in
a  final factor of $s_{n-1}$ (from the integral
$\int\e^{-V}\,\,P_{n-1}\,P_{n-1}$) which cancels the
$s_{n-1}$ on the left hand side of \enhn.
For the potential \eqns{\epex{,\ }\enhn} this gives
\eqn\egnpx{{g n\over N}=r_n+2r_n(r_{n+1}+r_n+r_{n-1})+
3 b(10 \ rrr \ {\rm terms})\ .}
(The 10 $rrr$ terms start with $r_n(r_n^2 +r_{n+1}^2+r_{n-1}^2+\ldots)$
and may be found e.g.\ in \rIYL.)

As mentioned before \enlnl, in the large $N$ limit the index $n$ becomes a
continuous variable $\xi$, and we have
$r_n\to r(\xi)$ and $r_{n\pm1}\to r(\xi\pm\varepsilon)$,
where $\varepsilon\equiv 1/N$ (as in \enlnl\ we assume that for $n,N$ large
$r_n$ becomes a smooth function of $n/N$). To leading order in $1/N$, \egnpx\
reduces to
\eqn\egxw{\eqalign{g \xi=r + 6 r^2+30 b r^3&=W(r)\cr
&=\gc+\half W''|_{r=r_c}\bigl(r(\xi)-r_c\bigr)^2+\cdots\ .\cr}}
In the second line, we have expanded $W(r)$ for $r$ near a critical point
$r_c$ at which $W'|_{r=r_c}=0$
(which always exists without any fine tuning of the parameter $b$),
and $\gc\equiv W(r_c)$. We see from \egxw\ that
$$r-r_c\sim(\gc-g \xi)^{1/2}\ .$$
%

To make contact with the 2d gravity ideas of the preceding section, let us
suppose more generally that the leading singular behavior of $f(\xi)$
$\bigl(=r(\xi)\bigr)$ for large $N$ is
\eqn\efx{f(\xi)/f_c-1 \sim K(\gc - g \xi)^{-\gamma}}
for $g$ near some $\gc$ (and $\xi$ near 1). (We shall see that
$\gamma$ in the above coincides with the critical exponent
$\gamma_{\rm str}$ defined in \elpoa.)
The behavior of \enlnl\ for $g$ near $\gc$ is then
\eqn\emc{\lbspace\eqalign{{1\over N^2}Z_{\rm sing.}
&\sim K\int_0^1\d \xi\,(1-\xi)(\gc- g \xi)^{-\gamma} \cr
&\sim-{K\over g(1-\gamma)}(1-\xi)(\gc-g \xi)^{-\gamma+1}\Big|_0^1
+{K\over g^2(1-\gamma)(2-\gamma)}(\gc- g \xi)^{-\gamma+2}\Big|_0^1\cr
&\sim{K\over g_c^2(1-\gamma)(2-\gamma)}(\gc- g)^{-\gamma+2}\sim \sum_n
{K\over g_c^\gamma\,\Gamma(\gamma)} n^{\gamma-3}(g/\gc)^n\ .}}
Comparison with \elpoa\ shows that the large area (large $n$) behavior
identifies the exponent $\gamma$ in \efx\ with the critical exponent
$\gamma_{\rm str}$ defined earlier.
We also note that the second derivative of $Z$ with respect to
$x=\gc-g$ has leading singular behavior
\eqn\elsbz{Z''(g)\sim K \gc{}^{-2}(\gc-g)^{-\gamma_{\rm str}}\sim
\gc{}^{-2}\Bigl(f(\xi=1)/f_c -1\Bigr)\ .}

{}From \efx\ and \emc\ we see that the behavior in \egxw\ implies a critical
exponent $\gamma_{\rm str}=-1/2$. From \epbl, we see that this corresponds
to the case
$D=0$, i.e.\ to pure gravity. It is natural that pure gravity should be
present
for a generic potential. With fine tuning of the parameter
$b$ in \epex, we can achieve a higher order critical point, with
$W'|_{r=r_c}=W''|_{r=r_c}=0$, and hence the r.h.s.\ of \egxw\ would instead
begin with an $(r-r_c)^3$ term. By the same argument starting from \efx, this
would result in a critical exponent $\gamma_{\rm str}=-1/3$.
\medskip
{\it General potential.}
%
%
For a general potential $V(M)=\sum_p v_p M^{2p}$ we find
$W(r)=2\sum_p{(2p-1)!\over ((p-1)!)^2} \, v_p r^p$, or equivalently in
more compact form:
\eqn\eWb{W(r)=\oint{\d z \over 2i\pi}V'\left(z+r/z\right)\ .}
(In the ``stairmaster'' interpretation following
\epex, we see that $z$ corresponds
to stepping up and $z\inv$ to stepping down.
The integral is non-vanishing only when there is an overall factor
$z\inv$, and correctly takes into account a factor of $r$ for each step down.)
Note that when $r$ is identified with $a^2/4$ (where $a$ is
the boundary of the eigenvalue distribution), eqs.~\egxw\ and \edetasq\
become identical (up to the change $g\xi\mapsto g$).
Eq.~\eWb\ also can be inverted to yield
\eqn\eVW{V(M)=\int^1_0{\d s \over s}W\bigl(s(1-s)M^2\bigr). }

With a general potential
$V(M)$ in \egpf, we have enough parameters to achieve an $m^{\rm th}$ order
critical point \rkazcon\ at which the first $m-1$ derivatives of $W(r)$ vanish
at $r=r_c$. The minimal potential has degree $2m$ and corresponds to a
function $W(r)$ of the form
\eqn\eWr{W(r)={g\dup_c\over r_c^m}\bigl(r_c^m-(r_c-r)^m\bigr)\ ,}
and $r(g)$ behaves as
\eqn\ebhav{{r_c-r(g)\over r_c}\sim\left(g\dup_c-g\over g\dup_c\right)^{1/m}.}
{}From \elsbz\ we then find
\eqn\efreesp{Z_{\rm sing}(g)\sim -{N^2 m^2\over(2m+1)(m+1)}
\left({g\dup_c-g \over g\dup_c}\right)^{2+1/m},}
identical, up to normalizations, to \eZmsg. The form \eWr\ for $W(r)$ is
thus associated with the critical exponent $\gamma_{\rm str}=-1/m$.
Expression \efreesp\ suggests the existence of a scaling region in which $N$
becomes large with fixed
\eqn\ezsc{z=(1-g/g\dup_c)N^{2m/(2m+1)}\ .}
\medskip
{\it The minimal critical model.} For the minimal critical model,
i.e.\ the model in which the polynomial $V$ has the smallest degree possible
at a fixed value of $m$, we find from the explicit form of $W$ in \eWr\
and the relation \eVW\ that the critical potentials $V_m(M)$ are
given by\foot{This expression was established by another method in \rNeu.
The $+$ subscript is defined as in \refsubsec\ssMp, meaning
the sum of terms of non-negative powers in the large $M$ expansion.}
\eqn\eVpm{\eqalign{V_m'(M) &=(-1)^{m-1}2m {\gc\over (4r_c)^m} M^{2m-1}
\int_0^1{\d s\over\sqrt{s}} \left(1-s-4r_c M^{-2}\right)^{m-1}\cr
&=(-1)^{m-1}2m {\gc\over(4r_c)^m} B(m,1/2)\left( M^{2m-1}\left(1-4r_c
M^{-2}\right)^{m-1/2}\right)_+\ .\cr}}
This also follows directly from the considerations of
\refsubsec\ssMp, since in the even case $\sigma(z)=(z^2-b^2)^{m-1/2}$.

\eVpm\ also shows that the term of highest degree of $V(M)/\gc$ in the
integral \eparta\ is positive for $m$ odd, and negative for $m$ even.
In the latter case, the minimal potential integrals can thus be defined only
by analytic continuation, as in the case of generic (not even) potentials.
This is the source of important differences between the two parities.

\subsec{The all genus partition function}
\subseclab\sstagpf

We now search for a solution to \egnpx\ that
describes the contribution of all genus surfaces to the partition function
\enlnl.
\medskip
{\it Pure gravity.} Let us for simplicity discuss first pure gravity.
We shall retain higher order terms in $1/N$ in \egnpx\ so that e.g.\
\egxw\ instead reads
\eqn\egxwp{\eqalign{g \xi
&=W(r)+ 2 r(\xi)\bigl(1+15br(\xi)\bigr)
\bigl(r(\xi+\varepsilon)+r(\xi-\varepsilon)-2r(\xi)\bigr) +\cdots \cr
&=\gc+\half W''|_{r=r_c}\bigl(r(\xi)-r_c\bigr)^2
+ 2r_c(1+15br_c)
\bigl(r(\xi+\varepsilon)+r(\xi-\varepsilon)-2r(\xi)\bigr)+\cdots\,  ,\cr}}
where the dots mean terms of order $(r-r_c)^3$,
$(r(\xi+\varepsilon)+r(\xi-\varepsilon)-2r(\xi)\bigr)(r-r_c)$ or terms which
approach $\varepsilon^4 r''''(\xi)$ in the small $\varepsilon$ limit.
We shall eventually verify that the omitted
terms are negligible.

As suggested at the end of \refsubsec\sdsl, we shall simultaneously let
$N\to\infty$ and $g\to\gc$ in a particular way.
Since $g-\gc$ has dimension [length]$^2$, it is convenient to introduce a
parameter $a$ with dimension length and let
\eqn\edefz{\gc-g=\gc a^2 z\ ,}
with $a\to0$. Our ansatz for a coherent large $N$ limit will be to take
$$\varepsilon\equiv 1/N=a^{5/2}\ ,$$
so that the quantity $\kappa\inv\equiv z^{5/4}=(g-\gc)^{5/4}N$ remains finite
as $g\to\gc$ and $N\to\infty$.

Moreover since the integral \enlnl\ is dominated by
$\xi$ near 1 in this limit, it is convenient to change variables from $\xi$ to
$x$, defined by
\eqn\edefx{\gc-g\xi=\gc\, a^2 x\ .}
Our scaling ansatz in this region is
\eqn\erescl{r(\xi)=r_c\bigl(1-a u(x)\bigr)\ .}
If we substitute these definitions into \egxw, the leading terms are of order
$a^2$ and result in the relation $u^2\sim x$.

To include the higher derivative terms, we note that
$$\eqalign{r(\xi+\varepsilon)+r(\xi-\varepsilon)-2r(\xi)&
= \varepsilon^2{\del^2 r\over \del \xi^2}+O\left(\varepsilon^4
{\del^4 r\over \del \xi^4}\right)\cr
&=-r_c a {\del^2\over \del x^2} a
u(x)+O\left(a^3\right) \mathop{\sim}_{a\to0}-r_c a^2 u''\ ,\cr}$$
where we have used
\eqn\edimder{\varepsilon(\del/\del \xi)=-(g/\gc) a^{1/2}(\del/\del x)
\sim -a^{1/2}(\del/\del x)\ ,}
which follows from the above change of variables from $\xi$ to $x$
and $g-\gc=O(a^2)$.
Substituting into \egxwp, the vanishing of the coefficient of $a^2$ implies
the differential equation
\eqn\eplv{ K x=u^2- \frac{1}{3} u''\qquad{\rm with}\quad
K^{-1}=\left(6r_c^2+90b r_c^3 \right)/\gc\,   .}
In the ordinary large $N$ limit, we have calculated the partition function in
eq.~\enlnl. It is however easy to verify that in the double scaling limit
the corrections to \enlnl\ are of order $\bigl(N(\gc-g)\bigr)^{-2}\propto
a^2$, and thus remain negligible.
After the changes of variables \eqns{\edefx{,\ }\edefz} and
with the definition \erescl, we then find
\eqn\elsbzsc{Z(z)\sim -\int_{a^{-2}}^z \d x\,(z-x)u(x)\ \Rightarrow\
Z''(z)=-u(z)\ ,}
generalizing \elsbz. In the double scaling limit, the
second derivative of the partition function
(the ``specific heat'') has leading singular behavior given by
$u(z)$ for $z=(1-g/\gc) N^{4/5}$.

The solution to \eplv\ characterizes the behavior of the partition function of
pure gravity to all orders in the genus expansion. (Notice that the leading
term is $u\sim z^{1/2}$, so after two integrations the leading term in $Z$ is
$z^{5/2}=\kappa^{-2}$, consistent with \elner.) Finally we note that the
change of normalization
\eqn\erescale{z\mapsto \rho z,\quad u(z)\mapsto \rho^2 u(\rho z)\ ,}
does not affect \elsbzsc\ but allows us to rewrite eq.~\eplv\ as
\eqn\eqpla{z=u^2 -{\textstyle{1\over3}}u''\ .}
The property that the r.h.s.\ of the equation is invariant is
directly related to the property that  $r-r_c=O(a)$ and
$\varepsilon(\del/\del \xi)= O(a^{1/2})$
(eqs.~\eqns{\erescl{,\ }\edimder}), which was essential in
selecting terms of the same order. This property will generalize to
higher order critical points.

\medskip
{\it The Painlev\'e I equation.}
Eq.~\eqpla\ is known in the mathematical literature as the
Painlev\'e I equation. One characteristic property is
that its only moveable singularities (in the complex plane) are double poles
\rBout. In the normalization
\eqpla\ they have residue 2, and, as eq.~\elsbzsc\ shows, correspond to
double zeros of $\exp\,Z$. Since it is a second order  differential
equation, its solutions are determined by two boundary conditions. We
are interested only in solutions that have an asymptotic expansion for $z$
large (the topological expansion) that begins with the
leading spherical result $u(z) \sim \sqrt{z}$.
The perturbative solution in powers of $z^{-5/2}=\kappa^2$ is then determined
and takes the form
\eqn\efasexp{u(z)=z^{1/2}(1-\sum_{k=1}u_k z^{-5k/2})\ ,}
where the $u_k$ are all positive.\foot{The first term, i.e.\ the
contribution from the sphere, is dominated by a regular part which has
opposite sign.  This is removed by taking an additional derivative of $u$,
giving a series all of whose terms have the same sign ---
negative in the conventions of \eqpla. The other solution, with leading term
$-z^{1/2}$, has an expansion with alternating sign
which is presumably Borel summable, but not physically relevant.}
This verifies for this model the claims made in eqs.~\eqns{\elolnhg{--}\elner}
of \refsubsec\sdsl.
For large $k$, the $u_k$ go asymptotically as $(2k)!$ (for details see
\refsubsec\ssPG), so the solution for
$u(z)$ is not Borel summable and thus does not define a unique function.
Our arguments in sec.~\sDsmmcl\ show only that the
matrix model results should agree with 2d gravity order by order in
perturbation theory. How to ensure that we are studying nonperturbative
gravity as opposed to nonperturbative matrix models is still an open question.
Some of the constraints that the solution to \eqpla\ should satisfy are
reviewed in \rfrdc. In particular it is known that real solutions to \eqpla\
cannot satisfy the Schwinger--Dyson (loop) equations for the theory.

\medskip
{\it Higher order critical points.}
In the case of the next higher multicritical point, with $b$ in \egxw\
adjusted so that $W'=W''=0$ at $r=r_c$, we have
$W(r)\sim \gc+{1\over6}W'''|_{r=r_c}(r-r_c)^3+\cdots$
and critical exponent $\gamma_{\rm str}=-1/3$.
In general, we take $g-\gc=\kappa^{2/(\gamma_{\rm str}-2)}a^2$, and
$\varepsilon=1/N=a^{2-\gamma_{\rm str}}$ so that the combination
$(g-\gc)^{1-\gamma_{\rm str}/2}N=\kappa\inv$ is fixed in the limit $a\to0$.
The value  $\xi=1$ now corresponds to $z=\kappa^{2/(\gamma_{\rm str}-2)}$,
so the string coupling $\kappa^2=z^{\gamma_{\rm str}-2}$.
The general scaling ansatz is
$r(\xi)=r_c(1-a^{-2 \gamma_{\rm str}}u(x))$, and the change of variables
from $\xi$ to $x$ gives
$\varepsilon(\del/\del \xi)\sim-  a^{-\gamma_{\rm str}}(\del/\del x)$.

For the case $\gamma_{\rm str}=-1/3$, this means in particular that
$r(\xi)=r_c(1-a^{2/3}u(x))$, $\kappa^2=z^{-7/3}$, and
$\varepsilon(\del/\del \xi)\sim-a^{1/3}{\del\over \del x}$.
Substituting into the large $N$ limit of \egnpx\ gives
(again after suitable rescaling of $u$ and $z$)
\eqn\etmcp{z=u^3-u u'' -\half(u')^2+\alpha\, u'''' ,}
with $\alpha={1\over10}$. Note again that since
$|r-r_c|^{1/2}=O\bigl(a^{-\gamma_{\rm str}}\bigr)
=O(\varepsilon(\del/\del\xi))$,
the r.h.s.\ of \etmcp\ is invariant under the
transformation \erescale. The solution to \etmcp\ takes the form
$u=z^{1/3}(1+\sum_k u_k\,z^{-7k/3})$. It turns out that the coefficients $u_k$
in the perturbative expansion of the solution to \etmcp\ are positive definite
only for $\alpha<{1\over12}$, so the $3^{\rm th}$
order multicritical point does not
describe a unitary theory of matter coupled to gravity. Although
from \epbl\ we see that
the critical exponent $\gamma_{\rm str}=-1/3$ coincides with that predicted
for the
(unitary) Ising model coupled to gravity,
it turns out \refs{\rIYL,\rising}\ that \etmcp\ with $\alpha={1\over10}$
instead describes the conformal field theory of the Yang--Lee edge singularity
(a critical point obtained by coupling the Ising model to a particular value
of imaginary magnetic field) coupled to gravity. The specific heat of the
conventional critical Ising model coupled to gravity turns out (see
\refsubsec\ssesgkf) to be as well determined by the differential equation
\etmcp, but instead with $\alpha={2\over27}$.

For the general $m^{\rm th}$ order critical point of the potential $W(r)$, we
have seen that the associated model of matter coupled to gravity has critical
exponent $\gamma_{\rm str}=-1/m$.
With scaling ansatz $r(\xi)/r_c=1-a^{2/m}u(x)$, we find leading behavior
$u(z)\sim z^{1/m}$ (and $Z\sim z^{2+1/m}=\kappa^{-2}$ as expected).
The differential equation that results from substituting the double scaling
behaviors given before \etmcp\ into the generalized version of \egnpx\ turns
out to be the $m^{\rm th}$ member of the KdV hierarchy of differential
equations (of which Painlev\'e I results for $m=2$). In the next section, we
shall provide some marginal insight into why this structure emerges.

In the nomenclature of \rBPZ, so-called ``minimal conformal field theories''
(those with a finite number of primary fields) are specified by a pair of
relatively prime integers $(p,q)$
and have central charge $D=c_{p,q}=1-6(p-q)^2/pq$.
(The unitary discrete series is the subset
specified by $(p,q)=(m+1,m)$.)
After coupling to gravity, these have critical exponent
$\gamma_{\rm str}=-2/(p+q-1)$, as calculated in the matrix model.
In general, the $m^{\rm th}$ order multicritical point of the one-matrix model
turns out to describe the $(2m-1,2)$ model (in general non-unitary) coupled to
gravity, so its critical exponent $\gamma_{\rm str}=-1/m$ happens to
coincide with that
of the $m^{\rm th}$ member of the unitary discrete series coupled to gravity.
The remaining $(p,q)$ models coupled to gravity can be realized in terms of
multi-matrix models (to be defined in sec.~\sqmomm).
We shall see that the interpretation of $\gamma_{\rm str}$ is slightly subtler
in the non-unitary case since the identity is no longer the lowest dimension
operator.

\subsec{Recursion formulae more generally}

In \refsubsec\sorth, we introduced orthogonal polynomials and used
them to calculate the matrix integral. We now introduce some additional
formalism related to this polynomials which will prove useful in what follows.
As an immediate application, it will allow us to discuss the
problem of general (not necessarily even) potentials.

Since the polynomials $P_n(\lambda)$ of \eopoly\ form a basis, we can write
(in matrix notation)
\eqn\eAB{P'=AP\ ,\qquad \lambda P=BP\ .}
In explicit component form, these relations are equivalently written
\eqna\erecur
$$\eqalignno{P'_n& =\sum_{m=0}^{n-1}A_{nm}P_m\ , & \erecur a \cr
\lambda P_n &= \sum_{m=0}^{n+1} B_{nm}P_m\ .&\erecur b\cr}$$
The normalization of $P_n$ (eq.~\eopoly) gives
$$A_{n,n-1}=n\ ,\qquad B_{n,n+1}=1\ .$$

Note that the matrices $A,B$ of \eAB\ necessarily form a representation of
the canonical commutation relations
$$[B,A]=1\ .$$
They can be related to the polynomial $V(\lambda)$ by the relations
\eqn\ereca{\int\d\mu(\lambda)\,\lambda\, P_m P_n= B_{nm} s_m =  s_n B_{mn}}
and
\eqn\erecb{0=\int\d\lambda\, {\d\over\d\lambda}\left(P_n P_m \,
\e^{-NV(\lambda)/g)}\right)
= A_{nm} s_m + s_n A_{mn} -{N \over g} V'_{nm} s_m\ ,}
where we employ the notation $V'_{nm}=[V'(B)]_{nm}$ and
$\d\mu(\lambda)\equiv\d \lambda\,\e^{-NV(\lambda)/g}$.
Introducing the matrix $S$ with matrix elements $S_{nm}=s_m\delta_{nm}$,
we can rewrite \ereca\ and \erecb\ in matrix form,
\eqn\eABmf{\eqalign{BS&=S B^T, \cr AS + SA^T&=(N/g)V'(B)S\ .\cr}}

Eq.~\ereca, together with the defining relation \erecur b,
shows that $B_{mn}$ is different from 0 only if $|m-n|\leq 1$.
If we no longer assume that the function $V(\lambda)$ is even, the recursion
relation \erecur b\ between orthogonal polynomials becomes
\eqn\erecurd{\lambda P_n =P_{n+1} + \tilde r_n P_n + r_n P_{n-1}\ .}
where $r_n, \tilde r_n$ are a short-hand notation for $B_{n,n-1},B_{n,n}$
respectively.

Specializing \ereca\ to $m=n-1$, we recover
$$\int\d\mu\, \lambda P_n P_{n-1}=s_n= r_n s_{n-1}\ ,$$
and consequently
\eqn\enorbn{r_n={s_n \over s_{n-1}}\ .}
Substituting into \egrt, we find
\eqn\efreen{Z=\ln\left(s_0^N N!\right)+\sum_{n=1}^{N-1}(N-n)\ln r_n\ .}

Specializing also \erecb\ to $m=n-1$, we recover \enhn:
\eqn\eqfund{g {n\over N} = V'_{n,n-1}\ .}
As we have already explained, since in the case of even potentials $V'_{mn}$
can be expressed entirely in terms of $r_n$, \eqfund\ leads to a recursion
relation for the coefficient $r_n$. For general potentials, to determine both
$r_n$ and $\tilde r_n$, we need instead another equation, in addition to
\eqfund, obtained by specializing \erecb\ to $m=n$,
\eqn\eqfundb{V'_{n,n}=0\ .}

\medskip
{\it Example.} Consider the potential
$$V(\lambda)=-\lambda+\lambda^3/3$$
(equivalent after translation to $\lambda^2+\lambda^3/3$).
The two equations \eqfund\ and \eqfundb\ in this case are
$$\eqalign{gn/N&= r_n\bigl(\tilde r_n+\tilde r_{n-1}\bigr)\ ,\cr
1&=\tilde r_n^2+r_n+r_{n+1}\ ,\cr} $$
with critical values
$$r_c=1/3\ ,\quad \tilde r_c=\pm\sqrt{r_c}\ ,\quad g\dup_c=2\tilde r_c^3\ .$$
The sign of $\tilde r_c$  is irrelevant (see however the remark at the end of
\refsubsec\sstlnl). Setting as in \erescl,
$r(g)/r_c=1-N^{-2/5}u\bigl(N^{4/5}(1-g/g\dup_c)\bigr)$,
we obtain the equation
\eqn\eqplab{{\textstyle{3\over 2}}u^2-{\textstyle{1\over 4}}u''=z\ .}

We see that eqs.~\eqpla\ and \eqplab\ are identical up to a rescaling of
the function $u$ and variable $z$.
Moreover the normalization of the partition function depends
only on the ratio of the coefficient of $u''$ to the coefficient of $u^2$.
We note here that this ratio is $1/6$ instead of $1/3$ in \eqpla.
The partition function for a potential of the form $-\lambda+\lambda^3$
from eq.~\elsbzsc\ is half of the previous partition function, and the double
poles of the solution of eq.~\eqplab\ have now residue equal to 1.  In a
direct saddle point technique, the interpretation of this result is that the
partition  function receives two identical contributions when the potential is
even.

\medskip
{\it General potentials in the spherical limit.} To leading order in large
$N$, eqs.~\eqfund\ and \eqfundb\ reduce to
\eqna\eqbbt
$$\eqalignno{g&=W(r,\tilde r)\equiv\oint{\d z \over 2i\pi}V'
\bigl(z+r/z+\tilde r\bigr)\ , &\eqbbt a\cr
0&=X(r,\tilde r)\equiv \oint{\d z \over 2i\pi z}
V'\bigl(z+r/z+\tilde r\bigr)\ . &\eqbbt b\cr}$$
We recognize eqs.~\econdcut{a,b}, with $r$ and $\tilde r$ related to the
end-points $a_1,a_2$ of the support of the matrix eigenvalue distribution by
$$\tilde r=(a_1+a_2)/2\ ,\quad r=(a_1-a_2)^2/16\ .$$

The criticality condition for eqs.~\eqbbt{a,b}\ reads
\eqn\eCc{{\del W\over\del r} {\del X\over\del \tilde r} -
{\del W\over\del \tilde r}{\del X\over\del r}=0\ .}
Note that
$${\del W\over\del r}={\del X\over\del \tilde r}\ ,$$
and
$${\del X\over\del r}=\oint{\d z \over 2i\pi z^2}
V''\left(z+r/z+\tilde r\right)=\ (z\mapsto {r\over z})\
{1\over r}{\del W\over\del \tilde r}\ ,$$
so that the criticality condition \eCc\ can be rewritten
$$r_c \left({\del X\over\del r_c}\right)^2 -
\left( {\del X\over\del \tilde r_c}\right)^2=0\ .$$
For $g$ close to $g\dup_c$, we parametrize $r$ and $\tilde r$ as
$$r=r_c(1-u),\quad \tilde r=\tilde r_c - \sqrt{r_c}\tilde u\ .$$
Eq.~\eqbbt b\ then implies
$$0=r_c\, u {\del X\over\del r_c}+\sqrt{r_c}\, \tilde u
{\del X\over\del \tilde r_c}\ ,$$
and thus
\eqn\erelimp{\tilde u=\pm u\ .}
This relation is indeed satisfied in the example considered above
(leading to \eqplab). 
Relation \erelimp\ will be useful in next section for the general
analysis of the one-matrix problem.

\subsec{Correlation functions}

We can also consider correlation functions of the form
$$\bigl< \tr F_1(M)\tr F_2(M)\cdots\bigr> \ .$$
For the simplest examples, we find
\eqna\ecorra
$$\eqalignno{\bigl<\tr F_1(M) \bigr>&=\sum_{n=0}^{N-1}
\bigl[F_1(B)\bigr]_{nn},&\ecorra a\cr
\bigl<\tr F_1(M)\tr F_2(M) \bigr>&=\sum_{n=0}^{N-1}\sum_{p=0}^{N-1}
\bigl[F_1(B)\bigr]_{nn}\bigl[F_2(B)\bigr]_{pp}&\cr
&\quad +\sum_{0\le n\le N-1 <p}
\bigl[F_1(B)\bigr]_{np}\bigl[F_2(B)\bigr]_{pn}\ .&\ecorra b\cr}$$
%
More generally, connected correlation functions may be
interpreted as averages of one-body operators in a free $N$-fermion state
\rbdss\ (for a review, see \rGiMo).

In the large $N$ limit, the sums are replaced by integrals. For example
$$\bigl<\tr F_1(M) \bigr>\sim N\int^{1}_{0}\d \xi\,
\bigl<\xi\bigl| F_1(B) \bigr|\xi \bigr>, \quad \xi=n/N\ .$$
In the scaling limit, the singularities come from the neighborhood
of $\xi=1$.
In terms of $xa^2=1-g/g\dup_c$ and $ya^2=1-g\xi/g_c$ ($1/N=a^{2+1/m}$),
the above trace involves an integral of the form
\eqn\etraceF{\bigl<\tr F_1(M) \bigr>\sim a^{-1/m} \int_{x}^{a^{-2}}\d y\,
\bigl<y| F_1(B) |y\bigr>\ .}
The upper  bound of the integral goes to $\infty$ in the large $N$ (double)
scaling limit.

In the large $N$ limit, away from the scaling region, the average of  a
product of traces is always dominated by the product of their  averages
because, as can be seen from the example \ecorra b,  the additional terms
are subleading by a factor $1/N^2$ due to the ``locality'' of $B$ and the
restrictions on the summation indices.

\subsec{Loop equations, Virasoro constraints}
\subseclab\ssLeVc

The loop equations \refs{\rkazcon,\rFDi,\rFDii}\
are obtained by performing the change of
variables $M\mapsto M+\varepsilon M^k$ in the matrix integral \eparta.
We find the identity
\eqn\eqmota{\sum_{l=0}^{k-1}\left< \tr M^l\, \tr M^{k-l-1} \right>
= (N/g)\left<\tr M^k V'(M) \right>\ ,}
where $\langle\cdots\rangle$ means
average with respect to the matrix integral.

It is convenient to rewrite these equations in terms of the operator
$G(z)=(M-z)\inv$. Multiplying eq.~\eqmota\ by $z^{-k-1}$ and summing on
$k$ gives
\eqn\eamz{{N \over g}\bigl<\tr G(z)V'(M)\bigr>
=-\bigl<\bigl(\tr G(z)\bigr)^2\bigr>\ .}
This expression can be further simplified by rewriting
$V'(M)=V'(z)+\bigl(V'(M)-V'(z)\bigr)$ and noting that
$\tr\bigl(V'(M)-V'(z)\bigr) G(z)$ is a polynomial in $z$.
Setting
$$R(z)={4g\over N}\left<\tr \bigl(V'(M)-V'(z)\bigr) G(z)\right>,$$
we obtain an equivalent form of \eamz
\eqn\eloopiv{{N \over g}V'(z) \bigl<\tr G(z)\bigr>+{N^2\over 4g^2}R(z)
=-\bigl<\bigl(\tr G(z)\bigr)^2\bigr>\ .}
Note that the loop equations, once expressed in terms of the matrix $B$, can
in principle be derived directly from eqs.~\eABmf, but the
derivation is not so straightforward. Introducing the loop average $L(s)$
(eq.~\eWloop), we can also write eq.~\eloopiv\ as
$$\int^s_0 \d t\,\bigl<W(t)\,W(s-t)\bigr>
={N\over g}V'(\d /\d s)\bigl<W(s)\bigr>\ .$$
The l.h.s.\ can be interpreted as the operation of gluing two boundaries
together.
\medskip
{\it Virasoro constraints.} Let us set $g=1$ for convenience here and
parametrize the potential $V(M)$ as
$$V(M)=\sum_{k=0}t_k\, M^k\ .$$
Calling the matrix integral $\vartheta(t)=\e^{Z}$, we can
rewrite the set of loop equations \eqmota\ in the form of a set of Virasoro
constraints (see e.g.\ \rdvvvir)
$${\cal L}_k \vartheta(t)=0\ , \qquad k \ge -1\ ,$$
with
$${\cal L}_k={1\over N^2}\sum_{m+n=k}{\del^2 \over \del t_m \del t_n}+
\sum_{m=0} m t_m {\del\over t_{m+k}}\ .$$
With respect to the trivial variable $t_0$, we have the additional
equation
$$\quad {\del \vartheta \over \del t_0} =-N^2 \vartheta\ .$$

\medskip
{\it Large $N$ limit.} In the large $N$ limit, \eloopiv\ can be solved by
noting from expression \ecorra b\ that the average of the product of traces
becomes asymptotically equal to product of averages. Setting
$$\omega(z)={1\over N}\bigl<\tr G(z)\bigr>\ ,$$
we find
$$\omega^2(z)+{1\over g}V'(z)\,\omega(z) + {1\over 4g^2}R(z)=0\ .$$
We recognize the form \ecolom\ of the saddle point equations.

\def\CP{{\cal O}_{(0)}}
\newsec{A general method: the canonical commutation relations}
\seclab\sAgm

It is possible
to study the recursion formulae \eqfund\ for general polynomials $V(\lambda)$.
In the continuum limit one finds a non-linear differential equation of more
general type for a scaling function $u$. A simpler algebraic method
has been found, however,
which easily generalizes to the several matrix problem.

It is convenient to introduce normalized orthogonal polynomials $\Pi_n$,
$$P_n=\sqrt{s_n}\Pit_n\ ,$$
satisfying
\eqn\eorthrl{\int\d\mu(\lambda)\,\Pi_m \Pi_n=\delta_{mn}\ .}
We now redefine matrices $A$ and $B$ in terms of the $\Pi_n$,
\eqn\erecurb{\Pi'_n=\sum_{m=0}^{n-1}A_{nm}\Pi_m,\quad
\lambda \Pi_n = \sum_{m=0}^{n+1} B_{nm}\Pi_m\ .}
With this new definition, the matrix $B$ is symmetric. In the even potential
case, in terms of the coefficients $r_n$ introduced in eq.~\erecurd\ the
recursion formula for the orthogonal polynomials becomes
\eqn\erfop{\lambda \Pi_n=\sqrt{r_{n+1}}\Pit_{n+1}+\sqrt{r_{n}}\Pit_{n-1}\ .}

Instead of as in \eABmf, the equation for $A$ now reads
\eqn\eAAd{A+A^T={N\over g}V'(B)\ ,}
while the commutator relation remains $[B,A]=1$.
It is convenient to shift $A$ and introduce the matrix
\eqn\esA{C\equiv A-{N \over2g}V'(B)=\ud\left(A- A^{T} \right)}
representing the operator $\d/\d\lambda$ acting on the orthogonal functions
$\e^{-NV(\lambda)/2g}\,\Pi_n$.
Then $C$ is antisymmetric and satisfies the same commutation relation as $A$,
\eqn\ecombc{[B,C]=1\ .}

\medskip
{\it A basic property.} A remarkable property can be proven (for details see
sec.~\sMccrda): Let $B$ be a Jacobi matrix $B$ (a symmetric matrix with
$B_{mn}=0$
for $|m-n|>1$) that satisfies the  commutation relation \ecombc, with $C$ an
antisymmetric local matrix (i.e.\ with $C_{mn}=0$ for $|m-n|> l$). Then
there exists a lower triangular matrix $A$ ($A_{mn}=0$ for $n\ge m$) and a
polynomial $V(\lambda)$ of degree $l+1$ such that
$${N \over g}V'(B)=A+A^T\ ,\qquad C=\half(A-A^T)\ \Rightarrow\ [A,B]=1\ .$$
The diagonal matrix elements of the last equation yield an equation equivalent
to the difference between \eqfund\ taken for two consecutive values of $n$.

Since the coefficients in $V$ are implicitly determined by the criticality
conditions, the original problem can thus be entirely reformulated in terms
of the  matrix $B$ satisfying the commutation relation \ecombc, from which the
singular part of the partition function $Z$ can be calculated. We now proceed
to take the large $N$ and scaling limits directly in these expressions.

\subsec{The large $N$ limit}
\subseclab\sstlnl

We now show that in the double scaling limit $B,C$ become differential
operators and that the commutation relation \ecombc\ determines the string
equation.

In the limit of large $N$ (and thus $n$), $\xi=n/N$ can be treated as a
continuous variable and then the matrix $B$ can be expanded for $r_n$ near
$r_c$,
$$\eqalign{(B\Pi)_n&=\sqrt{r_{n+1}}\Pit_{n+1}+\sqrt{r_{n}}\Pit_{n-1}\ ,\cr
&=\sqrt{r_c}\left(2\Pi_n+{r_n-r_c \over r_c}\Pi_n+\varepsilon^2{\del^2\Pi_n
\over(\del \xi)^2}\right)+O\left((r_n-r_c)^2\Pi_n,\varepsilon^{4}{\del^4\Pi_n
\over(\del \xi)^4}\right) ,\cr}$$
with $\varepsilon=1/N=a^{2-\gs}$. As we have discussed in \refsubsec\sstagpf,
for a general critical point we take $\gc-g=\gc z a^2$
so that the combination $(1-g/\gc)N^{2/(2-\gs)}=z$ is fixed in the
limit $a\to0$. The change of variables from $\xi$ to $x=a^{-2}(1-g\xi/\gc)$
gives $\varepsilon(\del/\del \xi)\sim- a^{-\gs }(\del/\del x)$.

The general scaling ansatz is then
$$r(\xi)=r_c\bigl(1-a^{-2\gs}u(x)\bigr)\ .$$
The leading term in the expansion of $B$, proportional to the identity, does
not contribute to the commutation relation \ecombc. The leading corrections
are of order $a^{-2\gs}$. Only two terms contribute which
together form  a second order differential operator $Q$:
\eqn\erelBQ{\bigl[(r_c^{-1/2}B-2)\,\Pi\bigr]_n\sim a^{-2\gs}\, Q\, \Pi_n\ ,}
with
\eqn\eQt{Q=\d^2 -u(x)}
(where $\d$ is a notation for $\d/\d x$). Note that the rescaling \erescale,
$$x\mapsto \rho x,\quad u(x)\mapsto \rho^2 u(\rho x)\ ,$$
transforms $Q$ into $\rho^2 Q$. The formal hermiticity of the
operator $Q$ follows directly from the symmetry of $B$.

\medskip
{\it General potentials.} The recursion relations between orthogonal
polynomials, generalizing \erfop, are
$$\lambda \Pi_n=\sqrt{r_{n+1}}\Pit_{n+1}+\tilde r_n \Pit_n
+\sqrt{r_{n}}\Pit_{n-1}\ .$$
In the large $N$ scaling limit, the matrix $B$ now becomes the differential
operator
$$(B\Pi)_n = \left(2\sqrt{r_c} + \tilde r_c\right) \Pi_n
+ \sqrt{r_c} \left\{\left({r_n -r_c \over r_c}
+ {\tilde r_n - \tilde r_c \over \sqrt{r_c}}\right)\Pi_n
+ {1\over N^2}\Pi''_n\right\} +\cdots\ .$$
Recalling now the relation $\tilde u=\pm u$ of eq.~\erelimp,
we see that one sign yields a trivial $Q$ operator and the
other replaces $u$ by $2u$ in $Q$. The sign which yields a trivial $Q$
has the following interpretation: In this case it is not the polynomials
$\Pi_n$ which are smooth functions of $n$ but instead $(-1)^n\Pi_n$.
Taking into account this property, one finds that in all cases $2u$ will now
be the solution of a differential equation described below, and the only
overall effect is to multiply $Z_{\rm sing}$ by a factor of $1/2$.

\medskip
{\it The matrix $C$.} In the double scaling limit, the matrix $C$ also becomes
proportional to a differential operator $P$, which moreover is formally
anti-hermitian because $C$ is antisymmetric. The degree of the operator
$P$ is seen to be at most $2l-1$ when the multicriticality conditions are met.
Let us prove this property for the minimal critical potential. At the critical
point we expect $C\propto \d^{2l-1}=P$, terms of lower degree in $\d$ arising
from deviations from criticality. At the critical point the dependence on $n$
of $B_{n,n}$ and $B_{n,n+1}$ can be neglected which means $B_{n,n+k}\sim b_k$.
Therefore powers of the matrix $B$ become very simple after Fourier
transformation. Let us introduce the variable $z=\e^{i\theta}$ and consider
the Fourier transform of $B$
$$\sum_k B_{n,n+k}z^k\sim\sum_k b_k z^k\equiv b(z)\ .$$
Then we have
$$\sum_k F(B)_{n,n+k}\,z^k\sim F\bigl(b(z)\bigr)\ .$$

In the same limit, the Fourier transform of the matrix $C$ has to be
proportional to $\theta^{2l-1}$ for $\theta$ small and thus to $(z-1)^{2l-1}$.
For the minimal potential, the matrix $C$ is then determined up to a
multiplicative constant:
$$\sum_{k=-l}^{l} C_{n,n+k}z^k \sim c(z)\propto z^{-l}(z-1)^{2l-1}(z+1)\ .$$
The function $V'(b(z))$ differs from  $c(z)$ only by the terms with
negative powers
$$\bigl[c(z)\bigr]_+=\ud \left[V'\bigl(b(z)\bigr)\right]_+\ ,$$
i.e.\ $V'$ is determined by the expansion of $c(z)$ for $z$ large.
Let us then set
$$t=\ud (z+1/z)\ \Rightarrow\ b(z)\sim 2B_{n,n+1}t+B_{n,n}\ .$$
Conversely we choose $z=t+\sqrt{t^2-1}$ such that $z$ large
corresponds to $t$ large. In the variable $t$, $V'(b)$ is a polynomial
which can be determined by expanding $c(z)$ for $t$ large. Noting
$\sqrt{2z}=\sqrt{t-1}+\sqrt{t+1}$, we find
$$c(z)\propto\left(\sqrt{z}-1/\sqrt{z}\right)^{2l-1}
\left(\sqrt{z}+1/\sqrt{z}\right) \propto 2^l (t+1)^{1/2}(t-1)^{l-1/2}\ .$$
It follows that
$$V'\bigl(b(t)\bigr)\propto  (t+1)^{1/2}(t-1)^{l-1/2}_+\ ,$$
which, up a shift of variables, is identical to \epotcrit.

In the double scaling limit, it will be convenient to normalize $P$ by
$$P=\d^{2l-1}+O\left(\d^{2l-2}\right)\ .$$
The operator $P$, obtained by collecting all terms of the same order
in $a$, must transform multiplicatively under \erescale. This implies
$P\mapsto \rho^{2l-1} P$. Moreover writing explicitly the powers of
$a$ in the relation between $C$ and $P$, we find
$$C\propto\ a^{-2+\gs} a^{-(2l-1)\gs}P\ .$$
The first factor $a^{-2+\gs}=N$ comes from the relation \esA\ between $C$
and $B$, and the second factor $a^{-(2l-1)\gs}$ from the change of variables
$\varepsilon \del/\del\xi\mapsto \d/\d x$. To check the consistency of the
scaling ansatz, note that the powers $a$ cancel in the
commutator $[B,C]$ as they should, $-2\gs-(2 -\gs)-(2l-1)\gs=0$
which indeed yields $\gs=-1/l$.

The commutator $[P,Q]$ has the form
$$[P,Q]=K\ ,$$
where $K$ is a constant which can be set to 1 after a rescaling
of the form \erescale,
i.e.\ of the variable proportional to  the deviation from the critical
coupling constant.

The problem of formulating the double scaling limit of matter systems
coupled to $2d$ gravity is thus
reduced to finding solutions of the canonical commutation relation
\eqn\ePQ{[P,Q]=1\ ,}
a problem we shall discuss in the next section.

Finally starting from eq.~\enlnl\ and introducing the quantities $x,z,u$, we
verify that the singular part of the partition function is always given by
eq.~\elsbzsc,
$${\d^2 Z_{\rm sing}\over \d x^2} =-u(x)\ ,$$
the result for the pure gravity case.
Note that this equation is left unchanged by a rescaling of the form
\erescale, a property which will be used systematically later on here.

\medskip
{\it Remark.} In this formulation, the argument
of the orthogonal polynomials, i.e.\ the variable $\lambda$,
becomes the eigenvalue of a Schr\"odinger-like
differential operator. The spectrum of $Q$ yields the asymptotic distribution
of the eigenvalues of the hermitian matrices.

\medskip
{\it Application: the resolvent in the large $N$ limit.} We can use
eq.~\etraceF, applied to $[M-z]^{-1}$, to compare the expression for the
singular part of the resolvent obtained from a steepest descent
calculation (eq.~\esigscii) with the resolvent of the operator $Q$ at leading
order. Considering eq.~\erelBQ, it is natural to set
$a^{2/l}\tilde z=2-r_c^{-1/2}z$. We then find
$$\omega(z)={1\over N}\tr[M-z]^{-1}\sim -r_c^{-1/2}a^{2/l}
\tilde\omega(\tilde z)\ ,$$
with
$$\tilde\omega(\tilde z)=\int_{x}^{a^{-2}}\d y\, \bigl<y| Q+\tilde z
|y\bigr>\ .$$
At leading order, the function $u(x)$ can be replaced by the leading term
$u(x)\sim x^{1/l}$ and the resolvent can be calculated in the semiclassical
limit, i.e.\ the non-commutation between $\d$ and $x$ can be neglected.
It follows that
$$\tilde\omega(\tilde z)=-\int_{x}^{a^{-2}}\d y\, \int{\d p \over 2\pi}
{1\over p^2+u(y)-\tilde z}={1\over 2}\int^{u(x)}_c{y'(s)\d s
\over \sqrt{s-\tilde z}}\ ,$$
with $u(y)=s$, and $c=u(a^{-2})$.
We recognize that the singular part of $\tilde \omega$,
for $c\to\infty$, is identical, up to normalizations, to expression
\eomegsc\ with $z\mapsto \tilde z$.

\medskip
{\it The scaling limit beyond the spherical approximation.} We now consider
the 
scaling limit of \ecorra{b}\ and find
\eqn\eloop{\tilde\omega^2(\tilde z)
+ \int_x \d t \int^x \d t'\, \langle t| (Q-\tilde z)\inv |t'\rangle
\,\langle t'| (Q-\tilde z)\inv |t\rangle
={\rm regular\ function\ of\ }\tilde z\ .}
It follows quite simply that the spectrum of $Q$ cannot be discrete \rFDi.
Indeed at an isolated eigenvalue $q_n$ of $Q$, corresponding to
a normalizable eigenvector $\varphi_n(x)$, the resolvent $(Q-\tilde z)\inv$
has a pole located at $\tilde z=-q_n$. The l.h.s.\ of \eloop\
thus has a double pole with residue
$$\left(\int_x \d t\, \varphi_n^2(t)\right)^2+\int_x\d t\int^x \d t'\,
\varphi_n^2(t)\,\varphi_n^2(t') = \int_x \d t\, \varphi_n^2(t)\ ,$$
where we have used the normalizability of $\varphi_n$.
This residue does not vanish except in the large $x$ limit,
where it decreases as $\exp(-2\int_x^{\infty}\d t\,\sqrt{u(t)})$.
The r.h.s.\ of \eloop, on the other hand, is
a regular function of $\tilde z$: a discrete spectrum for $Q$ thus
results in a contradiction.  Note, however,
that since the residue vanishes faster than any power for $x$ large, this
effect is invisible in perturbation theory.

\subsec{Construction of the differential equations}
\subseclab\sCofde

The differential equations following from \ecombc, $[P,Q]=1$, may be
determined directly in the continuum by relating this problem to the
corresponding KdV flows (see appendix A). The operator $P$ that can satisfy
this commutator is constructed as a fractional power of the operator
$Q=\d^2-u$ of \eQt\ (or more generally we can take $Q$ to be a $q^{\rm th}$
order differential operator in which case with $P$ a $p^{\rm th}$ order
operator, we construct the $(p,q)$ minimal models mentioned at the end of
\refsubsec\sstagpf\ coupled to gravity).
The differential equations describing the $(2l-1,2)$ minimal model are
given by $P=Q^{l-1/2}_+$:\foot{In \refsubsec\ssesgkf, we give a generalized
discussion of this ``string equation''.}
\eqn\ecom{\bigl[Q^{l-1/2}_+,\ Q\bigr]=1\ ,}
where $Q^{l-1/2}_+$ indicates the part of $Q^{l-1/2}$ with only
non-negative powers of $\d$. In terms of the Gelfand--Dikii polynomials
$R_l$, we find
\eqn\elcf{\bigl[\QT^{l-1/2}_+,\QT\bigr] =4R'_{l}\ .}
After integration and rescaling of $x$, the equation
$\bigl[\QT^{l-1/2}_+,\QT\bigr]=1$ thus takes the simple form
\eqn\ede{(l+\half)R_{l}[u]=x\ .}
The quantities $R_l$ can be for instance calculated from the recursion
relation
\eqn\erecR{R'_{l+1}=\frac{1}{4}R'''_{l}-uR'_{l}- \half u'R_l\ .}

\medskip
{\it An action principle.}
The $R_l$'s satisfy as well a functional relation
that allows us to write eq.~\ede\ as the variation of an action.
Indeed it is shown in \refapp{A.2}\ that
\eqn\egdp{{\delta\over\delta u}\int \d x\, R_{l+1}[u]
= -(l+\half)R_{l}[u]\ .}
The differential equation \ede\ therefore results as the variational
derivative with respect to $u$ of the action
\eqn\egenaci{{\cal S}=\int \d x \,\bigl(R_{l+1} + xu \bigr)\ .}
(We treat the above integral formally here and
ignore throughout that physically relevant boundary conditions on
$u$ typically preclude existence of such integrals.)

\subsec{Relevant perturbations and interpolation between multicritical models}
\subseclab\ssRpai

Instead of fixing to critical values
all but one parameter in the potential (i.e.\ the one
corresponding to the most singular perturbation),
it is possible to let them approach
criticality at a rate related to $N$ in the large $N$ limit.
The partition function then becomes a function of a set of new relevant
scaling parameters $t_{(k)}$ and this defines a general ``massive" model
interpolating between multicritical points.
The operator $P$ then takes the form
$$P=\sum_{k=1}^{l}
-(k+\half)t_{(k)}Q^{k-1/2}_+\ ,$$
and the corresponding string equation is
\eqn\estreq{\eqalign{0&=-x+\sum_{k=1}^{l}\
\bigl(k+\half\bigr) t_{(k)}\,R_k[u]\cr
&=\sum_{k=0}^{l}\bigl(k+\half\bigr) t_{(k)}\,R_k[u]\ ,\cr}}
with $t_{(0)}=-4x$. Note that the constant $t_{(l-1)}$ can be eliminated
by an irrelevant translation of $u(x)$ (adding a constant to $u$ means adding
a regular contribution $\propto x^2 t_{(l-1)}$ to the partition function).

Equivalently, using \egdp, eq.~\estreq\ can be
seen to follow from the action
\eqn\estreqa{\eqalign{{\cal S}&= 2 \int\d x
\Bigl(\sum_{k=0}^{l} t_{(k)} R_{k+1}[u]\Bigr)\cr
&=\tr\Bigl(\sum_{k=0}^{l} t_{(k)}\,Q^{k+1/2}\Bigr)\ .\cr}}
with $t_{(0)}\propto x$, the trace of a pseudo-differential operator being
defined in \refapp{A.3}. We shall see that the form
of the action \estreqa\ generalizes to $(p,q)$ models.

The scaling parameters $t_{(k)}$ can be thought of as dimensionful
couplings to certain operators ${\cal O}_{(k)}$ added to the pure gravity
action,
\eqn\purac{S=S_{\rm grav}+\sum_{k \geq 0} t_{(k)}
\int\d^2\xi\, \sqrt{g}\, {\cal O}_{(k)}}
(where the pure gravity action $S_{\rm grav}$ is as in \eZdo, suitably
defined by matrix model techniques as in sections \sDsmmcl,\stomm, or
by topological gravity or Liouville methods as in sections
\sTg,\stcalg).
One of these operators is the identity 
operator coupled to the (renormalized) cosmological constant. Other operators
are so-called gravitational descendants,\foot{This terminology is borrowed
from topological field theory, see sec.~\sTg\ for details.} arising from the
scaling limit of traces of certain polynomials of the matrix in the one
hermitian matrix formulation. The most general one-matrix model solution
corresponds to some perturbation of the pure gravity model by these operators:
eq.~\estreq\ describes the space of all these perturbations.
(See \rMSeSt\ for the transformation from the basis of the $t_{(k)}$'s to the
scaling operators of the theory.)

\medskip
{\it KdV flows.} 
As discussed in \refapp{A.4}, the dependence of the specific heat $u$ on
the parameters $t_{(k)}$ is given by the higher KdV flows
\eqn\ekdvf{{\del\over\del t_{(k)}}Q=-{\del\over\del t_{(k)}} u =
{\del\over\del x} R_{k+1}[u] =\bigl[Q^{k+1/2}_+,\,Q\bigr]\ .}
Using the commutativity of the higher KdV flows, it is straightforward to
verify consistency of \ekdvf\ with \estreq.

As usual in field theory, differentiation of the partition function with
respect to a parameter of the potential $t_{(k)}$ generates a correlation
function with the insertion of the corresponding operator ${\cal O}_{(k)}$. In
particular, the string susceptibility $-u=\del_x^2 \log Z$ is a
two-point correlator of the operator $\CP$:
$-u=\langle\CP\CP\rangle$.
We also have $(\del/\del t_{(k)})\langle\CP\CP\rangle=
\langle\CP\CP{\cal O}_{(k)}\rangle=(\del/\del x)\langle\CP
{\cal O}_{(k)}\rangle$,
where ${\cal O}_{(k)}$ is as above the appropriate scaling operator
that couples to $t_{(k)}$.
\ekdvf\ therefore identifies
$R_{k+1}[u] = \langle\CP {\cal O}_{(k)}\rangle$
as the 2-point function of the operator $\CP$ with
${\cal O}_{(k)}$,
and we can rewrite the string equation \estreq\ and the action \estreqa\ in
terms of these 2-point functions.

\medskip
{\it Scaling dimensions.} Previous relations can be used to calculate
correlation functions on the sphere. The string equation \estreq\ becomes an
algebraic equation for the string susceptibility $u$. The deviation $x\equiv
t_{0}$ from the critical coupling constant provides a scale to the theory and
we can assign it the dimension $1$.
Then $u=\langle {\CP}{\CP}\rangle \propto x^{-\gamma_{\rm str}}$ has dimension
$-\gamma_{\rm str}$, i.e.\ $1/l$ for the $l$-th critical point.
Each coupling constant $t_{(k)}$ carries a dimension $1-l/k$.
Correlation functions of the operator ${\cal O}_{(k)}$ are obtained
by differentiating  $u$ with respect to $t_{(k)}$ and integrating
twice with respect to $x$. It follows for example that the one-
and two-point functions  exhibit the scaling behavior
$$\langle {\cal O}_{(k)} \rangle \propto x^{1-\gs+k/l}\qquad
\langle {\cal O}_{(k)} {\cal O}_{(k')} \rangle \propto
x^{ -\gs+k/l+k'/l}\ldots\ .$$
This behavior is related to the KPZ \rKPZ\ scaling of conformal operators
``dressed" by gravity (see \refsubsec\ssDw).

\medskip
{\it The identity or puncture operator.} We see from the previous scaling
relations that the operator $\CP$ has the smallest dimension and is thus
the most singular.  As we have defined the coupling constant $g$,
i.e.\ multiplying the whole potential $V(M)$, it couples to all operators.
It is therefore not surprising that the most singular survives in the
double scaling limit. From the analysis of \refsec\sDsmmcl\ we would be tempted
also to identify $\CP$ as the puncture or identity operator coupled
to the cosmological constant, as in the pure gravity case. As we have
already seen in \refsubsec\ssMp, however, this identification is inconsistent
with the behavior of macroscopic loop averages. The behavior
in \elooplaw\ suggests instead that $t_{l-2}$ is coupled to the
identity, ${\cal O}_{(l-2)}=I$, because $t_{(l-2)}\equiv\mu\sim x^{2/l}$.
In terms of $\mu$, the partition function scales as $Z\sim\mu^{l+1/2}$
(correspondingly $\gs(\mu)=3/2-l$), and $\langle O_{(k)}\rangle
\sim \mu^{(l+k+1)/2}$, so we assign to $O_{(k)}$ the dimension
$\Delta_{(k)}=(l+k+1)/2-(l+1/2)+1=(k+2-l)/2$
in terms of $\mu$. This conclusion is
supported by a direct calculation in the continuum Liouville formulation
(see \refsec{\stcalg}).

In what follows, when we refer to {\it coupling\/} constant,
without other qualification, we shall mean
the parameter coupled to the operator of lowest dimension, here $\CP$,
which may or may not be the renormalized {\it cosmological\/} constant.

\def\qq{\nu}
\newsec{Multi-matrix models}
\seclab\sqmomm

\subsec{Solving the multimatrix models}

By a method of orthogonal polynomials, it is also possible to solve models
involving integration over several matrices $M^{(\alpha)}$. The basic identity
is \INT
\eqn\emulti{\eqalign{&\int\d M^{(1)}\, \ee{-\tr V_1\bigl(M^{(1)}\bigr)
+ c\,\tr M^{(1)} M^{(2)}}\cr
&\hskip3cm
\propto \int\d \Lambda^{(1)}\,\ee{\sum_i -V_1\bigl(\lambda^{(1)}_{i}\bigr)
+ c\lambda^{(1)}_{i}\lambda^{(2)}_{i}}
{\Delta\left(\Lambda^{(1)}\right)\over\Delta\left(\Lambda^{(2)}\right)}\ ,
\cr}}
where the $M^{(\alpha)}$  are $N\times N$ hermitian matrices, the
$\lambda_i^{(\alpha)}$ their eigenvalues, and $\Delta(\Lambda)$ is the
Vandermonde determinant $\det(\lambda_i^{j-1})$ (the proof of this identity
is much more involved than in the one-matrix case, see appendix~\smmaj\ for
a sketch of the proof).
When the action $S$ has the form
\eqn\acform{S\bigl(M^{(\alpha)}\bigr)=\sum_{\alpha=1}^{\qq-1}V_{\alpha}
\bigl(M^{(\alpha)}\bigr) -
\sum_{\alpha=1}^{\qq-2} c_{\alpha}\,M^{(\alpha)} M^{(\alpha+1)}\ ,}
it follows that
\eqn\eqmm{\eqalign{Z&=\int\prod_{\alpha=1}^{\qq-1} \d M^{(\alpha)}\
\ee{-\tr S\bigl(M^{(\alpha)}\bigr)}  \cr
&=\int \prod_{{\scriptstyle \alpha=1,\qq-1}\atop {\scriptstyle i=1,N}}
\!\!\d\lambda_i^{(\alpha)}\ \Delta\bigl(\Lambda^{(1)}\bigr)
\ee{-\sum_i S\bigl(\lambda_i^{(\alpha)}\bigr)}
\Delta\bigl(\Lambda^{(\qq-1)}\bigr)\ ,\cr}}
generalizing \eparta.

The result \eqmm\ depends on having $c_{\alpha}$'s that couple matrices
along a line (with no closed loops so that the integrations over the relative
angular variables in the $M^{(\alpha)}$'s can be performed).
Via a diagrammatic expansion, the matrix integrals in \eqmm\
can be interpreted to generate a sum over discretized surfaces, where
the different matrices $M^{(\alpha)}$ represent $\qq-1$ different matter
states that can exist at the vertices.
The quantity $Z$ in \eqmm\ thereby admits an interpretation as the partition
function of 2D gravity coupled to matter. Note that matter has only a finite
number of states (by taking $\qq\to\infty$, one can represent a $D=1$ model,
i.e.\ a single free boson, coupled to gravity).
Furthermore the only possible symmetry which can be implemented corresponds
to reflecting the line about its center.
Therefore these matrix models can only implement an
Ising-like $\IZ_2$ symmetry on a random lattice.

\medskip
{\it Generalized orthogonal polynomials.} To solve the matrix models we
define generalized orthogonal polynomials $\Pi_n(\lambda)$, satisfying
\eqn\eorth{\int\d\mu\bigl(\lambda^{(1)},\ldots,\lambda^{(\qq-1)}\bigr)\,
\Pi_m\bigl(\lambda^{(1)}\bigr)
\,\widetilde{\Pi}_n\bigl(\lambda^{(\qq-1)}\bigr)= \delta_{mn}\ ,}
(generalizing \eorthrl)
where the measure $\d\mu$ is defined by
\eqn\emeasq{\d\mu\bigl(\lambda^{(1)},\ldots,\lambda^{(\qq-1)}\bigr)=
\ee{-\sum_{\alpha}V_{\alpha}\bigl(\lambda^{(\alpha)}\bigr)+
\sum_{\alpha} c_{\alpha}\,\lambda^{(\alpha)}\lambda^{(\alpha+1)}}
\!\prod_{\alpha=1}^{\qq-1} \d\lambda^{(\alpha)}\ .}
To derive recursion formulae, we insert $\lambda^{(\alpha)}$ and
$\d/\d\lambda^{(\alpha)}$ respectively in the integral \eorth.

Let us denote by $\d\mu_{\alpha-1}$ the measure \emeasq\ in which the
integration is restricted to the first $\alpha-1$ variables.
It is then convenient to introduce a matrix  $B_{\alpha}$
associated with $\lambda^{(\alpha)}$ and defined by
\eqn\edefBa{\lambda^{(\alpha)}\int\d\mu_{\alpha-1}\,
\Pi_m\bigl(\lambda^{(1)}\bigr)=\bigl[B_{\alpha}\bigr]_{mn}
\int\d\mu_{\alpha-1} \Pi_n\bigl(\lambda^{(1)}\bigr)\ .}
We can also define  $\widetilde{B}_{\alpha}$, using a
similar expression but in which the roles of $\Lambda^{(1)}$ and
$\Lambda^{(\qq-1)}$ are exchanged, i.e.\ by integrating over
$\lambda^{(\alpha+1)},\ldots, \lambda^{(\qq-1)}$. Then, multiplying by
$\widetilde\Pi_n(\lambda^{(\qq-1)})$ and $\Pi_n(\lambda^{(1)})$ respectively,
and integrating over all $\lambda$'s, we obtain
$$\widetilde{B}_{\alpha}=B^T_{\alpha}\ .$$
Finally we define matrices $A_1$ and $\widetilde A_{\qq-1}$,
\eqn\edefAa{\Pi'_m=\bigl[A_1\bigr]_{mn}\Pi_n\,,\qquad
\widetilde{\Pi}'_m=\bigl[\widetilde A_{\qq-1}\bigr]_{mn}\widetilde\Pi_n\ ,}
and an additional matrix $A_{\qq-1}=\widetilde A^T_{\qq-1}$.

Inserting $\d/\d \lambda^{(\alpha)}$ in \eorth\ gives
\eqn\erecgen{\eqalign{A_1+c_1 B_2&=V'_1\left(B_1\right)\,,\cr
c_{\alpha-1}B_{\alpha-1}+c_{\alpha}B_{\alpha+1}&=V'_{\alpha}
\left(B_{\alpha}\right) \quad\qquad{\rm for}\ 1<\alpha<\qq-1\,,\cr
A_{\qq-1}+c_{\qq-2}B_{\qq-2}&=V'_{\qq-1}\left(B_{\qq-1}\right)\ ,\cr}}
with
\eqn\ecomab{\eqalign{[B_1,A_1]&=1\,,\cr
[\widetilde B_{\qq-1},\widetilde A_{\qq-1}]&=1\ .\cr}}
Eqs.~\erecgen\ imply that
\eqn\ecombb{c_{\alpha}\left[B_{\alpha+1},B_{\alpha}\right]=c_{\alpha-1}
\left[B_{\alpha},B_{\alpha-1}\right]=1\ ,}
and thus that $\left[A_{\qq-1},B_{\qq-1}\right]=1$, consistent with \ecomab.

Inspired by the one-matrix case \esA,
we can also introduce matrices $C_{\alpha}$ defined by
$$\eqalign{C_1&=A_1-\ud V'_1\left(B_1\right),\cr
C_{\alpha}&=c_{\alpha-1}B_{\alpha-1}-\ud V'_{\alpha}\left(B_{\alpha}\right)
\quad{\rm for}\ \alpha>1\ .\cr}$$
It follows from these definitions and eqs.~\eqns{\ecombb{,\ }\ecomab}\  that
\eqn\erecop{\eqalign{&\quad\left[B_{\alpha},C_{\alpha}\right]=1\,,\cr
&\quad C_{\alpha}=\ud V'_{\alpha}\left(B_{\alpha}\right)
- c_{\alpha}B_{\alpha+1} \qquad{\rm for}\ \alpha<\qq-1,\cr
&C_{\qq-1}=\ud V'_{\qq-1}\left(B_{\qq-1}\right)-A_{\qq-1}\ .\cr}}

Let us call $l_{\alpha}$ the degree of the polynomial $V_{\alpha}$. It is then
easy to verify from \erecgen\ that $[B_{\alpha}]_{mn}$
is non-vanishing only for
\eqn\ubound{\prod_{\beta=\alpha+1}^{\qq-1}\left(l_{\beta}-1\right) \le n-m\le
\prod_{\beta=1}^{\alpha-1}\left(l_{\beta}-1\right)\ .}
This uniform bound on the ``range" of the operators $B_{\alpha}$, henceforth
on the operators $C_{\alpha}$ through \erecop, is the heart of the result
obtained by Douglas \rD. In the large $N$ ``double-scaling'' limit (with all
but one of the couplings in \acform\ tuned to critical values), all
these operators become differential operators. The uniform
bound on the range above induces a bound on the order of the differential
operators.
\medskip

{\it $\IZ_2$ symmetry.} When $V_{\alpha}=V_{\qq-\alpha}$ and
$c_{\alpha}=c_{\qq-1-\alpha}$, the matrix problem has a $\IZ_2$ symmetry
corresponding to the mapping of matrices $M^{(\alpha)}\mapsto
M^{(\qq-\alpha)}$. We can then choose $\widetilde\Pi_n \equiv \Pi_n$. It
follows that $\widetilde{A}_{\qq-1}= A_1$, $\widetilde{B}_{\qq-1}=B_1$. In
addition this symmetry yields the relations
$$\widetilde{B}_{\alpha}=B^T_{\alpha}=B_{\qq-\alpha}\ ,
\quad\hbox{and hence}\quad
C^T_{\alpha}=-C_{\qq-\alpha}\ .$$

\medskip
{\it The free energy.} We now assume for simplicity $\IZ_2$ symmetry. We
normalize, as in the one-matrix case,
$\Pi_n(\lambda)=\lambda^n/\sqrt{s_n}+\ldots\,$.
We then find from eqs.~\eqns{\edefBa{,\ }\edefAa} the relations
$$\eqalign{[B_1]_{n,n+1}\equiv\sqrt{r_n}&
=\left(s_{n+1}/s_{n}\right)^{1/2},\cr
[A_1]_{n,n-1}&=n\left(s_{n-1}/s_{n}\right)^{1/2}\ .\cr}$$
As in \efreen, the matrix model free energy is given by
\eqn\energl{\eqalign{F=\ln Z&=\ln\left(N!\,s_0 s_1\ldots s_{N-1}\right),\cr
&=\ln\left(N!\,s_0^N\right)+\sum_{n=1}^{N-1}(N-n)\ln r_n\ .\cr}}
\medskip
{\it The two-matrix model.} Since it has been shown that it is sufficient to
consider the two-matrix model to generate the most general critical point
\rDoDKK, let us just write the previous equations in this special case
\eqn\etwomat{A_1+c B_2 =V'_1(B_1)\,,\quad A_2+c B_1 =V'_2(B_2),\quad
c[B_1,B_2]=1\ .}
In the $\IZ_2$ symmetric case the equations become particularly simple
$$A+c B^T=V'(B),\qquad [A,B]=1\ .$$
Note that in the various relations \etwomat\ the matrices $A_1,A_2$ can be
completely eliminated by writing them
\eqn\evrrew{c [B_2]_{mn} =[V'_1(B_1)]_{mn},\quad
c[B_1^T]_{mn}=[V'_2(B_2^T)]_{mn}
\quad {\rm for}\ n \ge m\ ,}
and $c[B_1,B_2]=1$.

Let us now briefly indicate, generalizing the arguments of  \refsubsec\sstlnl,
how in the large $N$ limit the potentials $V_1$ and $V_2$ can be chosen
(multicriticality conditions) such that the matrices $B_1,B_2$ become
linear combinations of two differential operators $P,Q$ of degree $p,q$
respectively. We recall that at criticality the matrix elements of the various
matrices depend only on the differences between their two indices, in such a
way that a Fourier transformation renders the previous relations particularly
simple. Let us introduce two functions $b_1(z), b_2(z)$
$$b_1(z)\sim \sum_{k=-1}[B_1]_{n+k,n}z^k\,,\qquad b_2(z)\sim
\sum_{k=-1}[B_2]_{n,n+k}z^k\ .$$
Both functions behave as $1/z$ for $z$ small. Then the relations \evrrew\
become
$$c[b_2(z)]_+=[V'_1(b_1(1/z))]_+ \,,\quad c[b_1(z)]_+=[V'_2(b_2(1/z))]_+\ ,$$
where as elsewhere the subscript $+$ means the sum of terms of non-negative
powers.

If we impose that in the large $N$ limit at criticality
$B_1$ and $B_2$ become proportional respectively to $\d^p,\d^q$ ($\d$ means for
example differentiation with respect to the index), then for $z\to 1$ we have
$b_1(z)\propto (z-1)^p$, $b_2(z)\propto (z-1)^q$. The simplest example is
$$b_1(z)=(z-1)^p/z\,,\qquad b_2(z)=(z-1)^q/z\ .$$
We introduce two new variables $t_1,t_2$ via
$$t_1=b_1(1/z)\,\qquad t_2=b_2(1/z)\ .$$
Inverting these relations, we choose the roots such that $z$ large corresponds
to $t_1,t_2$ large. Then the two critical potentials $V_1,V_2$ are determined
by
$$V'_1(t_1)=cb_2\bigl(z(t_1)\bigl)_+\,,\qquad
V'_2(t_2)=cb_1\bigl(z(t_2)\bigl)_+\ .$$

\subsec{The continuum limit}
\subseclab\sTcl

In this subsection, it is convenient to return to the normalization of
previous sections, i.e.\ multiply the whole potential by a factor
$N/g$ as in \eparta. To study the double scaling limit, we introduce the
renormalized deviation from the critical coupling constant
$x=a^{-2}(1-g/g_c)$ where $\varepsilon=1/N=a^{2-\gamma_{\rm str}}$.
The arguments which follow are then a simple generalization of those of
\refsubsec{\sstlnl}.
The matrices $B_1$ and $C_1$ generate differential operators
$Q$, $P$ of finite order say $q > p$, which still satisfy the ``string
equation":
\eqn\epqcc{\bigl[P,Q\bigr]=1\, ,}
(as in \ePQ, the commutator has been normalized for convenience to 1).
The operators $P$ and $Q$ are obtained by collecting the terms of
order $a^{-p\gs}$ and $a^{-2+\gs}a^{-q\gs}$ respectively. Since the commutator
is independent of $a$,
we obtain the value of the string susceptibility exponent:
\eqn\estrsce{-p\gs+(-2+\gs)-q\gs=0\ \Rightarrow\ \gs(p,q)
=-{2 \over(p+q-1)}\ .}
The operator $Q$ can always be written
\eqn\eQ{Q=\d^{q}+v_{q-2}(x)\d^{q-2} +\ \cdots\ + v_{0}(x)\ .}
(By a change of basis of the form $Q\to f\inv(x)Qf(x)$,
the coefficient of $\d^{q-1}$ may always be set to zero.)
The continuum scaling limit of the multi-matrix models
is thus abstracted to the mathematics problem of finding solutions of
\epqcc.

The various coefficient functions involved in eq.\eQ\ are scaling
functions of the coupling constant $x$.
In units of $x$, the ``grade'' of $\d=\partial_x$ is $-1$,
so that the grade\foot{This
notion of grade is related to the conventional
scaling dimensions of operators, see for instance the end of
\refsubsec\ssRpai.
It can also be used to determine the terms that may appear in many equations,
since these will only relate terms of overall equal grade.}
of $v_{q-\alpha}$ is $-\alpha$ for an operator $Q$ of overall grade $-q$.

The function $v_{q-2}$ can be identified (up to normalization)
in the continuum scaling limit
with the second derivative of the free energy with respect to
the coupling constant (here proportional to $x$). Equivalently
we can write $v_{q-2}\propto \partial^2_x \log Z
= \langle{\cal O}_{\min}{\cal O}_{\min}\rangle$
in terms of the 2-point function of the
operator coupled to $x$, which, according to the analysis of
\refsubsec\ssRpai, is the operator ${\cal O}_{\min}$
of smallest dimension (i.e.\ the most singular in the continuum limit).

\subsec{String equation solution and generalized KdV flows}
\subseclab\ssesgkf

The differential equations \epqcc\ may be constructed as follows.
Let $L=Q^{1 / q}$ denote the $1^{\rm st}$ order pseudo-differential
operator (i.e.\ a formal series with inverse powers of $\d$) whose
$q^{\rm th}$ power is equal to $Q$. Then, by a  theorem of \DS,
the most general form $P$ can have is
\eqn\forp{ P= \sum _{j=1}^p \mu_j  (L^j)_+\ ,}
where the subscript $+$, as before, indicates that we truncate the
pseudo-differential operators
to their differential part (non-negative powers of $\d$).
We will soon see that the $(p,q)$ minimal model  ($p$ and $q$
relatively prime) is described by the critical
equation with $\mu_j= \delta_{j,p}$:
\eqn\cripq{\bigl[(Q^{p / q} )_+,\,Q\bigr]=1\ .}

Let us concentrate on this case first. The equation \cripq\ is a coupled
differential system for the $q-1$ coefficients $v_{q-2},\ldots,v_0$ of $Q$.
Using the fact that
$Q=L^q$ and $L^p= (L^p)_+ +(L^p)_-$ commute, we write
\eqn\pqcri{ [P,Q]=- \bigl[(L^p)_-,L^q\bigr]\ ,}
and expand the commutator in formal powers of $\d$.
If $(L^p)_-=c_1(x) \d^{-1}+ \ \cdots $, then the leading term
$q\,c'_1\d^{q-2} + \cdots$ in \pqcri\
must vanish, giving rise to an integration constant
$c_1= \nu_{q-1}$.
We can now proceed by writing $(L^p)_--\nu_{q-1} L^{-1}=c_2 \d^{-2}+
\cdots\,$. By the same argument, we get a second integration constant:
$c_2 = \nu_{q-2}$, and so on, until we reach the $\d^0$ term in the
commutator \pqcri. Taking
$(L^p)_-= \sum_{i=1}^{q-2} \nu_{q-i} L^{-i} + c_{q-1}\d^{1-q}+ \cdots\,$,
$q c_{q-1}' =1$ integrates to
$c_{q-1} = x / q$, where we absorb the last integration constant
in a shift of $x$.

The occurrence of these integration constants is crucial for the following
reason: the dependence of $L$ (therefore of all  the coefficients of $Q$) on
these constants $\nu_i$ ($i=1,\ldots,q-1$) can be shown to be given by the
first
$q-1$ generalized KdV flows. Namely, in terms of the ``times"
$t_i = q \nu_i / i$  ($i=1,2,\ldots,q-1$), $L$ satisfies the evolution
equations:
\eqn\kdvpq{\partial_{t_i} L = \bigl[(L^i)_+, L\bigr]\ .}
We refer the reader to appendices A,B and \DIFK\ for a complete proof of this
property. It relies on the Jacobi identity for pseudo-differential operators.

Going back to the most general form \forp\ for $P$,  this result has a
beautiful generalization. Extending the set of KdV times to $t_i = - q
\mu_{i-q} / i$ for  $i=q+1,\ldots$,
the evolution of the solution to the string equation w.r.t.\ these variables
is given by the higher KdV flows \kdvpq, for $i=q+1,\ldots\,$.
This remarkable fact turns out to be very useful for studying the
physical content of the string equation \epqcc,
which now reads (see \refapp\ssEc)
\eqn\pqstreq{[P,Q]=1\,,\quad{\rm with}\quad P= \!
\sum_{\scriptstyle k\geq 1;\atop \scriptstyle k \neq 0\pmod{q}} \!
-(1 +{k / q})\, t_{k+q}\, Q^{k / q}_+ \ .}
Note that this general string equation can be derived as in the one-matrix
case from an action principle \rGGPZ, generalizing the result \estreqa, with
the action functional
\eqn\egenac{{\cal S}_q= \sum_{k=1}^{\infty} t_{k}\, \tr Q^{k / q}\ ,}
(and the trace operation defined in \refapp\sstlze). The direct proof is
tedious but the result also follows from the
corresponding action principle for the discrete matrix model which we present
in subsections {\it\sstomxc, \ssmmdis\/}.

\smallskip
{\it Remark.} It is obvious that in \egenac\ the terms proportional to
$t_{m}$, $m=0\pmod{q}$, give no contribution and can thus be omitted. This
seems to break the symmetry between $p$ and $q$. If we assume $p>q$, however,
only one term $t_p\,\tr Q^{p/q}$ is such that $m=0\pmod{p}$. It
is clear that this term can be eliminated by shifting $Q$ by an irrelevant
constant. More generally, in \refapp\sspqqp\ an equivalence is established
between the two actions expressed in terms of $Q$ and $P$ respectively.
(See also \rpqdual\ for more on $p$-$q$ duality in these models.)

To summarize, the solutions of the string equation are not unique, but
can be deformed in an infinite number of directions to reach any $(p,q)$ with
$p,q$ relatively prime. These directions are identified with KdV time
evolutions (see appendix~\sGkdv),
but from a field theoretical point of view they
correspond to flows along RG trajectories between various critical theories,
identified with $(p,q)$ minimal conformal models coupled to gravity. Such
trajectories are explored by adding an infinite number of relevant matter
operators dressed by gravity to the original critical action.  Therefore, the
KdV time variables provide
us with a natural definition of the operators in the theory: let the
insertions of operators $\phi\dup_{k}$ into correlators be defined dually by
differentiation w.r.t.\ the KdV flows $t_k$,
\eqn\defop{ \partial_{t_{k_1}} \ldots \partial_{t_{k_n}} \log Z =
\langle \phi\dup_{k_1} \ldots \phi\dup_{k_n} \rangle\ .}
The first $q-1$ operators have been singled out in our approach, as
dual to the integration constants introduced to rewrite the
string equation.  They play a special role in the $(p,q)$ model
picture: they correspond to the gravitational dressing of the order parameters
of the conformal model.

It is very useful to make an analogy with the
Landau--Ginzburg  (LG) picture \ZAMO\
of the minimal models in the particular case
where $|p-q|=1$ (unitary series).  From this point of view,
the order parameters
are normal ordered powers of the basic LG field, specifically
the first $q-2$ such powers
in the $(q+1,q)$ case (we exclude the identity).  On the other hand,
the LG action reads $S= (\grad\Phi)^2  + \Phi^{2q} +\ldots+
\Phi^{2q-2}+\Phi^{2q-4}$, and
$\Phi^{2q-4}$ is therefore the thermal operator $\Phi_{(1,3)}$,
which drives the
$(q+1,q)$ model to the $(q,q-1)$ model. We can see that a formal correspondence
$\Phi^{m} \mapsto L^{m+1}$ exists between the LG picture and its gravitational
counterpart in the framework of KdV flows
(with $m\in [0,q-2]$ \rMSeSt). We can also expect subtleties for
higher operators, since even the LG definition becomes ambiguous due to the
interplay with the equation of motion $\partial {\bar \partial} \Phi \propto
\Phi^{2q-1}$ (for which the corresponding flow, generated by $L^{2q}=Q^2$,
is trivial).

\medskip
{\it Scaling dimensions.} It is rather easy to extract directly some
information from the string equation \epqcc\ in the spherical limit.  Then
the string equations become a set of coupled algebraic equations for the
coefficients $v_i$. The solutions are scaling functions of the variables
$t_k$. Giving dimension 1 to the variable $x\propto t_1$ coupled to the
most singular operator, we find by a simple counting argument:
\eqn\edimtm{t_m \sim x^{(p+q-m)/(p+q-1)},\qquad
v_{q-\alpha}\sim x^{\alpha/(p+q-1)}\ .}
In particular for the string susceptibility $u \propto v_{q-2}$, we find
$u^{(p+q-1) /2} \propto x$, from which we
recover the string susceptibility exponent \estrsce.

Technically, once the string equation has been solved and  an operator $L$
is constructed, we have access to all the correlators \defop\
by repeated use of the flow equations \kdvpq.
Consider the evolution equation
\eqn\evolm{ \partial_{t_m} L =\bigl[ (L^m)_+,\,L\bigr]
= -\bigl[(L^m)_-,L\bigr]\ .}
Equating the coefficients of $\d^{-1}$ on both sides
of eq.~\evolm\ gives
\eqn\oneptpq{ {1 \over q} \partial_{t_m} v_{q-2}  = \partial_x \Res L^m\ ,}
where the symbol $\Res$ stands for the residue of a pseudo-differential
operator (see  \refapp\sstlze),
i.e.\ the coefficient of $\d^{-1}$ in its formal
power series. Integrating once w.r.t.\ $x$ gives an expression for the
2-point functions,
\eqn\twoptpq{ \langle \phi\dup_1 \phi\dup_m \rangle \propto \Res  Q^{m / q}\ .}

This gives in the spherical approximation the scaling behavior
$$\langle \phi\dup_m \rangle \sim x^{(p+q+m)/(p+q-1)}$$
(a result consistent with the dimension of the corresponding parameter $t_m$
in \edimtm). As we have already discussed in the one-matrix case, to compare
with the KPZ result \rKPZ\ for the gravitational dressing of conformal
dimensions we must first discuss the problem of the identification of the
identity operator and of the cosmological constant.


\noindent{\it Example 1: the critical Ising  model (4,3)}\par\nobreak
The Ising model has a natural realization as a two-matrix model \rKBK\
in which the two matrices represent the $+/-$ states of an Ising spin.
The two-matrix model has been first solved directly using recursion relations
\refs{\rIYL,\rising}. A simpler derivation follows from considering
the commutation relation \epqcc\ \refs{\rD, \rGGPZ} with
\eqn\eipqp{\eqalign{Q&=\d^3-{3\over4}\{u,\d\}+ {3\over2}w
=\left(\d^2-u\right)^{3/2}_+ + {3\over2}w\ ,\cr
-P&=Q^{4/3}_+ + {5 \over 3} t_5 Q^{2 / 3}_+
=\d^4 -\{u,\d^2\} +\{w,\d\}+{1 \over 2} u^2 -{1 \over 6}u'''
+ {5 \over 3} t_5 (\d^2 -u)\ .\cr}}
%
(we have set $t_7=3/7$).
The corresponding string action \egenac\ leads to two equations:
\eqn\isin{ \eqalign{& {1 \over 2}w''-{3 \over 2}uw
+{5 \over 2} t_5 w +t_2=0\,,\cr
&{1 \over 12}u^{(4)} - {3 \over 4}uu'' -{3 \over 8}(u')^2
+{1 \over 2} u^3 -{5 \over 12}t_5 (3u^2-u'')+{3\over 2}w^2+t_1=0\,, \cr}}
%
with $t_1\propto x$. We can identify the operators dual to the parameters
$t_1,t_2$, and we recover the identity ($\phi\dup_1 = {\cal P}$), spin
($\phi\dup_2= \sigma$, the order
parameter of the model)  and energy ($\phi\dup_5= \epsilon$) operators from
the $t_{1,2,5}$ flows respectively: $t_2$ is the exterior magnetic field and
$t_5$ the temperature shift from the critical value. We find the scaling
properties $t_5\sim[x]^{1/3}$, $t_2\sim [x]^{5/6}$. 

Note that, as explained above, the flows $t_3$ and $t_4$ have no significance
for the model.
Note also that in the limit $t_5 \to \infty$, the system \isin\ reduces to the
pure gravity equation. This is a perturbative manifestation of the well
known RG flow of the Ising model to pure gravity in the high temperature
limit.

\noindent{\it Example 2: The tricritical Ising model (5,4)}\par\nobreak
We parametrize now the two operators as ($t_9=4/9$)
\eqna\etcop
$$\eqalignno{Q&=(\d^2 - u)^2+\bigl\{w,\d\bigr\}+v\ ,&\etcop a\cr
-P&=Q^{5/4}_+=(\d^2 - u)^{5/2}_+ +{5\over4}\bigl\{w,\d^2\bigr\}
+ {5 \over 8}\bigl\{v,\d\bigr\}-{5\over4}uw\ , &\etcop b\cr}$$
where $w$ is again a $\IZ_2$ breaking field that results
in coupling to a magnetic field.
Again varying the corresponding action \egenac\ we find (with $t_9=4/9$)
\eqna\etci
$$\eqalignno{& 4R_3+{5\over8}v''-{5\over4}uv+{5\over4}w^2+{3\over2}t_3=0\
,&\etci a\cr
& {1\over2}w^{(4)} - {5\over4}(uw)'' - {5\over4}uw''
- {5\over2}vw + {5\over4}u^2 w -4 t_2 =0\ ,&\etci b\cr
& 8 R_4 + {1\over 16} v^{(4)}  + {5\over8} v^2
+ {15\over8} u^2 v - {5\over8} (u v'' + u' v' + v u'')\cr
&\qquad\qquad - {5\over4} w w'' + {5\over4} w^2 u -{3\over2}t_3u +t_1=0\
,&\etci c\cr}$$
where $x \propto t_1$.

We can also perturb the tricritical Ising model in the direction of
the next lower model, i.e.\ the Ising model. Instead of \etcop{b}, we use
\eqn\etcipi{\widetilde P=P - {7 \over 4} t_7 \,Q^{3/4}_+
= P - {7 \over 4} t_7 \,\Bigl( ({\d^2-u})^{3/2}_+ +{3\over2}w\Bigr) \ ,}
where $Q$ remains as in \etcop a. The equations that follow from
$\bigl[\widetilde P,Q\bigr]=1$ are the tricritical
Ising equations plus
$t_7$ times respectively the critical Ising equations \isin\ (with
critical temperature $t_5=0$).
(For $t_7$ ``large'' in some sense, the equations cross over to the Ising
equations.)
The most general perturbation of the tricritical Ising model also includes
the ``pure gravity" piece, and one simply has to consider ${\hat P}=
{\widetilde P} - {3 \over 2} t_6 (\d^2-u)$ as in \forp.

The genus zero equations for the tricritical Ising model
are given by ignoring the derivatives in the tricritical equations. {}From
these equations, we read off the scaling properties
$u\sim [x]^{1/4}$, $v\sim [x]^{1/2}$, $w\sim [x]^{3/8}$, so that
$t_2\sim [x]^{7/8}$, $t_3 \sim [x]^{3/4}$, $t_7\sim [x]^{1/4}$.
7/8 is the gravitationally dressed weight
of the spin field (the (2,2) operator with undressed conformal weight 3/80)
in the tricritical Ising model, 3/4 is the dressed weight of the energy
operator (the (3,3) operator with undressed conformal weight 1/10), and 1/4
is the dressed weight of the vacancy operator (the (3,2) operator with
undressed conformal weight 3/5). Derivatives of the free energy with
respect to the parameters $t_i$ generate correlation functions of the
associated operators.

More generally, the unitary minimal models $(q,p)=(n,n+1)$ flow onto each
other when one perturbs by the $\Phi_{(1,3)}$ ``thermal" operator generalizing
the  Ising energy. In our language, this is transparent in the string
equation
\eqn\strint{\Bigl[L^{n+1}_+ -{{2n-1}\over n}t_{2n-1} L^{n-1}_+
\,,\,L^n\Bigr] =1\ ,}
which interpolates between the $(n+1,n)$ and $(n,n-1)$ models, and
identifies the field $\phi\dup_{2n-1}$ dual to $t_{2n-1}$ with the
gravitationally dressed $\Phi_{(1,3)}$ operator.\foot{Note, however,
that this is a formal relation that can fail non-perturbatively \rbdss.}

\subsec{Solution of the unitary $(n+1,n)$ models on the sphere}
\subseclab\sssotumos

In this section, we carry out completely the program sketched above
in the case of the unitary series $p=n+1$, $q=n$.
By restricting to spherical correlation functions, we ignore the
possibly non-perturbative information contained in the string equation
(which anyway is subtle, as we will see in sec.~\stcalg), and concentrate on
their leading perturbative behavior.

This amounts to retaining in the solution $L=Q^{1 / n}$ only the leading
power-like behavior as a function of $x$, or equivalently rewriting the
correlators as
successive powers of the string susceptibility $u \propto v_{n-2}$.
Moreover the expected  $\IZ_2$ parity of the solution further restricts the
general form of $L$. In general, however,
the spherical solution is not completely determined by
\epqcc. Instead we have an algebraic equation (polynomial of fixed degree)
which admits several solutions for the coefficients $v_0,..,v_{n-2}$.
To resolve this discrete ambiguity, it is necessary to introduce
some physical constraint, as suggested by the results of the preceding
subsection: We impose the vanishing of all the order parameter one-point
functions on the  sphere. We now describe the particular solution
this determines.

Let us introduce for $m\geq 1$ the pseudo-differential operator
\eqn\qm{ {1 \over m} Q_m = \sum_{k=0}^{\infty} \left( { m-k-1 \atop k-1}
\right) {(-u/2)^k \over k} \d^{m-2k}\ ,}
where
\eqn\defcnp{ \left( { m \atop p} \right) = {{m(m-1)\ldots(m-p+1)} \over p!}\ ,}
and $u$ denotes the string susceptibility $u=-\partial_x^2 \log Z$. 
Then, up to terms involving two or more derivatives of $u$, we have
\eqn\propqm{[Q_m,Q_p]=0\ ,}
for all $m,p$. In addition the operators $Q_m$ satisfy by definition the
crucial identity:
\eqn\cruxid{ (Q_m)_- = -(u/2)^m \d^{-m} + O(\d^{-m-2})\ .}

In \DIFK, it was proved that the unique solution to the string equation with
vanishing order parameter one-point functions on the sphere takes the form
$P=(Q_{n+1})_+$ and $Q=(Q_n)_+$.
The properties \eqns{\propqm{,\ }\cruxid}\ suffice to prove that
this is indeed a solution. Uniqueness is a little more subtle and we refer the
reader to \DIFK\ for details.

The string equation at genus zero is
\eqn\sphstr{ (n+1)(u/2)^n = x\ ,}
giving the critical exponent $\gamma_{\rm str}(n+1,n)= -1 / n$. Correlators
are given as explained  above by the KdV flows:
\eqna\kdvcorr{ $$\eqalignno{
\langle \phi\dup_l\, \phi\dup_1 \rangle &= 2 \Res  L^l  &\kdvcorr a\cr
\langle \phi\dup_l\, \phi\dup_m\, \phi\dup_1 \rangle &= 2 \Res  [L^l_+,L^m_-]
&\kdvcorr b\cr
\langle \phi\dup_l\, \phi\dup_m\, \phi\dup_r\, \phi\dup_1 \rangle &=
2\Res \Bigl(\bigl[[L^l_-,L^m_+],L^r_-\bigr]-\bigl[L^m_+,[L^l_+,L^r_-]\bigr]
\Bigr)\ .&\kdvcorr c\cr }$$}
In the computations, we approximate the true $L=Q^{1 /n}$ by
$M=Q_n^{1/n}=Q_1$, which results in the systematic error
\eqn\errml{ M^k - L^k = -{k \over n}(u/2)^n \d^{n-2k} + \cdots\ ,}
so that for $k<n$ we can say that $L^k_-= -(u/2)^k \d^{-k} +\ldots\,$.

This suffices to calculate all the correlation functions
of the order parameters, which, up to a field redefinition
$\phi\dup_m \to (n+1)^{(m-1)/2n} \phi\dup_m$
and a partition function renormalization
$\log Z \to (n /2) (n+1)^{-1/n} \log Z$, can be written
\eqn\opresn{ \eqalign{
\langle \phi\dup_l \rangle &= \delta_{l,1} (n+1)x^{1+1/n}\ ,
\quad 1\leq l\leq 2n-2\cr
\langle \phi\dup_{2n-1} \rangle &= {1 \over 4}(n+1)(2n-1)x^2 \cr
\langle \phi\dup_l\, \phi\dup_m \rangle &= \delta_{l,m}\,nl\, x^{l/n}\ ,
\quad  1 \leq l,m \leq n-1 \cr
\langle \phi\dup_l\, \phi\dup_m\, \phi\dup_r \rangle &= lmr \,N_{lmr}\,
x^{-1+(l+m+r-1)/2n} \ , \quad 1 \leq l,m,r \leq n-1\ .\cr}}
We recover the KPZ scaling exponents $\Delta_l=(l-1)/2n$ \rKPZ\
(see sec.~\stcalg) attached to
each operator $\phi\dup_l$. Note the particularly simple form
of the three-point
function as compared with the rather cumbersome flat space expressions
\DOTFAT: coupling to gravity
has resulted in a dramatic simplification of the structure.
The coefficients $N_{lmr}\in \{0,1\}$ are the restrictions to the order
parameter space of the ordinary CFT fusion rules. They arise naturally in
the calculation as follows.

Using the symmetry of the three-point function under the interchange of $l$,
$m$, $r$, let us take $r \geq l,m$.
Then the first term in \kdvcorr c\ vanishes (the leading power of $\d$ is too
low to produce any residue), and we are left with
\eqn\leftwith{\langle \phi\dup_l\, \phi\dup_m\, \phi\dup_r \phi\dup_1 \rangle=
-2 \Res  (\bigl[L^m_+,[L^l_+,L^r_-]\bigr])\ .}
For the residue to be non-zero, we must have
\eqn\resnzc{ l+m \geq r+1\ ,}
and by symmetry, the cyclical permutations of \resnzc\ hold as well.

Comparing this with the restriction of the fusion rules of the CFT to
the diagonal of the Kac table, an additional condition, $l+m+r \leq 2n-1$,
must be satisfied for the three-point correlator not to vanish.
To derive this, let us take again some general $l,m,r$ with the
restrictions \resnzc\ and permutations.
We still have to evaluate \leftwith, where we can drop the $+$ subscripts
from $L^m,L^l$, since their negative part does not contribute to the residue.
The expression is further simplified by noticing that, on the sphere,
the derivative $\d$ acts only once, so that
$[L^l,L^r_-]=lL^{l-1} [ L,L^r_-]$,
and $\bigl[L^m,[L^l,L^r_-]\bigr]=ml\, L^{m+l-2} \bigl[L,[L,L^r_-]\bigr]$.
This gives the general form for the three-point correlator by symmetry,
$$\langle \phi\dup_l\, \phi\dup_m\, \phi\dup_r \rangle
= mlr\, F(l+m,r)=mlr\,G(l+m+r)\ .$$

It is now easy to compute the function $G$ for some
convenient values of $l,m,r$. In the case $r=n$ for example,
the $t_n$ flow is
trivial so the correlator vanishes. This means that $G$ vanishes whenever
its argument $l+m+r >2n$ (recall $l+m >r$). This is exactly the relevant fusion
rule.  To calculate the value of $G$ when it is non-zero, let us
take $l+m=r+1$ (the boundary of the fusion rules).
The leading contribution to \leftwith\ reads
\eqn\leadcont{ 2\bigl[\d^m,[\d^l, (u/2)^r\d^{-r}]\bigr]'
= 2lm \,\bigl((u/2)^r\bigr)'''\,\d^{-1}\ ,}
which, after integration over $x$, yields $G$ and \opresn\ follows.

A slight adaptation of this line of arguments yields the general $N$-point
function of order parameters on the boundary of the fusion rules,
i.e.\ $\langle \phi\dup_{m_1}\ldots \phi\dup_{m_N}\rangle$,
with $m_1+\ldots+m_{N-1} = m_N +N-2$.
We evaluate
\eqn\nptfr{ \eqalign{
\langle \phi\dup_{m_1} \ldots \phi\dup_{m_N} \rangle &= -2\int \d x\,
\Res\bigl[L^{m_1}_+,[L^{m_2}_+,[\ldots,L^{m_{N-1}}_+,L^{m_N}_-]\ldots\bigr] \cr
&= -2m_1 \ldots m_N \int \d x \Res  (L^{j_N-1}\bigl[L,[L \ldots [L,
L^{m_N}_- / m_N] \ldots \bigr])\cr
&= m_1 \ldots m_N \,G_N(m_N)\ .\cr}}
Let us now take $m_1=\cdots =m_{N-3}=1$. Then we use the three-point function
result
\opresn\ to identify $G_N= \partial_x^{N-3} x^{2-N+(m_N/n)}$, so that
\eqn\resnptkv{ \langle \phi\dup_{m_1}  \ldots \phi\dup_{m_N} \rangle =
m_1 \ldots m_N \partial_x^{N-3}  x^{s+N-3}\ ,}
where we recognize the KPZ scaling exponent
$s=\sum_{i=1}^N \Delta_{m_i}
-\gamma_{\rm str} -N+2= 2-N+(m_N / n)$ for the $N$-point function
(see sec.~\stcalg).

\subsec{An alternative method for solving the $(p,q)$ models on the sphere}
\subseclab\ssaamfs

The tree level solution for the $Q$ differential operator
of the $(n,n+1)$ models of previous subsection can be easily expressed
in terms of a variant of the Chebychev polynomials of the first kind
(see also \rMSeSt), as
\eqn\extche{ Q=v^{n/2}T_n\left(\d/\sqrt{v}\right)\ ,}
where
\eqn\tcheby{T_n(2 \cos \theta)=2 \cos (n \theta)\ ,}
and we have set $v(x)=-u(x)/2$.
This can be proven simply and compactly,
leading to interesting possible generalizations.

At tree level, the differential operators $P=P\bigl(v(x),\d\bigr)$ and
$Q=Q\bigl(v(x),\d\bigr)$, with respective degrees $p$ and $q$, can be rewritten
$$ P=v^{p/2}\, P(1,\d/\sqrt{v})\ ,\qquad Q=v^{q/2}\, Q(1,\d/\sqrt{v})\ ,$$
since we can neglect the non-commutation of $\d$ and $v(x)$ at this order.
In the spherical (i.e.\ semiclassical) approximation, the commutator
in the string equation \epqcc\ can be replaced by Poisson brackets
and thus becomes
$$\left({\partial P \over \partial \d}{\partial Q \over \partial v}
-{\partial Q \over \partial \d}{\partial P \over \partial v} \right)v'=1\ .$$

Let $P(z)\equiv P(1,z)$ and $Q(z)\equiv Q(1,z)$ denote the
two corresponding polynomials, with respective degrees $p$ and $q$.
The string equation now reads
\eqn\stpq{qP'(z)\, Q(z)-p Q'(z)\, P(z) = 2pq\ .}
Suppose $p>q$, then the polynomial $P$ is given by
$P(z)=\bigl(Q(z)^{p/q}\bigr)_+$, where again the subscript $+$
means we retain only the
polynomial part of the formal series expansion in $z$.
On the other hand, due to the differential equation \stpq,
we can also write
\eqn\intpasq{ P=2p\,Q^{p/q}(z) \int_0^z \d t\,Q^{-1-p/q}(t)\ .}

To find solutions to \stpq, let us try $Q=T_q(z)$
where the Chebychev polynomial $T_q(z)$
is defined in \tcheby. The formula \intpasq\ yields a polynomial
expression for $P$ if and only if $p$ is of the form
$p=(2m+1)q \pm 1$, for some integer $m$, in which case we have
\eqn\finvalp{P(z)=\sum_{l=0}^m {p/q \choose l} T_{p-2lq}(z)\ .}
We recover the unitary
solution of previous subsection when $m=0$, $p=q \pm 1$.
In addition, we see a very particular pattern emerge for the theories
$\bigl(q,(2m+1)q \pm 1\bigr)$ \refs{\rEyZJ,\rEyZJii}.
A general study of the equation \stpq\ remains to be performed.

\medskip
{\it The resolvent in the spherical limit.} A surprising and useful identity
can be proven in the spherical limit. Let us calculate the trace of the
resolvent $\omega(z)$ of one of the differential operators, e.g.\ $Q$:
$$\omega(z,x)=\tr (Q-z)^{-1} \sim -\int^x\d y\int{\d k\over 2\pi}
{1\over Q(ik,y)-z}\ .$$
The integral over $k$ is given by a residue. We thus find the relation
\eqn\eomQ{{\del \omega(z,x)\over \del x}{\del Q(ik,y) \over\del \d}=1\,,\quad
{\rm with}\quad  Q(ik,y)=z\ .}
The derivative of $\omega$ with respect to $x$
is taken at $z$ fixed. We can transform it into a derivative at $\d=ik$
fixed,
$$\eqalign{ \left.{\del \omega(z,x)\over \del x}\right|_z &= \left.{\del
\omega(z,x)\over \del x}\right|_{\d}+{\del \omega(z,x)\over \del \d}
\left.{\del \d \over \del x}\right|_z \cr &= \left.{\del
\omega(z,x)\over \del x}\right|_{\d}-{\del \omega(z,x) \over\del \d}
{\del Q\over\del x}\left({\del Q\over\del \d}\right)^{-1}\ .\cr}$$

Using this relation, eq.~\eomQ\ can be rewritten
$${\del \omega\over\del x}{\del Q\over\del \d}
- {\del \omega\over\del \d}{\del Q\over\del x} =1\ .$$
We recognize the Poisson brackets of $\omega$ and $Q$. We conclude that the
Poisson brackets of $\omega+P$ and $Q$ vanish. Therefore $\omega+P$ is a
function of $Q$, i.e.\ of $z$. Comparing the analytic
properties in $z$ of $\omega$ and $P$ we conclude that they can differ only
by analytic terms. We thus obtain the curious relation:
\eqn\eomegP{\omega_{\rm sing}(z,x)=-P(\d,x)\quad{\rm with}\quad Q(\d,x)=z\ .}

\medskip
{\it The loop average and the problem of the cosmological constant.}
In the multimatrix models, a surface with a boundary of length $s$ can be
created by considering any non-vanishing average of a trace of the product
of $s$ matrices. In the double scaling limit, all matrices become a linear
combination of the identity and the two matrices $P$ and $Q$. If we assume
$p>q$, then $Q$ is the leading operator and in the generic situation,  the
renormalized loop average for a loop of macroscopic length $s$ is thus
proportional to $\tr \e^{-s_{\rm ren} Q}$ with $s_{\rm ren}=sa^{-q\gs}$
(see the arguments of the end of \refsubsec\ssMp). In the spherical
approximation, the dimension of $Q$ is $Q\sim x^{q/(p+q-1)}$. Therefore the
coupling constant coupled to the area, i.e.\ the cosmological constant,
should scale as $x^{2q/(p+q-1)}$. From the scaling relations
\edimtm, we thus conclude that the parameter $t_{|p-q|}\equiv \mu$ plays the
role of the renormalized cosmological constant. In terms of $\mu$, the
partition function in the spherical approximation behaves as
$Z\sim \mu^{1+p/q}$ and thus the associated string susceptibility exponent is
\eqn\egscosc{\gs(\mu)=1-p/q\qquad {\rm for}\ p>q\ .}
The parameters $t_m$ have a dimension $t_m\sim
\mu^{(p+q-m)/2q}$. The dimension $\Delta_m$ of the corresponding operators
$\phi_m$ is then
\eqn\edimopcc{\Delta_m=1-(p+q-m)/2q=(q-p+m)/2q\ .}
These conclusions are consistent
with analysis of the continuum Liouville equation of \refsec\stcalg.

Note however that it is as well
possible to create boundaries corresponding to the
second operator $P$, and their interpretation is less clear.

\def\D{{\rm D}\,}
\def\KK{M. Kontsevich}
\def\Gs{{\sigma}}
\def\Gl{{\lambda}}

\newsec{Topological gravity}
\seclab\sTg

We saw in \refsubsec\ssRpai\ how operators could be defined in the framework
of the one-matrix model description of some two dimensional quantum
gravity theories. They should correspond to polynomials in traces
of powers of the hermitian matrix of the model, with leading
behavior ${\cal O}_n \propto {\tr}(M^{2n})$.
In a pioneering paper \WSTR,
Witten compared the structure of correlators of such operators,
obtained from the KdV flows
and string equation of one-matrix models, to his reinterpretation of
Donaldson's cohomological theory in terms of topological field theory,
referred to as topological gravity.
The latter has a precise mathematical definition as the
theory of intersection of classes (topological invariants)
of the moduli space of (punctured) connected
Riemann surfaces with arbitrary genus which we describe below.
The Virasoro constraints in the context of these topological models were
studied in \refs{\rDijkW,\DVV}\ and the addition of minimally coupled matter
was first provided in \KEKE\ (where the topological system that corresponds
to the multi-matrix models was presented). In this section we shall not 
directly follow this historical approach, but rather we shall focus on
the analysis of these models allowed by the later work of Kontsevich \KON.

\subsec{Intersection theory of the moduli space of punctured Riemann
surfaces}

More precisely, let ${\cal M}_{g,n}$ denote the moduli space of
Riemann surfaces $\Sigma$ of genus $g$ with
$n$ marked points (or punctures)
$x_1,\ldots,x_n$.
The basic operators of the theory are built from
exterior powers of the first Chern class $c_1({\cal L}_{(i)})$
of the holomorphic
line bundle ${\cal L}_{(i)}$, whose fiber is the cotangent space to
$\Sigma$ at the point $x_i$, considered as a complex one dimensional
vector space.
The intersection numbers attached to these operators are just integrals
over some compactification ${\bar{\cal M}}_{g,n}$ of
${\cal M}_{g,n}$ of their exterior product, and can be
thought of as  correlators,
\eqn\intersec{ \langle \Gs_{d_1} \Gs_{d_2} \cdots \Gs_{d_n} \rangle=
\int_{{\bar{\cal M}}_{g,n}} c_1({\cal L}_{(1)})^{\wedge d_1}
\cdots c_1({\cal L}_{(n)})^{\wedge d_n}\ .}
These numbers are positive rationals, and vanish unless the total
degree of the integrated form equals the dimension of the moduli space
i.e.\ $6g-6+2n=2\sum d_i$.
The puncture operator is ${\cal P}=\Gs_0$.
The effect of its insertion in a correlator is not trivial, since it
corresponds to fixing a point $x_i$ (i.e.\ in generally
covariant terms, one divides out by the reparametrization group
which leaves $x_i$ fixed).
The topological definition \intersec\ leads to a number of recursion
relations between correlators at various genera, which make their
computation a finite exercise.
Actually all correlators can be expressed
in terms of correlators involving only the puncture operator $\Gs_0$.
In this sense, the other operators $\sigma_n$, $n>0$, are
(gravitational) ``descendents"
of the puncture operator.

Due to striking resemblances to the one-matrix model correlators
as defined through KdV flows and string equation, Witten was lead to
conjecture the equivalence between topological gravity and two
dimensional quantum gravity in the one-matrix model form \Wun.

\subsec{The Kontsevich matrix model}

Although the abovementioned recursion relations enabled
computation of all the intersection numbers in principle, it seemed very
difficult to prove the relation to the KdV framework.
Dramatic progress emerged from M. Kontsevich's combinatorial treatment
of the problem, in which, upon introducing a ``cell decomposition" of the
moduli space ${\cal M}_{g,n}$, he was able to compute the
intersection numbers by explicitly performing the integrals of
exterior products of chern classes.
Skipping details of the mathematical construction \KON, we shall write just
the final answer for the ``free energy" generating function
$F(t_0,t_1,\ldots)$ of the intersection numbers \intersec, with
arbitrary genus and punctures, as follows.
For a given correlator as in \intersec, define
$n_i=\# \{ d_k,\ \big| \ d_k=i \}$, and rewrite
$\langle \prod {\Gs_i}^{n_i} \rangle=\langle \prod {\Gs}_{d_i}\rangle$.
The generating function $F$ reads
\eqn\genftg{F(t_0,t_1,\ldots)= \sum_{n_i \geq 0}
\prod_{i \geq 0} {t_i^{n_i} \over n_i!}
\left \langle \prod {\Gs_i}^{n_i}\right \rangle\ ,}
for arbitrary real parameters $t_0,t_1,\ldots$ .
The result of Kontsevich allows $F$ to be rewritten in the form
\eqn\konres{ F(t_0,t_1,\ldots)=\sum_{{\rm fatgraphs}\ \Gamma}
{(i/2)^{v(\Gamma)} \over |{\rm Aut}(\Gamma)|} \prod_{{\rm lines}
\ (ij)} {2 \over {\Gl_i+\Gl_j}} \ ,}
where the sum is over connected fatgraphs, or ribbon graphs
(i.e.\ a set of vertices joined by non-intersecting double lines, or ribbons,
each line carrying a ``color"
index $i=1,2, \cdots$, also summed over, as described after \esmft),
$v(\Gamma)$ and $|{\rm Aut}(\Gamma)|$ denote respectively
the number of vertices and the order of the automorphism group
of the connected fatgraph $\Gamma$, and
the product extends over all the double-lines in $\Gamma$,
carrying colors $i,j$.
The collection of parameters $\Gl_i$ is related to the $t_i$ by
\eqn\eclopti{t_i = -(2i-1)!!\, \sum_{k \geq 1} \Gl_k^{-2i-1}\ ,}
where $(2i-1)!!=(2i-1)\cdot(2i-3)\cdots3\cdot1$, and $(-1)!!=1$.
The expression \konres\ is of course very suggestive of the
Feynman graphical expansion of the one hermitian matrix integral discussed
in sec.~\sDsmmcl.

Suppose the number $N$ of colors is fixed, and as earlier take $N$ to
be the size of the hermitian matrix. Introduce the (real) diagonal
matrix $\Lambda$ with diagonal entries $(\Gl_1,\ldots,\Gl_N)$.
Kontsevich proved that the $N \times N$ hermitian matrix integral
\eqn\xikon{ \Xi_N(\Lambda)= {\int \d M\, \e^{\tr(i{M^3 / 6}-
{\Lambda M^2 / 2})}\over \int \d M\, \e^{-\tr({\Lambda M^2/ 2})}}}
has a well-defined $N\to \infty$ limit $\Xi(\Lambda)$
as an asymptotic series of the
variables $t_i=-(2i-1)!!\, \tr \Lambda^{-2i-1}$,
coinciding with $\e^{F(t_{\scriptstyle .})}$.
Moreover it is a so-called $\tau$--function for the
KdV flows, i.e.\ its evolution in terms of the parameters
$\theta_{2i+1}=\tr(\Lambda^{-2i-1})/(2i+1)=-t_i/(2i+1)!!$
is given by the KdV flows
\eqn\kdevol{
\eqalign{&Q=\d^2 -u \qquad u=-2 \partial_{\theta_1}^2
\log \Xi(\theta_{\scriptstyle .}) \cr
&\partial_{\theta_{2i+1}}Q = \bigl[(Q^{i+{1/ 2}})_+,Q\bigr]\ ,\cr}}
and it satisfies in addition the one-matrix model string equation
$$[P,Q]=1\ ,\quad {\rm with}\quad P=\sum_{k \geq 1} -(k+ \half)\theta_{2k+1}
(Q^{k-{1 / 2}})_+ \ .$$

These various statements will be proved in next subsections.
This enables the identification $t_{(k)}=\theta_{2k+1}$
between the scaling constants of the one--dimensional matrix model
and the $\theta$'s, which are proportional to the couplings
$t_k$ of the topological gravity theory.

Note that the fatgraphs involved in the summation \konres\ are
connected, hence the ``free energy" $F$ is related to the matrix
model partition function \xikon\ through the usual logarithm,
which extracts only the connected fatgraph contribution
to the Feynman expansion.
Of course analogous disconnected correlators can be defined, as
counting intersection numbers of the moduli space of
possibly disconnected Riemann surfaces, an interesting object
of study for mathematicians.
The disconnected free energy is then just the
Kontsevich integral $\Xi$.

Note also the important
distinction between the matrix integrals considered earlier
(e.g.\ \ecmm, \egpf\ or \eqmm), in which a double
scaling limit had to be taken (tuning the overall coefficient of the
polynomial potential {\it and\/} sending $N \to \infty$), and
the integral \xikon,
in which the potential is fixed and only cubic, and only $N\to\infty$.
Despite these apparent differences, Witten's conjecture is equivalent
to identifying the operator content of the double--scaled one
matrix hermitian model with that of Kontsevich's integral \xikon.

\subsec{Computing the Kontsevich integral}

The numerator of the Kontsevich integral \xikon\ is best reexpressed
after a change of variable $M \to M-i\Lambda$ as\foot{We summarize here
various techniques applied to the computation
of the Kontsevich integral. For more details, see for instance
especially \CJB, and \refs{\WIT\DIJ{--}\KMM}.}
\eqn\yyk{A_N(\Lambda)=
\int \d M\, \e^{i \tr({M^3 / 6}+{\Lambda^2 M / 2})}\ ,}
up to a multiplicative factor $\e^{\tr(\Lambda^3/3)}$, independent of $M$.
The now customary integration over angular variables leaves us
with an integral over the eigenvalues $m_i$ of $M$.
A direct computation of the denominator of \xikon, which is just
a trivial Gaussian integral, enables us to rewrite
\eqn\onedimk{\Xi_{N}(\Lambda)= \prod_{i=1}^N \int \d\mu_{\Gl_i}(m_i)
\prod_{1 \leq i <j \leq N}
{{m_i - m_j} \over {i(\Gl_i -\Gl_j)}}\ ,}
with the measure
$$\d\mu_{\Gl}(m)= \d m \left({ \Gl \over {2 \pi }}\right)^{1 \over 2}
\e^{i({m^3 / 6}+{\Gl^2 m / 2}+{(i\lambda)^3/ 3})}\ .$$

Let us use the notation
$$|x^0,x^1,\ldots,x^{N-1}|=\left\vert\matrix{1&x_1&\ldots &x_1^{N-1} \cr
1&x_2&\ldots&x_2^{N-1} \cr
.&.&\ldots&. \cr
1&x_N&\ldots&x_N^{N-1} \cr}\right\vert
=\Delta(x)$$
for the Vandermonde determinant \eVand.
By rebuilding a Vandermonde determinant out of integrals over the $m_i$'s,
the multiple integral \onedimk\
can be recast in a single determinantal expression
\eqn\finresk{\Xi_N(\Lambda)= {|\D^0z,\D^1z,\ldots,\D^{N-1}z| \over
|\Gl^0,\Gl^1,\ldots,\Gl^{N-1}|} }
where the function $z(\lambda)=\Xi_1(\Lambda\equiv \Gl)$ is the
one-dimensional Kontsevich integral, and $\D$ denotes the differential
operator $\Gl+{1 \over 2 \Gl^2}-{1 \over \Gl}\partial_{\Gl}$.
In particular, the function $z(\Gl)$ satisfies Airy's equation
\eqn\airy{(\D^2-\Gl^2) z(\Gl)=0\ .}
The function $z$ has the well known asymptotic expansion
\eqn\aircoef{z(\Gl)= \sum_{k=0}^{\infty} c_k\, \Gl^{-3k}\ ,
\quad{\rm with}\quad c_k=(-1/36)^k {(6k-1)!! \over (2k)!}\ .}
Let us define also the function
${\bar z}(\Gl)=\Gl^{-1}\D z=\sum d_k\, \Gl^{-3k}$ with
$d_k = c_k (6k+1)/(1-6k)$.
In the determinant appearing in the numerator
of \finresk, we can replace $\D^{2k}z$ and $\D^{2k+1}z$ respectively
by $\Gl^{2k}z$ and $\Gl^{2k+1}{\bar z}$, up to linear combinations
of columns of lower rank. When reexpressed in terms of the variables
$x_i=\Gl_i^{-1}$, this results in
$$\Xi_N(\Lambda)= {|x^{N-1}z,x^{N-2}{\bar z},\ldots| \over
|x^{N-1},x^{N-2},\ldots|}\ .$$
Expanding both functions, we get the final result in
terms of the coefficients $a_n^{(2p+1)}=c_n$, $a_n^{(2p)}=d_n\,$:
\eqn\kondet{ \Xi_N(\Lambda)= \sum_{n_1,\ldots,n_N \geq 0}
\left(\prod_{i=1}^N a_{n_i}^{(i)}\right)
{|x^{3n_1+N-1},x^{3n_2+N-2},\ldots,x^{3n_N}| \over
|x^{N-1},x^{N-2},\ldots,x^0|}\ .}

Recall now the definition of the characters of irreducible
representations of GL($N$), indexed by Young tableaux $(l_1,l_1,\ldots,l_N)$,
with the $i^{th}$ row of length $l_i$  ($l_1>l_2>\ldots>l_N$).
When evaluated on a matrix $X \in {\rm GL}(N)$, the characters read
\eqn\schupol{\chi\dup_{l_1,\ldots,l_N}(X)
={|x^{l_1+N-1},x^{l_2+N-2},\ldots,x^{l_N}|
\over |x^{N-1},x^{N-2},\ldots,x^0|}
=\det\bigl[p\dup_{l_i+j-i}(X)\bigr]_{1 \leq i<j \leq N}\ ,}
where $x={\rm diag}(x_1,\ldots,x_N)$ is the diagonalized version of $X$.
The result has been reexpressed in terms of the Schur polynomials $p_m(X)$,
themselves characters corresponding to the totally symmetric
representations of GL($N$), $p_m(X)=\chi\dup_m(X)$ (one row of $m$ boxes).
These can also be expressed in terms of traces of powers of the matrix $X$,
\eqn\schur{p_m(X) = \sum_{\scriptstyle{n_i \geq 0\ , \  i=1,2,\ldots} \atop
\sum \scriptstyle{i n_i=m}}\ \prod_{i \geq 1}
{\bigl(\tr(X^i)/i\bigr)^{n_i} \over n_i!}\ ,}
and $p_m\equiv 0$ if $m<0$. The definition of
these characters can be extended
to any ordering of the $l_i$'s by using the
determinantal expression in terms of Schur polynomials \schupol. By a slight
abuse of notation, we still call these functions
$\chi\dup_{l_1,\ldots,l_N}(X)$.
We readily see that $\Xi_N$ of eq.~\kondet\ is expressed in terms of these
generalized characters
\eqn\charsum{\Xi_N(\Lambda)= \sum_{n_i \geq 0, \ i=1,\ldots,N}
\left(\prod_{i=1}^N a_{n_i}^{(i)}\right)
\chi\dup_{3n_1,3n_2,\ldots,3n_N}(\Lambda^{-1})\ .}
In this form, the result is obviously a function of the
traces $\tr \Lambda^{-i}$, $i=1,2,\ldots$ (see the definition of
$\chi$ \eqns{\schupol{,\ }\schur}, with $X=\Lambda^{-1}$).

{\it Large N limit.}
The first question one may ask concerns the large $N$ limit: is
it well defined? The answer is surprisingly simple.
Note first that there is a natural gradation of the expansion
of $\Xi_N(\Lambda)$ in terms of the traces $\tr \Lambda^{-i}$,
to which we assign degree $i$. The
sum is over the characters $\chi\dup_{3n_1,\ldots,3n_N}$, which have
total degree $3 \sum n_i$.
For instance the first few terms in \charsum\ up to degree
$3$ read (recall that
$c_0=d_0=1$ and $c_1=-5/24$, $d_1=7/24$)
\eqn\examk{
\eqalign{ \Xi_N(\Lambda)&= c_0+c_1|p_3|
+c_0d_1 \left\vert\matrix{p_0&p_1\cr p_2&p_3\cr}\right\vert
+c_0d_0c_1\left\vert\matrix{p_0&p_1&p_2\cr
0&p_0&p_1\cr p_1&p_2&p_3\cr} \right\vert+\cdots\cr
&=1-{\tr(\Lambda^{-3})\over 24} - {\tr(\Lambda^{-1})^3 \over 6}
+ \cdots\cr}}
for $N \geq 3$ (we omitted the remainder of the $N \times N$ determinants
because they are just formed by some upper diagonal blocks
with diagonal elements $p_0=1$ and do not alter the result).
Note also that no  determinant of size $4$ or more with $p_3$ as last
diagonal element contributes to the degree $3$ piece of $\Xi_N$, because
one would have two identical lines in the corresponding matrices.
Note finally that the terms of the form
$\tr(\Lambda^{-1})\,\tr(\Lambda^{-2})$ have been cancelled out automatically,
leaving us with only negative odd powers of $\Lambda$ in the traces.

More generally, it is clear that the degree $3k$ contribution to
$\Xi_N(\Lambda)$ is independent of the value of $N$, provided
$N \geq 3k$: the matrix determinants which actually contribute must have
$p_0$ as $i^{th}$ diagonal term for $i >3k$, otherwise they would have
two identical lines.
This means that as $N$ grows, the contributions of given degree $3k$ to
$\Xi_N(\Lambda)$ stabilize as soon as $N \geq 3k$.
This enables to define
order by order the $N \to \infty$ limit $\Xi$
of $\Xi_N(\Lambda)$,
now considered as a function of the traces $\tr \Lambda^{-i}$, as
\eqn\finxi{\Xi=\sum_{n_i \geq 0, \ i=1,2,\ldots}
\left(\prod_{i \geq 1} a_{n_i}^{(i)}\right)
\chi\dup_{3n_1,3n_2,\ldots}(\Lambda^{-1})\ ,}
where only finitely many $n$'s are non zero for each term.

{\it Dependence on odd traces only.}
As already evident in the above example \examk, the traces of even powers
of $\Lambda^{-1}$ cancel out of $\Xi$.
This is a general fact.
To see why, introduce the variables $\theta_i=\tr(\Lambda^{-i})/i$,
and compute the derivative
$\partial_{\theta_{2i}}\Xi(\theta_{\scriptstyle .})$ of $\Xi$, considered
as a function of the $\theta$'s.
{}From the definition \schur, it is easy to see how the Schur polynomials
behave under differentiation:
$$ \partial_{\theta_{2i}} p_m = p_{m-2i}\ .$$
Let us differentiate eq.~\finxi\ w.r.t.\ $\theta_{2i}$, term by term.
Each  determinant yields a sum of terms of the form
$$\left\vert\matrix{p_{3n_1}&p_{3n_1+1}&p_{3n_1+2}&\ldots&p_{3n_1+N-1}\cr
.&.&.&\ldots&.\cr .&.&.&\ldots&. \cr p_{3n_s-s+1-2i}&p_{3n_s-s+2-2i}
&p_{3n_s-s+3-i}&\ldots&p_{3n_s-s+N-2i}\cr
.&.&.&\ldots&.\cr
p_{3n_N-N+1}&p_{3n_N-N+2}&p_{3n_N-N+3}&\ldots&p_{3n_N}\cr}\right\vert\ ,$$
for any $N\geq 3k=3\sum n_i$ (the degree of the term considered).
Now take $N\geq s+2i$ (and $N\geq 3k$). In that case, the $(s+2i)^{\rm th}$
line of this determinant is obtained from the $s^{\rm th}$ line by
exchanging the roles of $n_s$ and $n_{s+2i}$; but  the prefactor
$a_{n_1}^{(1)}\cdots a_{n_s}^{(s)}\cdots a_{n_{s+2i}}^{(s+2i)}\cdots$
is the same for both terms, and symmetric in the interchange
$s \leftrightarrow s+2i$
(the corresponding $a$'s are both $c$'s or $d$'s), whereas
the determinant is antisymmetric under the interchange of lines
$s$ and $s+2i$. Therefore the total contribution vanishes,
and $\Xi$ does not depend on the traces of even powers of
$\Lambda^{-1}$.
That the step be $2i$ is crucial for the prefactor to be
symmetric and not mix $c$'s and $d$'s: this is why only odd powers survive.

The compact expression \finxi\ certainly leads to a straightforward
computation of any intersection number \intersec\ (see \CJB\ for even
more powerful variants, and many examples).
Still the link with one--matrix model quantum gravity is far from
transparent at this point.

\subsec{Equivalence between topological gravity and one-matrix model}

{\it The KdV flows again.}
We now turn to the derivation of the KdV evolution equations for
the large $N$ limit of the Kontsevich integral
$\Xi(t_{\scriptstyle .})$, as a function of the
$t_i = -(2i-1)!!\ \tr \Lambda^{-2i-1}$.
First notice that eq.~\finxi\ allows us to rewrite $\Xi_N(\Lambda)$
as a single Wronskian determinant
\eqn\xwron{\Xi_N(\Lambda)=\left\vert\matrix{f_1&f_1'&\ldots&f_1^{(N-1)}\cr
f_2&f_2'&\ldots&f_2^{(N-1)}\cr
.&.&\ldots&.\cr
f_N&f_N'&\ldots&f_N^{(N-1)}\cr}\right\vert\ ,}
where $f'$ denotes $\partial_{\theta_1}f=\partial_{\tr\Lambda^{-1}} f$ and
\eqn\fdefk{ f_i=\sum_{n \geq 0} a_n^{(N+1-n)} p_{3n+i-1}\ .}

This form suggests that we introduce a differential operator $\Delta_N$,
whose action on a function $f$ is just the normalized $(N+1)\times(N+1)$
Wronskian
\eqn\wronsk{ \Delta_N f =\sum_{r=0}^N w_r\, \d^{N-r}(f)=
{1 \over \Xi_N(\Lambda) }\left\vert\matrix{
f&f'&\ldots&f^{(N)}\cr f_1&f_1'&\ldots&f_1^{(N)}\cr
.&.&\ldots&.\cr f_N&f_N'&\ldots&f_N^{(N)} \cr }\right\vert\ ,}
where $\d=\partial_{\theta_1}$, and we used the fact that
\eqn\schuprop{
\partial_{\theta_i}p_m=p_{m-i}=\partial_{\theta_1}^i p_m\ .}
Write $\Delta_N=W_N\, \d^N$, $W_N=\sum_{r \geq 0} w_r\, \d^{-r}$.
The functions $w_r$ are just minors of the determinant \wronsk.
In particular, $w_0=1$ by the choice of normalization, and
we find
\eqn\wone{ w_1= -\partial_{\theta_1} \log \Xi_N(\Lambda)\ ,}
as a direct result of the action of $\partial_{\theta_1}$
on the columns of the determinant \xwron.
Each of the coefficients $w_r$ enjoys for obvious reasons the
same property as $\Xi_N$ to be stabilized order by order as $N$ grows.
As the limit $N \to \infty$ of $W_N$, the pseudo--differential operator
$W=\sum w_r \d^{-r}$ of degree $0$ is therefore well-defined.

Anticipating the final result, we plan to show that
the degree $1$ pseudo-differential operator $L=W\d W^{-1}$ satisfies
the KdV evolution equations \kdevol\ with $Q=L^2$, in the
variables $\theta_{2i+1}=\tr\Lambda^{-2i-1}/(2i+1)$.
Let us show first that
\eqn\partresk{
\partial_{\theta_i} \Delta_N= Q_{i,N}\Delta_N-\Delta_N \d^i,}
with $Q_{i,N}$ some differential operator of order $i$.
For the right--hand side of this equation to be a differential operator
of degree at most $N-1$, it is necessary that
$$Q_{i,N}=(\Delta_N\, \d^i \Delta_N^{-1})_+\ .$$
Moreover for the equation \partresk\ to hold, it is necessary and sufficient
that it be satisfied when acting on $N$ linearly independent functions.
Choosing the basis of the $f_j$'s of eq.~\fdefk, which are by definition
in the kernel of $\Delta_N$, gives
$$\Bigl(\partial_{\theta_i}(\Delta_N)+\Delta_N \partial_{\theta_1}^i
-Q_{i,N} \Delta_N \Bigr)f_j=\partial_{\theta_i}(\Delta_N f_j)=0\ ,$$
where we used again the property \schuprop\ to trade
$\partial_{\theta_1}^i$ for $\partial_{\theta_i}$. This completes the proof
of the statement \partresk.

Translating eq.~\partresk\ in terms of $W$ and $L=W \d W^{-1}$,
and taking $N \to \infty$, results in
\eqn\kdvkonw{ \partial_{\theta_i} L = \bigl[(L^i)_+,L\bigr]\ .}
For the same reasons as above, the coefficients $w_r$ of $W$
do not depend on the even traces $\theta_{2i}$, which amounts to the
KdV restriction that $Q=L^2$ is itself differential.\foot{To see why,
one can restrict the expressions of $f_j$ to depend only on the
odd $\theta$'s, by setting $\theta_{2s}=0$.
This does not alter the result \xwron, but allows writing
$\partial_{\theta_{2s}}(\Delta_N f_j)=\partial_{\theta_{2s}}(\Delta_N) f_j=0$,
which implies that $\partial_{\theta_{2s}}\Delta_N=0$ and consequently
that $\partial_{\theta_{2s}}W=0$.}
Writing $Q=\d^2 -u$, and using the values of $w_0=1$ and
$w_1=-\partial_{\theta_1} \log \Xi$ of eq.~\wone, with $N\to \infty$,
we finally identify
$$ u=-2\, \partial_{\theta_1}^2 \log \Xi\ .$$
This completes the proof of eq.~\kdevol.
The evolution of the generating function $F$ as a function
of the $\theta$'s is therefore dictated by the KdV flows.

{\it The string equation.}
Let us start from the expression \yyk\ of the matrix
Airy function $A_N(\Lambda)$, numerator of the
Kontsevich integral, after the change of variables $M \to M -i \Lambda$.
The string equation will appear as a particular case of equations
of motion for this integral.
We first have
$$\eqalign{
0 &= 2 \int \d M {\partial \over \partial M_{kk}} \e^{i\tr({M^3 / 6}
+{M \Lambda^2 / 2})} \cr
&=\int \d M \bigl((M^2)_{kk}+\Gl_k^2\bigr)\,
\e^{i\tr({M^3 / 6}+{M \Lambda^2 / 2})}\cr
&\equiv \bigl< (M^2)_{kk} + \Gl^2_k \bigr>\ .\cr}$$
On the other hand, expressing the invariance of the integral under the
infinitesimal change of variables $M \to M+i\epsilon[X,M]$,
with $X_{ij}=\delta_{i,k}\,\delta_{j,l}\, M_{kl}$, gives to first order
in $\epsilon$,
$$ 0= \Bigl< M_{ll} -M_{kk}
+{i \over 2}(\Gl_k^2-\Gl_l^2)M_{kl}M_{lk} \Bigr>\ ,$$
where the first term is the contribution of the Jacobian
$J=1+ i \epsilon(M_{ll}-M_{kk})+O(\epsilon^2)$, and the second
comes from the exponential.
Collecting both equations, we deduce
$$0= \Bigl< \Gl_k^2 +(M_{kk})^2-2i \sum_{l \neq k}
{M_{kk} -M_{ll} \over \Gl_k^2 -\Gl_l^2} \Bigr>\ .$$

An insertion of $M_{kk}$ in $A_N(\Lambda)$ is generated by
differentiation w.r.t.\ $\Gl_k$.
More precisely, we have
\eqn\matairy{\Bigl( \Gl_k^2 -(\Gl_k^{-1}\partial_{\Gl_k})^2
-2\sum_{l \neq k}{1 \over{\Gl_k^2 -\Gl_l^2} }
(\Gl_k^{-1}\partial_{\Gl_k} - \Gl_l^{-1}\partial_{\Gl_l}) \Bigr)
A_N(\Lambda)=0\ ,}
referred to as matrix Airy equations, generalizing the ordinary Airy
equation \airy.
This results straightforwardly in a differential equation
\eqn\consvir{
\eqalign{\Bigl( \Gl_k^{-2}(\sum_l \Gl_l^{-1})^2 +&{1\over4}\Gl_k^{-4}+
2 \sum_{l \neq k}{1 \over{\Gl_k^2 -\Gl_l^2} }
(\Gl_k^{-1}\partial_{\Gl_k} - \Gl_l^{-1}\partial_{\Gl_l}) \cr
&-2(1+\Gl_k^{-2}\sum_l {1 \over \Gl_k+\Gl_l})\partial_{\Gl_k}
+(\Gl_k^{-1} \partial_{\Gl_k})^2 \Bigr) \Xi =0}}
for the quantity
$$\Xi=\lim_{N \to \infty} {\e^{\tr(\Lambda^3/3)}
A_N \over \int \d M\, \e^{ -\tr (\Lambda M^2/2)}}\ .$$

Using $\Gl_k$ as expansion parameter gives an infinite set
of constraints, order by order in $\Gl_k$, of the form
$$2 \sum_{m=-1}^{+\infty} \Gl_k^{-2(m+1)} L_m\, \Xi =0\ ,$$
with $L_m$ some differential operators of the $\Gl$'s, expressible
in terms of the $\theta_{2i+1}=\tr \Lambda^{-2i-1}/(2i+1)$.
These operators were found to form a piece of a representation
of the Virasoro algebra, and the above constraints are often
called Virasoro constraints.
Let us stress again that they express nothing but the equations of motion
(reparametrization invariance) of the matrix model, and as we have
seen in \refsubsec\ssLeVc, their
appearance is quite general in the framework of matrix models

Concentrate on the first constraint $L_{-1}$, obtained as the
leading order term in \consvir,
\eqn\lminone{ L_{-1} \Xi = \left[\frac{1}{4}\theta_1^2
-\half \partial_{\theta_1} - \textstyle{\sum_{k=1}^{\infty}}
(k+ \half) \theta_{2k+1} \partial_{\theta_{2k-1}}\right]\Xi=0\ .}
Dividing by $\Xi$ and differentiating twice w.r.t.\ $\theta_1$ gives
$$ \Bigl(- \half \partial_{\theta_1}
-\textstyle{\sum_{k=1}^{\infty}}(k+ \half)
\theta_{2k+1} \partial_{\theta_{2k-1}}\Bigr)
(-2 \partial_{\theta_1}^2 \log \Xi)=1\ , $$
which after the trivial translation $\theta_3 \to \theta_3 -1/3$ can be
recast by application of the evolution equations \kdvkonw\
into the familiar one-matrix model string equation \ePQ: $[P,Q]=1$, with
$$P= \sum_{k=1}^{\infty} -(k+ \half) \theta_{2k+1}
\bigl(Q^{k-{1 \over2}}\bigr)_+\ .$$

This completes the proof of the equivalence between topological gravity
as intersection theory of the moduli space of connected punctured
Riemann surfaces and one hermitian matrix models
in the double scaling limit. We stress
that the equivalence at this point holds {\it only at a perturbative level\/}
since we demonstrated only an identity between two perturbative expansions,
namely that of the free energy interpolating between multicritical points
of the one-matrix model and the generating function $F$
for intersection numbers \intersec.

\subsec{Polynomial averages and observables}

The Kontsevich integral \xikon\ possesses a number of other remarkable
properties. Among those, two concern interpretations of
polynomial averages taken with the Gaussian part of the
Kontsevich weight.
Such quantities are the natural counterparts of the operators
usually defined in the context of matrix models, as insertions of sources
in the Feynman expansion of the model. Typically, inserting $\tr M^k$
amounts to restricting to ``triangulations" surrounding a given
polygon with $k$ edges. The net effect of this is interpreted
in the double scaling limit as an operator creating a microscopic
hole in the Riemann surfaces.
It is therefore interesting to identify such operators
in the framework of the Kontsevich integral.

{\it Gaussian polynomial averages.}
We consider the average
\eqn\konmap{ \bigl< P(M) \bigr> =
{\int \d M\, P(M)\, \e^{-\tr({\Lambda M^2 / 2})}
\over \int \d M\, \e^{-\tr({\Lambda M^2 / 2})}}\ ,}
where $P(M)$ is some polynomial of the traces of {\it odd\/}
powers\foot{The restriction to odd powers is crucial in the following.
It is indirectly related to the underlying KdV structure induced
by the Gaussian weight, which only retains traces of odd powers
of $\Lambda^{-1}$.} of $M$.

In \KON, Kontsevich proved the following statement
(see also \PCJB\ for an alternative algebraic proof).
For any polynomial $P$ of odd traces of $M$, the quantity
$Q(\Lambda^{-1})=\langle P \rangle$ defined in eq.~\konmap\
is itself a {\it polynomial}\/ of the odd traces of $\Lambda^{-1}$.
This defines a map
$$\eqalign{ K&: \pi[X] \to \pi[X] \cr
&P \to Q(\Lambda^{-1})=\bigl< P(M) \bigr> \cr}$$
from the set $\pi[X]$ of polynomials of odd traces of $X$ to itself.
In \PCJB, this mapping was constructed explicitly using a
convenient basis of $\pi[X]$, formed by generalized Schur polynomials
(see also \PPK\ for another presentation).
Attaching the degree $i$ to the variables
$\theta_i(X)=\tr X^i / i$, the polynomial $Q$ has a total degree
${\rm deg} (Q) = {\rm deg} (P) /2$.

Now we see that the Kontsevich integral \xikon\  may be expressed as
$$\Xi_N(\Lambda)= \sum_{n=0}^{\infty} {(-1/36)^n \over (2n)!}
\bigl< (\tr M^3 )^{2n} \bigr>\ .$$
Order by order in $M$, we therefore find an asymptotic expansion
with degree $3n$ polynomials in odd traces of $\Lambda^{-1}$
(the third order term is computed in eq.~\examk). This gives
an alternative proof of the fact that \xikon\ only depends
on odd traces of $\Lambda^{-1}$.

{\it Observables versus polynomial averages in the Kontsevich model.}
{}From our knowledge of the usual matrix models, we might expect that
the actual correlators of topological gravity be realized
as polynomial averages within the framework of the Kontsevich integral.
The final picture is a little more involved, however, since
this turns out to be true only in the disconnected case
(i.e.\ where the free energy of the theory is directly
the Kontsevich integral \xikon\ and not its logarithm).
We have the following result, conjectured in \WIT\ and proved
constructively in \PCJB, using the same basis as above:
For any polynomial $R(\partial_{\theta_{.}})$ of derivatives
w.r.t.\ $\theta_{2i+1}\equiv \tr\Lambda^{-2i-1}/(2i+1)$, there
exists a polynomial $P(M)$ of odd traces of $M$ such that
\eqn\eRdxi{\eqalign{
R(\partial_{\theta_{.}})\,\Xi(\theta_{.})
&= \bigl< P(M) \e^{i\tr({M^3 / 6})} \bigr> \cr
&={\int \d M\,  P(M) \e^{\tr(i{M^3 / 6}-
{\Lambda M^2 / 2})}\over \int \d M \,\e^{-\tr({\Lambda M^2 / 2})}}
\ .\cr}}

This shows that the intersection numbers of the moduli space of possibly
disconnected (punctured) Riemann surfaces have a representation in
the Feynman diagrammatic expansion of their generating function $\Xi$
as insertion of polynomial sources. This renders even deeper the
connections between the double--scaled one-matrix model and the
Kontsevich matrix model. In particular, it was pointed out
in \WIT\ that these polynomial representations of observables
lead to an interesting generalization of short distance operator products
in the topological framework.

\subsec{Generalization: multi-matrix models and topological field theory}

The Kontsevich integral may be generalized \KON\ to include higher
degree potentials. Actually one gets a $\tau$ function for the
$q^{th}$ reduction of the KP hierarchy by taking
\eqn\genekon{ \Xi_N^{(q)}(\Lambda) ={ \int \d M\, \ee{{i(q^2+1)\over
2(q+1)} \tr(M+i^{q+1} \Lambda)^{q+1}\big\vert_{\geq 2} } \over
{\int \d M \,\ee{{i(q^2+1)\over 2(q+1)}
\tr (M+i^{q+1} \Lambda)^{q+1}\big\vert_{= 2} }}}\ ,}
where the subscript $\geq 2$
in the numerator means we omit the constant and linear
terms in $M$ in the expansion of the polynomial, and the
subscript $=2$ in the denominator means we only keep the  terms
quadratic in $M$ in the expansion of the polynomial.
The logarithm of this integral, considered as a function
of the traces of negative powers of $\Lambda$,
is the generating function of a set of correlators generalizing the
intersection numbers \intersec.
These form a topological field theory coupled to topological gravity
as axiomatized by Witten \refs{\WSTR,\DWIT}.
Instead of just one ``primary" operator $\Gs_0$ and its gravitational
descendents $\Gs_n$, $n>0$, the theory possesses a set of
$q-1$ ``primaries" ${\phi}_1,\ldots,{\phi}_{q-1}$, the first of
which is the puncture operator $\phi_1={\cal P}$, and their
``gravitational descendents", whose correlators are expressed
through recursion relations in terms of
correlators involving only primaries.
An explicit realization of these theories was given  \refs{\KEKE,\DVV}
in the framework
of $N=2$ superconformal theories, made topological (i.e.\ correlators
do not depend any longer on the points of insertion) by a twist of the
stress tensor \EY.

Remarkably, these gravitational primaries can be identified with
the $q-1$ order parameter fields of the minimal $(p,q)$ models
coupled to gravity in the framework of the double scaling limits of
$q-1$ matrix models discussed in sec.~\sqmomm. Actually the whole set of
correlators defined through $F=\lim_{N \to \infty}\log\Xi_N^{(q)}$
coincide with that of the gravitationally dressed operators of the
$(p,q)$ minimal theories coupled to gravity.
The proof of this equivalence goes essentially
as in the previous $q=2$ case.

By computing the integral \genekon\ in terms of generalized
GL($N$) characters \schupol, or alternatively
as a Wronskian determinant, one can establish the following results.

First of all, the $N \to \infty$ limit of \genekon\ is well defined
since the same phenomenon occurs as in the $q=2$ case:
terms with given degree in the variables
$\theta_i =\tr \Lambda^{-i} /i$ stabilize as $N$ grows (they become
independent of $N$ for sufficiently large $N$).

{\it q--reduced KP flows.}
The integral \genekon\ is a $\tau$--function for the
$q^{th}$ reduction of the KP hierarchy: i.e.\ suitably defined,
the differential operator $Q=\d^q -{q \over 2} u \d^{q-2} +\ldots$
with $u=-2 \partial_{\theta_1}^2 \log \Xi^{(q)}$ is
found to evolve with the
$\theta_i=\tr \Lambda^{-i} /i$ as
$$ \partial_{\theta_i} Q = \bigl[(Q^{i/q})_+,Q\bigr]\ .$$
This immediately implies that the generalized Kontsevich integral
\genekon\ does not depend on the $\theta_{qi}$, $i=1,2,\ldots$

{\it q--string equation.}
Moreover, the equations of motion for the matrix model \genekon\
can be recast into Virasoro constraints, the first of which
imply the $q-1$ matrix model string equation $[P,Q]=1$, as in
eqs.~\eqns{\epqcc{,\ }\forp{,\ }\pqstreq}, with
$$ P=\sum_{\scriptstyle{k\geq 1;}\atop \scriptstyle{k \neq 0 \pmod{q}}}
\!-(1+{k / q})   \theta_{k+q} (Q^{k / q})_+\ .$$

There is therefore an exact perturbative equivalence
between $(p,q)$ models
coupled to gravity within the framework of double--scaled
$q-1$ matrix models and the $q^{th}$ theory of topological field theory
coupled to topological matter realized through the
generalized Kontsevich integral \genekon.

There are still many open questions regarding these generalizations.
In particular, no generalization of the disconnected case observables
as polynomial averages is known yet, nor any equivalent of
Kontsevich's observation about the Gaussian averages in the case $q=2$.
More topological field theories are known, providing possible candidates
for the coupling of minimal models with W--symmetry (larger
symmetry algebras, including the Virasoro symmetry, for a review
see e.g.\ \rpoperev) to W--gravity
(a higher tensorial form of ordinary gravity, which is a metric
theory), and no matrix models have been found yet to represent
the corresponding free energies.


\def\ap{{\alpha_+}}
\def\am{{\alpha_-}}
\def\ao{{\alpha_0}}

\newsec{The continuum approach: Liouville gravity}
\seclab\stcalg

Thus far we have solved some toy examples of 2D quantum gravity formulated as
matrix models. The original 2D string theory with a target space comprised of a
$D<1$ conformal field theory (coupled to a Liouville mode)
is not guaranteed to admit such a formulation.
In the pure gravity ($D=0$), Hard Dimer ($D=-22/5$) \rstaud, Ising ($D=1/2$)
\refs{\rIYL,\rising}, and tricritical Ising \rtric\
cases, there is a complete line of arguments leading directly
from the discrete string
formulation of the statistical models coupled to gravity
to a matrix model
solvable by the orthogonal polynomial techniques of sections \stomm--\sqmomm.
In general, however, the link between the matrix model solution and
the continuum string is not necessarily proven.

In support of this identification, however, a number of features already known
from continuum calculations,
such as scaling exponents, initially helped to identify which matrix model
was a good candidate to describe which string theory.
But experience showed that a limited set of exponents was not sufficient to
distinguish even between solutions to different matrix models (recall
that the Ising model
and the Hard Dimer model share the same string susceptibility exponent).

In this section, we will derive general spherical
correlators for the operators
of the continuum theory (a string in $D<1$ dimension),
using the framework of Liouville theory.
Comparison with the results of \refsubsec\sssotumos\ will show a complete
agreement for the unitary minimal series coupled to gravity,
further establishing the link between 2D strings and double-scaled
matrix models.

\subsec{Liouville gravity and conformal matter}

Starting from the original Polyakov string action \eString\ in flat
$D$-dimensional Euclidean space,
\eqn\polst{{\cal S}_M(X,g)=
{1\over2\pi}\int\sqrt{g}\,g^{ab}\partial_a X^i\partial_b X^i}
($i=1,\ldots,D$), the most convenient prescription to quantize
this generally covariant two dimensional system is to fix a conformal gauge
$g_{ab}=\e^\varphi\hat{g}_{ab}$. The system is then described by the
Liouville mode $\varphi$ and space coordinates $X^i$,
living in the background metric $\hat g$.
The gauge fixing also introduces reparametrization ghosts
$b,c$ with spins $2, -1$ respectively, which we henceforth omit since we
are interested only in zero ghost number operators.
The effective action is a sum of three pieces: the ghost action, the matter
action \polst\ in the background metric ($g \to \hat g$), and the Liouville
action:
\eqn\liouac{ {\cal S}_L= { 1 \over 2 \pi}
\int \sqrt{\hat g}\left( {\hat g}^{ab}
\partial_a \varphi \partial_b \varphi
- {Q \over 4}{\hat R} \phi+ 2 \mu\, \e^{\ap \varphi}\right)\ .}

In the following we will concentrate on peculiar matter theories, made of one
free boson, with action:
\eqn\matter{ {\cal S}_M={1 \over 2\pi}\int
\sqrt{\hat g} \left( {\hat g}^{ab}
\partial_a X \partial_b X +{{i \ao} \over 2} {\hat R} X \right). }
\noindent{}This is nothing but the Feigin--Fuchs \FF\
representation of a conformal
field theory with central charge $c=1-12 \ao^2$, therefore meriting the
description as $D=c \leq 1$ conformal matter.
For comparison with the results of matrix models of \refsubsec\sssotumos,
$\ao$ has to
take the discrete values $1/\sqrt{2n(n+1)}$. In general, minimal matter is
obtained by taking $\ao^2$ rational. (Note that the above representation
also requires an explicit truncation of the spectrum to treat only the
self-contained set of states of interest.)

The various parameters in eqs.~\eqns{\liouac{,\ }\matter} are fixed by
requiring that the total action
${\cal S}={\cal S}_L+{\cal S}_M+{\cal S}_{\rm gh}$
be independent of
the choice of $\hat g$. This can be recast into a BRST invariance condition
on ${\cal S}$, with the BRST charge $\oint c \,T$, where $T$ is the total
energy-momentum
tensor of the system, and $c$ the dimension $-1$ ghost. The result is:\foot{For
recent review, see \refs{\rGiMo,\rdhok}.}
\eqn\param{ Q= \sqrt{{25-D} \over 3} \quad ; \quad \ap=-{Q \over 2}+
\sqrt{{1-D} \over 12}\ .}

The center of mass of the string is described by the ``tachyon" field
\eqn\tach{ {\cal T}_k = \e^{ikX+\beta(k)\varphi}\ , }
where the Liouville momentum $\beta$ is related to the matter momentum $k$
by BRST invariance (i.e., that the 2D integral $T_k=\int {\cal T}_k$ not depend
on the choice of background metric $\hat g$). This translates to the condition
that ${\cal T}_k$ is a dimension (1,1) operator:
\eqn\dimone{ {1 \over 2}k(k-2\ao) -{1 \over 2} \beta(\beta+Q) =1\ , }
resulting in the mass shell condition:
\eqn\lioumom{ \beta(k)= -{Q \over 2} + |k-\ao|\ . }

Note that the quadratic equation \dimone\ admits two branches of solutions.
In the semi-classical approach of \NATI, the choice with the plus
sign in front of the absolute value is chosen as follows.
The Liouville coordinate can be considered as a time variable,
and the (Wheeler--DeWitt)
wavefunction reads $\Psi \propto \exp(Q\varphi/2) \, {\cal T}_k$.
In this language, we have retained only the states of positive energy
$E=\beta+Q/2$.\foot{Analogously,
the matter part of the wave function has momentum $p=k-\ao$,
so the mass-shell condition \lioumom\ implies
$\bigl(\beta(k)+Q/2\bigr)^2-(k-\ao)^2=E^2 - p^2=0$, and hence describes
massless propagation as appropriate for a ``tachyon'' ${\cal T}_k$ at $D=1$.
For $D<1$ the gravitational dressing as well admits an interpretation
as a target space on-shell condition, with $E^2-p^2=m^2=(1-D)/24$.
By abuse of terminology, we refer to the massless mode at $D=1$ as the tachyon
since it becomes tachyonic for $D>1$.}
The insertion of such operators in a correlator results in local disturbances
of the surface, due to the infinite peak of the wave function at small
geometries, $\varphi\to \infty$. The negative energy states do not correspond
to
local disturbances of the surface, and it was argued in \NATI\
that they cannot exist. (For a recent survey of the situation, see e.g.\
\rGiMo. The truncation is as alluded to after \matter.)
As an example, the cosmological constant or ``identity" operator present
in the Liouville action, is ${\cal T}_{k=0}$, with $\beta(0)=\ap$.

\subsec{Dressed weights}
\subseclab\ssDw

It is now straightforward to derive some of the continuum results
that compare with the matrix model results of sections \sDsmmcl--\sqmomm.
Note that in \liouac, $\ap$ is determined by the requirement that the
physical metric be $g={\hat g}\,\e^{\ap\ph}$. Geometrically, this means that
the area of the surface is represented by
$\int \d^2\xi\,\sqrt {\hat g}\,\e^{\ap\ph}$.
$\ap$ is thereby determined by the requirement that $\e^{\ap\ph}$ behave
as a (1,1) conformal field (so
that the combination $\d^2\xi\,\e^{\ap\ph}$ is conformally invariant). For
the energy-momentum tensor $T=-\half\del\ph\del\ph-{Q\over2}\del^2\ph$
derived from \liouac, the conformal weight\foot{Recall that $\Delta$
is given by the
leading term in the operator product expansion $T(z)\,\e^{\ap\ph(w)}\sim
{\Delta\,\e^{\ap\ph}/(z-w)^2}+\ldots\ $.
Recall also that for a conventional
energy-momentum tensor $T=-\half\del\ph\del\ph$, the conformal weight of
$\e^{ip\ph}$ is $\Delta=\overline \Delta=p^2/2$.} of $\e^{\ap\ph}$ is
\eqn\ecw{\Delta(\e^{\ap\ph})
=\overline\Delta(\e^{\ap\ph})=-\half\ap(\ap+Q)}
(as in \dimone\ for $\beta$).
Requiring that $\Delta(\e^{\ap\ph})=\overline \Delta(\e^{\ap\ph})=1$
determines that
\eqn\eQgam{Q={-2/\ap}-\ap\ .}
Substituting $Q=\sqrt{(25-D)/3}$ from \param\ and solving for $\ap$ gives
\eqn\ealph{\ap=-{1\over\sqrt{12}}\bigl(\sqrt{25-D}-\sqrt{1-D}\bigr)=
-{Q\over 2}+\ha\sqrt{Q^2-8} \ ,}
%
verifying \param.

A useful critical exponent that can be calculated in this formalism
is the string susceptibility $\gs$ of \elpoa.
We write the partition function for fixed area $A$ as
\eqn\eZA{Z(A)=\int\CD\ph\,\CD X\ \ee{-S}\
\delta\Bigl({\textstyle\int} \d^2\xi\,\sqrt {\hat g}
\,\e^{\ap\ph} - A\Bigr)\ ,}
where for convenience we group the ghost determinant and integration over
moduli into $\CD X$. We define a string susceptibility $\gs$ by
\eqn\eZainf{Z(A)\sim A^{(\gs-2){\chi/2}-1}\ ,\quad A\to\infty\ ,}
and determine $\gs$ by a simple scaling argument. (Note that for
genus zero,
we have $Z(A)\sim A^{\gs-3}$.)
Under the shift $\ph\to\ph+\rho/\ap$ for $\rho$ constant, the measure in
\eZA\ does not change. The change in the action \liouac\ comes from the term
$${Q\over 8 \pi}\int \d^2\xi\,\sqrt {\hat g}\, \hat R\,\ph\to
{Q\over 8 \pi}\int \d^2\xi\,\sqrt {\hat g}\, \hat R\,\ph +
{Q\over 8 \pi}{\rho\over\ap}\int \d^2\xi\,\sqrt {\hat g} \hat R\ .$$
Substituting in \eZA\ and using the Gauss-Bonnet formula
${1\over4 \pi}\int \d^2\xi\,\sqrt {\hat g} \hat R=\chi$ together with the
identity $\delta(\lambda x)=\delta(x)/|\lambda|$ gives
$Z(A)=\e^{+{Q \rho \chi/ 2\ap}-\rho}\,Z(\e^{-\rho}A)$.
We may now choose $\e^\rho=A$, which results in
$$Z(A)=A^{+Q\chi/2\ap-1}\,Z(1)= 
A^{(\gs-2){\chi/2}-1}\,Z(1)\ ,$$
and we confirm from \param\ and \ealph\ that
\eqn\egam{\gs=2+{Q\over\ap}
={1\over12}\bigl(D-1-\sqrt{(D-25)(D-1)}\,\bigr)\ .}

Recall from the comments at the end of \refsubsec\sstagpf\ that
minimal conformal field theories are specified by a pair of
relatively prime integers $(p,q)$ and have central charge
$D=c_{p,q}=1-6(p-q)^2/pq$.
The unitary discrete series, for example,
is the subset specified by $(p,q)=(m+1,m)$.
After coupling to gravity, the general $(p,q)$ model has critical exponent
$\gs=-2|p-q|/(p+q-|p-q|)$ (as calculated in Liouville theory, i.e.\ with
respect to the area dependence of the partition function).
Notice that $\gs=-1/m$ for the values $D=1-6/m(m+1)\,)$
of central charge in the unitary discrete series.
Notice also that \egam\
ceases to be sensible for $D>1$, an indication of a ``barrier'' at
$D=1$.\foot{The ``barrier'' occurs when coupling gravity to $D=1$ matter in
the language of non-critical string theory, or equivalently in the case of
$d=2$ target space dimensions in the language of critical string theory.
So-called non-critical strings (i.e.\ whose conformal anomaly is compensated by
a Liouville mode) in $D$ dimensions can always be reinterpreted as critical
strings in $d=D+1$ dimensions, where the Liouville mode provides the additional
(interacting) dimension. (The converse, however, is not true since it is not
always possible to gauge-fix a critical string and artificially disentangle the
Liouville mode.)}

\smallskip
\noindent{\it Dressed operators / dimensions of fields}
\par\nobreak
Now we wish to determine the effective dimension of fields after coupling
to gravity. Suppose that $\Phi_0$ is some spinless primary field in a
conformal field theory with conformal weight $\Delta_0=\Delta(\Phi_0)
=\overline\Delta(\Phi_0)$ before coupling to gravity.
The gravitational ``dressing'' can
be viewed as a form of wave function renormalization that allows $\Phi_0$
to couple to gravity. The dressed operator $\Phi=\e^{\beta\ph}\Phi_0$ is
required to have dimension (1,1)
so that it can be integrated over the surface $\Sigma$ without breaking
conformal invariance. (This is the same argument used prior to \ealph\ to
determine $\ap$). Recalling the formula \ecw\ for the conformal weight of
$\e^{\beta\ph}$, we find that $\beta$ is determined by the condition
\eqn\ehdr{\Delta_0-\half \beta(\beta+Q)=1\ }
(which becomes the mass shell condition eqns.~\eqns{\dimone{,\ }\lioumom}\
in the string target space interpretation).

We may now associate a critical exponent $\Delta$ to the behavior of the
one-point function of $\Phi$ at fixed area $A$,
\eqn\eopt{F_\Phi(A)\equiv {1\over Z(A)}\int\CD\ph\,\CD X\ \ee{-S}\
\delta\Bigl({\textstyle\int} \d^2\xi\,\sqrt {\hat g}\,\e^{\ap\ph} - A\Bigr)
\ {\textstyle\int} \d^2\xi\,\sqrt{\hat g}\, \e^{\beta\ph}\,\Phi_0
\sim A^{1-\Delta}\ .}
This definition conforms to the standard convention that $\Delta<1$ corresponds
to a relevant operator, $\Delta=1$ to a marginal operator, and $\Delta>1$ to an
irrelevant operator (and in particular that relevant operators tend to dominate
in the infrared, i.e.\ large area, limit).

To determine $\Delta$, we employ the same scaling argument that led to \egam.
We shift $\ph\to\ph+\rho/\ap$ with $\e^\rho=A$ on the right hand side of
\eopt, to find
$$F_\Phi(A)
={A^{Q\chi/2\ap-1+\beta/\ap}\over A^{Q\chi/2\ap-1}}\,F_\Phi(1)
=A^{\beta/\ap}\,F_\Phi(1)\ ,$$
where the additional factor of $\e^{\rho\beta/\ap}=A^{\beta/\ap}$ comes
from the $\e^{\beta \ph}$ gravitational dressing of $\Phi_0$.
The gravitational scaling dimension $\Delta$ defined in \eopt\ thus satisfies
\eqn\ehba{\Delta=1-\beta/\ap\ .}
Solving \ehdr\ for $\beta$ with the same branch used in \ealph,
\eqn\embeta{\beta=-\half Q+\sqrt{{\textstyle {1\over4}}Q^2-2+2\Delta_0}
=-{1\over\sqrt{12}}\bigl(\sqrt{25-D}-\sqrt{1-D+24 \Delta_0}\bigr)}
(for which $-\beta\le Q/2$, and $\beta\to0$ as $D\to-\infty$).
Finally we substitute the above result for $\beta$ and the value \ealph\ for
$\ap$ into \ehba, and find\foot{We can also substitute
$\beta=\ap(1-\Delta)$ from \ehba\ into \ehdr\ and use
$-\half\ap(\ap+Q)=1$ (from before \ealph) to rederive the result
$\Delta-\Delta_0=\Delta(1-\Delta){\ap^2/2}$ for the difference between the
``dressed weight'' $\Delta$ and the bare weight $\Delta_0$ \rKPZ.}
\eqn\ehf{\Delta
={\sqrt{1-D+24\Delta_0}-\sqrt{1-D}\over\sqrt{25-D}-\sqrt{1-D}}\ .}

Assurance that the procedure of this section for identifying operators 
in these theories is consistent comes from calculations of the
toroidal partition functions \rtorPF, which essentially just count the
states.

\subsec{Tachyon amplitudes}

We shall now evaluate tachyonic amplitudes of the form
\eqn\tachamp{A(k_1,k_2,\ldots,k_N)= \langle  T_{k_1} T_{k_2}\ldots  T_{k_N}
\rangle\ ,}
where the expectation value is taken with respect to the total action $\cal
S$. In the rational $\ao^2$ case, the spectrum of $k$'s in the CFT can be
restricted to a finite set of
degenerate representations of the Virasoro algebra, closed
under operator product. Following \DOTFAT,
the flat space correlators for these minimal CFT are computed
by inserting a suitable number of screening operators of dimension 1,
$\exp(ia_{\pm} X)$, where $a_{\pm}=\ao \pm \sqrt{\ao^2+1}=\mp \alpha_{\pm}$.
We see that these
are pure matter operators, since their Liouville momentum vanishes.
In principle the computation of the most general amplitude \tachamp\
involves the insertion of say $n$ $T_{a_+}$'s and $m$ $T_{a_-}$'s.

The first step in computing \tachamp\ consists of the
integration over the zero modes of the fields.
Splitting $\ph=\ph\dup_0+{\hat \ph}$
and $X=X_0+{\hat X}$,
with $\int \sqrt{\hat g} {\hat \ph}=\int \sqrt{\hat g} {\hat X}=0$,
and integrating over $\ph\dup_0$ and $X_0$ along the
real line, gives
\eqn\amptach{ A(k_1,k_2, \ldots ,k_N)=
\left(\mu \over \pi \right)^s \Gamma(-s)\biggl< T_{k_1}T_{k_2}\cdots T_{k_N}
\Bigl( \int \e^{\ap\ph} \Bigr)^s \biggr>_{\mu=0}\ ,}
where the scaling factor $s$ is defined by
\eqn\scafac{ \ap s + \sum_{i=1}^N \beta(k_i) =-Q (1-h)\ .}
$h$ denotes the genus of the world sheet surface, and the above average is
performed over the non-zero modes of the fields with the free action
\eqn\acfree{ {\widehat {\cal S}} = {1 \over 2 \pi} \int \sqrt{\hat g}\,
 {\hat g}^{ab} \bigl( \partial_a {\hat \ph} \partial_b {\hat \ph} +
\partial_a {\hat X} \partial_b {\hat X} \bigr) }
(from now on, we drop the hat symbol on the free fields).
The integration over the zero
mode of $X$ yields the electric neutrality condition,
including the screening operators:
\eqn\elecneut{ n a_+  + m a_- + \sum_{i=1}^N k_i = 2 \ao (1-h). }

What has been gained in this process?
First of all, we have reobtained the KPZ \rKPZ\ scaling exponent $s$ for the
dependence on the cosmological constant $\mu$ of the correlators as in the
preceding subsection. Namely to each tachyon operator $T_k$,
we associate a scaling dimension
$\Delta_k= 1-\beta(k)/\ap$, and on genus $h$ surfaces
the bare partition function
scales like $\mu^{2-\gs^{(h)}}$, with the string susceptibility
exponent $\gs^{(h)}=2 + Q(1-h) / \ap$. Notice that the screening
operators
of the matter sector do not contribute. For the $(p,q)$ minimal CFT
coupled to gravity ($\ao^2=(p-q)^2 / 2pq$), we find
\eqn\gmastr{ \gs^{(h)}(p,q)=\gs(p,q)
+h(2-\gs)\ ,}
where the genus zero string susceptibility exponent reads
\eqn\strsph{\gs(p,q)= 2+{Q \over \alpha_+}=
-{2|p-q| \over {p+q-|p-q|}}\ ,}
confirming eq.~\egam.
Recall that the linear behavior of the string susceptibility exponent
\gmastr\ is the basis of the double scaling limit procedure of
\refsubsec\sdsl.

Note also that this result is in general different from that of the
matrix models eq.~\estrsce\
(in the non-unitary case $|p-q|\neq 1$), because the Liouville calculation
always selects a definition of the ``cosmological constant'' $\mu$
as the coupling to the ``dressed identity'' operator, different from that of
the matrix models (in which the scaling variable couples to the dressed
lowest dimension operator, see eqs.~\eqns{\egscosc{,\ }\edimopcc}).
These two definitions coincide only in the unitary case $|p-q|=1$,
where we will be able to compare the Liouville results
directly to those of the matrix model.

We can moreover find the dressed scaling dimensions of the tachyon operators
$T_{k_{r,s}}$, where each momentum is quantized in units of $a_{\pm}$
according to the number of matter screening operators of each kind,
$$\eqalign{ k_{r,s}&= {1 \over 2}(1-r)a_+ + {1 \over 2}(1-s)a_- \cr
&={1 \over 2 \sqrt{2pq}}((r+s-2)|p-q|+(s-r)(p+q)) \cr
\Delta_{r,s}&= 1-{\beta(k_{r,s}) \over \alpha_+}\cr
&={(\half (r+s)-1)|p-q|+\half(s-r) (p+q) \over p+q -|p-q|}\ .\cr}$$
Taking into account the restrictions over $r,s$, the last expression
can be rewritten 
\eqn\edimLiou{\Delta_{r,s}={-|p-q|+|ps-qr| \over p+q-|p-q|}\ ,}
with $1\le s\le q-1$ , $1\le r\le p-1$, and with the symmetry
$\Delta_{r,s}=\Delta_{q-s,p-r}$.
These coincide with the matrix model result for unitary theories,
because the identity $r=s=1$ is then the operator of minimal dimension, while
for non-unitary theories the minimal operator which corresponds to $r,s$ such
that $|ps-qr|=1$ differs from the identity operator.

Beyond this, we would like to compute the factor of $\mu^s$ in the
general amplitude. Our strategy is as follows:
\item{(i)} We perform the computation for non negative integer $s$, where
\amptach\ becomes a simple free
field correlator, with $s$ insertions of the tachyon $T_0$, integrated
over all positions of the operators. Although tractable in principle,
the free correlators have an ugly form for genus $h \geq 1$, and
it is not evident how to perform explicitly the integrations over the positions
of the tachyons, and the moduli of the surface. We will only present the
spherical results ($h=0$) here.
This last integral is particularly well-behaved only in the $D \leq 1$ case:
the kinematics allow for a finite region of convergence
in momentum space. We will calculate \amptach\ in this region,
where the result is a simple polynomial in $s$.
\item{(ii)} analytically continue this result to arbitrary (real) $s$.\foot{As
originally suggested in \rGLi. In \rAoDhok, it is argued that this
analytic continuation agrees with the asymptotic behavior for complex $s$.}
Here we have to resort to a more physical argument: the polynomial
amplitudes above will be interpreted as locally described
by a two dimensional effective field theory, which
for large momenta gives an algebraic growth of the amplitudes.
Requiring that all amplitudes be polynomial will fix them uniquely.
\item{(iii)} knowing the expressions for $A(k_1,k_2,\ldots,k_N)$
for arbitrary $s$ and $k_i$ in the convergence domain of the integral,
we still have to analytically continue them to arbitrary momenta.
This last step is not necessary in the $D<1$ case, since the
momenta eventually take discrete values inside the convergence domain.
We will comment later on the $D=1$ case,
where this last step becomes crucial.

An alternative (and sometimes more powerful) approach to these calculations
is to use the ground ring structure \grndrng, as implemented in \rmgrng.

\subsec{Three-point functions}

We first want to evaluate the three-point function without
screening:
\eqn\threept{ A(k_1,k_2,k_3) = (-\pi)^3 \left(\mu \over \pi
\right)^s\Gamma(-s)\bigl< {\cal T}_{k_1}(0) {\cal T}_{k_2}(\infty)
{\cal T}_{k_3}(1) T_0^s \bigr>\ ,}
where we have used the $SL(2,\IC)$ invariance to fix the positions
of three tachyons, and rescaled the partition
function by a constant
factor for technical convenience. The momenta are subject to
eqs.~\eqns{\scafac{,\ }\elecneut} with $m=n=h=0$.
Performing first the free field Wick contractions,
with the propagators $\langle \phi(z)\, \phi(0) \rangle =
\langle X(x)\, X(0) \rangle = -\log|z|^2$, we are left with the integral
\eqn\inthree{
\eqalign{\langle T_{k_1} T_{k_2} T_{k_3} T_0^s \rangle &=
\prod_{j=1}^s \int \d^2w_j|w_j|^{2\alpha} |1-w_j|^{2\beta}
\prod_{1\leq i<j\leq s} |w_i-w_j|^{4\rho} \cr
&= {\cal I}(\alpha,\beta;\rho)\ ,\cr}}
where
\eqn\paramthr{ \alpha=-\ap \beta(k_1)\ , \qquad
\beta=-\ap \beta(k_3) \ , \qquad \rho =-{\ap^2 \over 2}\ .}

The integral \inthree\ was first computed by Selberg \SEL\
by analytic continuation from integer $\rho$. Introducing
the function $\Delta(x)= \Gamma(x) / \Gamma(1-x)$, the result reads
\eqn\intthree{\eqalign{ {\cal I}(\alpha,\beta;\rho)&=s!
\left(\pi \Delta(-\rho)\right)^s \cr
&\cdot\prod_{i=0}^{s-1}\Delta((i+1)\rho)\Delta(1+\beta+i\rho)
\Delta(-1-\alpha-\beta -(s+i-1)\rho)\ .\cr}}
With no loss of generality,
we can take $k_1 \geq \ao$, $k_2 \geq \ao$ and $k_3 \leq \ao$.
This enables us to solve for $k_3$ using \scafac\ and \elecneut, or
equivalently solve for $\beta$,
\eqn\valbethr{ \beta = \cases{ \rho(1-s) & for $\ao >0$\cr -1-\rho s &
for $\ao <0 $\ .\cr} }
This expression for $\beta$ implies many cancellations
in \intthree, and we find
\eqn\threeres{\eqalign{\ao >0 \ \ \ &A(k_1,k_2,k_3)=0 \cr
\ao<0 \ \ \ &A(k_1,k_2,k_3)=-\pi \Delta(-s)
\left(\mu \Delta(-\rho)\right)^s\prod_{i=1}^2 -\pi\Delta(m_i)\ ,\cr}}
where
\eqn\mastach{ m_i= {{\beta_i^2 -k_i^2} \over 2}\ .}
In the latter case, $\ao <0$, we have $m_3=-s$,
so the result \threeres\ can be put in a more symmetric form,
\eqn\resthree{ A(k_1,k_2,k_3)=
\bigl( \mu \Delta(-\rho)\bigr)^s \prod_{i=1}^3
-\pi \Delta(m_i)\ . }

A few remarks are in order:
\item{(i)}
The apparent divergence of the $\ao <0$ result due to poles of
the $\Gamma$-function at negative integers should be understood
as a finite contribution for fixed area;
the area $A$ and the cosmological constant $\mu$
are related by a Laplace transformation
\eqn\laptrans{ \mu^s \Delta(-s) = {1 \over s!}
\int_0^{\infty} \d A\, \e^{-\mu A} A^{-s-1}}
(recall that the KPZ scaling is derived for fixed area amplitudes).
\item{(ii)}
This explains why we seem to get two qualitatively
different results for the two signs of $\ao$.
Taking a closer look at \resthree, we see that all the factors
are finite for $\ao >0$, whereas $m_3=-s$ yields a divergence
for $\ao<0$. Therefore \resthree\ gives a vanishing finite area
amplitude when $\ao>0$ and a finite one for $\ao<0$.
Both answers are correct, but the $\ao>0$ vanishing is just
an artifact of the resonance condition imposed on the
momenta. \resthree\ is thus the general result,
independent of the sign of $\ao$.
\smallskip

The factorized form of the result suggests defining ``renormalized"
tachyon operators:
\eqn\renormtach{ {\tilde T}_k = {T_k \over -\pi
\Delta\bigl((\beta^2-k^2)/2\bigr)}\ .}
(This includes the cosmological constant operator, and we redefine $\mu$
as the coefficient of ${\tilde T}_0$ in the Liouville action \liouac.)
In terms of the operators \renormtach, the amplitude is simply
\eqn\ampthreefin{ \langle {\tilde T}_{k_1} {\tilde T}_{k_2}
{\tilde T}_{k_3} \rangle = \mu^s\ . }
The situation here seems to be much better than in ordinary
string theory, where tachyon amplitudes contain poles
corresponding to the massive modes of the string and integrating them
out leads to a highly non-local effective action. Here the
tachyon interacts with an infinite set of massive modes,
existing only at discrete values of the momentum (the poles of the
renormalization factor $\Delta(m)$), and integrating out these modes
has the mild effect of renormalizing the tachyon field, therefore
described by a 2D effective field theory.
This suggestion is compatible with higher amplitudes growing
algebraically with the momenta, i.e.\ polynomial in momenta.
Under this assumption, \ampthreefin\ holds trivially
for arbitrary $s$ ($1$ is the analytic continuation of
the polynomial $P(s)=1$).

If we include arbitrary numbers $n$ and $m$ of screening operators,
\intthree\ is replaced by generalizations of the Selberg integral
computed by Dotsenko and Fateev \DOTFAT. For the special kinematics
chosen above, correlators simplify owing to many cancellations that occur.
After some algebra, the result is
\eqn\threescr{ A_{mn}(k_1,k_2,k_3) =
\bigl( \mu \Delta(-\rho) \bigr)^s \bigl(-\pi \Delta(-\rho_+)\bigr)^n
\bigl( -\pi \Delta(-\rho_-) \bigr)^m \prod_{i=1}^3 -\pi \Delta(m_i)\ ,}
where $\rho_{\pm}=a_{\pm}^2/2$.
Performing the field redefinitions \renormtach\ also for the
screening operators $T_{a_{\pm}}$, we find the same
result \ampthreefin, trivially extended to arbitrary $s$.

This answer is not quite final, however, since a subtlety arises in the
discussion of the selection rules for the three-point amplitudes.
In the present case, we have $k_i=k_{r_i,s_i}=
{1 \over 2}(1-r_i)a_+ + {1 \over 2}(1-s_i)a_-$ for $i=1,2$, and
$k_3=k_{\overline {r_3,s_3}}={1 \over 2}(1+r_3)a_+ +{1 \over 2}(1+s_3)a_-$,
with $a_-^2 =2p/q$, $p<q$. We have introduced the Kac indices
$1\leq r_i <p$, $1 \leq s_i <q$, such that $r_i q>s_i p$.
The only constraint on the integers $r_i,\,s_i$ arises from \elecneut,
and amounts to $r_1+r_2>r_3$, $s_1+s_2>s_3$, and
$\sum r_i=1\ {\rm mod}\  2$, $\sum s_i = 1 \ {\rm mod} \ 2$.
This cannot be the only constraint on the Kac indices for the
three-point coupling to be non-zero, because it would
violate the symmetry under permutation of the tachyons.
Moreover we seem to have lost the truncation of the CFT fusion
rules: $\sum r_i <2p$, $\sum s_i <2q$. In fact the
correct result must be a {\it consistent projection} of
\ampthreefin. In the flat space CFT, this projection is given
by the factorization properties of the four-point functions
onto three-point correlators. This result does not
arise directly from
the Feigin--Fuchs integrals for the three-point correlators,
which in addition have to be truncated by allowing any flip of two
vertex operators $V_{r,s}$($=\exp ik_{r,s}X$)$\to V_{p-r,q-s}$.
We expect a similar phenomenon here, therefore leading
to the same fusion rules as in the ordinary CFT case (recall
that those arose naturally in the matrix model solution of sec.~\sqmomm).
The final result should then read:
\eqn\threfus{ \langle {\tilde T}_{k_{r_1,s_1}}
{\tilde T}_{k_{r_2,s_2}} {\tilde T}_{k_{r_3,s_3}} \rangle
= {\rm N}_{(r_1,s_1)(r_2,s_2)(r_3,s_3)} \ \mu^s}
with the CFT fusion numbers ${\rm N}_{(r_i,s_i)} \in \{0,1\}$,
and reproduces the KdV results of \refsubsec\sssotumos,
apart from different field normalization factors.

\subsec{$N$-point functions}

We now turn to the computation of arbitrary
$N$-point functions of the tachyon field.
We believe the screening to be just a decoration of the
amplitude, responsible only for a global renormalization factor
and for implementing the CFT fusion rules. We will therefore
concentrate on the $N$-point functions without screening.
The conservation laws \eqns{\scafac{,\ }\elecneut} now read:
\eqn\consln{ \eqalign{&\quad\sum_{i=1}^N k_i = 2 \ao \cr
&s \ap + \sum_{i=1}^N |k_i-\ao|=({N \over 2}-1)Q \ .\cr}}

Fixing three tachyon positions by $SL(2,\IC)$ invariance, the
$N$-point amplitude reads for integer $s$:
\eqn\nptamp{ \eqalign{A(k_1,k_2,\ldots,k_N) &= -\pi^3 \left( {\mu \over \pi}
\right)^s \Gamma(-s) \prod_{a=1}^s \int \d^2 w_a
\prod_{i=1}^N \int \d^2 z_i \cr
&\quad\Bigl< {\cal T}_{k_1}(0) {\cal T}_{k_2}(\infty) {\cal T}_{k_3}
(1) \prod_{a=1}^s {\cal T}_{0}(w_a) \prod_{i=4}^N
{\cal T}_{k_i} (z_i) \Bigr>\ ,\cr}}
and the free field integral reads:
\eqn\nffint{ \eqalign{
\Bigl< T_{k_1} T_{k_2} T_{k_3} &\prod_{i=4}^N T_{k_i} T_0^s
\Bigr> = \prod_{a=1}^s \int \d^2\ w_a  \prod_{i=4}^N \d^2 z_i
|w_a|^{2 \delta_1} |1-w_a|^{2 \delta_3} \cr
&\cdot |z_i|^{2 \theta_{1,i}} |1 - z_i|^{2 \theta_{3,i}}
\prod_{a<b} |w_a-w_b|^{4 \rho} \prod_{4\leq i < j}
|z_i -z_j|^{2\theta_{i,j}}   \prod_{i,a} |z_i -w_a|^{2 \delta_i}\ ,\cr}}
with
\eqn\defsn{ \delta_i=-2 \ap \beta(k_i)\ ,\quad
\theta_{i,j}=k_i k_j - \beta(k_i) \beta(k_j)\ ,
\quad \rho= -{ \ap^2 \over 2}\ .}

We now restrict to the $(N-1,1)$ kinematics (it is
possible to show that the amplitude vanishes for any other choice
$(n,m)$ just as in the three-point case):
\eqn\kinN{ k_1 > \ao \ , \ k_2>\ao \ , \ \ldots\ ,\ k_{N-1}>\ao \ ,\
{\rm and} \ k_N<\ao\ ,}
which enables us to solve for $k_N$ using \consln,
\eqn\kknN{ k_N= {{N+s-3} \over 2} \ap + {1 \over 2} \am\ . }
It is natural to trade the momenta $k_i$ for the variables
$m_i=\bigl(\beta(k_i)^2 - k_i^2\bigr) / 2$, in terms of which we have
\eqn\kinematics{\eqalign{\delta_i &= \rho - m_i\ , \quad i<N \cr
\delta_N &= -1- (N+s-3) \rho \cr
\theta_{i,j} &= -m_i - m_j \ ,\quad i<j<N \cr
\theta_{i,N} &= -1+(N+s-3)m_i\ . \cr }}

The complexity of the integral \nffint\ resides in its pole 
structure.\foot{In cases where the pole structure is insufficient to resolve 
ambiguities, one can refer as well to the ground ring structure 
possessed by the operators (see \refs{\grndrng,\rmgrng}).}
As in general Veneziano string amplitudes, poles arise when
intermediary channels have momenta in certain discrete sets.
{}From the point of view of the integral,
those correspond to integration regions where some number
of $z_i$ approach each other. We can make this more precise in an example:
suppose we want to investigate the behavior
of \nffint\ when $z_4$, $z_5$, $\ldots$, $z_p$ approach $0$ simultaneously.
Then we can redefine
\eqn\pptsto{ z_4=\epsilon\,,\ z_5=\epsilon\, y_5\,,\ \ldots\,,
\ z_p=\epsilon\,y_p\ .}
Performing the integral over $\epsilon$ in a small disk, a pole arises
at the intermediate state of energy $E={Q \over 2}+\sum_{i=1}^p \beta$,
and momentum $k=\sum_{i=4}^p k_i$, such that
$E^2 - (k-\ao)^2 = 2 l $, where $l$ a non-negative integer labelling the
excitations ($l=0$ is a tachyon intermediate state, $l=1$ a graviton, etc.).
What are the residues at these poles?  It is easy to see that the answer is
given by
the factorization of the amplitudes. For instance the tachyon residue is:
\eqn\tapol{ \langle T_{k_1 }T_{k_2}\ldots  T_{k_N} \rangle \propto  {
{\langle T_{k_1} T_{k_4} \ldots  T_{k_p} T_{2 \ao -k} \rangle\,
\langle T_{k} T_{k_2}T_{k_3}T_{k_{p+1}} \ldots T_{k_N} \rangle}
\over { E^2 -(k- \ao)^2 }}\ ,}
and a straightforward generalization holds for non-zero $l$.

The miracle  which occurs in our case is the vanishing of most of these
residues at intermediary states,  leaving us with almost completely factorized
$n$-point amplitudes.
Concentrating first on the tachyonic poles ($l=0$),
we are going to show that at least one of the two pieces of the residue
\tapol\ vanishes identically, due to the fact that the intermediate state is a
special state in the wrong branch. Namely, one has in this case
$m(k)=\bigl(\beta(k)^2 - k^2\bigr)/2 = p-3$, and $\beta = {-Q/2}  - |k-\ao| $.
Note that in this case the momentum takes the value $k=(\am - p \ap) /2$,
and therefore corresponds, from the matter CFT point of view, to a
representation
with Kac indices $(0,p-1)$, outside of the minimal table.

We proceed by induction. Let us suppose that all the $M$-point functions with
$M \leq  N-1$ have the form:
\eqn\rechyp{ \langle T_{k_1} T_{k_2} \ldots T_{k_{M}} \rangle =
P(k_1,k_2,\ldots,k_M) \prod_{i=1}^M \Delta(m_i)\ ,}
with some polynomial $P$ in the momenta.
This is certainly satisfied for all $M \leq 3$.
Plugging this form into the residue of \tapol, we find that the first bracket
is finite, whereas the second one vanishes
due to the factor $\Delta\bigl(m(k)\bigr)=\Delta(p-3)=0$,
and therefore the residue vanishes.
More generally, one can consider the situation in which any subset of
$z_4,\ldots,z_{N-1}$ is taken simultaneously to $0$. The corresponding residue
can be shown to vanish as well using the recursion hypothesis.
More interesting are the poles obtained from integration regions where
a subset of $z_4,\ldots,z_{N-1}$, together with $z_N$, approach $0$.
In that case, one finds that the second bracket  is finite, but the first one
vanishes,
due to the factor $\Delta\bigl(m(2\ao -k)\bigr)=0$, enforced by the kinematics.
We are therefore left only with poles where $z_N$ approaches other vertices.
But as a consequence of the relation \kknN, these turn out to be factorized
poles  in the individual momenta, involving $s$ only. The residues of the
latter poles do not vanish, due to the simultaneous presence of a special
state of the wrong branch (which would in principle induce the vanishing of
the residue), and of a special state of the right branch (which would
give a divergent contribution), whose competing effects cancel each other.

The poles at excited intermediary states ($l>0$)  have the same fate.
The latter can be seen as gravitational descendents (Liouville dressing of the
Virasoro descendent fields) of the (wrong branch) tachyons.
Typically such an operator is built by multiplying  an operator ${\cal
T}_k^{(-)}$ with $\beta$ in the wrong branch, by a differential polynomial of
the matter field $X$. Starting from a momentum $k=k_{r,s}$ of the minimal
conformal grid, one obtains the first excited state at level $rs$, etc.
The point is again that any insertion of such an operator in a generic
correlator yields a vanishing result.  The proof proceeds by induction,
and we leave it as an exercise to the reader.
This shows, just as in the $l=0$ case, that all residues vanish
except for a set of factorized poles in the individual momenta.

Where are these last poles located? The poles in $m_1$ come from the region
where $z_N$ approaches $0$. The poles in $m_2$ come from $z_N \to \infty$,
but can be recast into poles in $m_1$, using the kinematic  relation:
\eqn\kinn{ \sum_{i=1}^N m_i = 1 + \rho s\ .}
In principle poles in $m_1$ occur whenever $\theta_{1,N}=-l$,
$l=1,2,\ldots$, but
it is easy to see that among those only the ones corresponding to $m_1=-n$,
$n=1,2,\ldots,$
have a non-vanishing residue: we know already that the locations of
the poles only depend on $s$, so that we can take $k_3,k_4,\ldots,k_{N-1}
\to 0$
and be left with the three-point function \resthree,  differentiated
$N-3$ times w.r.t.\ $\mu$ (corresponding to $N-3$
insertions of the cosmological constant operator $T_0$), from which the poles
in $m_1$ are obvious.
Similarly,  we find poles at $m_2=-p$, $p=1,2,\ldots$, or  equivalently at
$m_1= 1 + \rho s -\sum_{1=3}^{N-1} + p$, $p=1,2,\ldots,$ .

We now form the ratio
\eqn\ration{ f_s(m_1; m_3,\ldots,m_{N-1})
= {{\langle T_{k_1} \ldots  T_{k_{N}}
\rangle} \over {\prod_{i=1}^{N} \Delta(m_i) }}\ .}
As a function of $m_1$, $f_s$ can have poles only  at the zeros of the
denominator, i.e.\ at $m_1=l, \ m_2=n$, $l,n=1,2,\ldots,$ .
To show that the numerator also vanishes at these points, we
apply the recursion hypothesis \rechyp\ to the residue of the individual pole
at $m_{N-1}=-n$, which vanishes for $m_1=l$ or $m_2=n$.
Therefore $f_s$ is an entire function of $m_1$.
Moreover we find that $f_s$ is bounded when $|m_1| \to \infty$.
For large $|m_1|$,  the integral  \nffint\  is dominated by the region where
all the points are close to $1$.  To blow up this region, we perform the
change of variables $z_i = \exp (x_i /m_1)$, $w_a= \exp( w_a / m_1)$,
and estimate the integral. The result is simply
$\lim_{m_1 \to \infty} f_s = {\rm const}$.
Repeating the argument for the other $m_i$, we find that
$f_s$ is only a function of $N$ and $s$.
Again we can send $k_3,\ldots,k_{N-1} \to 0$, and read the result from the
three-point function \resthree, differentiated $N-3$ times w.r.t.\ $\mu$.
This gives
\eqn\resnn{ \langle T_{k_1} T_{k_2} \ldots T_{k_N} \rangle =(- \pi
\Delta(-\rho))^s \Bigl[\prod_{i=1}^N -\pi \Delta(m_i)\Bigr]
(\partial_{\mu})^{N-3} \mu^{s+N-3}\ ,}
or after the redefinitions \renormtach\ of the tachyon and
cosmological  constant operators,
\eqn\resnpt{ \langle {\tilde T}_{k_1} {\tilde T}_{k_2} \ldots {\tilde
T}_{k_{n}} \rangle= (\partial_{\mu})^{N-3}  \mu^{s+N-3}\ . }
Assuming, as explained above, that the result is polynomial in the momenta,
\resnpt\ still holds for arbitrary $s$, not necessarily integer.

This concludes our calculation of the $N$-point amplitudes
on the sphere without screening, showing a perfect agreement with the matrix
model (KdV) results of \refsubsec\sssotumos\ (eqn.~\resnptkv).

The general amplitudes of the rational case involve insertions of screening
operators, enforcing the fusion rules of the CFT, but we believe the form of
\resnpt \ is unaffected by these decorations.

As indicated earlier,
we have to be careful with the problems of convergence of
the defining integral \nffint.
It is easy to see that the convergence domain is simply: $m_i >0$,
$i=1,2,\ldots,N-1$, with the kinematical constraint \kinn.
Therefore, taking the momenta at discrete values corresponding to the
conformal grid
$k_i = k_{(r_i,s_i)}$, $i=1,2,\ldots,N-1$ and $k_N= k_{(p-r_N,q-s_N)}$ does not
violate convergence, and we do not have to worry about analytic continuation
of the results in momentum space.
However,  in the interesting limiting case $\ao \to 0$ ($ c \to 1$), all real
values of the momenta are permitted, and we have to worry about this problem.
The structure of the amplitudes turns out to be much richer in that case,
although identical in the $(N-1,1)$ kinematical regime (see \DKU\ for details).

\newsec{Large order behavior and Borel summability}
\seclab\sLOBBS

In this section we show how one can determine the large order behavior of
the topological expansion of the $d<1$ models by a straightforward analysis
of the differential equation satisfied by their partition functions. After
recalling some standard facts about divergent series, we discuss in detail the
simplest case of pure gravity, and sketch the generalization to other cases.

\subsec{Divergent series and Borel transforms}

We start by recalling some standard features of divergent series,
Borel summability, and summation methods (see, for example, pp.~840--842 of
\rZJ\ for a recent treatment with physical
applications). Consider a function $f(w)$, analytic in some sector $S$
(say $|{\rm Arg}\,w|\le\alpha/2$, $|w|\le|w_0|$) in which it
has an asymptotic expansion
\eqn\easex{f(w)\approx\sum_0^\infty f_k\,w^k\ .}
This means that the series diverges for all non-vanishing $w$, but in $S$
there is a bound of the form
\eqn\ebd{\left|f(w)-\sum_{k=0}^N f_k\,w^k\right| \le C_{N+1}\,|w|^{N+1}
\qquad\hbox{for all }N\ ,}
and for definiteness we assume that $C_N= M\,A^{-N}\,(\beta N)!$.
Though the series diverges, it can be used to estimate $f(w)$ for $|w|$
small by taking the truncation of the series \easex\ at a value of $N$ that
minimizes the bound \ebd. This gives the best possible estimate of $f(w)$,
with a finite error
$\varepsilon(w) =\min_{\{N\}} C_N\,|w^N|\sim \exp-(A/|w|)^{1/\beta}$.
An asymptotic series does not in general define a unique function since we
can always add to it any function analytic and smaller than $\varepsilon(w)$
in the sector $S$.

When the angle $\alpha$ defining the sector $S$ above
satisfies $\alpha>\pi\beta$, however, a classical theorem of
analytic functions applies to show that a function analytic in $S$ and
bounded there by $\varepsilon(w)$ must vanish identically.
This is the only case in which the asymptotic series defines
a unique function $f(w)$, and for which
there exist methods to reconstruct the function from the series.
One such method is based on the Borel transform
$B_f(w)$ of $f(w)$, defined from the expansion \easex\ by
\eqn\ebxfm{B_f(w)=\sum_0^\infty b_k\,w^k\equiv
\sum_0^\infty {f_k\over(\beta k)!} w^k\ .}
%
According to the assumptions following \easex, we have
$|f_k/(\beta k)!|< M\,A^{-k}$,
so $B_f(w)$ is analytic at least in a circle of radius $A$ and uniquely
defined by the series. Then the integral representation
\eqn\einre{f(w)=\int_0^\infty \d t\ {\rm e}^{-t}\,B_f(w\,t^\beta)}
converges in the sector $|{\rm Arg}\,w|\le\alpha/2$
for $|w|$ small enough, and
yields the unique function which has the asymptotic expansion \easex\
in the domain $S$.

In general, the function $B_f(w)$ may have poles and
cuts running from its singularities off to infinity on the complex plane.
If there is a singularity on the positive real axis, we say that the
original series
for $f(w)$ is not Borel summable since we cannot run the contour in \einre\
along the real axis.  Cuts in $B_f(w)$ are indicative of
possible ``non-perturbative'' effects (i.e.\ exponential in an inverse
string coupling, $\kappa\inv\sim x^{(2l+1)/2l}$ in the notation of \ede),
and generally the choice of contour in \einre\ from the
origin to $\infty$ in the Re$\,t>0$ half-plane reflects possible
``non-perturbative'' ambiguities.

We now recall how the large order behavior of $f(w)$ may be extracted
from such non-perturbative behavior.
The coefficients of the series defining $B_f(w)$ satisfy
\eqn\ebk{b_{k}
={1\over 2i\pi}\oint_C {\d s\over s^{k+1}}\,B_f(s)
\ \mathop{\propto}_{k\to\infty}\
{1\over2i\pi}\int_r^\infty \d s\,{1\over s^{k+1}}\,{\rm disc}\,B_f(s)\ ,}
where the contour $C$ encloses the origin.
For $k$ large, the behavior of $b_k$ is related to values
of $B_f(s)$ near the point on the contour where $|s|$ is minimal, so by
deforming the contour to run along the cuts of $B_f(s)$ to infinity,
we see that the integral above is dominated by the discontinuity of
$B_f(s)$ along the cut corresponding to the singularity $r$
closest to the origin. After Borel transformation, it follows
that the large order behavior of the original series is given by
\eqn\elor{\eqalign{f_k &= (\beta k)!\, b_k
 =\int_0^\infty \d t\,\e^{-t}\,t^{\beta k}\, b_k
={1\over2i\pi} \int_0^\infty \d t\,\e^{-t}
\oint {\d s\over s^{k+1}}B_f(st^\beta)\cr
&\mathop\propto_{k\to\infty}\ \int_0^\infty \d t\,\e^{-t}
\int_{r/t}^\infty {\d s\over s^{k+1}}\,{\rm disc}\,B_f(st^\beta)
\propto\int_0^\infty {\d s\over s^{k+1}} \int_r^\infty \d t\,\e^{-t}\,
{\rm disc}\, B_f(st^\beta)\cr
&\qquad=\int_0^\infty {\d s\over s^{k+1}} \bigl(f_+(s)-f_-(s)\bigr)\ ,\cr}}
where $f_\pm(s)$ are the Borel transforms
corresponding to integrations in \einre\ on opposite sides of
the cut.

In the cases of interest to follow here, both $f_\pm$ will satisfy
the same differential equation, and the (exponentially small) difference
\eqn\eeps{\epsilon(w)=f_+(w) - f_-(w) =
\int_r^\infty \d t\,\e^{-t}\,{\rm disc}\, B_f(w\,t^\beta)}
will be determined by the corresponding linearized equation.
Knowledge of the leading behavior of
$\epsilon(w)$ can be used to infer a great deal about
$f(w)$ and $B_f(s)$.
For example, when $\epsilon(w)$ has leading behavior
\eqn\eepsb{\epsilon(w)\sim w^{-b/\beta}\ \ee{-(A/w)^{1/\beta}}\ ,}
we find from \elor\ that $f_k\sim \Gamma(\beta k+b)$ for $k$ large.
Moreover we see from \eeps\ that the above behavior for $\epsilon(w)$
results from the singular behavior $B_f(s)\sim (1-s/A)^{-b}$ near $s=A$,
where $A$ is the singularity nearest the origin in the Borel plane.
For large $k$ this means that $b_k\sim A^{-k}k^{b-1}$ and hence we have the
refined estimate $f_k\sim (\beta k)!\,A^{-k}k^{b-1}$. This is typical of
large order behavior of perturbation theory in quantum mechanics and field
theory, where $A$ is a classical instanton action.\foot{As stressed by
Shenker \rScar, however, the value of $\beta$ provides an important
distinction between string theory and field theory. In field theory,
$\beta=1$ and the non-perturbative effects in the exponential
in \eepsb\ go as the inverse loop coupling, $w\inv=1/g^{2}$.
In string theory, on the other hand, we shall see that $\beta=2$,
leading to much larger non-perturbative effects in the exponential
that go as the inverse square root of the loop coupling, $w^{-1/2}=1/\kappa$.}
In what follows, we can now bypass the intermediate steps and
use eqs.~\eqns{\elor{--}\eepsb} directly to determine
the asymptotics of $f_k$ and the locations of singularities of $B_f(s)$.

\subsec{Pure gravity}
\subseclab\ssPG

For pure gravity, the differential equation satisfied by the second
derivative of the partition function is \eqnn\eai%
(\ede\ with $l=2$, after suitable rescaling)\foot{We exchange $x$ for $z$
in what follows since we continue to the complex plane.
Our normalization in \eai\ corresponds to a matrix model with an even
potential; for an odd potential the second term is instead $-{1\over6}u''$.}
$$u^2(z)-{1\over3}u''(z)=z\ .\eqno\eai$$
%
If $u(z)$ has an asymptotic expansion for $z$ large, it satisfies
$u(z)=\pm \sqrt{z}+O\left(z^{-2}\right)$.
The solution that corresponds to pure gravity has a $z$ large expansion of
the form
\eqn\eaii{u(z)=z^{1/2}\Bigl(1-\sum_{k=1}u_k\,z^{-5k/2}\Bigr)\ ,}
where the $u_k$ are all positive.\foot{The first term, i.e.\ the
contribution from the sphere, is dominated by a regular part which has
opposite sign.  This is removed by taking an additional derivative of $u$,
giving a series all of whose terms have the same sign ---
negative in the conventions of \eai.
The other solution, with leading term $-z^{1/2}$,
has an expansion with alternating sign
which is presumably Borel summable, but not physically relevant.}

To determine the large order behavior of the expansion we argue as in
\eqns{\ebxfm{--}\elor}.
We consider the Borel transform of the expansion, defined by
\eqn\eBor{B(s)=\sum_{k=1}^\infty {u_k \over (\beta k)!} s^k \ ,}
in which $\beta$ is chosen so that the series \eBor\ is
convergent in a circle of finite radius.
Then a solution (in general complex) to eq.~\eai\ is obtained from
the integral
\eqn\eBortr{\tilde u(z)=z^{1/2}\Bigl(1-\int_{0}\d t\ \e^{-t}
B\bigl(t^\beta\,z^{-5/2}\bigr) \Bigr)\ ,}
provided there exists a suitable
contour of integration from the origin to infinity in
the $\Re t>0$ half-plane on which the integral converges.


The functions $u_\pm(z)$ defined respectively by integration in \eBortr\
above and below the cut both satisfy eq.~\eai.
For $z$ large, their difference $\epsilon\equiv u_+ - u_-$
is exponentially small compared to their average
$u_0\equiv(u_+ + u_-)/2$. $\epsilon(z)$ is therefore a solution
of the equation obtained by linearizing \eai.
Taking the differences of the equations \eai\ satisfied by
$u_+$ and $u_-$, we find that $\epsilon$ satisfies
\eqna\eaiii
$$\epsilon''(z)-6 u_0(z)\,\epsilon(z)=0\ ,\eqno\eaiii a$$
where $u_0$ is determined by
$$u_0^2(z)-{1\over3}u_0''(z)+\epsilon^2(z)=z\ \eqno\eaiii b$$
(and $\epsilon^2(z)$ in \eaiii b\
can be ignored to leading order in large $z$). To leading order,
the function $\epsilon$ is also proportional to the difference
between any Borel sum of the series and the exact non-perturbative
solution of the differential equation (up to even smaller exponential
corrections corresponding to multi-instanton like effects).

Eq.~\eaiii a\ can easily be solved by the WKB method for $z$ large.
Substituting the ansatz $\epsilon'/\epsilon=ru_0^{1/2}+b\,u_0'/u_0$,
we find $r^2=6$ and $b=-1/4$.
Dividing by the leading term $u_0\sim z^{1/2}$ to remove the overall
factor $z^{1/2}$ in \eaii\ gives
\eqn\ews{{\epsilon(z) \over z^{1/2}}\ \propto\
z^{-5/8}\,\ee{-{4\sqrt{6}\over5}z^{5/4}}\bigl(1+\cdots\bigr)\ .}
In terms of the expansion parameter (string loop coupling)
$\kappa^2=z^{-5/2}$, the ratio \ews\ reads
\eqn\ewsp{\sigma\bigl(z(\kappa)\bigr)
\equiv{\epsilon(z)\over z^{1/2}}\ \propto\
\kappa^{1/2}\,\ee{-{4\over5}(\sqrt6/\kappa)}\ .}
The above solution is valid for $z$ large, i.e.\ $\kappa$ small, so we may
apply \elor\ to find that the large order behavior in \eaii\ is given by
\eqn\elop{u_k
\mathop{\propto}_{k\to \infty}\ \int_0 {\d\kappa\over\kappa^{2k+1}}
\,\sigma(\kappa)\ \propto\ \left({5\over4\sqrt6}\right)^{2k}
\Gamma(2k-\half)\ .}
(The constant of proportionality in the above
cannot be determined by this method.) The asymptotic $\Gamma(2k-\half)$
behavior is a slight refinement of the $(2k)!$ behavior determined in
\refs{\rDS,\rGM,\rBIZ}.

{}From the discussion following \eepsb, we can see directly from \ewsp\ that
$\beta=2$, and the reality of $r^2$ has implied a singularity on the real
axis in the Borel plane. This obstruction to Borel summability is
consistent with the large order behavior in \elop, in which all terms have
the same sign.

{\it Remarks.} Note that other singularities of the Borel transform are
related to higher order corrections in the expansion of $\epsilon$. Because
the coupled equations \eaiii{a,b}\ for $u_0$ and $\epsilon$ have a
well-defined parity in $\epsilon$, the exponential behavior of successive
non-perturbative corrections to $\epsilon$ will be of the form
$\exp(-n\,C\,z^{5/4})$, where $C=(4/5)\sqrt6$ and $n$ is an odd integer.
This follows from iterating \eaiii{a,b}:
$\exp(-n\,C\,z^{5/4})$ terms in $\epsilon$ in \eaiii b,
with $n$ odd (including the leading $\exp(-C\,z^{5/4})$ piece),
result only in $\exp(-m\,C\,z^{5/4})$ terms in $u_0$ with $m$ even,
and vice versa, in \eaiii a.

In \rFDii, it is confirmed that the exponential in \ews\
coincides with the action for a single eigenvalue climbing to the top of
the barrier in the matrix model potential, allowing us to interpret the
exponential piece of the solution to \eai\ as an instanton effect. In
\refsubsec\ssIalobom\ here,
we shall reproduce and generalize this observation to
arbitrary one-matrix models.

\subsec{Ising / Yang--Lee}

We now consider the
fourth order differential equations \refs{\rIYL,\rising,\rD}\ for the
Yang--Lee edge singularity and the critical Ising model. After
suitable rescaling (different for the two cases, \ede\ with $l=3$ and
\isin\ with $t_2=t_5=0$), the equations are written
\eqn\eiyle{u^3- u u''-\half u'{}^2 + a u^{(4)}=z\ ,}
where $a={1\over10},{2\over27}$ respectively for Yang--Lee and critical Ising.

The asymptotic expansion takes the form
$$u(z)=z^{1/3}\Bigl(1+\sum_{k=1}u_k\,z^{-7k/3}\Bigr)\ .$$
Substituting $u=u_0+\epsilon$ (where now $u_0\sim z^{1/3}$), we find that
the linearized equation for the discontinuity $\epsilon(z)$, at leading
order for $z$ large, reads
$$a\epsilon^{(4)}+3z^{2/3}\epsilon-z^{1/3}\epsilon''-{1 \over 3}z^{-2/3}
\epsilon' =0\ .$$
The ansatz $\epsilon'/\epsilon=r u_0^{1/2}+bu_0'/u_0$ now gives $b=-3/4$
and yields a solution with the asymptotic form
\eqn\ewsi{\sigma\equiv{\epsilon(z)\over z^{1/3}}
\ \propto\ z^{-7/12}\,\ee{-{6\over7} r z^{7/6}}\ ,}
where $r$ satisfies
\eqn\emter{ar^4-r^2+3=0\ .}
The solutions are $r^2=9/2,9$ in the case of the Ising model and
$r^2=5\pm i\sqrt{5}$ in the Yang--Lee edge case.

In terms of the expansion parameter $\kappa=z^{-7/6}$, we have
$\sigma(\kappa)\propto\kappa^{1/2}\exp\bigl(-{6\over7}(r/\kappa)\bigr)$,
and the large order behavior according to \elor\ is
\eqn\elobiyl{u_k\ \mathop{\propto}_{k \to \infty}\
\int_0 {\d\kappa\over\kappa^{2k+1}}\,\sigma(\kappa)\ \propto\
\int_0 {\d\kappa\over \kappa^{2k+1/2}}\,\ee{-{6\over7}(r/\kappa)}
\ \propto\ \left({7\over6r}\right)^{2k}\Gamma(2k-\half)\ .}
Remarkably enough, the large order behavior takes the same form,
$\Gamma(2k-\half)$, as in \elop.

{}From \emter, we see that $r^2$ is complex when $a>1/12$,
and from \elobiyl\ we see that
this is related to the non-unitarity of the Yang--Lee edge case,
since it leads to asymptotic coefficients
$u_k$ that are not positive definite. This is also related to the Borel
summability in the Yang--Lee edge case, since from the discussion following
\eepsb\ we see that the
poles nearest the origin are a finite distance off the real axis in the
Borel plane. There thus exists a real and physically acceptable Borel sum,
presumably equal to the solution of \rBMP.
In the (unitary) Ising case, on the other hand (with $r^2$ real), the
terms of the series are all positive, there is a singularity on the real
axis in the Borel plane, and the series is not Borel summable.

We shall shortly generalize the conclusion concerning Borel summability in
the ($l=3$) Yang--Lee case to all the $l$ odd one-matrix models. We can
also show that it is not affected by higher order exponential corrections.
If we look for such corrections by expanding in $\epsilon$, we find that
the coefficient $r$ in the exponential is replaced by $n_+ r + n_- r^*$,
but because the equations have a well-defined parity in $\epsilon$, $n_+
+n_-$ must necessarily be odd.\foot{The solution $\epsilon_1$
of the linearized equation is a linear combination of the two exponentials
involving $r$ and $r^*$, each multiplied by power series.  The leading
order corrections to $u_0$ involve $\epsilon_1^2$ and thus correspond
to exponentials involving $2r$, $2r^*$, and $r+r^*$.
The argument then proceeds by iteration, just as remarked at the end of
\refsubsec\ssPG\ concerning the analogous properties of solutions to
eqs.~\eaiii{a,b}.}
All such allowed terms will not occur (some of them can correspond to
singularities of the Borel transform $B$ in other sheets of the complex
plane), but we see in any event that no singularity will appear on the
positive real axis and an integral like \eBortr\ may be performed to define
a unique real function.  This function solves the differential equation of
interest and is therefore the natural candidate for the partition function
of the original matrix problem.

\subsec{The tricritical Ising model}
\subseclab\ssttIm

In the case of the tricritical Ising model,
we perform the analogous computation by substituting $u\to u+\epsilon_u$,
$v\to v+u\epsilon_v$ into eqs.~\etci{a\hbox{--}c}\ with $T=h=0$.
We introduce the ansatz $\epsilon_u'/\epsilon_u=r u^{1/2}$ and
$\epsilon_v'/\epsilon_v=r u^{1/2}$, and remember that at leading order $v\sim
-u^2/2$. We then find a system of two linear equations for $\epsilon_u$ and
$\epsilon_v$, \eqna\etrisys
$$\eqalignno{\left(-10+5r^2-\half r^4 \right) \epsilon_u
+\left(5r^2-10\right)\epsilon_v &=0\ ,&\etrisys a\cr
\left(40-30 r^2+7r^4-\half r^6 \right) \epsilon_u
+\left(20-10r^2+r^4\right)\epsilon_v&=0\ .&\etrisys b\cr}$$
Writing that the determinant of the $2\times2$ matrix vanishes yields an
equation for $r^2$. Actually two roots are obvious because in
eq.~\etrisys{b}\ the polynomial $20-10r^2+r^4$ can be factorized.
The equation for the other roots is then simply $r^4-5r^2+5=0$,
and the four solutions for $r^2$ are
$$r^2=5\pm\sqrt{5},\ \half(5\pm\sqrt{5})\ .$$
Since these give real values of $r$,
the theory is not Borel summable. This is again
as expected for a unitary theory, in which
the coefficients in the asymptotic expansion have fixed sign.
The large order behavior is, up to the value of $r^2$, the same as
found in the earlier cases, eqns.~\eqns{\elop{,\ }\elobiyl},
\eqn\elotci{u_k\ \propto\
\left(9\over8r \right)^{2k}\Gamma\left(2k-\half\right)\ .}

\subsec{The general problem: Preliminary remarks}

In this subsection we shall explain the structure of the large order
behavior in the general case. In the next section we shall then discuss
explicitly the Borel summability of a large class of $(p,q)$ models.
\medskip
{\it The one-matrix models.} We consider the string equation \ede, $R_l[u]\sim
z$. To examine the leading large order behavior of perturbation theory,
it is only necessary to know the terms in $R_l[u]$ of the form
\eqn\erld{R_l[u]= {1\over l}\,A_{ll}\,u^l +
\sum^{l-1}_{j=1} A_{lj}\,u^{j-1}u^{(2l-2j)} + \cdots}
%
(i.e.\ that contain at most one derivative of $u$ factor.
The next leading contribution is given by terms such as
$u^{j-2}u^{(2l-2j-1)}u'$, i.e.\ with a single factor of $u'$ as well).
{}From the recursion relation
\erecR, we have for example that
the coefficient of the highest derivative term $u^{(2l-2)}$ in \erld\
is given by $A_{l1}=-4^{-l}$ and the coefficient of $u^l$
is $A_{ll}/l=(-1)^l (2l-1)!!/(2^{l+1} l!)$.

Denoting the discontinuity of $u(z)$ by $\epsilon(z)$
and substituting $u=u_0+\epsilon$ as before
(now $u_0\sim x^{1/l}$), to leading
order for $z$ large we find that $\epsilon$ satisfies
\eqn\elwkb{0=\sum^{l}_{j=1} A_{lj}\,u_0^{j-1}\epsilon^{(2l-2j)}\ .}
Substituting the WKB ansatz
\eqn\ewkbana{{\epsilon'\over\epsilon} = r u_0^{1/2}}
then gives
\eqn\esa{0=\sum^{l}_{j=1} A_{lj}\,r^{2l-2j}\ ,}
an $(l-1)^{\rm st}$ order equation for $r^2$ with real coefficients.
We shall determine the coefficients $A_{lj}$ and discuss the solutions of the
equation in the next subsection.

The subleading terms in $R_l[u]$ mentioned after \erld\
are immediately deduced from the leading terms by noting that since $R_l[u]$
is derived from an action (see eq.~\egdp),
the operator acting on $\epsilon$ is hermitian.
Therefore the operator $u^{j-1}\d^{2l-2j}$ should be replaced by the
symmetrized form $\half\{u^{j-1},\d^{2l-2j}\}$, correcting \elwkb\ to
\eqn\elwkbp{0=\sum^{l}_{j=1} A_{lj}\bigl(u^{j-1}\epsilon^{(2l-2j)}
+\half(2l-2j)(j-1)u^{j-2}\,u'\,\epsilon^{(2l-2j-1)}\bigr)\ .}
To characterize more precisely the large order behavior, to next
order we set
\eqn\ewkbanb{{\epsilon'\over\epsilon} = ru^{1/2} + b{u'\over u}\ ,}
from which it follows, to the same order, that
\eqn\ewkbno{{\epsilon^{(k)}\over\epsilon}
= r^k u^{k/2} + r^{k-1} u^{(k-3)/2}\,u'\,k\Bigl(b+{1\over4}(k-1)\Bigr)\ .}
Substituting into \elwkbp, we find
\eqn\elwk{0=\sum^{l}_{j=1} A_{lj} \Bigl(r^{2l-2j}
+ u^{-3/2}\,r^{2l-2j-1}\, u' (l-j)
\bigl[j-1 + 2\bigl(b+{\textstyle{1\over4}}(2l-2j-1)\bigr)\bigr]\Bigr)\ .}
We see that $r$ remains a solution to \esa\ and $b=(3-2l)/4$,
independent of $r$. Dividing $\epsilon$ by
the leading term $u\ \propto\ z^{1/l}$ results in
\eqn\egeps{\sigma\equiv{\epsilon(z)\over z^{1/l}}
\ \propto\ z^{-(2l+1)/4l}\,\ee{-{2l\over2l+1} r z^{(2l+1)/2l}}\ ,}
generalizing \ews\ and \ewsi.
In terms of the expansion parameter $\kappa=z^{-(2l+1)/2l}$, we find
\eqn\elobg{\eqalign{
&\qquad\sigma(\kappa)\ \propto\
\kappa^{1/2}\,\ee{-{2l+1\over2l} (r/\kappa)},\cr
&u_k\ \mathop{\propto}_{k \to \infty}\ \int_0
{\d\kappa\over \kappa^{2k+1/2}}\,\ee{-{2l+1\over2l} (r/\kappa)}
\ \propto\ \left({2l+1\over2lr}\right)^{2k}\Gamma(2k-\half)\ .\cr}}
We see that the $\Gamma(2k-\half)$ factor in
\eqns{\elop{,\ }\elobiyl{,\ }\elotci}
is general, owing in the case of one-matrix models to the special form
of the subleading term in the equation \elwkbp\ satisfied by $\epsilon$
(due to the fact that the original equations descended from an action
principle).

\medskip
{\it General $(p,q)$ model.} In the case of the general $(p,q)$ model
(eqs.~\epqcc\ and \egenac) there results a system of coupled
linear differential equations for the variations $\epsilon_u$,
$u^{\delta_i}\epsilon_{v_i}(x)$ associated with the functions $u(x)$,
$v_i(x)$ (the power $\delta_i$ of $u$ is determined by the grading).
As in the case of the tricritical Ising model considered in \refsubsec\ssttIm,
at leading order we set $\epsilon'_u/\epsilon_u=r u^{1/2}=\epsilon'_{v_i}
/\epsilon_{v_i}$. We obtain, taking into account the leading relations
between $u$ and the $v_i$, a linear system for $\epsilon_u$ and
$\epsilon_{v_i}$. Expressing again that the determinant of the linear system
vanishes, provides an equation for the coefficient $r$.

To determine more precisely the behavior of $\epsilon_u$ we have to
consider subleading terms. As in the one-matrix case they can be
determined by a hermiticity argument.
Since the equations for $u,v_i$ derive from an action \egenac, the linear
equations for $\epsilon_u$, $\epsilon_{v_i}$ define a hermitian operator.
Eliminating for example all the $\epsilon_{v_i}$ yields an equation to next
leading order for $\epsilon_u$ which can be expressed as a hermitian
operator acting on $\epsilon_u$ (as was the case leading to \elwkbp).  The
coefficient of the subleading term that led to the $\Gamma(2k-\half)$
behavior found in \elobg\ for the one-matrix models, since it only depended
on the hermiticity of the operator acting on $\epsilon$, is thus universal
for all the $(p,q)$ models.
\foot{The $(2k)!$ large order behavior
is also the generic behavior for $D=1$ models coupled to gravity.
More precisely\rGZ, we found the contribution from genus $k$ surfaces to go
as $f_{2k}\sim\Gamma(2k-1)$. This was based on an instanton
analysis that allowed an understanding of this behavior as a result of barrier
penetration effects, typically of the form $k!/ A^k$,
where the $A$ is an instanton action given by the integral
$\int\d x\,\sqrt{V(x)-E}$ between the turning points.
The perturbative expansion by itself for $D=1$ does not fully determine the
partition function and instead misses some essential non-perturbative
feature of the problem (see also \rGiMo).}

\subsec{Borel summability}

Having shown that perturbation theory at large order has for all $(p,q)$
models the generic $A^k\,\Gamma(2k- \half)$ behavior, we shall discuss for a
class of models the existence of real positive values $A$, relevant for the
Borel summability of the theory. For this purpose we need all equations only
at leading order. We will show here that the relevant equations can be
directly derived from the canonical commutation relations $[P,Q]=1$ in the
semiclassical limit \rEyZJii. The
important remark which simplifies the analysis is that no derivative of $u$
contributes to the equation for $\epsilon$ at leading order
(e.g.\ \elwkb). Therefore
$u$ appears only as a scale parameter and can be eliminated from the
equations.

We have seen that in the semiclassical limit, the operators $P,Q$ take the
form\foot{Note that in this subsection the normalization of $u(x)$
corresponds to generic potentials, i.e.\ the minimum residues of double poles
is 1.}
$$P(\d,u)=u^{p/2}P\left(\d u^{-1/2}\right) ,\qquad Q(\d,u)
=u^{q/2}Q\left(\d u^{-1/2}\right)\ .$$
{}From now on we call $P(z)$, $Q(z)$ the two polynomials $P(z=\d u^{-1/2},1)$,
$Q(z=\d u^{-1/2},1)$.
We recall that in \refsubsec\ssaamfs\ we have determined $P,Q$ for all models
such that $p=(2m+1)q\pm1$.

\medskip
{\it One-matrix models.}
Before discussing the general $p=(2m+1)q\pm1$  models, let us return to the
one-matrix models. From the analysis of the
corresponding non-linear differential equations, we have learned that the
variation $\epsilon$ of the specific heat $u(x)$ has
for $x$ large the asymptotic form
\eqn\eepscom{\epsilon'/\epsilon\sim r \sqrt{u}\ ,}
where $r$ is a constant determined by an algebraic equation. Since
the function $u$ can be treated at leading order as a constant, we can rescale
$\d$, i.e.\ set $u$ to 1. Equation \eepscom\ can then be written as a
commutation relation
\eqn\eepscomii{\d \epsilon=\epsilon(\d+r)\quad \Longrightarrow\quad
f(\d)\epsilon=\epsilon f(\d+r)\ .}
Then the operators $P,Q$ are simply
$$Q=\d^2-2\ ,\qquad P\equiv P_{2l+1}(\d)=\left(\d^2-2\right)^{l+1/2}_+\ .$$

The equation  for $\epsilon$ is obtained by expanding at first order in
$\epsilon$ the commutation relation $[P,Q]=1$. Setting
$$\delta P=\bigl\{\epsilon,R(\d)\bigr\}\equiv\sum_{k=0} R_k \{\epsilon,
\d^{2l-1-2k}\}$$
($\{,\}$ means anticommutator and this form takes into account the
antihermiticity of $P$), we obtain
$$\bigl[\{\epsilon,R(\d)\}\,,\,\d^2-2\bigr]+[P,-2\epsilon]=0\ .$$
Using the commutation relation \eepscomii\ to commute $\epsilon$ to the left,
we find the equation
$$-\left(2r\,\d+ r^2\right)\bigl(R(\d)+R(\d+r)\bigr)-2
\bigl(P_{2l+1}(\d+r)-P_{2l+1}(\d)\bigr)=0\ .$$

The first term vanishes for $\d=-r/2$, so this must give as well
a zero of the second term.
Taking into account the parity of $P_{2l+1}$, we obtain
\eqn\eonemat{P_{2l+1}(r/2)=0\quad
\Longleftrightarrow\quad (r^2-8)^{l+1/2}_+=0\ .}
The polynomial $R(\d)$ is then determined by division.
The function $(z^2-1)^{l+1/2}_+$ is also proportional to
$C^{-l}_{2l+1}(z)$ where $C_{2l+1}^\nu$ is a Gegenbauer polynomial defined by
analytic continuation in $\nu$  \rGM. Note that the number of zeros is exactly
the same as the number of operators in a $(p=2l-1,2)$ minimal conformal model
\rKPZ.
This is a property we shall meet again in the general case. In the
one-matrix case there is a natural explanation: the steepest descent
analysis shows that the number of different instantons (see the next
subsection) is
related to the degree of the minimal potential corresponding to a critical
point. This degree in turn is also related to the number of relevant
perturbations.\foot{We thank F.~David for this remark.}

The l.h.s.\ of the equation has a useful integral representation:
\eqn\eodra{\left(r^2-8\right)^{l-1/2}_+={\Gamma(l+1/2)\over
\Gamma(l)\Gamma(1/2)}\int_0^1 {\d s\over\sqrt{s}}
\bigl(r^2(1-s)-8\bigr)^{l-1}\ .}
For $l$ even, eq.~\eonemat\ is an odd--order equation that will have at least
one real solution for $r^2$, positive as is obvious from the integral
representation \eodra. The series therefore cannot be Borel summable.

For $l$ odd, on the other hand, the equation \eonemat\ for $r^2$ has no real
solutions and therefore we expect the solution of the differential equation
to be determined by the perturbative expansion.

\medskip
{\it General $(p,q)$ problem.} In the general $(p,q)$ case, in
classical limit as above and after the same rescaling, we have:
$$Q=Q(\d),\quad P=P(\d)=Q^{p/q}_+(\d),\quad \delta Q=\{S(\d),\epsilon\},\quad
\delta P=\{R(\d),\epsilon\}\ ,$$
where $P$, $Q$ are polynomials of degrees $p$, $q$, respectively,
and $R$, $S$ are polynomials with the same parity as $P$, $Q$
but with degrees $p-2$, $q-2$.

The equation for $\epsilon$ then leads to
$$[P,\delta Q]+[\delta P,Q]=0\ \Longleftrightarrow \
\eqalign{&\bigl(P(\d+r)-P(\d)\bigr) \bigl(S(\d)+S(\d +r)\bigr)\cr
&-\bigl(Q(\d+r)-Q(\d)\bigr) \bigl(R(\d)+R(\d +r)\bigr)=0\ .\cr}$$
The polynomial $P(\d+r)-P(\d)$ has degree $p-1$ in $\d$, while $R$ only has
degree $p-2$. An equivalent property is true for $Q,S$. Thus the polynomials
$P(\d+r)-P(\d)$ and $Q(\d+r)-Q(\d)$ must have at least one common root.
Note that the first polynomial has $p-1$ roots and the second $q-1$. Moreover
these roots are symmetric in the exchange $\d\mapsto -r-\d$.
The existence of a common root thus leads to $(p-1)(q-1)$ values
of $r$, up to the symmetry. Note that the number of zeros is again exactly
the same as the number of relevant operators in a $(p,q)$ minimal conformal
model \rKPZ\ of gravitationally dressed weights $\Delta_{m,n}$ (see
eq.~\edimLiou) with $1\le n\le q-1$ , $1\le m\le p-1$ , and possessing the
symmetry $\Delta_{m,n}=\Delta_{q-n,p-m}$. The explanation of this relation is
probably again
that the number of different instanton actions is related to the degree
of the minimal potential needed to generate a critical point in the
multi-matrix model, and thus to the number of different relevant operators.
Also we note that we are studying a general deformation of a critical
solution and therefore the appearance in some form of the relevant operators
should be expected.

The above condition determines the possible values of $r$ when the polynomials
$P$ and $Q$, i.e.\ the differential operators, are known in the classical
limit. Examples are provided by the models $p=(2m+1)q\pm 1$ where
these polynomials have been determined explicitly. The simplest examples
are provided by the $(q+1,q)$ models, i.e.\ the unitary models which we
examine below.

Finally we verify that we can indeed find the polynomials
$R,S$. We call $\alpha$ the common root and assume first that $\alpha\neq
-r/2$. Then the parity properties imply that $-r-\alpha$ is also a common
root. Setting
$$\eqalign{\bigl(P(\d+r)-P(\d)\bigr)&=(\d-\alpha)(\d+r+\alpha)
\,\tilde P(\d) \cr
\bigl(Q(\d+r)-Q(\d)\bigr)&=(\d-\alpha)(\d+r+\alpha)
\,\tilde Q(\d)\ , \cr}$$
we find that $R$ and $S$ are solutions of
$$R(\d)+R(\d +r)=(\d+r/2)\tilde P(\d)\ ,\quad S(\d)+S(\d +r)=(\d+r/2)\tilde
Q(\d)\ .$$
Note that these equations satisfy both the degree and parity requirements.

If $\alpha=-r/2$, the situation is even simpler
$$\eqalign{ R(\d)+R(\d +r)&=\bigl(P(\d+r)-P(\d)\bigr)/(\d+r/2) \cr
S(\d)+S(\d +r)&=\bigl(Q(\d+r)-Q(\d)\bigr)/(\d+r/2)\ .\cr}$$

\medskip
{\it Application: The unitary models.} In \refsubsec\ssaamfs\
we have seen that the differential operators $P,Q$ may be written in the
classical limit as
$$P= 2\, T_p(\d/2)\ , \qquad Q=2\, T_q(\d/2)\ ,$$
where $T_p$ is the $p$-th Chebychev's polynomial:
$$T_p(\cos\varphi)=\cos{p\varphi}\ .$$
%
As explained above, taking into account the degrees of the polynomials $R$
and $S$, we conclude that the polynomials $T_q((r+\d)/2)-T_q(\d/2)$ and
$T_p((r+\d)/2)-T_p(\d/2)$  must have a common root $\alpha=2\cos{\varphi_0}$.
Let also set $\alpha+r=2\cos{\psi_0}$.
We have
$$\cos{p\psi_0}=\cos{p\varphi_0} \quad {\rm and} \quad
\cos{q\psi_0}=\cos{q\varphi_0}\ .$$
The solution is
$$\psi_0 = \pm \varphi_0 + {2m\pi\over p} = \mp \varphi_0 + {2n\pi\over
q}\ .$$
Since $r=2\cos{\psi_0}-2\cos{\varphi_0}$, excluding the solutions $r=0$
which is not acceptable, we have the different solutions:
$$ r=\pm 4\sin{m\pi/p}\sin{n\pi/q},\quad 0< 2m  \le p\ ,\quad 0< 2n \le q
\ .$$

It is easy to verify that these results agree with the explicit solutions of
the $(2,3)$, $(4,3)$ and $(4,5)$ models. The results also show that,
as expected, all
unitary models lead to non-Borel summable topological expansions because all
terms of the series have the same sign. These models thus suffer from the same
disease as the pure gravity model. Note finally that the number of
different values of $r$ is indeed the same as the number of operators in the
minimal $(p,q)$ conformal model.

\medskip
{\it The general $p=(2m+1)q\pm 1$ models.}
For $m\neq 0$, $r$ is a solution to more complicated algebraic equations.
In the notation of previous subsection, we still have
$$\psi_0=\pm \varphi_0+{2n \pi \over q}\ .$$
We set
$$\alpha=\ud (\psi_0+\varphi_0),\qquad \beta=\ud (\psi_0-\varphi_0)\ ,$$
so that, making a choice of signs, we have
$$\beta={n \pi \over q},\qquad r=4\sin\alpha\sin \beta
=4\sin\alpha\sin(n\pi/q)\ ,$$
where
$$\sum_{l=0}^m {p/q\choose l}\sin\bigl((p-2ql)\alpha\bigr)
\sin\bigl((p-2ql)\beta\bigr)=0\ .$$
We note that $\sin\bigl((p-2ql)\beta\bigr)
=\sin(n\pi p/q)$, which can be factorized.
We thus find an equation for $\alpha$,
$$A(\alpha)\equiv \sum_{l=0}^m {p/q\choose l}\sin\bigl((p-2ql)\alpha\bigr)
=0\ .$$
The function $A(\alpha)$ satisfies the differential equation
$$p A(\alpha)(\cos q\alpha)'-q A'(\alpha)\cos q\alpha= K(p,q)\cos \alpha\ ,$$
where $K$ is a constant. This equation implies that
$A(\pi/2q)\,A(3\pi/2q)\le 0$
and thus $A(\alpha)$ vanishes at least once in the interval $(0,\pi)$. We
conclude that  for all these models the topological expansion is not Borel
summable.

\subsec{Instantons and large order behavior for one-matrix models}
\subseclab\ssIalobom

Following \rFDii, we now show that the large order behavior analysis
coincides with an instanton calculation of barrier penetration effects
for the one-matrix models.
For $g$ negative and $N$ finite, instanton effects, corresponding to a
single eigenvalue of the matrix model climbing to the top of the barrier,
are responsible for the divergence of perturbation theory at large orders.
In the large $N$ limit this effect is suppressed as $\e^{-K(g) N}$ and thus
the tree-level free energy is analytic for $0>g>g_c$.
Near $g_c$, however, we shall show that the function $K(g)$ vanishes as
$(g_c-g)^{(2l+1)/2l}$ so that in the double scaling limit the tunneling
amplitude remains finite and moreover given identically by \egeps.
In what follows we use the notation of \refsubsec\sSD.

In the large $N$ limit, the free energy is given by
\eqn\esaddF{F=\ln Z=N^2\left(\int\d\lambda\,\d\mu\,
\rho(\lambda)\rho(\mu)\ln|\lambda-\mu| -{1\over g}\int
\d\lambda\,\rho(\lambda)V(\lambda)\right)\ ,}
where $\rho(\lambda)$ is the eigenvalue density.
We wish to calculate the variation of the action
in the large $N$ limit when one eigenvalue is
displaced from position $\lambda_i$ to $\lambda_f$.
This variation of the distribution $\rho(\lambda)$ is of order $1/N$
and therefore can be obtained by the first order variation of \esaddF,
\eqn\eFvar{\delta F=N^2\int\d\lambda\ \delta\rho(\lambda)
\left(2\int\d\mu\,\rho(\mu)\ln|\lambda-\mu|-{1\over g} V(\lambda)\right)}
(which is non-vanishing since $\lambda$ is outside the
support of $\rho$). For
$$\delta\rho(\lambda)=
N\inv\bigl(\delta(\lambda-\lambda_f)-\delta(\lambda-\lambda_i)\bigr)\ ,$$
eq.~\eFvar\ can be written (using eq.~\esigsol)
\eqn\edeltaF{\eqalign{\delta F&=-2 N\int^{\lambda_f}_{\lambda_i}
\d\lambda\,\Bigl(\omega(\lambda)+{1\over 2 g}V'(\lambda)\Bigr)\ ,\cr
&=-{N\over g}\int^{\lambda_f}_{\lambda_i}
\d\lambda\,\sqrt{\bigl(V'(\lambda)\bigr)^2+ R(\lambda)}
=-2N \int^{\lambda_f}_{\lambda_i}\d\lambda\ \omega_{\rm sing}(\lambda)\ ,\cr}}
where $\omega_{\rm sing}(\lambda)$ is the singular part (which
scales in the continuum limit) of
$\omega(\lambda)=N\inv\bigl<\tr(M-\lambda)\inv\bigr>$.

{\it The case $l=2$}.
In the special case $l=2$, the explicit form \erhodist\
of $\omega_{\rm sing}(\lambda)$ gives
\eqn\edFlt{\delta F=-{N\over g}\int^{\lambda_f}_{\lambda_i}\d\lambda \,
(\lambda^2+1+\half a^2) \sqrt{\lambda^2-a^2}\ ,}
with $a^2={2\over3}\bigl(-1+\sqrt{1+12g}\bigr)$.
It is easily verified (and intuitively clear) that
the variation of the action is minimized by moving the eigenvalue
$\lambda_i=\pm a$ at the edge of the distribution.
The final position $\lambda_f$ is taken as the top of the barrier,
$\lambda_f^2=-1-a^2/2$, at which $(\d/\d\lambda_f)\delta F=0$.
For $g$ small and negative, it is easy to verify
that $\delta F$ is strictly negative and
therefore tunneling is suppressed in the large $N$ limit. Let us now examine
what happens as $g\to g_c=-1/12$. Then $\lambda_f \to \lambda_i$,
and substituting $a^2={2\over3}(-1+\sqrt x)$ into \edFlt, and evaluating
in the limit of small $x\equiv 1-g/g\dup_c$, gives
$$\delta F=-{\textstyle{4\over5}}\sqrt6 N x^{5/4}\ ,$$
in agreement with the result \ews\ (with $z=N^{4/5}x$).
The conclusion is simple:
instanton effects are suppressed in the large $N$ limit for fixed $g>g_c$ but
remain finite in the double scaling limit.
The scaling limit of the $l=2$ one-matrix model
thus corresponds to a {\it complex\/} solution of the Painlev\'e I
equation, presumably the complex Borel sum of the perturbative expansion.

{\it The case of general $l$}.
Returning now to the expression \edeltaF\ for the case of general $l$,
we take $\lambda_i$ to be the cut end-point $\lambda_i=a$ and $\lambda_f$
again the top of the barrier.
In the scaling limit, we know that from the steepest descent analysis of
\refsubsec\ssMp\ that $\omega_{\rm sing}(\lambda)$ has the form (eq.~\eomegsc)
\eqn\eomsing{\omega_{\rm sing}(\lambda)
=b^{-1/2}\int^a_\lambda\d s\,{\del x\over \del s}(s-\lambda)^{-1/2}\ , }
where $x=1-g/g_c$ is considered as a function of the cut end-point position
$a$ (as in \eomegsc). $\delta F$ then becomes
\eqn\egdF{\delta F = -2Nb^{-1/2}\int^{\lambda_f}_{a}\d \lambda
\int^{a}_{\lambda}\d s\, x'(s)\,(s-\lambda)^{-1/2}\ ,}
where $\lambda_f$ is determined by the condition
$$0={\d\over\d\lambda_f}\delta F =-2N\omega_{\rm sing}(\lambda_f)\ .$$
Changing variables in \eomsing\ to $s=\lambda+t(a-\lambda)$, this condition
can be rewritten
$$\int^{1}_{0}{\d t\over \sqrt t}
\,\bigl(\lambda_f+t(a-\lambda_f)\bigr)^{l-1} =0\ .$$
Comparing with eq.~\eodra, we find the solution $\lambda_f=a(1-r^2/8)$.
Substituting in \egdF, we derive
$${\d\over\d x}\delta F=
-2N b^{-1/2}\int_a^{\lambda_f}\d\lambda\,(a-\lambda)^{-1/2}
=Nr(2a/b)^{1/2}\ .$$
{}From \eZab\ we finally conclude that
\eqn\eagrlob{{\d\delta F\over \d x}=r\sqrt{u}={\epsilon'\over\epsilon}
\quad\Longrightarrow\quad\delta F\sim \ln\epsilon(x)
\sim-{2l\over2l+1} r z^{(2l+1)/2l}\ ,}
yielding a result in agreement with the result \egeps\ of the large order
behavior analysis.

\medskip
{\it Remarks.} The interpretation of the large order behavior calculation in
terms of instantons in the steepest
descent method allows us to understand the direct correspondence between the
property of Borel summability and the existence of the original integral. It
has been noted \rBMP\ that according to whether
$l$ is odd or even, the original minimal matrix integral is well-defined or
not because the integrand goes to zero in the first case while in the
latter case it blows up for $M$ large (see subsections {\it\ssMp\/},
{\it\sstgzpfr\/}). The direct
calculation given here of the instanton action
\rGZaplob, using steepest descent,
confirms that when the potential is unbounded from below the
instanton action is indeed real and the series therefore non-Borel summable.
Note moreover that replacing the minimum potential by a potential of
higher degree which would be bounded from below does not solve the problem.
Indeed the instanton result depends only on universal properties and thus
instantons would still appear. In the latter case they would reflect the
existence of another minimum of the potential, lower than the one in which one
assumes the eigenvalues are contained. Such a minimum would also invalidate
the direct calculation.

\subsec{$l=m$ perturbed by $l=m-1$}
\subseclab\sYLp


The calculation of the previous subsection can be adapted to relate
the loss of Borel summability when flowing from a model with $l$ odd
to another model with $l$ even. For definiteness we consider here the case
of flowing from $l=m$ to $l=m-1$, with $m$ odd. The general case is
treated identically.

To describe a perturbation of an $l=m$
model in the direction of an $l=m-1$ model we write, in the notation of
sec.~\stomm,
\eqn\eWR{W(b) = -(1-b)^m - \xi(1-b)^{m-1}}
(with $W$ as in \egxw).
A particular case would be the $l=3$ Yang--Lee edge singularity perturbed
by $l=2$ pure gravity, and the reader might derive further intuition by
substituting this particular case in the more general formulae that follow.
(The matrix model potential $V(M)$ associated to \eWR\ is easily reconstructed
by recalling that a potential $V(M)=\sum_p g_p\, M^{2p}$ in general leads to
$W(b)=2\sum_p {(2p-1)!\over ((p-1)!)^2} \, g_p\, b^p$, but the explicit form
of $V$ but will not be necessary in what follows here.)
%
%

The double scaling limit in this case involves taking
$b=1+N^{-2/(2m+1)}u$, and $\xi=N^{-2/(2m+1)}T$. The resulting
(all-genus) string equation that describes the flow of the $l=m$ model
to the $l=m-1$ model is
\eqn\efltt{R_m[u]+T\,R_{m-1}[u] = z\ .}
The scaling limit corresponds to $z\sim (g-g\dup_c)N^{2m/(2m+1)}$,
$T\sim u\sim z^{1/m}\sim N^{2/(2m+1)}$.
Thus, at leading order (genus zero),  $z$ and $T$
are large with $z/T^m$ fixed in such a way that the leading order equation
becomes
\eqn\elead{u^m+ Tu^{m-1}=z\ .}
To treat this (mixed) case more easily, we change variables to
$\tilde z\equiv z/T^m$, $\tilde u(\tilde z)\equiv u/T$,
and $\kappa^2\equiv T^{-(2m+1)}$. Eq.~\efltt\ reads with these notations
$$\tilde R_m[\tilde u]+\tilde R_{m-1}[\tilde u]=\tilde z\eqno\efltt'\ ,$$
where the notation $\tilde R$ indicates the substitution $\d\to\kappa\d$,
and perturbation theory corresponds to an expansion in small $\kappa$.
Eq.~\elead\ becomes
$$\tilde u^m+\tilde u^{m-1}=\tilde z\ .\eqno\elead'$$

By the same linearization procedure ($\tilde u\mapsto\tilde u+\epsilon$)
as used before, we find at leading order (in the notation of \elwkb)
$$0=\sum^{m}_{j=1} A_{m,j}\,\tilde u^{j-1}\kappa^{2m-2j}\epsilon^{(2m-2j)}
+\sum^{m-1}_{j=1} A_{m-1,j}
\,\tilde u^{j-1}\kappa^{2m-2-2j}\epsilon^{(2m-2-2j)}\ .$$
Substituting the WKB ansatz $\epsilon'/\epsilon=\kappa^{-1}\tilde u^{1/2}r$
gives an algebraic equation for $r$,
\eqn\efleq{0=\sum^{m}_{j=1} A_{m,j}\,r^{2m-2j}
+{1\over\tilde u}\sum^{m-1}_{j=1} A_{m-1,j}\,r^{2m-2-2j}\ .}
$r^2$ depends on $\tilde z=z/T^m$ implicitly through $\tilde u=u/T$.
In the limits $\tilde u$ large and small
(and hence via $\elead'$ $\tilde z$ large and small), we
recover the cases $l=m$ and $l=m-1$ respectively. Since we are assuming
$m$ odd, for some large enough value of
$\tilde z$ (and hence of $\tilde u=u/T$ classical),
the solution of eq.~\efleq\ goes from complex to real.
Thus for $T$ large enough, the equation for $r^2$
has real roots and the Borel summability is lost.

Now consider the value of $\xi$ for which the interpolating
potential $V(\lambda)$ corresponding to \eWR\
begins to admit real instantons. Employing the same steps leading
from this potential to the result \eagrlob\ for the instanton action,
we find that the instanton action
becomes real at precisely the value of $T$ for which the Borel summability is
lost, since both are determined by when the same equation \efleq\ has
real solutions $r$. Thus
above some value of $\xi$, which corresponds to the critical value
of $\tilde u\inv=T/u$
in \efleq\ for which the roots in $r^2$ become real, we see that the
interpolating potential substituted in \esaddF\
allows real instantons, giving a physical interpretation for
the loss of Borel summability.\foot{For $D=1$ models coupled to gravity,
``multicritical'' models, determined by
a potential $V(\lambda)$ that has its first
$s-1$ derivatives vanishing at the critical point, were considered in \rGZ.
These models have string
susceptibility $\gamma=-(s-2)/(s+2)$, and their physical interpretation is
in general unclear (but see \rGiMo).
They do however provide a useful arena for studying how
perturbations from one model to another can destroy Borel summability.
We expect again a mod$\,2$ grading that determines which models are Borel
summable, and as well which models may flow to one another.}

\subsec{Other properties of the equations. Moveable singularities}

We have investigated earlier in this section the large order behavior of
perturbation theory and concluded that in many cases of interest, in
particular pure gravity, the series is not Borel summable. In this subsection,
we investigate a few other properties of these equations, again first
considering the Painlev\'e equation.

\medskip
{\it The Painlev\'e I equation.} Eq.~\eai,
\eqn\ePagain{u^2(z)-{1\over3}u''(z)=z\ ,}
is a second order differential equation. The general solution depends on two
parameters fixed by the boundary conditions. It has as moveable singularities
double poles of residue 2, as one can immediately verify, and it can be proven
that these are the only moveable singularities \rBout.
Moreover it is easy to show that
any real solution of the equation has an infinite number of double poles on
the negative real axis. This leads to logarithmic singularities of the free
energy with weight 2 and thus to double zeros of the partition function.
(If we consider a general potential (not even), the free energy is
divided by 2 and thus the partition function has simple zeros.)

A more quantitative analysis can be obtained by transforming the Painlev\'e
equation. We set
$$u(z)=z^{1/2}v(y)\quad {\rm with}\quad y=\frac{4}{5}z^{5/4}\ ,$$
and then approximate the equation by keeping only the leading order
terms for $z$ large. The approximation
is also valid when $v'\gg v$, e.g.\ near the double poles.
It is finally very easy to write the resulting equation: we simply replace
$u\mapsto v$, $z\mapsto 1$, in \ePagain\ to find
$$v^2(y)-\frac{1}{3}v''(y)=1\ .$$
The equation can be integrated once
$$\frac{1}{3}v^3(y)-\frac{1}{6}v'{}^2(y)=v(y)+{\rm const}\ .$$
The solution has in general the form of a Weierstrass elliptic function
${\cal P}(y/\sqrt{2})$, with a doubly periodic lattice of double poles in the
complex plane.

\medskip
{\it Boundary conditions.} To understand the role of boundary conditions, we
calculate the variation of a solution $u_0$ of eq.~\eai\ when the boundary
conditions are infinitesimally changed,
$$u(z)=u_0(z)+\epsilon(z)\ .$$
$\epsilon$ satisfies the linearized equation \eaiii{a}\
$$\epsilon'' - 6u_0\epsilon=0\ .$$
It follows that $\epsilon$ is a linear combination of two independent
solutions $\epsilon_{\pm}$ which can be obtained, for $z$ large, by a WKB
ansatz (see eq.~\ews),
$$\epsilon_{\pm}\mathop{\propto}_{z\to+\infty}z^{-1/8}
\e^{\pm(4\sqrt{6}/5)z^{5/4}}\ .$$
This form shows that the equation is unstable for $z$ large both when solved
with $z$ increasing or decreasing 
since a small change in the boundary
conditions produces an exponentially increasing change in the solution.
Moreover solutions with the asymptotic
expansion \efasexp\ form only a one-parameter subset of solutions, because the
coefficient of the growing exponential must vanish. The meaning of this
remaining parameter is, at this point, obscure.
This result is related to the property that the topological expansion
\efasexp\ is non-Borel summable (see \refsubsec\ssPG),
and thus that its sum is ambiguous.

{}From these properties the following conclusions can be drawn:

(i) For the solutions with the correct asymptotic behavior, the existence of
double poles implies that the operator $H= -(\d/\d z)^2 +f(z)$ has a
discrete spectrum. This property is surprising since the operator $H$ is a
representation of the multiplication operator $\lambda$
(see \refsubsec\sstlnl).
It seems to indicate that, in contrast to what is seen in perturbation
theory, the
limiting distribution of eigenvalues of hermitian matrices is discrete.

(ii)  A careful analysis of the loop equations in the scaling limit
\rFDi, however, has shown that the spectrum of $H$ cannot be
discrete (see \refsubsec\ssLeVc).

(iii) Real zeros of the partition function, which is a sum
of positive terms, are not expected.

(iv) The initial matrix integral can only be defined by analytic
continuation since the critical value $g_c$ is negative. It can be argued
\rFDii\ that, as a consequence of this continuation, the function $f(z)$
is actually a complex solution of Painl\'eve equation, most likely the
complex Borel sum of the series, which for $z\to +\infty$ behaves like
$\sqrt{z}$ and like $\pm i \sqrt{-z}$ for $z\to -\infty$.

The conclusion is that the matrix integral does not define the sum over
topologies beyond perturbation theory.

\medskip
{\it The $l=3$ model.} In the case of the $l=3$ model, the situation is more
favorable. Since a real solution exists \rBMP,
the perturbation series is presumably Borel-summable. The initial
matrix integral is convergent in the scaling region. A solution has been
found numerically that behaves like $z^{1/3}$ for $z\to\infty$,
$-z^{1/3}$ for $z\to-\infty$ (for $N$ finite, the matrix model is defined
both for $g<g_c$ and $g>g_c$ and therefore the solution should have an
asymptotic expansion both for $z\to+\infty$ and $z\to -\infty$), and
which has no real double poles. The spectrum of
$H$ is thus continuous. The solution thus satisfies all requirements
and is presumably the correct solution to the initial problem. The
non-positivity of the terms in perturbation theory, however, is itself a
pathology of the model, and indeed it has been found \rIYL\
that the $l=3$ model corresponds to the Yang--Lee edge singularity, i.e.\
to the critical point of an Ising model in an imaginary magnetic field.

Note that as in the preceding case, some information about the structure
of the solution in the $l$ model, in the complex plane for $z$ large, can be
obtained by setting
$$u(z)=z^{1/l}v(y)\ ,\qquad y=z^{1+1/2l}/(1+1/2l).$$
Again the effect of the transformation is to yield an equation for $v$ which
at leading order is the same as for $u$ except that $z$ in the r.h.s.\ has
been replaced by 1. Because the equation follows from an action principle
a first integration can be performed.

\medskip
{\it General models.}  Although there is no general proof, there is very
good evidence for a systematic difference between the odd and even cases in
the one-matrix problem: Before taking the large $N$ continuum limit, the
operators $A$ and $B$ have a continuous spectrum and form a regular
representation of the canonical commutation relations. This property still
holds for the operator $\d^2-u$ when $l$ is odd, while for $l$ even the
operator has a discrete spectrum.\foot{In \rDSS,
it was further shown that it is
impossible to flow by means of the perturbation of eq.~\eWR\ from
the real solution of the $l=3$ model found in \rBMP\ to a
solution of the $l=2$ pure gravity model. This result generalizes to show
that flows from real solutions of arbitrary $l$-odd models to solutions of
$l$-even are impossible \rMoore\
(see also \rFDii), further distinguishing the $l$-odd and $l$-even cases.}

\medskip
{\it Moveable singularities.} All the differential equations of the $(p,q)$
models (subsections {\it\sCofde\/}, {\it\sTcl\/})
have as moveable singularities
double poles. For all the equations investigated so far, the residues of
these double poles are even integers. Although it was shown in \rFDi\
that solutions to these equations with poles on the real axis are not
physically relevant, for completeness we briefly recall here some previous
results and add results for the Ising and tricritical Ising models.

For the one-matrix models, we set
$$u(x)\sim{a/x^2}\ ,\qquad
R_l[u]\sim \rho_l(a)/x^{2l}\ ,$$
and using the recursion relation between $R_l$'s gives
$$\rho_{l+1}(a)={2l+1 \over2(l+1)}\bigl(a-l(l+1)\bigr)\rho_l(a)\ .$$
The solutions of the equation $R_l[u]=x$ have double poles with
residues belonging to the set ${2,\ldots,j(j+1),\ldots,l(l-1)}$ (all of which
are even integers).

For the critical and tricritical Ising models, a similar analysis shows that
the possible residues are $2,10$ and $2,6,8,14,30$, respectively ---
again even integers. In general for a $(p,q)$ model, it can be shown that the
residues are integers
whose smallest and largest values are respectively 2 and
${1\over12}(p^2-1)(q^2-1)$.
These results are calculated using a normalization for the differential
equation generated by an even matrix model potential, and the
poles above all correspond to double zeroes of the partition
function. For generic models with non-even potentials, on the other hand,
the residues are all divided by two and the partition function has only
simple zeroes. 

\newsec{Matrix canonical commutation relations, discrete action
principle and discrete KdV flows}
\seclab\sMccrda

As we shall show, many of the algebraic properties of the differential
equations of the continuum limit considered in previous
sections here are consequences of the properties of the
recursion formulae when $N$ is finite. We shall examine the one-matrix case
first because it is simpler and more explicit results are obtainable.
In what follows an essential role is played by the matrix
equivalent of differential operators, what we shall call {\it local\/}
matrices: matrices $X$ that have non-vanishing matrix elements only in a
strip of finite width, i.e.\ satisfying $X_{mn}=0$ for $|m-n|>l_X$.
(For other consideration of the discrete case, and in particular the
connection to Toda theory, see e.g.\ \rMardis.)

\subsec{The one-matrix case}
\subseclab\sstomxc

We have seen that all perturbative properties of the models can be obtained
from the commutation relations $[P,Q]=1$ of the continuum differential
operators which in turn are direct consequences of the matrix commutation
relation \ecombc\ of two matrices representing the operations of
differentiation and multiplication.

We shall therefore assume that we are given two local matrices $B$ and $C$,
i.e.\ such that the matrix elements $B_{mn}$, $C_{mn}$ are non-vanishing
only for $|m-n|\le 1$ and $|m-n|\le l-1$ respectively, which satisfy
\eqn\ecombcp{[B,C]=1\ .}
Moreover we assume $B$ symmetric and $C$ antisymmetric, and
$B_{n,n+1}$ non-vanishing for all values of $n$.
\medskip
{\it Reconstruction.}
With the hypothesis $B_{n,n+1}\ne 0\ \forall n$, we can construct by induction
an algebraic basis of polynomials $P_n$  such that
\eqn\edefPn{\lambda P_m(\lambda)=B_{mn}P_n(\lambda),\quad P_0=\
{\rm const}\,,\quad P_{-1}=0\ .}
With the $P_n$ defined, we can introduce a matrix $A$ representing
$\d/\d \lambda$,
\eqn\eAPn{P'_m(\lambda)=A_{mn}P_{n}(\lambda)\ .}
$A$ is by definition a lower triangular matrix ($A_{mn}=0$ for $n\ge m$)
which also satisfies
\eqn\ecomba{[B,A]=1\ .}

The matrix $A$, however, is not necessarily local. Setting
$X=\half (A-C)$, from \ecombcp\ and \ecomba\ we see that
$$[B,X]=0\ .$$
Because $B$ is symmetric, both the symmetric and antisymmetric part of $X$
commute with $B$. But an antisymmetric matrix commuting with $B$
necessarily vanishes, as can be proven by a double induction
on $m$ and $k$ on the matrix elements $X_{m,m+k}$ (see \refapp\ssDccr).
Thus $X$ is symmetric, and
$$X=A+A^T\ ,\quad C=\half(A-A^T)\ .$$

Because $A$ is lower triangular with vanishing diagonal elements, it is
uniquely defined  by the second equation above and is thus local.
Then $X$ is also local, $X_{mn}=0$ for $|m-n| > l-1$.
Again the commutation relation $[B,X]=0$ translates into recursion relations
for the matrix elements. It is shown in \refapp\sAfur\ that the commutation
relation has a general solution of the form $X=V'(B)$ where $V'(B)$ is an
arbitrary polynomial of degree $l-1$ in $B$. We thus find
\eqn\eAAt{A+A^T=V'(B)\ ,}
a formula analogous to \eAAd. Note that $C$ is then a representation of the
operator $\d/\d\lambda$ on the functions $\e^{-V(\lambda)/2}P_n(\lambda)$.

Equation \eAAt\ however does not embody the full content of eq.~\ecombcp.
Let us indeed assume the existence of a matrix $A$ with $A_{mn}\ne 0$ only for
$0< m-n \le l-1$, and satisfying \eAAt. We consider the commutator
$$J=[B,A]\ .$$
Because $A$ is lower triangular ($A_{mn}=0$ for $n\ge m$), $J$ satisfies
$J_{mn}=0$ for $n>m$. Moreover using \eAAt\ we have
$$J^T=\bigl([B,A]\bigr)^T=[A^T,B]=J\ .$$
Demanding $[B,A]=J=1$  then leads to an additional equation
\eqn\eJmn{J_{mm}=B_{m,m+1}A_{m+1,m}-B_{m-1,m}A_{m,m-1}=1\ ,}
with solution
\eqn\eabm{m=B_{m-1,m}\,A_{m,m-1}\ .}
The property that eq.~\eJmn\ is the difference of the two successive equations
\eabm\  is undoubtedly related to the property that in the continuum limit
the relation $[P,Q]=1$ has yielded the derivative of the string equation.

Of course eq.~\eabm\ can be directly obtained by comparing the terms of
highest  degree in eqs.~\edefPn\ and \eAPn. Indeed setting $P_n(\lambda)=p_n
\lambda^n +O(\lambda^{n-1})$, we obtain
$$ p_m=B_{m,m+1}p_{m+1}\ ,\qquad m p_m =A_{m,m-1}p_{m-1}\ ,$$
from which follows \eabm.
It is not proved however from this argument that this is the only missing
equation.

We have thus recovered the
recursion relations \eqns{\eqfund{,\ }\eqfundb} of the one-matrix case starting
only from the commutation relation and some conditions on the matrices $C$,
$B$.
The recursion relations determine the matrix $B$, however, only as a function
of its first $l-2$ matrix elements, whereas it is completely determined in the
matrix model or when it is related to orthogonal polynomials. It follows
that even when the integral $\int\d\lambda\,\e^{-V(\lambda)}$ exists,
the polynomials $P_n(\lambda)$ are not in
general the orthogonal polynomials corresponding to the measure
$\d\lambda\,\e^{-V(\lambda)}$. In particular the symmetric matrix $S$,
$$S_{mn}=\int\d\lambda\,\e^{-V(\lambda)}P_m(\lambda)P_n(\lambda)\ ,$$
is not in general diagonal.
\medskip
{\it Remarks.} (i) It is easy to verify that $S$ commutes
with $B$ and $A$. Therefore, as shown in \refapp\sAfur,
if $S$ is local, it is a
polynomial in $B$: $S=S(B)$. Then expressing that it commutes with $A$ we find
$S'(B)=0$, i.e.\ the polynomial $S(B)$ is a constant and $S$ is a multiple of
the identity. Therefore, when the first $l-2$ elements of $B$ are not those
given by the matrix model, the matrix $S$ is not local.

(ii) Based on considerations concerning
the instability of the sequences generated
by the recursion relations (analogous to the instability of the continuum
differential equations), we believe that when the $l-2$ coefficients of $B$ do
not correspond to orthogonal polynomials, the continuum limit does not exist
because the matrix elements $B_{n,n}, B_{n,n+1}$ do not have a smooth behavior
for $n$ large.
In particular when the integral $\int\d\lambda\,\e^{-V(\lambda)}$ does
not converge,
no real set of matrix elements has a smooth behavior and the continuum limit
can only be defined when the $P_n$ are taken orthogonal with respect
$\e^{-V}$, with the integral calculated along a complex path on which it
converges. This argument applies to the even critical models. It emphasizes
the importance of the boundary conditions for the differential equations
representing the string equations in the continuum limit.

(iii) We recognize in the decomposition of $V'(B)$ into the sum of a lower and
upper triangular matrix a structure very  similar to the one encountered in
the continuum case (sec.~\sAgm) in which we introduced $Q^{k-1/2}_{\pm}$.

\medskip
{\it An action principle.} As in the continuum, the discrete recursion
relations \eqns{\eqfund{,\ }\eqfundb} can be obtained from a variational
principle by varying the discrete action
\eqn\edaS{{\cal S}(B)=-\sum_n {gn\over N}\ln B_{n-1,n} +\tr V(B)\ .}
%

\subsec{Discrete form of KdV flows, one-matrix case}

The solutions of the commutation relations depend on the coefficients of the
polynomial $V'(B)$. Let us explore what happens when these coefficients vary.
We denote a parameter characterizing $V$ by $t$. Since the polynomials
$P_n$ also depend on $t$, we set
\eqn\edefSig{{\del P_m \over \del t}=\Sigma_{mn}P_n\ .}
It follows that $\Sigma_{mn}=0$ for $n>m$.
Differentiating eqs.~\edefPn\ and \eAPn\ with
respect to $ t$ and using the linear independence of the $P_n$'s, we
find \Wun
\eqna\eKdVdis
$$\eqalignno{{\del A \over \del t}& =\bigl[\Sigma,A\bigr]\ ,
&\eKdVdis a \cr
{\del B\over\del t}& =\bigl[\Sigma,B\bigr]\ .&\eKdVdis b \cr}$$
For a set of parameters $ t_i$ and corresponding matrices $\Sigma_i$,
\edefSig\ implies
$$\del_i \Sigma_j-\del_j \Sigma_i +\bigl[\Sigma_j,\Sigma_i\bigr]=0\ .$$
Eqs.~\eKdVdis{a,b}\ thus define a set of commuting discrete KdV flows
analogous to the flows \ekdvf\ of the continuum differential equations.

Since $B$ is symmetric, \eKdVdis{b}\ implies that
\eqn\esigmab{\bigl[B,\,\Sigma +\Sigma^T\bigr] =0\ .}
Differentiating \eAAt\ with respect to $ t$ and using \eKdVdis{a},
we find after transposing:
\eqn\ecomAS{\bigl[A,\ \Sigma +\Sigma^T\bigr]={\del V'(B)\over\del t},}
where $\del V/ \del t$ in the r.h.s.\ means derivative with respect to $t$ at
$B$ fixed. From \esigmab\ and $[B,A]=1$ we see that a particular solution is
\eqn\esigmas{\Sigma +\Sigma^T =-{\del V(B)\over\del t}
+ \hbox{ const}\ .}
The constant corresponds to a trivial rescaling of all polynomials $P_{n}$.
Since $\Sigma$ is lower triangular, it is entirely determined by \esigmas\
--- in particular we see that it is local.

A general solution differs from \esigmas\ by a matrix which commutes both
with $B$ and $A$.  A local matrix which commutes with $B$ and $A$, however,
must be a
multiple of the identity (see remark (i) above). Therefore if $\Sigma$ is
local, eq.~\esigmas\ yields the most general solution of
eqs.~\eqns{\esigmab{,\ }\ecomAS}.

More general solutions are necessarily non-local. It can be shown
that a general symmetric matrix $X$ that commutes with both $A$
and $B$ vanishes only when its matrix elements $X_{mn}$ with $m+n\le l-3$
vanish. The reason is easily understood: Variations of the matrix $B$ can
be separated into variations induced by a variation of the function $V$,
corresponding to local matrices $\Sigma$ of the form \esigmas, and
variations induced by variations of the $l-2$ first matrix elements of $B$ not
determined by the recursion relations, corresponding to non-local matrices
$\Sigma$. This result indicates that KdV
flows generated by local and non-local matrices play a very different role.

The result \esigmas\ can be easily verified in the matrix model. Let
$ t_i$ be for example the coefficient of $B^i$ in the potential $V(B)$.
Differentiating the orthogonality relations \eorthrl, we find
\eqn\esigmay{\Sigma_i +\Sigma_i^T ={\del V(B) \over \del t_i}=B^i\ .}
Note that the commutation relation $[B,A]$ together with eq.~\eAAt\
constitute a
particular case of this result: They correspond to an infinitesimal variation
of $V(M)$ resulting from a translation of $M$: $M\mapsto M+ t 1$, which
induces an identical translation of $B$.

\medskip
{\it Remarks}

(i) Since $C$ is also a representation of a fixed operator, we expect a flow
equation for it. One verifies indeed that $\half(\Sigma-\Sigma^T)$
(an antisymmetric matrix) generates a flow for $C$ and $B$.
We denote by $B^i_+$ the antisymmetric matrix defined by
$$\Sigma_i+\Sigma_i^T=B^i \ ,\qquad B^i_+=\half (\Sigma_i-\Sigma_i^T)\ .$$
As above we parametrize the potential $V(B)$ as $V(B)=\sum_i t_i B^i$.
With these definitions, we can write
$${\del B\over \del t_i}=[B^i_+, B]\ ,\qquad C=\sum_i i \, t_i B^{i-1}_+\ ,$$
in complete analogy with \pqstreq.

\smallskip
(ii) The flow equations reflect the property that the matrices $B$ and $A$ are
the representation in a basis which depends on some parameters of a fixed
operator corresponding to a multiplication by $\lambda$. Indeed two
matrices $B$ corresponding to two
sets of parameters $ t_i$ and $ t'_i$ of the initial integrand are
related by a linear transformation $R$,
$$B\left( t_i\right)=R\left( t_i, t'_i\right)
B\left( t'_i\right) R^{-1}\left( t_i, t'_i\right)\ .$$
Differentiating with respect to $ t_i$, one finds \eKdVdis b\ with
$$\Sigma_i={\del R \over \del t_i}R^{-1}\ .$$
This definition of $\Sigma_i$ may differ from the
previous one by the addition of a matrix commuting with $B$.

Conversely, let us assume \eKdVdis{b}\ and in addition that $B$ is
diagonalizable. In our problem, this condition is satisfied when $B$
is defined via polynomials orthogonal with respect to the measure
$\d\lambda\,\e^{-V(\lambda)}$, ensuring that the functions
$\e^{-V(\lambda)/2}P_n(\lambda)$ are the
eigenvectors of $B$ corresponding to the eigenvalue
$\lambda$. The other cases require a separate investigation of the behavior
of $B_{mn}$ for $m,n$ large.

We can then express $B$ in terms of its eigenvalues $\lambda_i$ (we write
them as a discrete set although they could as well be continuous),
$$B=R^{-1}\Lambda R\ ,$$
which transforms \eKdVdis{b}\ into
$${\del\Lambda\over\del t}=\bigl[\tilde\Sigma, \Lambda\bigr]\ ,$$
with
$$\tilde\Sigma=R\Sigma R^{-1}+{\del R\over\del t}R^{-1}\ .$$
Since $\Lambda$ is diagonal, the equation implies that $\tilde\Sigma$ is
also diagonal. It follows that
$${\del\Lambda\over\del t}=0\ .$$
The spectrum of $B$ is thus independent of $ t$. This confirms indirectly
that in the large $N$ limit the spectrum cannot become discrete.

\medskip
{\it The resolvent.} It is easy to verify that, in direct analogy with the
continuum case, the diagonal matrix elements of the resolvent
$G(z)=(B-z)\inv$ satisfy a four term recursion equation. Defining
$$\gamma_n =\bigl[G(z)\bigr]_{nn}\ ,$$
one finds (for an even potential) 
$$z^2\left(\gamma_n-\gamma_{n-1}\right)=r_{n+1}
\gamma_{n+1}+r_n\left(\gamma_n-\gamma_{n-1}\right) -r_{n-1}
\gamma_{n-2}\ .$$
(In the continuum limit, this equation becomes eq.~\eresolv\ of
\refapp\ssTrLcq.)

\subsec{Multi-matrix case}
\subseclab\ssmmdis
The general case is not in the same satisfactory state. We shall therefore
give a few general results valid for the multi-matrix case  and then consider
separately the two-matrix case for which more detailed results can be
obtained.

In the continuum limit, all the matrices reduce to two differential operators.
It would be useful to derive, as in the one-matrix case, all
properties entirely from consideration of two local matrices satisfying
canonical commutation relations. Only partial results can be easily obtained.
The major reason for this new difficulty is that while in the continuous case
only two operators appear, the matrix model is originally defined in terms of
several matrices forming equivalent representations of the commutation
relation.  As we shall see, the general problems we would like to solve are
the following: (i) Characterize all pairs of local matrices $B$ and $A$
satisfying the commutation relations $[B,A]=1$.
(ii) Understand whether all solutions
correspond to a matrix model, or whether additional conditions
are necessary to insure such a correspondence.
(iii) Find the subclass of matrices $B$ which
lead to a continuum limit. Characterize all flows generated by local
matrices $\Sigma$:
$${\del B\over \del t} =\bigl[\Sigma,B\bigr]\ ,\quad
{\del A\over \del t} =\bigl[\Sigma,A\bigr]\ ,$$
such that $B$ and $A$  remain in the class defined above if they belong to
this class for $t=0$. Unfortunately we shall only give partial answers to
these questions.

{\it An action principle\/}.
It is, however, easy
to verify that eqs.~\erecgen\ can also be  directly derived
from a more complicated variational principle than \edaS, taking for action
$$\eqalign{{\cal S}\left(B_{a}\right)& =\sum_{a=1}^{\nu-1}\tr
V_{a}\left(B^{(a)}\right)
-\sum_{a=1}^{\nu-2} c_{a}\,\tr B^{(a)} B^{(a+1)} \cr &\quad
- \sum_n n\bigl(\ln [B_1]_{n-1,n}+\ln[B_{\nu-1}]_{n,n-1}\bigr).\cr}$$

\noindent{\it Discrete canonical commutation relations}\par\nobreak

Thus let $B$, $A$ be two {\it local}\/ matrices  be such that $B_{mn}\ne 0$
only for $1-r\le n-m\le 1$ and $A_{mn}\ne 0$ for $m\ge n$ with $[B,A]=1$.
(In the notation of
sec.~\sqmomm, we have in mind here the matrices $A_1$ and $B_1$.)
We again assume $B_{m,m+1}\ne 0$ for all $m\ge0$.

\medskip
(i) It is proven in \refapp\ssDccr\ that the matrix $A$ is uniquely defined in
terms of $B$. The matrix $A$ can then be explicitly determined by first
constructing inductively an algebraic basis of polynomials $P_n$ such that
\eqn\edefPna{\lambda P_m(\lambda)=B_{mn}P_n(\lambda)\ ,}
and thus
\eqn\eAder{P'_m(\lambda)=A_{mn}P_n(\lambda)\ .}

(ii) The preceding results do not involve the locality of $A$ and $B$.
Expressing that $A$ and $B$ are local matrices yields $r+1$ recursion
relations  for the matrix
elements $B_{m+k-1,m}$, $0\le k\le r$. To prove this result, one solves
$[B,A]=1$ following the method of \refapp\ssDccr. This leads
to recursion relations for $A_{m+k,m}$ which depend on the $A_{m+k-n,m}$,
$n=1,...,r$ which have already been determined. Expressing that
$A_{m+k,m}$ vanishes for all $k>s$ leads then to the corresponding equations
for $B$.

\subsec{Generalized loop equations}

Let us write the equations which follow from the commutation relation
$[B,A]=1$. We first specialize to $m=n$. Then
$$B_{n,n+1}A_{n+1,n}-A_{n,n-1}B_{n-1,n}=1\ ,$$
and thus
$$B_{n,n+1}A_{n+1,n}=n+1\ .$$
Writing the other equations and after some tedious algebraic manipulations,
one finds generalized loop equations 
\eqn\eloopg{\eqalign{&\sum_{n=0}^{N-1}
\left[A(z-B)^{-1}+(z-B)^{-1}A\right]_{nn}=
\left(\sum_{n=0}^{N-1}(z-B)^{-1}_{nn}\right)^2\cr &\qquad +\sum_{0\le m\le
N-1<n} (z-B)^{-1}_{mn}(z-B)^{-1}_{nm}\ .\cr}}
This equation should be understood as a generating function for the set of
equations obtained by expanding in powers of $1/z$.

\medskip
{\it Matrix model.} Let us verify this result in the matrix model. The
ingredients we need are the transformation law of the measure when
$\lambda_i^{(1)}\mapsto \lambda^{(1)}_i+\varepsilon
\bigl(\lambda^{(1)}_i\bigr)^k$. We find
$$\eqalign{&\prod_i \d\lambda_i
\prod_{i<j}\left(\lambda_i-\lambda_j\right)\cr
&\qquad\qquad\mapsto
\prod_i \d\lambda_i \prod_{i<j}\left(\lambda_i-\lambda_j\right)
\left(1+\varepsilon \Bigl(\sum_i k\lambda_i^{k-1}+\sum_{l=0}^{k-1}
\sum_{i<j} \lambda_i^{l}\lambda_j^{k-l-1}\Bigr)\right)\ .\cr}$$
It is convenient to rewrite the last factor, using
$$\sum_i k\lambda_i^{k-1}+\sum_{l=0}^{k-1}
\sum_{i<j} \lambda_i^{l}\lambda_j^{k-l-1}=
\ha\Bigl(\sum_i k\lambda_i^{k-1}+\sum_{l=0}^{k-1}
\sum_{i,j} \lambda_i^{l}\lambda_j^{k-l-1}\Bigr)\ .$$
The corresponding variation of the integrand is the sum over eigenvalues, and
we need only the variation for one eigenvalue. Let us call $\rho(\lambda)$
the integrand concentrating only on the dependence on $\lambda\equiv
\lambda^{(1)}$, then
$$\delta \rho(\lambda)=\varepsilon \lambda^{k}\rho'(\lambda)\ .$$

We wish to calculate the integral of $\delta\rho$ multiplied by a
polynomial $P_n(\lambda)$. Integrating by parts, we find
$$\eqalign{\int\d\lambda\,\lambda^{k}\rho'(\lambda)P_n(\lambda) &=
-\int\d\lambda\,\rho(\lambda)\left(\lambda^{k}P_n(\lambda)\right)'\cr
&= -\int\d\lambda\,\rho(\lambda)\,B^k_{nm}\,A_{ml}\,P_l(\lambda)\ .\cr}$$
Gathering all terms, we recover the expansion in powers of $1/z$ of the loop
equations \eloopg. Note that the proof starting from the matrix
model does not depend on the potentials being polynomials, in the same way as
the proof starting from the commutation relations does not depend on the
locality of $B$ and $A$. On the other hand, the existence of the resolvent
appearing in the loop equations implies some non-trivial topological
properties of the matrix $B$.
As we have seen in the one-matrix case, these have non-perturbative
consequences and may be connected with the existence of
a continuum limit.

\subsec{Discrete form of KdV flows}

We assume now that the matrices $B$, $A$ depend on a parameter $t$. It
follows that the polynomials $P_n$ depend also on $t$. We again set
\eqn\edefSiga{{\del P_m \over \del t}=\Sigma_{mn}P_n\ .}
It follows that $\Sigma_{mn}=0$ for $n>m$. Differentiating eq.~\edefPna\ with
respect to $t$ and using the linear independence of the $P_n$'s, we
find
\eqn\eKdVdisc{{\del B \over \del t}=\bigl[\Sigma,B\bigr]\ ,\quad
{\del A \over \del t}=\bigl[\Sigma,A\bigr]\ .}
For a set of parameters $t_i$ and corresponding matrices $\Sigma_i$
\edefSiga\ implies that
$$\del_i \Sigma_j-\del_j \Sigma_i +\bigl[\Sigma_j,\Sigma_i\bigr]=0\ .$$
Eq.~\eKdVdisc\ thus defines a set of commuting discrete KdV flows
analogous to the flows \ekdvf\ of the continuum differential equations,
and thus far the argument has been identical to the one-matrix case
eq.~\edefSig\ and following.
Unlike the one-matrix case, however, it is no longer obvious from these
considerations when the matrices $\Sigma$ are local. It is easy to verify
that the flows associated with the matrix model are all local. We shall show
below that in the multi-matrix model, as in the one-matrix case, non-local
flows correspond instead to modifications of the boundary conditions of the
recursion equations satisfied by the matrix elements of $B$, generating
matrices $B$ which no longer correspond to a matrix model.  However, an
interesting problem remains: Can all flows generated by local matrices which
preserve the locality of the matrices $A$ and $B$ be associated with a matrix
model?

\medskip
{\it Explicit form of the flow generators in the multi-matrix models.} If we
assume again the set of
equations \erecgen\ then we can find an explicit expression for the matrices
$\Sigma$.  Note that in what follows, to simplify
notations, we shall set $c_a=1$ (this corresponds just to a rescaling of the
matrices $M_a$ or $B_a$). Let us call $\Sigma^{(1)}$ the matrix generator
acting on $A_1,B_1$,
\eqn\eKdVdisi{{\del B_1 \over \del t}=\bigl[\Sigma^{(1)},B_1\bigr]\ ,\quad
{\del A_1 \over \del t}=\bigl[\Sigma^{(1)},A_1\bigr]\ .}
In the same way, starting from the variation of the polynomials
$\widetilde P_n$  associated with the matrix $\tilde B_{q-1}$,
we can define a
lower triangular matrix $\widetilde \Sigma^{(q-1)}$
which generates the flow of $\tilde A_{q-1},\tilde B_{q-1}$.

We now differentiate eqs.~\erecgen\ with respect to the parameter
$t$. The first equation
$$A_1+B_2=V'_1(B_1)$$
yields
$${\del B_2\over\del t}=\bigl[\Sigma^{(1)},B_2\bigr]
+{\del\over\del t} V'_1(B_1)\ .$$
We introduce a matrix $\Sigma^{(2)}$,
$$\Sigma^{(2)}=\Sigma^{(1)}-{\del \over \del t}V_1(B_1)\ .$$
Using the commutation relation $[B_2,B_1]=1$, we find
\eqn\eKdVdisii{{\del B_1 \over \del t}=[\Sigma^{(2)},B_1],\quad
{\del B_2 \over \del t}=[\Sigma^{(2)},B_1]\ .}

This argument can be repeated after introducing matrices $\Sigma^{(a)}$
$a=3,\ldots, q$. The result is
\eqn\eKdVdisq{{\del B_{q-1} \over \del t}=[\Sigma^{(q)},B_{q-1}]\ ,
\quad  {\del A_{q-1} \over \del t}=[\Sigma^{(q)},A_{q-1}]\ .}
Comparing with the definition of $\widetilde\Sigma^{(q-1)}$ above, we also have
$$\Sigma^{(q)}=(\widetilde\Sigma^{(q-1)})^T+X\ ,$$
where $X$ is a matrix that commutes with $A_{q-1}$ and $B_{q-1}$.
When $X$ vanishes, we can write
\eqn\esigsigen{\Sigma^{(1)}+(\widetilde\Sigma^{(q-1)})^T=
{\del\over\del t} \sum_{a=1}^{q-1} V_a(B_a)\ ,}
which determines the matrices $\Sigma_{(1)}$ and $\widetilde\Sigma^{(q-1)}$
because they are both lower triangular, except for
the diagonal elements for which only the sum for the two matrices is fixed.
This was to be expected since an opposite change in the diagonal elements of
these two matrices corresponds to a trivial change in the normalization of the
polynomials $P_n$ and $\widetilde P_n$:  $P_n \mapsto \mu_n P_n$, $\widetilde
P_n \mapsto \mu_n^{-1} \widetilde P_n$. It is easy to verify equation
\esigsigen\ in the matrix model.

As in the one-matrix case, a flow generated by a matrix $X$ commuting with $A$
and $B$ corresponds to a variation of the boundary conditions of the
recursion relations. It remains to be shown that, as in the one-matrix case,
such a matrix is necessarily non-local. This follows from the result
proven in the \refapp\sAfur.

Therefore the main difference from the one-matrix case is that we have not been
able to characterize all local solutions of the commutation relations and all
flows generated by local matrices as being related to a multi-matrix model.
Note finally that the commutation relation $[B,A]=1$ does not yield
additional constraints on $\Sigma$ because the commutator $[B,A]$ is a flow
invariant.


\def\pp{r}
\def\qq{s}
\def\rr{\pp}
\def\Om{\Omega}

\newsec{The $O(n)$ matrix model}
\seclab\stOnmm

In this section we employ the techniques developed in earlier sections
to the case of the $O(n)$ matrix model coupled to gravity.
Many of the results for the latter model provide interesting analogs for
the other cases considered earlier here. (For early work on the $O(n)$ matrix
model, see \rIK.)

The partition function of the $O(n)$ matrix model is given by an integral over
$n+1$ hermitian $N\times N$ matrices, $n$ matrices $A_i$ and a matrix $M$:
\eqn\eZOn{Z=\int \d M\,\d A_1\ldots \d A_n\, \ee{-(N/ g)\tr
[M(A^2_1+\cdots+ A^2_n)+V(M)]}\ ,}
with $V(M)$ a general polynomial potential.

The free energy of the $O(n)$ matrix model $F=\ln Z$ can be interpreted as the
partition function of a gas of loops, each indexed by an integer $i$,
$i=1,...,n$,  drawn on a random lattice of the form of a Feynman diagram
\refs{\rIK,\rEyZJ}. In the special case $n=1$ and
for a specific class of cubic potentials $V(M)$, the model can be shown to be
equivalent to a two-matrix  model of the form considered in sec.~\sqmomm,
representing an Ising model on a random triangulated lattice.
The model cannot be solved exactly in the general case;  we shall
show however that
it can be solved in the large $N$ limit, i.e.\ on the sphere, by
steepest descent.

The corresponding model on regular lattices can become critical only for
values $-2\le n\le 2$ of $n$
mainly associated with non-physical symmetry groups
(here the integral is only even defined for $n\le 2$). It is thus
convenient to set $n=-2\cos{\theta}$. Although
we could restrict to the interval $0\le \theta
\le \pi$, it is convenient for book-keeping purposes to consider all positive
values of $\theta$. Note that the case $n=0$,
$\theta=(2m+1)\pi/2$ reduces to the standard one-matrix model.

The integral over the matrices $A_i$ is gaussian and can be performed to give
\eqn\eZIii{Z=\int\d M\, [\det(M\otimes 1+1\otimes M)]^{-1/2}\e^{-(N/g)\tr
V(M)}\ .}
We can then parametrize $M$ in terms of a unitary transformation and its
eigenvalues $\lambda_i$. After integration over unitary matrices,
the integral \eZIii\ becomes
\eqna\eZlambda
$$\eqalignno{Z&=\int  \Delta^2(\Lambda)
\prod_{i,j}(\lambda_i+\lambda_j)^{-n/2}\prod_i \d\lambda_i\,
\ee{-(N/ g) V(\lambda_i)}& \eZlambda{a} \cr
&=\int \d\lambda\, \ee{-N\, \Sigma[\lambda]}\ ,& \eZlambda{b} \cr} $$
with the effective action
$$ \Sigma[\lambda] =\sum_i {1\over g}V(\lambda_i)-{1\over N}\sum_{i\neq j}
\ln{\vert\lambda_i-\lambda_j\vert}+{n\over 2N}\sum_{i,j}
\ln(\lambda_i+\lambda_j). $$

\subsec{The saddle-point equation}

In the planar limit $N\to\infty$, $Z$ can be calculated by the steepest
descent method. The saddle point equation is
\eqn\eesaddle{{\del \Sigma\over\del\lambda_i}=0={1\over
g}V'(\lambda_i)-{2\over N}\sum_{j\neq i} {1\over\lambda_i-\lambda_j}+{n\over
N}\sum_j{1\over\lambda_i+\lambda_j}\ .}
We introduce the density of eigenvalues $\rho(\lambda)=
{1\over N}\sum_i \delta(\lambda-\lambda_i)$, and its Hilbert's transform,
$$ \omega_0(z)={1\over N}\sum_i {1\over z-\lambda_i}=\int \d\lambda
{\rho(\lambda)\over z-\lambda}\ ,$$
the trace of the resolvent.
In the large $N$ limit, $\rho(\lambda)$ becomes a continuous function and
$\omega_0$ becomes
a function analytic  except when $z$ belongs to the spectrum of
$M$, i.e.\ has a cut on a segment $[a,b]$ of the real positive axis.

Eq.~\eesaddle\  may be written in terms of $\omega_0$ as
\eqn\eqwze{{\omega_0(\lambda+i0)+\omega_0(\lambda-i0)+n\omega_0(-\lambda)
={1\over g}V'(\lambda)\ ,\qquad (\lambda\in [a,b])}\ .}
This linear equation has a polynomial solution,
\eqn\eqwr{\omega_r(z)={1\over g}{1\over 4-n^2}\bigl(2V'(z)-nV'(-z)\bigr)\ .}
Note that the cases $n=\pm 2$ are special and must be examined separately.
The function $\omega(z)$, defined by
\eqn\eomreg{\omega_0=\omega_r+\omega/g\ ,}
then satisfies the homogeneous equation
\eqn\egenRH{\omega(\lambda+i0)+\omega(\lambda-i0)+n\omega(-\lambda)=0\ .}
Since $\omega_0(z)$ behaves as $1/ z$ for $z$ large, $\omega(z)$ has the
large $z$ expansion
\eqn\ezlarge{\omega(z)
=-{1\over 4-n^2}\bigl(2V'(z)-nV'(-z)\bigr) +{g \over z}
+O\left(z^{-2}\right)\ .}

\medskip
{\it A quadratic relation.} We introduce the following function:
\eqn\equad{\rr(z)=\omega^2(z)+\omega^2(-z)+n\omega(z)\omega(-z)\ ,}
and verify that the
discontinuity on the cut of $\rr(z)$ vanishes as a consequence of equation
\egenRH :
$$\eqalign{\rr(z+i0)-\rr(z-i0)&=\left[\omega(z+i0)-\omega(z-i0)\right] \cr
&\quad \times\bigl(\omega(z+i0)+\omega(z-i0)+n\omega(-z)\bigr)=0\ .}$$
Therefore $\rr$ is an even function, analytic  in the whole complex plane. The
behavior of $\omega$ for $z$ large implies that $\rr$ is a polynomial.
As in the usual one-matrix case, equation \equad\ can be directly derived from
the saddle point equation or the loop equation. An expression for $\rr(z)$ in
terms of the potential follows.
\medskip
{\it The one-cut solution: a useful representation.}
We introduce two auxiliary functions:
\eqn\edefwpm{\left\lbrace\eqalign{
\omega_{+}(z)&={i \over 2\sin\theta}\left(\e^{i\theta/2}\omega(z)-
\e^{-i\theta/2}\omega(-z)\right)\cr
\omega_{-}(z)&=-{i \over 2\sin\theta}
\left(\e^{-i\theta/2}\omega(z)-\e^{i\theta/2}\omega(-z)\right)\ ,  \cr}
\right.}
such that $\omega_{+}(-z)=\omega_{-}(z)$, and we have
$$\omega_{+}(z)\,\omega_{-}(z)=\pp(z)\ .$$
Conversely, $\omega(z)$ is given in terms of $\omega_+,\omega_-$  by
\eqn\esolw{\omega(z)=-\left(\e^{i\theta/2}\omega_{+}(z)
+\e^{-i\theta/2}\omega_{-}(z)\right)\ ,}
and eq.~\egenRH\ is equivalent to the simple relations
\eqn\edisc{\omega_{\pm}(z-i0)=\e^{\pm i\theta}\omega_{\mp}(z+i0)\ , }
which, by themselves, imply that $\omega(z)$ has cuts only on the positive
axis.

For generic values of $n$, the Riemann surface of $\omega(z)$ has an infinite
number of sheets. Only for the exceptional values $n=-2\cos(\pi p/q)$,
i.e.\ for $\theta=\pi p/q$, $p,q$ being two
relatively prime integers (and thus $\e^{iq\theta}=\pm1$), is this number
finite. Then the function  $\omega(z)$ is the solution of an algebraic
equation of degree $q$ with polynomial coefficients.

\medskip
{\it The case $\e^{iq\theta}=1$.} This implies $\theta=\pi p/q$,
where $p$ is an even integer. The function
\eqn\edefq{\qq(z)=\ud (\omega_{+}^q+\omega_{-}^q)}
has no discontinuity on the cut, and is analytic in the whole complex
plane. It is therefore an even polynomial of a degree determined by the
degree of the potential .
We have thus the two following algebraic equations:
\eqn\eompqk{\omega_{+}\omega_{-}=\pp(z),\qquad
\omega_{+}^q+\omega_{-}^q =2\qq(z)\ . }
The solution of these equations is  $\omega_{\pm}^q=\qq\pm\sqrt{\Delta}$,
with
\eqn\edisq{\Delta=\qq^2-\pp^q\ ,\qquad
\sqrt{\Delta}=\ud\left(\omega^q_+-\omega^q_-\right)\ . }
Note that $\sqrt{\Delta}$ is thus an odd function.
The function $\omega$ is given by eq.~\esolw.

\medskip
{\it The case $\e^{iq\theta}=-1$.}  This implies $\theta=\pi p/q$,
where $p$ is now an odd integer. The expressions are quite similar, but the
role of $\qq(z)$ and $\sqrt{\Delta}$ are formally exchanged. It is now the
function
\eqn\edefqm{\qq(z)={1\over2i}(\omega_{+}^q-\omega_{-}^q)}
which has no discontinuity on the cut, and is analytic in the whole complex
plane. It is therefore an odd polynomial.
We have the two algebraic equations,
\eqn\eompqkm{\omega_{+}\omega_{-}=\pp(z), \qquad
\omega_{+}^q-\omega_{-}^q=2i\qq(z)\ ,}
with solution $\omega_{\pm}^q=\sqrt{\Delta}\pm i\qq$, where
$$\Delta=\pp^q-\qq^2\ ,\qquad
\sqrt{\Delta}=\ud\left(\omega^q_+ +\omega^q_-\right)\ .$$
Here $\sqrt{\Delta}$ is now even, and the function $\omega$ remains given
by equation \esolw.

\medskip
{\it One-cut solution.}
We still have to determine the coefficients of the polynomials
$\pp,\qq,\Delta$. They can be found from the additional condition that
$\omega$ has only one cut $[a,b]$ on the positive real axis.
We can see from \esolw\ that the singularities of $\omega$ are the single
roots of $\Delta$. We demand  that
except for $a$ and $b$ (and $-a$,$-b$ by parity), all the roots of $\Delta$
are double, such that  $\Delta$ can be written:
$$ \Delta=-(z^2-a^2)(z^2-b^2)R^2(z)\ ,$$
where $R(z)$ is an odd or even polynomial depending on the different cases.
Due to the special form of the conditions \edisc, all one-cut solutions
$\omega_{\pm}$ can be factorized:
$$\omega_{\pm}=\Omega_{\pm}(z)\left(\pm z A(z)+
B(z)\sqrt{(z^2-a^2)(z^2-b^2)}\right)\ ,$$
where $A$ and $B$ are even functions,  in general rational fractions because
$\Omega{\pm}(z)$ may have zeros. The function $\Omega$, which has
only singularities at $\pm a$, $\pm b$, is a ``minimal" solution of:
$$\Omega_+(z)=\Omega_-(-z)\,,\qquad
\Omega_{\pm}(z-i0)=-\e^{\pm i\theta}\Omega_{\mp}(z+i0)\ .$$
This factorization property is a consequence of the algebraic equation
satisfied by $\omega(z)$.

\subsec{Critical points}

A critical point is generated by the confluence of two
different zeros of $\Delta$ or (and this is new with respect to the usual
one-matrix model) when a cut endpoint (called  $a$ in what follows)
approaches the origin where the integrand is singular. Let us examine the
different possible situations:

(i) a non-vanishing zero of $\Delta$ coalesces with the cut endpoint $a$:
This is the
case of an ordinary critical point of the one-matrix model.
the determinant
coming from the integration over $A_i$ plays no special role, and just
modifies the form of the potential. From the point of view of the statistical
model, this is the low temperature phase in which
all matter degrees of freedom are frozen.

(ii) $a=0$: this is a new critical point specific to the structure of the
integral \eZIii, and the only case we shall consider from now on. The condition
$a=0$ implies the divergence of $N\langle\tr A^2_i\rangle$,
which characterizes the matter
fluctuations. Indeed this quantity is proportional to
$\sum_{i,j}1/(\lambda_i+\lambda_j)$ and diverges only when some eigenvalue of
$M$ vanishes \rIK. This argument is confirmed in the continuum limit
by a determination of its scaling properties.

Finally a general critical model in the continuum limit is obtained when both
confluences occur simultaneously:  a cut endpoint and some zeros of $\Delta$
approach the origin. Note, however, that in this limit the eigenvalue
distribution approaches a singularity of the integrand. In such a situation
the validity of the steepest descent method is questionable. The Ising model
provides a useful test of the method.

\medskip
{\it The resolvent.}
We thus consider only critical points for which $a=0$. The function
$\omega(z)$ at a critical point has a cut for $0\le z\le b$. The general
form of such a solution is
\eqn\eompqcrg{\omega_{\pm}(z)=\left(\sqrt{1-b^2/z^2}\pm ib/z\right)^{-l/q}
\left(A(z)\sqrt{1-b^2/z^2} \pm i b B(z)/z\right)\ ,}
where $l,q$ are relatively prime integers with $0<l<q$, and
$A, B$ are polynomials which can be chosen even without loss of generality.
Indeed the situation $A,B$ odd is equivalent to $A,B$ even with the
change $l\mapsto q-l$.

It follows immediately
that $\pp(z),\qq(z)$ and $\Delta(z)$ are polynomials of a form consistent
with a one-cut solution, provided $A$ and $B$ vanish at $z=0$.

A minimal realization of a critical point with polynomial potentials is:
\eqna\ecrpq
$$\eqalignno{\omega_{\pm}(z)&=\mp i(z/2b)^{2m+1} \left(\sqrt{1-b^2/z^2}\pm
ib/z\right)^{-l/q}\ ,& \ecrpq{a} \cr
\omega_{\pm}(z)&=(z/2b)^{2m+2} \left(\sqrt{1-b^2/z^2}\pm
ib/z\right)^{1-l/q}\ ,& \ecrpq{b} \cr} $$
where in both cases we have
\eqn\ethetag{\omega_+(z-i0)=\e^{i\pi(1-l/q)}\omega_-(z+i0)
\quad\Longrightarrow\quad \theta=\pi(1-l/q)\ .}
The two cases $q-l$ even and odd  correspond to
the two situations $\e^{iq\theta}=\pm 1$.

For $z\to 0$ we find that
$$\omega(z)\propto z^{p/q}\ ,$$
where in the two cases \ecrpq{a,b}\ we parametrize respectively
$$\eqalign{{\rm case}\ (a):\qquad &p=(2m+1)q-l\cr
{\rm case}\ (b):\qquad &p=(2m+1)q+l\ .\cr}$$
A simple scaling argument shows that the same result would be obtained
in the $(p,q)$ string model of sec.~\sqmomm\ for the trace of the resolvent
of the operator $Q$.

Note that the values of $p$ are such that $m$ can also be defined as the
integer part of $p/2q$ since $p/(2q)-1<m<p/(2q)$. Finally we see that for
book-keeping purposes it is convenient to assign the angle $\theta=\pi p/q$ to
the critical point characterized by the integers $(p,q)$.

%

\subsec{Scaling region}

We now wish to derive $\omega(z)$ in the scaling region, where the variable
$x=1-g/g_c$, which characterizes the deviation of the coupling constant
from its critical value, is small. Functions that satisfy equation \egenRH\
and are singular only at $z=\pm a$ and $z=\infty$ have the general form:
\eqn\eomscgen{\omega_{{\rm sc,}\pm}
=\left(\sqrt{a^2-z^2}\pm iz\right)^{-l/q}
\left(C(z) \sqrt{a^2-z^2}\pm iz D(z)\right)\ ,}
where again $C$ and $D$ are even polynomials.
A comparison between the large $z$ behavior of $\omega_{\rm sc}$ and the
small $z$ behavior of $\omega$ at the critical point  yields the degrees of
$C$ and $D$.
This determines them completely only for the minimal critical points $m=0$,
for which we find
$$\omega_{{\rm sc,}\pm}\propto \left(\sqrt{a^2-z^2}\pm
iz\right)^{p/q}\ .$$
To obtain the relation between $x$ and $a$ and completely
determine the form of the polynomials $C,D$ for multicritical points ($m>0$),
we calculate the deviation from the critical form at leading order for $x$
small.

\medskip
{\it Deviation from the critical form at leading order.}
We now calculate the deviation from the critical form at leading order
in the variable $x=1-g/g_c$. We normalize the potential in
such a way that for $z$ large, the variation $\delta\omega_{\pm}$ is
$$\delta\omega_{\pm}\sim \mp ixb/z\ .$$
Introducing the function
$$\Omega(z)={\del\over\del g}\bigl(g\omega_0(z)\bigr)\ ,$$
we find that eq.~\eomreg\  implies
\eqn\eOmdefg{\Omega(z)= {\del\omega(z)\over\del g}\ .}
It thus satisfies the homogeneous equation \egenRH.

We also introduce the decomposition
\eqna\eOmpmg
$$\eqalignno{
\Omega_{\pm}(z)&=\pm{i \over 2\sin\theta}\left(\e^{\pm i\theta/2}\Omega(z)-
\e^{\mp i\theta/2}\Omega(-z)\right)&\eOmpmg{a}\cr
\Omega(z)&=-\left(\e^{i\theta/2}\Omega_+(z)
+\e^{-i\theta/2}\Omega_-(z)\right)\ .&\eOmpmg{b}\cr}$$
{}From the definition of $\omega_0$, we infer the behavior of $\Omega$ for
$z$ large, $\Omega(z)\sim 1/z$. Moreover since $\Omega$ is the derivative
of a function which has singularities of the form $(z-z_0)^{1/2}$, $z_0=\pm
a,\pm b$, it can have
a stronger singularity of the form $(z-z_0)^{-1/2}$. These conditions determine
$\Omega(z)$ uniquely as a function of the location of the singularities. The
variation $\delta\omega$ is proportional to $\Omega$
at the critical point. At the critical point, $\Omega$ has the form \eompqcrg.
To obtain its complete form, we need its small $z$ behavior which
must be consistent with the leading correction to the large $z$ behavior of
$\omega_{\rm sc}$. Since $\Omega$ is independent of the potential, we can
compare it to the form of $\omega_{\rm sc}$ in the case $(a)$
(see preceding subsection) for $m=0$.
The leading correction to $\omega_{\rm sc}$ is then of order $z^{l/q-1}$,
and the unique solution is
$$\delta\omega_{\pm}=\mp {ixb \over z\sqrt{1-b^2/z^2}}
\left(\sqrt{1-b^2/z^2}\pm ib/z\right)^{1-l/q}\ .$$
With this information we can now explicitly calculate the
scaling functions for all critical points.

\medskip
{\it Scaling function.} From the preceding analysis, we conclude that for $z$
large $\omega_{\rm sc}$ satisfies
\eqn\econd{\omega_{\rm sc}(z)-{\rm const\ }z^{p/q}
=O\left(x z^{l/q-1}\right)\ .}
It is easy to verify that this fixes the polynomials $C,D$. We now show
that the scaling
function $\omega_{\rm sc}$ can be expressed in terms of the function
$\vartheta(z)$ given by the integral representation
\eqn\eVint{\vartheta(z)= \int_{-iz}^{\sqrt{a^2-z^2}}\d
t\,(t+iz)^{p/q-m-1}(t-iz)^m\ ,}
the proof relying on a verification of condition \econd. Calculating the
integral \eVint\ we first verify that the function
$\vartheta(z)$ has a form consistent with expression \eomscgen:
$$\vartheta(z)=\left(\sqrt{a^2-z^2}+ iz\right)^{\pm l/q}\left(C(z)
\sqrt{a^2-z^2}+ iz D(z)\right)\ ,$$
where $C,D$ are two even polynomials of degree $2m$ and $\pm l/q=p/q-2m-1$.

Another representation of the function is useful. Setting $z=a\cos(q\varphi)$
and integrating over $t$, we find
\eqn\evartphi{\vartheta(z)={(-1)^m\over p/q-m}{ (-ia)^{p/q}\over{p/q\choose
m}} \sum_{r=0}^m {p/q\choose r}\e^{-i\varphi(p-2rq)}\ .}
Moreover if we introduce the parametrization of case $(a)$ of the preceding
subsection, $p=(2m+1)q-l$, we obtain
$$\vartheta(l,z)= \int_{-iz}^{\sqrt{a^2-z^2}}\d t\,
(t+iz)^{(p-q-l)/(2q)}(t-iz)^{(p-q+l)/(2q)}\ .$$
It follows that
$$\vartheta(l,z)-\vartheta(-l,-z)=(2z)^{p/q}
\,\e^{i\pi(q-l)/(2q)}\,\sigma_{pq}\ ,$$
where we have set
\eqn\esigmax{\sigma_{pq}=B(m+1,p/q-m)={\Gamma\bigl((p+q+l)/(2q)\bigr)
\,\Gamma\bigl((p+q-l)/(2q)\bigr) \over \Gamma\bigl((p+q)/q\bigr)}\ .}
Now we need to expand $\vartheta(\pm z)$ for $z$ large.

\medskip
{\it Large $z$ expansion.} For $z=-i\lambda$ large, we find
\eqna\eVzlarge
$$\eqalignno{\vartheta(-i\lambda)&= (-1)^m
(2\lambda)^{p/q}{\Gamma(m+1)\Gamma(p/q-m)
\over \Gamma(p/q+1)}+(2\lambda)^{p/q-2m-2}{a^{2m+2}\over
m+1} & \cr &\quad +O\left(\lambda^{p/q-2m-4}\right)\ ,&\eVzlarge{a}\cr
\vartheta(i\lambda) &\sim {(2\lambda)^{2m-p/q} a^{2p/q-2m} \over p/q-m}\ .
&\eVzlarge{b}\cr}$$

\medskip
{\it Case $(a)$.} If $p=(2m+1)q-l$, and thus $p/q<2m+1$, $\vartheta(-z)$ is
asymptotically larger than the correction to $\vartheta(z)$. Moreover
$2m-p/q=l/q-1$. The solution is then
\eqn\eomscpqa{\omega_{{\rm sc},\pm}(z)=\vartheta_0\vartheta(\pm z)\ .}
Moreover comparing the expansion \eVzlarge{}\ with the expansions of
the critical functions $\omega_{\pm}$ and $\delta\omega_{\pm}$,
$$\eqalign{\omega_+(-i\lambda)&\sim(-1)^{m+1} (\lambda/2b)^{p/q}\cr
\delta\omega_-(-i\lambda)&\sim - x 2^{1-l/q}(\lambda/b)^{l/q-1}\ ,\cr}$$
we obtain the normalization constant $\vartheta_0$ and the relation between
$a$ and $x$:
\eqna\evartab
$$\eqalignno{\vartheta_0 &=-(4b)^{-p/q}\sigma_{pq}^{-1},&\evartab{a}\cr
\left(a\over b\right)^{(p+q-l)/q}&= 2^{(2p+q-2l)/q}
{(p+q-l)\over q}\sigma_{pq}\,x\ ,&\evartab{b}\cr}$$
where we have used the definition \esigmax.

\medskip
{\it Case $(b)$}. If $p=(2m+1)q+l$, and thus $p/q>2m+1$, the correction to
$\vartheta(z)$ is asymptotically larger than the correction to
$\vartheta(-z)$. Moreover $p/q-2m-2=l/q-1$.
The solution is then
\eqn\eomscpqb{\omega_{{\rm sc},\pm}(z)=\vartheta_0\vartheta(\mp z)\ ,}
%
where,
in our normalizations and in the set of variables $p,q,l$,
the same relations \evartab{}\ continue to determine $\vartheta_0$ and the
relation between $a$ and $x$.

\subsec{The singular free energy}

We can find the singular part of the free energy. We have shown that
\eqn\efreeder{ {\del\over \del g}
\left(g^3{\del F\over \del g}\right)
={N^2\over 2i\pi}\oint \d z\,V(z)\,\Om(z)\ ,}
where $\Omega(z)$ is the function \eOmdefg. Using the decomposition
\eOmpmg{b}, we can rewrite equation \efreeder\ as
\eqn\eFrb{{\del\over \del g}\left(g^3{\del F\over \del
g}\right) =-{N^2\over 2i\pi} \oint \d z\,\Om_{+}(z)\left(\e^{i\theta
/2}V(z)-\e^{-i\theta /2}V(-z)\right) \ .}
The critical function for $a=0$ is
\eqn\eOmcrq{\Om_{\pm}=\pm{i\over 2\sin(\theta/2) z\sqrt{1-{b^2/ z^2}}}
\left(\sqrt{1-{b^2/ z^2}} \pm i{b/ z}\right)^{1-l/q}.}

\medskip
{\it The scaling region.}
Let us now consider the case $a\neq 0$ but small. For $z$ small, the function
\eOmcrq\ behaves as $z^{l/q-1}$. This, together with the other properties,
determines the scaling form of $\Omega(z)$:
$$\Omega_{{\rm sc},\pm}(z)={\vartheta_0\over\sqrt{a^2-z^2}}
\left(\sqrt{a^2-z^2}\mp iz\right)^{l/q} , \qquad
\vartheta_0={2^{-2l/q}b^{-l/q} \over \sin(\theta/2)}\ .$$
Conversely, the next-to-leading term in the large $z$
expansion of $\Omega$ provides the additional information needed to completely
determine the first correction to the critical function for $a$ small. This
correction behaves like $z^{-1-l/q}$, and therefore
$$\Omega_+(a)-\Omega_+(a=0)\propto {1 \over z^2 \sqrt{1-b^2/z^2}}
\left(\sqrt{1-b^2/z^2}+ ib/z\right)^{-l/q}\ .$$
The leading correction to $\Omega_{{\rm sc},+}(z)$ is
$$\Omega_{{\rm sc},+}(-i\lambda)\sim \vartheta_0 2^{-l/q}a^{2l/q}
\lambda^{-1-l/q}\ ,$$
and thus
$$\Om_{+}(a)-\Omega_{+}(0)\sim -{2^{-4l/q}\over \sin(\theta/2)}\left({a \over
b}\right)^{2l/q}  {b \over z^2 \sqrt{1-b^2/z^2}}
\left(\sqrt{1-b^2/z^2}+ ib/z\right)^{-l/q}\ .$$
The identity
$${\d \over \d z} \left(\sqrt{1-b^2/z^2}+ ib/z\right)^{-l/q}=
ib{l\over q} {1 \over z^2 \sqrt{1-b^2/z^2}}\left(\sqrt{1-b^2/z^2}+
ib/z\right)^{-l/q}\ ,$$
allows us to cast this expression into the form
$$\Om_{+}(a)-\Om_{+}(0)=-i{q \over l\sin(\theta/2)} \left({a\over
4b}\right)^{2l/q}{\d \over\d z}\left(\sqrt{1-{b^2/
z^2}}+i{b/ z}\right)^{-l/ q}.$$

With this expression we can integrate by parts in the integral \eFrb,
giving the second
derivative of the singular part of the free energy as
$$\eqalign{g_c^2 F''_{\rm sg}={N^2\over 2i\pi g_c}{i q\over l\sin{\theta /2}}
\left({a \over 4b}\right)^{2l/q} &\oint\d z \left(\sqrt{1-{b^2/
z^2}}+i{b/ z}\right)^{-l/ q}\cr
&\quad \cdot \left(
\e^{i\theta/2}V'(z)+\e^{-i\theta/2}V'(-z)\right)\ .\cr}$$
We now substitute the identity
$$\eqalign{\e^{i\theta/2}V'(z)+\e^{-i\theta/2}V'(-z)&
=2ig\sin\theta\left(\e^{-i\theta/2}\,\omega_0(z)
-\e^{i\theta/2}\omega_0(-z)\right)\cr&\qquad +4\sin^2\theta\,
\omega_-(z)\ ,\cr}$$
where we need $\omega_-$ only at leading order,
as given by the expressions \ecrpq{}.
We see that the contribution to the integral coming from $\omega_-$ vanishes.
The contribution due to $\omega_0$ can be calculated by taking the residue at
infinity. Then only the leading behavior of $\omega_0$ for $z$ large is
relevant,
$$2ig\sin\theta\left(\e^{-i\theta/2}\omega_0(z)
-\e^{i\theta/2}\omega_0(-z)\right)\sim {4ig \sin\theta\cos(\theta/2)\over
z}\ .$$

In terms of the variable $x=1-g/g_c$ we finally obtain
$$g_c^2 {\d^2 F_{\rm sg} \over (\d g)^2}=F''_{\rm sg}(x)
=-N^2 (q/l)(2-n)\, 2^{1-4l/q}\left({a\over b}\right)^{2l/q}\ .$$
%
%
{}From this expression, we derive the scaling of the free energy for all
critical points:
\eqn\eFrscpq{F''_{\rm sg}(x)\propto x^{2l/(p+q-l)}\,,
\quad\Longrightarrow\quad \gamma_{\rm str}= -{2l\over p+q-l}\ .}

{\it Discussion.} We find a result in agreement with the $(p,q)$ string models
only in the case $l=1$, i.e.\ when $p$ is of the form $p=(2m+1)q\pm1$.
We note that this is the particular class of models for which we know the
operators $P,Q$ at leading order (see \refsubsec\ssaamfs). The trace of the
resolvent  of $Q$ can be easily calculated. In the semiclassical limit
we have indeed (eq.~\eomegP)
$$\omega(z,x)=P(\d,x),\qquad Q(\d,x)=z\ .$$
In the case $p=(2m+1)q\pm1$, $Q$ is a Chebychev polynomial,
$$Q(\d,x)=u^{q/2}(x)T_q\left(\d/\sqrt{u}\right)\ ,$$
and $P$ is given by (eq.~\finvalp)
$$P(\d,x)=u^{p/2}\sum_{r=0}^m {p/q \choose r} T_{p-2rq}(\d/\sqrt{u})\ .$$
Setting $z=2u^{q/2}\cos(q\varphi)$, we see that $\omega$, up to
normalizations, agrees with the
results in eqs.~\eqns{\eVint{,\ }\eomscpqa{,\ }\eomscpqb},
where $\vartheta$ is replaced by the form
\evartphi, obtained for the  corresponding $(p,q)$ critical points of the
$O(n)$ model.

We conclude that on the sphere we have obtained in the case $p=(2m+1)q\pm1$
complete agreement between the results of the $(p,q)$ string models and those
of the $O(n)$ models. For other cases they seem to differ.
One possible interpretation is that the most relevant operator of multimatrix
models is not present here. And indeed one can find in the
$(p,q)$ model another relevant operator which has the proper dimension to
yield the value of $\gamma_{\rm str}$ found here. Introducing the parameter
$t$ coupled to this operator (which has the dimension $x^{(p+q-1)/(p+q-l)}$),
one finds that the resolvent $\omega(z,x,t)$ of the $(p,q)$ string model
coincides indeed for $x=0$ with the resolvent of the $O(n)$ model \rEyZJiii.

\newsec{Open problems}

Many questions that require further investigation have been pointed out
in the text. For example, the solutions of
the differential equations which have been derived should be studied further.
The perturbation expansion has been shown to be non-Borel summable
in many cases
of interest, and the solutions of the differential equations, having
unphysical properties, do not lead to a solution beyond perturbation theory.
In the pure gravity case, a candidate for a non-perturbative solution has been
proposed based on the solution of a Fokker--Planck equation \rMPsup.

It is possible to integrate over the relative unitary matrices for multi-matrix
models in which the interactions among the matrices
involves no closed loop. The method of
orthogonal polynomials, however, has only been applied to
the case of matrices interacting along a line.

It will be useful to understand the underlying relation between the KdV
and Liouville approaches to these theories in all generality, rather than
the current successful comparison of partial results for low genus and specific
correlation functions.

Finally, the matrix models considered here
representing $d<1$ matter coupled to gravity are quantum mechanical models
with a finite number of degrees of freedom. For application to more realistic
string theories, it will be necessary to extend these techniques to the
case of $d>1$ matter coupled to gravity, which would correspond to a
quantum field theory of matrices.  The borderline case of $d=1$ matter
coupled to 2D gravity, realizable as a solvable
quantum mechanical matrix model, admits an interpretation as a critical string
theory with a two dimensional target space (for review, see \rGiMo).
This model already provides many
tantalizing hints of interesting physics that emerges
as we approach the field theory case. The spacetime interpretation of these
theories and the role of the Liouville ``time'' will greatly enhance our
understanding of physical realizations of string theory.

\bigbreak\bigskip\bigskip\centerline{{\bf Acknowledgements}}\nobreak
We thank especially F.~David, B.~Eynard, and I.~Kostov for discussions.
PG is supported by DOE contract W-7405-ENG-36.

\vfill\eject

\def\tr{{\rm tr}\,}
\def\ud{\half}
\def\ee#1{{\rm e}^{^{\textstyle#1}}}
\def\d{{\rm d}}
\def\e{{\rm e}}
\def\CP{{\cal P}}
\def\Res{{\rm\ Res}\,}
\def\QT{Q}
\def\LT{L}

\appendix{A}{KdV flows and KdV hierarchy}

We consider the Schr\"odinger operator $-L$
$$L=\d^2-u(x,t)\ ,$$
where in what follows $\d$ means $\d/\d x$, and $t$ parametrizes the
potential. We introduce a second (anti-hermitian) differential operator
$M$ of order $2l-1$, and look for the compatibility condition between the
Schr\"odinger equation
$$-L(t)\psi(t)=E\psi(t)\ ,$$
where $E$ is assumed to be $t$ independent, and the condition
\eqn\eflowi{\del_t \psi =-M(t)\,\psi\ ,}
which then implies that the Schr\"odinger operators corresponding to different
values of $t$ are unitary equivalent.

Note that $M$ can be written
$$M=\sum_{k=1}^l\bigl\{m_k(x,t),\,\d^{2k-1}\bigr\}\ ,$$
where the symbol $\{,\}$ means anticommutator, and thus {\it a priori}\/
depends on $l$ independent functions.

The compatibility condition for this linear system,
$$\bigl[\del_t +M,\, L+E\bigr]\psi=0\ ,$$
implies that
$$\bigl\{\del_t L -[L,M]\bigr\}\psi=0$$
for all eigenfunctions $\psi$. It follows that
\eqn\eflowii{\del_t L=[L,M]\ .}

\subsec{The resolvent. Local conserved quantities}
\subseclab\ssTrLcq

Since all operators $L(t)$ are unitary equivalent, any function of
the eigenvalues is conserved in the flow  \eflowi\ or \eflowii. Let us
consider the resolvent
$$G(z)=(z+L)^{-1}\ .$$
Then we have
$$\eqalignno{\del_t G(z)&=-(z+L)^{-1}\del_t L (z+L)^{-1}& \cr
&=(z+L)^{-1}\,\bigl[M,L\bigr]\, (z+L)^{-1}=\bigl[G(z),M\bigr]  \ ,& \cr}$$
and thus taking the trace of both sides, we find
\eqn\econserv{\del_t \tr G(z) =0\ .}
By expanding $\tr G(z)$ for $z$ large,
\eqn\elargez{\tr G(z)=-\sum_{k=0} \int\d x\, R_k[u] (-z)^{-k-1/2}\ ,\qquad
R_0=\ud\ ,}
we generate an infinite number
of conserved quantities which we explicitly construct below and
show to be the space integrals
of {\it local\/} differential polynomials in $u$.

\medskip
{\it Explicit construction of the conserved quantities.}
Let us consider the Schr\"odinger equation for the resolvent
$G(z)=(z+L)\inv$,
\eqn\eresol{\bigl(z+\d^2-u(x)\bigr)G(z;x,y)=\delta(x-y)\ ,}
where $z$ does not belong to the spectrum of $-L$.

Let us recall how $G(z;x,y)$ can be expressed in terms of two independent
solutions of the homogeneous equation
\eqn\ehom{\bigl(\d^2-u(x)+z\bigr)\varphi_{1,2}=0\ .}
If we partially normalize by
$$\varphi'_{1}\varphi_2-\varphi_1 \varphi'_2=1\ ,$$
and moreover impose the boundary conditions
$$\varphi_1(x)\to 0\ {\rm for}\ x\to -\infty,\quad
\varphi_2(x)\to 0\ {\rm for}\ x\to +\infty\ ,$$
then it is easily verified that $G(z;x,y)$ is given by
$$G(z;x,y)=\varphi_1(y)\varphi_2(x)\,\theta(x-y)+
\varphi_1(x)\varphi_2(y)\,\theta(y-x)\ .$$
%
%
The diagonal matrix elements
$r(x)\equiv G(z;x,x)=\varphi_1(x)\varphi_2(x)$ satisfy
$$\eqalign{r'(x)&=\varphi'_1(x)\varphi_2(x)+\varphi_1(x)\varphi'_2(x)\ ,\cr
r''(x)
&=2\bigl(\varphi'_1(x)\varphi'_2(x)+\bigl(u(x)-z\bigr)r(x)\bigr)\ ,\cr
r'''(x)&=2\bigl(u'(x)r(x)+2\bigl(u(x)-z\bigr)r'(x)\bigr)\ ,\cr}$$
where we have systematically used the Schr\"odinger equation \ehom.

The last equation, involving only $r(x)$, is usefully written
\eqn\eresolv{-z r'(x)= \frac{1}{4}r'''(x)-u(x)r'(x)- \half u'(x)r(x)
=\frac{1}{4}\bigl(\d^3-2\{u,\d\}\bigr)r\ .}
Let us now use this equation to expand $r(x)$ for $z$ large and negative.
We can obtain $r(x)$ at leading order for $z$ large, directly starting from
the definition \eresol, and applying the WKB method (in this limit the
non-commutation between $\d$ and $x$ can be neglected). We find
$$\eqalignno{r(x)&\sim{1 \over 2\pi}\int {\d p\over z-p^2-u(x)}=-\ud{1\over
\sqrt{-z+u(x)}}&\cr& =-{1\over2} (-z)^{-1/2}+{u(x)\over
4}(-z)^{-3/2}+O\left((-z)^{-3/2}\right)\ .&\cr}$$
%
It follows that
\eqn\ezlargea{G(z;x,x)=-\sum_{k=0}R_{k}[u] (-z)^{-k-1/2}}
for any smooth potential $u(x)$, where $R_k[u]$ is a local functional of
$u$. Equation \eresolv\ leads to the recursion relation \erecR\ for the
$R_k$'s,
$$R'_{l+1}= \frac{1}{4}R'''_{l}-uR'_{l}- \half u'R_l\ .$$

\medskip
{\it Remark.} The conservation equation \econserv\ shows that as a consequence
of the flow equation \eflowii,
$\del_t R_l$ can be written as a total derivative.

\medskip
\noindent{\it Two useful properties.}\par\nobreak
(i) $G(z)$ defined by \eresol\ satisfies
\eqn\erestr{{\delta \over \delta u(x)}\tr G(z) =
\langle x| (L-z)^{-2} |x\rangle = -{\del \over \del z}G(z;x,x)\ .}
Eq.~\ezlargea\ relates the quantities $R_l$ to the expansion of $G(z;x,x)$
for $z$ large. An expansion of \erestr\
for $z$ large yields immediately
\eqn\egdpa{{\delta\over\delta u}\int \d x\, R_{l}[u]
= -(l-\half)R_{l-1}[u]\ .}
%

\smallskip
(ii) Let us denote by $\del/ \del u$ the derivative corresponding to
constant variations of $u$. Then the resolvent satisfies:
$${\del\over \del u} G(z)=-{\del\over \del z} G(z)\ .$$
Again an expansion of this identity for $z$ large yields
$${\del\over \del u} R_l[u]=-(l-\ud)R_{l-1}[u].$$

\subsec{The flow equation: discussion}

Let us examine the content of the flow equation  \eflowii. The
l.h.s.\ is no longer a differential operator, while the r.h.s.\
is at most a differential operator of degree $2l$. Identifying the
coefficients of all independent operators yields $l$
differential equations and one partial differential equation (hermiticity
being taken into account) for the $l$ coefficients $m_k$ and $u$.
The operators $M$ can then be constructed using the algebraic technique of
pseudo-differential operators.

\medskip
{\it Pseudo-differential operators.}
One considers non-integer powers of the operator $L$. Formally,
the operator $L^{\sigma}$ may be represented within an algebra of formal
pseudo-differential operators as
\eqn\eLrg{L^{\sigma}=\d^{\sigma}+\sum_{i=1}^{\infty}e_i(x)\,\d^{\sigma-i}\ .}
The choice of putting all pseudo-differential operators on the left is
arbitrary. The equivalent expansion with operators on the right can be
obtained by using the commutation relation:
\eqn\ecomRd{e(x)\d^{\rho}=\sum_{n=0}(-1)^n{\Gamma(\rho+1)\over
\Gamma(\rho-n+1) n!}\d^{\rho-n}e^{(n)} .}
The successive terms of the expansion of $L^{\sigma}$ can be
obtained for instance by writing $L^{\sigma}$ as a Laplace transform.

\medskip
{\it Explicit construction of the flow equations.}
The formal expansion of $L^{l-1/2}$
(an anti-hermitian operator) in powers of $\d$ is given by
\eqn\eDl{\LT^{l-1/2}=\d^{2l-1}- {2l-1\over4}\left\{u,\d^{2l-3}\right\}
+ \cdots\ }
(where only symmetrized odd powers of $\d$ appear in this case).
We now decompose $\LT^{l-1/2}= \LT^{l-1/2}_+ + \LT^{l-1/2}_-$, where
$\LT^{l-1/2}_+=\d^{2l-1}+\cdots$ contains only non-negative powers of $\d$,
and the remainder $\LT^{l-1/2}_-$ has the expansion
\eqn\eLR{\LT^{l-1/2}_-=\sum_{i=1}^\infty\
\bigl\{e_{2i-1},\d^{-(2i-1)}\bigr\}
=\left\{R_{l},\d^{-1}\right\}+O(\d^{-3})+\cdots\ .}
Here we have identified $R_l\equiv e_1$ as the first term in the expansion
of $\LT^{l-1/2}_-$.
For $\LT^{1/2}$, for example, we find $\LT^{1/2}_+=\d$ and $R_1=-u/4$.
We justify below that $R_l$ defined by equation \eLR\
is also the functional of $u$ which appears in the large $z$ expansion of
$\tr G(z)$ in \elargez.

Since $\LT$ commutes with $\LT^{l-1/2}$, we have
\eqn\eKKc{\bigl[\LT^{l-1/2}_+,\LT\bigr]=\bigl[\LT,\LT^{l-1/2}_-\bigr]\ .}
But since $\LT$ begins at $\d^2$, and since from the l.h.s.\ above
the commutator can have only positive powers of $\d$,
only the leading ($\d\inv$) term from the r.h.s.\ can contribute, which
results in
\eqn\elc{\bigl[\LT^{l-1/2}_+,\LT\bigr]
=\,{\rm leading\ piece\ of}\ \bigl[\LT,2 R_{l}\,\d^{-1}\bigr]
=4R'_{l}\ .}
If we then take for $M$ a linear combination of the operators $L^{l-1/2}_+$,
$$M=\sum_{k=0}^{l-1} \mu\dup_k\, \LT^{k+1/2}_+ \ ,$$
the equations for the coefficients $m_k$ are automatically satisfied and
we find an equation for the potential $u$:
$$\del_t u =-4{\del \over \del x}\sum_{k=0}^{l-1}
\mu_k R_{k+1}[u]\ .$$

\medskip
{\it A recursion relation.}
The quantities $R_l$ in \elc\ are easily seen to satisfy a simple
recursion relation.
{}From $\LT^{l+1/2}=\LT\LT^{l-1/2}=\LT^{l-1/2}\LT$, we find
%
$$\LT^{l+1/2}_+=\ha\left(\LT^{l-1/2}_+ \LT
+ \LT\LT^{l-1/2}_+\right)+\bigl\{R_l,\d\bigr\} \ .$$
%
Commuting both sides with $\LT$ and using \elc, simple algebra gives
$$R'_{l+1}={1\over4}R'''_{l}-uR'_{l}-{1 \over 2}u'R_l\ ,$$
where  we recognize equation \erecR.
While this recursion formula only determines $R'_{l}$,
by demanding that the $R_{l}$ ($l\ne0$) vanish at $u=0$, as implied by the
definition, we obtain
\eqn\erex{\eqalign{R_0&=\ha\ ,\qquad\qquad
R_1=-{1\over4}u\ ,\qquad\qquad
R_2={3\over16}u^2-{1\over16}u''\ ,\cr
R_3&=-{5\over32}u^3+{5\over32}\bigl(uu''+\half u'{}^2\bigr)
-{1\over64}u^{(4)}\ ,\cr
R_4&= {35\over256}u^4 - {35\over128}\bigl(u u'{}^2 + u^2 u''\bigr)
+ {7\over256}\bigl(2 u u^{(4)} + 4 u' u''' + 3 u''{}^2\bigr)
- {1\over256} u^{(6)} \ .\cr}}
We summarize as well the first few $\LT^{l-1/2}_+$,
\eqn\ekex{\eqalign{\LT^{1/2}_+&=\d\ ,\qquad\qquad
\LT^{3/2}_+=\d^3-{3\over4}\{u,\d\}\ ,\cr
\LT^{5/2}_+&=\d^5-{5\over4}\{u,\d^3\}+
{5\over16}\left\{(3u^2+u''),\d \right\}\ ,\cr
\LT^{7/2}_+&=\d^7 -{7\over4}\{u,\d^5\}
+{35\over16}\left\{(u^2 + u'')\ ,\d^3\right\}\cr
&\quad\qquad-{7\over64}\Bigl\{\bigl(13 u^{(4)} + 10 u u'' + 10 u^3
+ 25 u'{}^2\bigr),\d \Bigr\}\ .\cr}}

\medskip
{\it Examples of flow equations.} The simplest equation only involves $R_1$
and is thus linear
$$\del_t u=\mu_1 u'\ .$$
the unitary transformation simply corresponds to the translation of the
coordinate $x$. The second equation is non-linear instead and is actually the
original KdV equation
$$\del_t u={\mu_2\over4}\,\left(u'''-6uu'\right)\ .$$

\subsec{Large $z$ expansion of the resolvent: residue and trace of
pseudo-differential operators}
\subseclab\sstlze

We now relate  the large $z$ expansion of the resolvent to two
operations called ``residue'' and ``trace'' of pseudo-differential operators.
To calculate the coefficients of the expansion of $G(z;x,x)$, we assume for
convenience that the spectrum of $-L$ is strictly positive and consider the
integral
\eqn\eIsz{I(s,x)=-\ud\oint_{C}\d z\, z^{s-1} G(z;x,x)\ ,}
where the integration contour  can be taken for example going from
$-i\infty+i0$ to $+i\infty+i0$. The residue of $I(s,x)$ at its poles
located at $s=k+1/2$ is the coefficient of $z^{-k-1/2}$ in the expansion
of $G(z;x,x)$. After the change of variables $z\mapsto -Lz$ and integrating
over $z$, $\eIsz$ becomes
\eqn\eIszb{I(s,z)=i\pi\left<x|(-L)^{s-1}|x\right>\ ,}
where $\langle x|(-L)^{s-1}|x\rangle$ is defined by analytic continuation from
$s<\ud$. The poles of the expression are related to the short distance,
large momentum singularities.

If we expand $L^{s-1}$ as a formal series
in $\d$, poles in $s$ will appear when one power of $\d$ approaches $-1$,
the usual logarithmic divergent term. A short calculation then shows
that the coefficient of $(-z)^{-k-1/2}$ in the expansion of $G(z;x,x)$ is
given  by the {\it residue\/} of $L^{k-1/2}$, where
the residue of the pseudo-differential operator $A=\sum_{i=-\infty}^{k}
a_i(x)\,\d^i$, is defined by
\eqn\eresid{\Res\,A \equiv a_{-1}\ .}
Then we have
$$I(s,z)\mathop{\sim}_{s\to k+1/2}{\ud \Res(-L)^{k-1/2}\over s-k-1/2}\
\Rightarrow\ G(z;x,x)=-\ud \sum_{k=0}(-z)^{-k-1/2}\Res L^{k-1/2}\ .$$

Similarly the {\it trace\/} of $A$ is defined by
\eqn\ert{\tr A \equiv \int \d x\,\Res\,A=\int \d x\,a_{-1}\ . }
The trace $\tr L^{k+1/2}$ is thus the coefficient of  $(-z)^{-k-1/2}$ in the
expansion of $\tr G(z)$,
$$\tr G(z)=-\ud \sum_{k=0}(-z)^{-k-1/2}\tr L^{k-1/2}\ ,$$
justifying the nomenclature and in particular
the cyclicity property\foot{The cyclicity property can also be verified
directly by considering basis elements $a(x)\d^m$ and $b(x)\d^n$.
We find\hfill\break
$\tr a\d^m b\d^n
=\int\sum_{j=0}^\infty\left({m\atop j}\right){\rm Res}\,a b^{(j)}\d^{m+n-j}
=\int \left({m\atop m+n+1}\right)a b^{(m+n+1)}$.
Integration by parts then gives\hfill\break
$\tr a\d^m b\d^n=\int\left({n\atop m+n+1}\right)ba^{(m+n+1)}
=\tr b\d^n a\d^m$,
where we have used the identity $(-1)^{m+n+1}\left({m\atop m+n+1}\right)=
\left({n\atop m+n+1}\right)$, following from the definition
$\left({k\atop j}\right)\equiv k(k-1)\ldots(k-j+1)/j!$.}
$\tr AB = \tr BA$ for any two differential operators $A,B$.

We now give two simple examples illustrating the usefulness of these
operations.\par\nobreak
(i) Since
$$\del_t L^{l-1/2}=[L^{l-1/2},M]\ ,$$
we can take the trace of both sides. The cyclicity of the trace implies
$$\delta_t \tr L^{l-1/2}=0\ ,$$
and we thus recover the conserved quantities \elargez.

(ii) Since $R_{l+1}=\ha\,{\rm Res}\,L^{l+1/2}$, we see that
$$\eqalign{{\delta\over\delta u}\int \d x\, R_{l+1}[u]
&={\delta\over\delta u}\ha\tr \LT^{l+1/2}
=-(l+\half)\ha\Res\,\LT^{l-1/2}\cr
&= -(l+\half)R_{l}[u]\ ,\cr}$$
recovering the identity \egdpa.
Finally, introducing a polynomial $\Pi(L)$ in the variable $L^{1/2}$,
such that $M=\Pi_+(L)$, we can write the flow equation \eflowii\ as
\eqn\eflowiii{\del_t L=\bigl[L,\Pi_+(L)\bigr]
=-2{\del\over\del x} \Res \Pi(L)\ .}

\subsec{Commutation of higher KdV flows.}

Let us consider the set of flows associated with the operators
$M=L^{l+1/2}_+$ and call $t_l$ the corresponding parameters. We show now that
all these flows commute. We have to prove that the linear system
$${\rm D}_l L\equiv {\del L\over \del t_l}-[L,L^{l+1/2}_+]=0$$
is compatible. This is again a zero curvature condition
$$0=[{\rm D}_k,{\rm D}_l]={\del_k}L^{l+1/2}_+ -{\del_l}L^{k+1/2}_+
+\bigl[L^{k+1/2}_+,L^{l+1/2}_+ \bigr]\ .$$
The equation can be rewritten
$$\left( {\del_k}L^{l+1/2} -{\del_l}L^{k+1/2}
+\bigl[L^{k+1/2}_+,L^{l+1/2}_+ \bigr]\right)_+ =0\ ,$$
and thus, using the flow equations, we can write
$$\left(\bigl[L^{l+1/2},L^{k+1/2}_+\bigr]-\bigl[L^{k+1/2},L^{l+1/2}_+\bigr]
+\bigl[L^{k+1/2}_+,L^{l+1/2}_+ \bigr]\right)_+ =0\ .$$
Replacing $L^{j+1/2}_+$ in the above by $L^{j+1/2}-L^{j+1/2}_-$
for $j=k,l$, we find that the above equation is identically satisfied,
and hence the flows commute.

\subsec{The canonical commutation relations}

It is now very easy to relate the previous considerations to the solutions
of the string equation $[P,Q]=1$. Let us examine the compatibility of the
equation, where we take $P$ of the form
%
$$P=\bigl[K(Q)\bigr]_+ \ ,$$
$K(Q)$ being a polynomial in $Q^{1/2}$, with the KdV flow
\eqn\eflowQ{{\del Q\over \del t_k}=\bigl[Q,Q^{k+1/2}_+\bigr]\ .}
Differentiating the string equation with respect to $t_k$ and using the flow
equation \eflowQ, we find
$$[\del_k P, Q]+\bigl[P,[Q,Q^{k+1/2}_+]\bigr]=0\ .$$
We rewrite the second term by using the Jacobi identity and the string
equation and find, as expected, the compatibility condition
$${\del P\over \del t_k}=\bigl[P,Q^{k+1/2}_+\bigr]\ .$$

Let us now calculate explicitly the l.h.s.:
%
$${\del P\over \del t_k}={\del K_+\over \del t_k}+\left({\del Q\over\del t_k}
{\del K \over \del Q}\right)_+\ ,$$
where, as in the verification of commutation of KdV flows, we have used the
property that differentiation with respect to $t_k$ and truncation of the
differential part commute. Thus we have
%
$${\del P\over \del t_k}={\del K_+\over \del t_k}+
\bigl([K(Q),\,Q^{k+1/2}_+]\bigr)_+\ ,$$
and it follows that
$${\del P\over \del t_k}-[P,Q^{k+1/2}_+]= {\del K_+\over \del t_k}+
\bigl([K_-(Q),\, Q^{k+1/2}_+]\bigr)_+\ .$$
%
In the r.h.s., $Q^{k+1/2}_+$ can be replaced by $Q^{k+1/2}$. Moreover,
%
$$\bigl[K_+(Q),\,Q\bigr]=1=-\bigl[K_-(Q),\,Q\bigr]
\quad\Longrightarrow\quad
\bigl[K_-(Q),Q^{k+1/2}\bigr]=-(k+\half)Q^{k-1/2}\ .$$
%
The compatibility condition between the KdV flows and the
string equation is thus satisfied with the choice
$${\del K\over \del t_k}=(k+\half)Q^{k-1/2}\quad\Longrightarrow\quad
P=\sum_k (k+\half)t_k  Q^{k-1/2}_+\ ,$$
justifying the equations which appear in \refsubsec\ssRpai. Note also
the relation between the action \estreqa\ and the form of $P$
\eqn\erelSP{{\cal S}(Q)=\tr S(Q),\qquad P\propto [S'(Q)]_+\ .}

\appendix{B}{Generalized KdV flows}
\applab\sGkdv

Most of the preceding algebraic results can be generalized to higher order
differential operators
\eqn\eLgen{L=\d^{q}+\ud\sum_{j=0}^{q-1}\bigl\{u_j,\d^{j}\bigr\}\ .}
The explicit form of the equations is of course slightly different. In this
appendix we provide a few of these expressions.

\medskip
{\it The large $z$ expansion of the resolvent.} From the analysis of
\refapp\sstlze, we
know that the large $z$ expansion of the resolvent is related to the
residues of $L^{s-1}$. The residues do not vanish only when $s$ is of the
form $s=1+m/q$.
We thus find
\eqn\elargezg{\eqalign{G(z;x,x)&=\sum_{m=-1}(-z)^{-1-m/q}\Res L^{m/q}\ ,\cr
\tr G(z) &=\sum_{m=-1}(-z)^{-1-m/q}\tr L^{m/q}\ ,\cr}}
showing in particular that the traces $\tr L^{m/q}$ form a
complete set of local conserved quantities in the flow.
Note that relations \elargezg\ can be verified in the WKB approximation in
which
$$\left<x | (z+L)^{-1}|x\right>\equiv G(z;x,x)\mathop{\sim}_{\rm WKB}{1
\over 2\pi}\int{\d p \over\left(ip\right)^q+ \sum_{j=0}^{q-1}
u_j\left(ip\right)^j+z}\ .$$
Rescaling $p$, we see that $G(z;x,x)$ has an expansion of the form
\elargezg.

\subsec{Explicit construction}
\subseclab\ssEc

The new feature is that $\del_t L$ is now a differential operator of
order $q-1$.  The operators  $M$ however can again be expanded on a
basis formed by the set of differential parts of rational powers of $L$.
Indeed the relation
$$[L,L^{m/q}_+]=-[L,L^{m/q}_-]$$
shows that the commutator on the r.h.s.\ is a hermitian differential operator
of order $q-1$ as required.

Note that since the proof of commutation of KdV flows did not depend on the
specific form of the $L$ operator, the property remains true in this more
general case. Also since the problem has become purely algebraic
through the use of the pseudo-differential operator formalism
(at least as long as we do
not discuss the existence of solutions of the equation), we can immediately
generalize previous considerations to general differential operators $L$ of
arbitrary (not necessarily even) degree $q$. We then expand the second
operator $M$ in the basis $L^{m/q}_+$.

\medskip
{\it $(p,q)$ string equations.} We now consider the string equations
$[P,Q]=1$ with
\eqn\eLgena{Q=\d^{q}+\left\{v_{q-2}(x),\d^{q-2} \right\}+\ \cdots\ +
2v_{0}(x)\ .}
(By a change of basis of the form $Q\to f\inv(x)Qf(x)$,
the coefficient of $\d^{q-1}$ may always be set to zero.)
We then expand $P$ on the basis $Q^{l/q}_+$. Let $K(Q)$ be a polynomial
in the variable $Q^{1/q}$. We can write
$$P=\bigl[K(Q)\bigr]_+ $$
(in $K(Q)$, integer powers of $Q$ can of course be omitted).
Again we can study the compatibility  of the string equation with general
KdV flows generated by
$${\del Q\over\del t_k}=[Q, Q^{k/q}_+]\ .$$
The calculation is the same as in the case $q=2$. One finds the solution
$${\del K\over \del t_k}=(k/q)Q^{k/q-1}\quad\Longrightarrow\quad
P=\sum_k (k/q) t_k\,  Q^{k/q-1}_+\ ,$$
%
consistent, up to sign conventions, with \eqns{\kdvpq{,\ }\pqstreq}.
Comparing with eq.~\egenac, we also see that the relation \erelSP\
is valid for all $(p,q)$ models.

\subsec{$(p,q)$ and $(q,p)$ actions}
\subseclab\sspqqp

In \refsubsec\ssesgkf, we have indicated that the string equations could be
derived from an action principle (eq.~\egenac). This action seems to break the
initial symmetry between the two operators $P$ and $Q$. Let us first examine
this problem here in the simplest case, i.e.\ with  $t_{m}=0$ only for $m\ne
1$ or $m\ne p+q$.

The action $\tr Q^{p/q+1}$ was constructed from a $q^{\rm th}$ order
differential operator $Q$. We assume $p>q$, and discuss what happens if
we consider instead the action $\tr P^{q/p+1}$,
where $ P$ is a $p^{\rm th}$ order differential operator.
We shall show that the two actions
$\tr Q^{p/q+1}$ and $\tr  P^{q/p+1}$
give rise to the same equations of motion
(modulo extra integration constants that arise
because $ P$ contains higher derivatives).

We expand
\eqn\eQp{ P=\d^{p}+\left\{\tilde v_{p-2},\d^{p-2} \right\}
+\ \cdots\ + 2\tilde v_{0} \ .}
Since $p>q$, we can take $Q= P_+^{q/p}$, where the
coefficients $v_\alpha$ in an expansion of $Q$ of the form \eQ\ may be
expressed in terms of the $\tilde v_\alpha$'s of \eQp.
Note however that $Q$ will only depend on the $q-1$ quantities
$\tilde v_{p-2},\ldots,\tilde v_{p-q}$.

First we eliminate $\tilde v_0,\ldots,\tilde v_{p-q-1}$
by imposing the equations of motion
\eqn\ecii{{\delta\over\delta \tilde v_\alpha}
\tr  P^{q/p+1} = 0 \qquad (\alpha=0,\ldots,p-q-1)\ .}
These imply that
${\rm Res}\,\bigl\{\d^\alpha, P^{q/p}\bigr\} = 0$
for $\alpha=0,\ldots,p-q-1$, and consequently that
\eqn\edi{ P_-^{q/p} = \bigl\{e_{p-q+1},\d^{-(p-q+1)}\bigr\}
+\ \ldots \ ,}
where the additional terms involve only derivatives $\d^{-k}$ with $k>p-q+1$.

We now show that the actions $\tr  P^{q/p+1}$ and
$\tr Q^{p/q+1}$ are proportional after imposing \ecii,
\eqn\eQq{\eqalign{\tr Q^{p/q+1}
&= \tr \left ( P_+^{q/p}\right )^{p/q + 1} =
\tr \left ( P^{q/p} -
 P_-^{q/p}\right )^{p/q+1} \cr
&=\tr  P^{q/p+1}
-({p/q}+1)\tr\bigl( P\, P_-^{q/p}\bigr)\cr
&\qquad\qquad+\half({p/q}+1){p\over q}\,\tr \Bigl( P^{1-q/p}
\bigl( P_-^{q/p}\bigr)^2\Bigr) + \ \ldots\ ,\cr}}
where the neglected terms are of lower order in $\d$ either because they
involve at least one commutator $[P^{q/p},\,P_-^{q/p}]$ or a higher power
of $P_-^{q/p}$ (see the appendix of \rGGPZ).
{}From \edi, we see that only the first two terms on the r.h.s\ above
can contribute a nonvanishing $\d^{-1}$ term.  But since
$\tr  P\, P_+^{q/p}=0$
(both contain only positive powers of $\d$), we also have
$\tr  P\, P_-^{q/p}=\tr  P\, P^{q/p}=\tr  P^{q/p+1}$.
Substituting in \eQq\ gives the result
\eqn\eApqqp{\tr Q^{p/q+1}=-{p\over q}\,\tr  P^{q/p+1} \ ,}
which shows that the two actions generate the same equations of motion.

\medskip
{\it A general action.} We shall see that if we had restricted ourselves
to operators $P$ of the form $P=Q^{p/q}_+$, the argument would have been even
simpler. Let us consider a more general case where the action
${\cal S}(Q)$ is the trace of polynomial $S(Q)$ of degree $p+q$ in $Q^{1/q}$,
normalized by its term of highest degree:
$${\cal S}(Q)=\tr S(Q),\qquad P=[K(Q)]_+\ {\rm with}\ K(Q)\equiv
S'(Q)=Q^{p/q}+ \cdots\ .$$

\smallskip
{\it A first order calculation.} To explain the idea we first restrict to
\eqn\eKQm{{\cal S}(Q)={q \over p+q}\tr \left(Q^{1+p/q}
+ t_m \tr Q^{m/q}\right)\ ,}
and calculate to first order in $t_m$.  We find
$$\eqalign{\tr P^{1+q/p}&=\tr\left(K-K_-\right)^{1+q/p} \cr
&=\tr K^{1+q/p}-(1+q/p)\tr K^{q/p}K_-+ \half(1+q/p)(q/p)\tr K^{q/p-1}K_-^2+
\cdots\ ,\cr}$$
where again only the two first terms contribute because $q<p$.
Then we have
$$K^{q/p}=Q+(m/p)t_m Q^{(m-p)/q}\ ,$$
and thus, since $\tr QK_-=\tr Q(K-K_+)=\tr QK$
($QK_+$ has only positive powers) there results
$$\tr P^{1+q/p}=-(q/p)\left(\tr Q^{1+p/q}
+(m/ p)t_m\,\tr Q^{(m-p)/q}Q^{p/q}_-\right)\ .$$

Let us now calculate also $\tr P^{m/p}$ at leading order:
$$\eqalign{\tr P^{m/p}&=\tr \left(Q^{p/q}-Q^{p/q}_-\right)^{m/p} \cr
&=\tr Q^{m/q}-(m/p)\tr Q^{p/q}_-Q^{(m-p)/q}\ .\cr}$$
We can eliminate the term $\tr Q^{p/q}_-Q^{(m-p)/q}$ between the two
equations. Introducing the notation
$$\tilde{\cal S}(P)={p\over p+q}\left(\tr P^{1+q/p}-(q/p)t_m \tr
P^{m/p}\right)\ ,$$
we obtain
\eqn\equivPQ{\tilde {\cal S}(P)=-{\cal S}(Q)+O\left(t_m^2\right)\ .}
This shows, at first order in $t_m$, that there exists an action $\tilde
{\cal S}(P)$ equivalent to ${\cal S}(Q)$.

\medskip
{\it General proof.} Let us now consider a general polynomial $\tilde S(P)$ of
degree $p+q$ in $P^{1/p}$. For convenience, we normalize it by
$\tilde S'(P) \sim P^{p/q}$ for $P$ large.
The general proof relies on the identity
$$\eqalign{\tr \tilde S(P)&=\tr \tilde S(K)- \tr \tilde S'(K)K_-+\half \tr
\tilde S''(K)K_-^2+\cdots \cr
&=\tr \tilde S(K)- \tr \tilde S'(K)K_- \ ,\cr}$$
because $p>q$. Again since $\tr QK_-=\tr Q(K-K_+)=\tr QK$, we have
$$\tr \tilde S'(K)K_-=\tr QK+\tr \bigl(\tilde S'(K)-Q\bigr) K_- \ .$$
We then choose the polynomial $\tilde S$ such that
$$Q=\bigl[\tilde S'(K)\bigr]_+ \ .$$
This equation is satisfied by inverting $Q=\tilde S'\bigl[K(Q)\bigr]$
in the functional
sense, expanding for $Q$ large, and taking the polynomial part in $K^{1/p}$
of the solution. It follows that
$$\tr\tilde S(P)=\tr \tilde S(K)-\tr QK \ .$$
Differentiating the r.h.s.\ with respect to $Q$, we obtain $-K=S'(Q)$, and
therefore
$$\tr\tilde S(P)=-\tr S(Q)\ .$$
The quantity $\tr\tilde S(P)$ is thus the action expressed in terms of the
$P$ operator. In the functional sense, $\tilde S(P)$ is the Legendre transform
of $S(Q)$.


\appendix{C}{Matrix models and jacobians}
\applab\smmaj

In sec.~\stomm, we gave a justification for the appearance
of the Vandermonde determinant \eVand\
in the one-matrix integral. As usual a more
powerful method to calculate Jacobians relies on determination of a metric
tensor.  Let $M$  be a hermitian matrix
which we parametrize in terms of a unitary matrix $U$ and eigenvalues
$\lambda_k$: $M=U^\dagger \Lambda U$. Let us then calculate the square of the
line element in the new variables. We express the variation $\d U$ of $U$ in
terms of a hermitian matrix $\d T$, $\d U=i \d T\, U$. Then
\eqn\emetric{\tr (\d M)^2=\tr\Bigl(U^\dagger
\bigl(\d \Lambda +i [\Lambda,\d T]\bigr)U\Bigr)^2
= \sum_k (\d\lambda_k)^2+\sum_{i,j}\left(\lambda_i-\lambda_j\right)^2
|\d T_{ij}|^2\ .}
Note that the independent variables are the variation of the eigenvalues
$\d\lambda_k$ and $\Re\d T_{ij}$, $\Im\d T_{ij}$, for $i<j$. From eq.~\emetric,
we immediately obtain the Jacobian $M\mapsto \Lambda,U$ which is proportional
to $\sqrt{G}$, where $G$ is the determinant of the metric tensor.
In the notation of sec.~\stomm, we have explicitly
$$G=\prod_{i\ne j}\left(\lambda_i-\lambda_j\right)^2\ \Rightarrow\ \sqrt{G}=
\prod_{i<j}\left(\lambda_i-\lambda_j\right)^2=\Delta^2(\Lambda)\ ,$$
and we thus recover the identity \egpf.

The expression \emetric\ has another useful application here: it
yields the laplacian
$${\cal L}\,\psi(\Lambda)\equiv\tr \left(\del \over \del M\right)^2
\psi(\Lambda)=G^{-1/2}\sum_k {\del
\over \del \lambda_k} G^{1/2} {\del \over \del \lambda_k}
\psi(\Lambda)\ .$$
applied to symmetric functions of the eigenvalues of the matrix $M$.
It is easy to verify that this laplacian is equivalent to a sum of free terms.
Setting $\Delta(\Lambda)\psi(\Lambda)=\phi(\Lambda)$, we have
$$\Delta(\Lambda){\cal L}\,\Delta^{-1}(\Lambda)=\Delta^{-1}(\Lambda)
{\del \over \del \lambda_k}\Delta^2(\Lambda){\del \over \del \lambda_k}
\Delta^{-1}(\Lambda) \ .$$
After a series of commutations, this expression reduces to
$$\Delta(\Lambda){\cal L}\,\Delta^{-1}(\Lambda)=\sum_k \left(\del \over \del
\lambda_k\right)^2 - \Delta^{-1}(\Lambda)\left[ \sum_k\left(\del \over \del
\lambda_k\right)^2 \Delta(\Lambda)\right]\ ,$$
where the square brackets are meant to indicate
that the second term is no longer a differential operator,
the derivatives acting only on $\Delta$. The quantity $\sum_k \left(\del/
\del \lambda_k\right)^2 \Delta(\Lambda) $ is a totally antisymmetric
polynomial in the $\lambda_i$'s.  Any antisymmetric polynomial, however,
is necessarily proportional to $\Delta$. Since the quantity has a degree
smaller than $\Delta$, it must vanish identically so that
\eqn\efrLap{\Delta(\Lambda){\cal L}\,\Delta^{-1}(\Lambda)
=\sum_k \left(\del \over \del \lambda_k\right)^2\ .}

This result has several applications. It provides a partial solution
to the $D=1$ matrix problem. As long as unitary excitations can be neglected,
the quantum mechanics of hermitian matrices can be reduced to a free fermion
problem: The ground state $\psi$ is a completely symmetric function of the
eigenvalues. Because the Vandermonde determinant is completely antisymmetric,
the transformed function $\phi$ must also be antisymmetric.

\medskip
{\it The Itzykson--Zuber integral.}
The same result can be used to prove the basic
identity \emulti\ for the multi-matrix model.

The free hamiltonian $-\cal L$ has the plane-wave solutions:
\eqn\efhpws{-{\cal L}\,\e^{i\tr KM}=\tr K^2 \e^{i\tr KM}\ ,}
where $K$ is an arbitrary hermitian matrix. We diagonalize $K$,
$$K=V^\dagger \Gamma V\ ,$$
and integrate both sides of \efhpws\ over the unitary matrix $V$.
The eigenfunction
$$\psi(\Lambda)=\int \d V\, \e^{i\tr V^\dagger \Gamma V M}$$
then remains a function only of the eigenvalues $\lambda_i$ of the matrix $M$.
Using the result \efrLap, we can set
$\Delta(\Gamma)\Delta(\Lambda)\psi(\Lambda)=\phi(\Lambda)$ and the function
$\phi$ satisfies
$$- \sum_{k} \left(\del \over \del \lambda_k\right)^2 \phi(\Lambda)=
\tr K^2  \phi(\Lambda)\ ,$$
where $ \phi(\Lambda)$ in an antisymmetric function of the $\lambda_i$'s
and thus is superposition of Slater determinants of free solutions. We can
therefore write
$$\phi(\Lambda)=\int\d\rho(\mu_k)\,\det\e^{i \lambda_i \mu_j}\quad {\rm
with}\quad \sum_i \mu_i^2=\sum_i\gamma_i^2\ .$$

We now note that $\phi(\Lambda)$ is symmetric in the exchange
$\Lambda\leftrightarrow \Gamma$. This fixes the solution up to a global
normalization, with the result
$$\int\d U\,\e^{i\tr U^\dagger K U M}\propto \Delta^{-1}(\Gamma)\,
\Delta^{-1}(\Lambda)\, \det \e^{i \gamma_i \lambda_j}\ .$$
If, instead of integrating over a unitary matrix, we integrate $\e^{i\tr KM}$
over $K$, then, because $\Delta^{-1}(\Gamma)$ is antisymmetric in the
$\gamma_i$'s, we can replace the determinant by only one term. Moreover
the jacobian for $K\mapsto V, \Gamma$ generates as we have seen a factor
$\Delta^2(\Gamma)$. We thereby recover the identity \emulti.

\appendix{D}{Discrete canonical commutation relations: A few additional
results}

In this appendix, we consider the matrix solutions to certain problems
posed as matrix commutation relations.

\subsec{A uniqueness theorem}
\subseclab\ssDccr

Let $A$ be a lower triangular matrix with vanishing diagonal elements,
$B$ a matrix such that $B_{n,n+1}\ne 0$ for all $n\ge 0$ and $B_{mn}=0$
for $n>m+1$, satisfying $[B,A]=1$. Then we shall show that
$${\rm given}\ B\,,\ A {\rm\ is\ unique\ }.$$

Indeed let
us assume that we have found two solutions $A_1,A_2$ of this linear equation,
then $X=A_1-A_2$ commutes with $B$: $[B,X]=0$.
We consider the diagonal elements of this commutator,
$$[B,X]_{mm}=0=B_{m,m+1}X_{m+1,m}-X_{m,m-1}B_{m-1,m}\ .$$
An induction over $m$ thus shows that $X_{m+1,m}$ vanishes. Let us then assume
that $X_{m,m-k}$ vanishes for all $k\le n$. It follows that
$$[B,X]_{m,m-n}=B_{m,m+1}X_{m+1,m-n}-X_{m,m-n-1}B_{m-n-1,m-n}=0\ .$$
Again an induction over $m$ shows that $X_{m,m-n-1}$ vanishes, and thus
finally $X=0$. The equation $[B,A]=1$ for $A$ has a unique solution.

\medskip
{\it Application.} If $B$ is symmetric, as in the one-matrix problem, $X$ an
antisymmetric matrix, then $[B,X]=0$ implies $X=0$. The proof follows simply
from the previous result. We set $X=A-A^T$ where $A$ is a lower triangular
matrix with vanishing diagonal elements. Then
$$[B,A]=[B,A^T]\ .$$
The first commutator has vanishing matrix elements $[B,A]_{mn}$ for $n>m$,
the second $[B,A^T]_{mn}$ for $n<m$. Thus $J=[B,A]$ is a diagonal matrix and
$$J=[B,A]=[B,A^T]=J^T=[A,B]=0\ .$$
According to the preceding result, $A$ thus vanishes as well as $X$.

\subsec{Another useful result}
\subseclab\sAfur

Let $X$ be a matrix such that $X_{mn}=0$ for $n\ge m+s$ ($s>0$),
and let $X$ commute
with $B$, where $B$ has the form of the matrix $B_1$ of the multi-matrix model
(i.e.\ $B$ is local with $B_{mn}=0$ for $n>m+1$, and $B_{m,m+1}\ne 0$).
We shall show that
\eqn\eXVB{X=V'(B)\ ,}
where $V(B)$ is some polynomial in $B$.

Let us write the
set of  equations $[B,X]=0$. We start from the matrix element $n=m+s$ of the
commutator. Only one term contributes to each product of matrices
$$B_{m,m+1}X_{m+1,m+s}-X_{m,m+s-1}B_{m+s-1,m+s}=0\ ,$$
which we rewrite
$${X_{m+1,m+s}\over X_{m,m+s-1}}={B_{m+s-1,m+s}\over B_{m,m+1}}\ .$$
Multiplying the numerator and the denominator in the r.h.s.\ by the
product $B_{m+1,m+2}\ldots B_{m+s-2,m+s-1}$, we write
$${X_{m+1,m+s}\over X_{m,m+s-1}}
={B^{s-1}_{m+1,m+s}\over B^{s-1}_{m,m+s-1}}\ ,$$
and therefore
$$X_{m,m+s-1}=K_{s-1}B^{s-1}_{m,m+s-1}\ .$$

Next we write the matrix element $n=m+s-1$. Two terms contribute and we
consider the equation as an inhomogeneous recursion relation for
$X_{m,m+s-2}$,
$$\eqalign{ & B_{m,m+1}X_{m+1,m+s-1}+B_{m,m}X_{m,m+s-1}
-X_{m,m+s-2}B_{m+s-2,m+s-1} \cr
&\quad -X_{m,m+s-1}B_{m+s-1,m+s-1}=0\ .\cr}$$
Because $[B^{s-1},B]=0$, $X_{m,m+s-2}=K_{s-1}B^{s-1}_{m,m+s-2}$ satisfies this
equation. The general solution is thus the sum of this special solution and
the general solution of the homogeneous  equation
$$B_{m,m+1}X_{m+1,m+s-1}-X_{m,m+s-2}B_{m+s-2,m+s-1}=0\ .$$
We now use the same argument as for the earlier equation and conclude
$$X_{m,m+s-2}=K_{s-2}B^{s-2}_{m,m+s-2}\ .$$

Iterating the argument, we find that $X$ can be written
\eqn\eXsol{X=V'(B)+Y\ ,}
where $V(B)$ is an arbitrary polynomial of degree $s$ and $Y$ is a
lower triangular matrix with vanishing diagonal elements that commutes
with $B$,
$$[B,Y]=0\ .$$
In \refapp\sAfur, however,
we have shown that this equation has the unique
solution $Y=0$, thus establishing \eXVB.

\medskip
{\it Remarks.}

(i) It follows in particular that
if in addition $[X,A]=0$, with $[B,A]=1$, then $V''(B)=0$, and thus $X$ is
proportional to the identity.

(ii) If with the same conditions on the matrices $B$ and $X$ we look for a
solution of the equation $[X,B]=A$, with $A_{mn}=0$ for $n>m$,
we find from the
previous arguments that $X$ can be written in the form \eXsol\
$$X=V'(B)+Y\ ,$$
where $Y$ is an lower triangular matrix with vanishing diagonal elements
satisfying $[Y,B]=A$.

(iii) Of course all results remain true if we transpose all matrices.

\subsec{The two-matrix model}

Some problems unsolved in the case of the general $q$-matrix model
can be solved in the two-matrix
case. For example, let $B$ be a local matrix such that
$$B_{mn}\ne 0\quad{\rm for}\ m-r+1\le n \le m+1\ (r\ge 2)\ .$$
We look for $B$ such that there exists
another local lower triangular matrix $A$ such that
$$A_{mn}\ne 0 {\rm\ only\ for\ } m>n\ ,$$
and which satisfies
$$[B,A]=1\ .$$
Note that if $B$ and $A$ are
solutions to this problem, $S^{-1} BS$ and $S^{-1} AS$ are also solutions if
$S$ is a diagonal matrix, $S_{mn}=s_m \delta_{mn}$. Therefore we can fix some
of the matrix elements $B_{mn}$, for example $B_{m,m+1}$ or $B_{m,m-r+1}$.

We shall construct a set of solutions to this problem by adding one assumption
and show that the solutions then correspond to the general two-matrix model:
we shall assume that we can find another local matrix $\tilde B$ such that
$X=A+\tilde B$ commutes with $B$,
$$X=A+\tilde B,\quad [X,B]=0\quad\Longrightarrow\quad [\tilde B, B]=1\ ,$$
with
$$A_{mn}= 0\ {\rm for}\ m\le n,\quad \tilde B_{mn}\ne 0
\hbox{ only for}\ m-1\le n \le m+s-1\ .$$
The decomposition is ambiguous for the elements $n=m-1$, so we leave
$A_{m,m-1}$ arbitrary. The matrix $X$ is also local.  Then according to the
result of \refapp\sAfur\ we have
$$A+\tilde B=V'(B)\ ,$$
where $V$ is an arbitrary polynomial of degree $s$. It follows that
$B_{mn}=0$ for $n\le m+s$, and therefore
\eqn\eJcom{[B,X]=0\quad\Longrightarrow\quad [B,A]=[\tilde B,B]=J\ .}

The commutators $[B,A]$ and $[\tilde B,B]$ have non-vanishing matrix
elements
$J_{mn}$ only for $m \ge n$, $n\ge m-r$, respectively. Therefore the matrix
$J$ has the same width as $B$. The equation $J=1$ then yields just the right
number of equations, i.e.\ $r$ recursion relations, which determine the
matrix $B$. For $m= n$, we use the commutator $[B,A]$ to find
$$B_{n,n+1}A_{n+1,n}-A_{n,n-1}B_{n-1,n}=1\ ,$$
and thus
$$B_{n,n+1}A_{n+1,n}=n+1\ .$$

If we write the other equations in terms of $A$, we get a subset of the loop
equations \eloopg. If instead we use the commutator $[\tilde B,B]=1$,
we can use the result (ii) of \refapp\sAfur, the role of $B$
played by $\tilde B$. It follows that we can introduce a polynomial
$\tilde V(\tilde B)$ of degree $r$ such that
$$B_{mn}=\bigl[\tilde V'(\tilde B)\bigr]_{mn}\quad {\rm for} \ n \le m\ ,$$
which can be rewritten
$$B+\tilde A=\tilde V'(\tilde B)\,,\quad \tilde A_{mn}=0\ {\rm for}\
n \le m\ .$$
We recognize exactly the structure of the two matrix model. In particular
$$[\tilde B^T,\tilde A^T]=1\ ,$$
and
$$\widetilde B_{n+1,n}\tilde A_{n,n+1}=n+1\ .$$
Note that we could have stated the initial problem more symmetrically by
starting directly from the commutator $[\widetilde B,B]=1$.

\medskip
{\it Remarks.}

(i) In the two-matrix case we have proven with one additional
assumption 
eqs.~\erecgen\ and thus we know the form of the
generators of the KdV flows of the two-matrix model.

(ii) The $\IZ_2$ invariant two-matrix problem is even simpler because then
the matrices satisfy $B_2=B_1^T$, and the model can described in terms of
only two matrices $B$ and $A$ satisfying
\eqn\etwom{A+B^T=V'(B)\ ,}
and thus
\eqn\esigmaii{\Sigma +\Sigma^T={\del\over\del t}
\bigl(V(B)+V(B^T)-B^T B\bigr)\ .}

\subsec{The general $\IZ_{2}$ invariant model}

In the case of the multi-matrix model
with $\IZ_{2}$ invariance, the general problem of finding solutions to
the canonical commutation relations
can be reduced to that of finding
the solution of commutation relations $[B,C]=1$ where $B$ and $C$ are two
local matrices, respectively symmetric and antisymmetric, satisfying
$$B_{mn}=0\quad {\rm for}\  | m-n| \ge q\,,\quad C_{mn}=0\quad {\rm for}
\ | m-n| \ge r\ .$$
Let us introduce a matrix $D$, in general non-local, such that
$$B=D^{q-1}\quad {\rm and}\ D_{mn}=0 \quad {\rm for }\ n>m+1\ .$$
Note that although $D$ is not local, the calculation of matrix elements of
powers of $D$ involves only finite sums.

It is easy to verify that the matrix elements of $D$ can be systematically
calculated from the matrix elements of $B$. The first equation is
$$B_{m,m+q-1}=\prod_{j=m}^{m+q-2}D_{j,j+1}\ .$$
We assume that the elements $B_{m,m+q-1}$ never vanish. This equation
determines all matrix elements $D_{m,m+1}$ except for
the first $q-2$ which remain arbitrary. The next equation is
$$B_{m,m+q-2}=\prod_{j=m}^{m+q-3}D_{j,j+1}\sum_{i=m}^{m+q-2}D_{ii}\ .$$
This equation determines $D_{mm}$ again when the $q-2$ first elements are
given.

More generally for $k\ge 1$,
$$B_{m,m+k-1}=\prod_{j=m}^{m+k-2}D_{j,j+1}\sum_{i=m}^{m+q-2}D_{i,i+k-q+1}
\prod_{l=i+k-q+1}^{i-1}D_{l,l+1}+\ {\rm known\ terms}\ .$$
We always find $q-1$ terms recursion equations; 
however for the first values
of $m$ the situation is different and only the first $k-1$ elements are
undetermined. For $k<1$, the equations are similar but all elements are
determined.

Let us now consider the equation $[B,C]=1$. We first specialize to the matrix
element $n=m+q+r-2$ because only two terms contribute,
$$C_{m,m+r-1}\prod_{j=m+r-1}^{m+r+q-3}D_{j,j+1}-
\prod_{j=m}^{m+q-2}D_{j,j+1}C_{m+q-1,m+q+r-2}=0\ .$$
Setting
$$C_{m,m+r-1}=[D^{r-1}]_{m,m+r-1}C'_{m,m+r-1}\ ,$$
we obtain
$$C'_{m,m+r-1}-C'_{m+q-1,m+q+r-2}=0\ .$$
This equation determines all $C'_{m,m+r-1}$ except the first $q-1$. However
we remember that the $q-2$ elements of $D$ are not determined. We can use
them to reduce $C'_{m,m+r-1}$ to a constant. Also if $D$ has a continuum
limit, then we expect that only the constant solution  corresponds to a
matrix $C$ which also has a continuum limit.

Generalizing this argument to $n-m\ge q-1$, we conclude 
that $C$ can be written
$$C=W(D)-L\ ,$$
where $W$ is a polynomial of degree $r-1$ and $L$ is a lower triangular matrix
with vanishing diagonal elements, $L_{mn}=0$ for $n\ge m$. Since $C$ is
antisymmetric, we have also
$$C^T=-C\ \Rightarrow\ L+L^T=W+W^T\ ,$$
which determines $C$ and $L$ in terms of $W$. Moreover we have
$$[B,C]=[L,B]=1\ .$$

Now we consider the equations corresponding to $m\le n<m+q-1$. The
other equations are satisfied by symmetry. A first equation is
$$L_{m,m-1}\,B_{m-1,m+q-2}-B_{m,m+q-1}\,L_{m+q-1,m+q-2}=0\ .$$
The difference between this equation and the previous ones is that the first
term vanishes for $m=0$. Therefore it seems likely that the only solution
compatible with a continuum limit is $L_{m,m-1}=0$. The argument extends to
all non-diagonal matrix elements of the commutator and thus
$$L_{m,m-k}=0\quad {\rm for}\ k<q-1\ .$$

Finally the last equation is
$$L_{m,m-q+1}\,B_{m-q+1,m}-B_{m,m+q-1}\,L_{m-q+1,m}=1\ ,$$
which leads with the same arguments to
$$L_{m,m-q+1}\,B_{m-q+1,m}={m\over q-1}\ .$$
In terms of $W$, these equations read
\eqn\eWsol{\eqalign{W_{m,n}+W_{n,m}&=0\quad {\rm for}\ |n-m| <q-1\ ,
\cr W_{m,m-q+1}+W_{m-q+1,m}&={2m\over(q-1)B_{m-q+1,m}}\ . \cr}}
Note that the number of equations we have obtained is equal to the number of
coefficients of $B$.

\medskip
{\it An action principle.} Eqs.~\eWsol, which determine $B$, can be
derived from an action principle with action
\eqn\eactD{\eqalign{S(B) &=\tr V(B)-\sum_m {m\over q-1}\ln B_{m-q+1,m}\ ,\cr
V(B)&=\sum_{k=0}^{r-1} V_k\,\tr B^{1+k/(q-1)}\ .\cr}}

\medskip
{\it KdV flows.} KdV flows are generated by antisymmetric matrices $\Sigma$
which can be reconstructed from analogous considerations. One finds
$$\Sigma_{mn}={\del \over \del t}\bigl [V(B)\bigr]_{mn}
\quad {\rm for}\ n>m\ ,$$
where $V(B)$ is defined in eq.~\eactD. 

\listrefs\bye